\documentclass[11pt,a4paper,pagesize,BCOR0mm,DIV15,oneside,headinclude]{scrbook}
\usepackage[automark]{scrpage2}
\usepackage{natbib}  
   
\usepackage{helvet}
\usepackage{xcolor}

\usepackage{amsmath,amssymb,empheq}
\usepackage[retainorgcmds]{IEEEtrantools}

\usepackage{aas_macros}

\usepackage[margin=0.2in,labelfont=bf,labelsep=space,format=plain]{caption}

\usepackage[english]{babel}

\usepackage[Lenny]{fncychap}

\newcommand{\bv}[1]{\boldsymbol{#1}}

\newcommand{\settocdepth}[1]{%
  \addtocontents{toc}{\protect\setcounter{tocdepth}{#1}}} 

\usepackage[bookmarksnumbered,
colorlinks,
linkcolor=black,
citecolor=black,
filecolor=black,
urlcolor=blue,
breaklinks=true,
]{hyperref}

\clearscrheadfoot
\lohead{\rightmark}
\rehead{\leftmark}
\ohead[\thepage]{\thepage}
\setheadsepline{.5pt}[\vspace*{4pt}]

\usepackage{booktabs}
 
\author{Christian Knobel}

\renewcommand{\cleardoublepage}{\clearpage}

\newcommand{\kfrac}[2]{\frac{\displaystyle{#1}}{\displaystyle{#2}}}

\newcommand{\supeq}[2]{\:\stackrel{#1}{#2}\:}

\begin{document}
\frontmatter

\begingroup
\fontsize{12pt}{14pt}\selectfont


\begin{titlepage}
\begin{center}

\vspace*{1cm}
\begin{huge}
\sffamily\textbf{An Introduction into the Theory of Cosmological Structure Formation}\\[2cm]
\end{huge}

\begin{LARGE}
Christian Knobel\\[0.3cm]
\end{LARGE}

Institute for Astronomy, ETH Zurich, Zurich 8093, Switzerland\\[1cm]

January 2013\\[0.2cm]
\begin{small}
(Second Edition)\\[2.5cm]
\end{small}

\begin{minipage}{0.7\textwidth}
This text aims to give a pedagogical introduction into the main concepts of the theory of structure formation in the universe. The text is suited for graduate students of astronomy with a moderate background in general relativity. A special focus is laid on deriving the results formally from first principles. In the first chapter we introduce the homogeneous and isotropic universe defining the framework for the theory of structure formation, which is discussed in the three following chapters. In the second chapter we describe the theory in the Newtonian framework and in the third chapter for the general relativistic case. The final chapter discusses the generation of perturbations in the very early universe for the simplest models of inflation.
\end{minipage}

\end{center}
\end{titlepage}

\endgroup


\chapter*{Preface}
\addcontentsline{toc}{chapter}{Preface}

This text aims to give a pedagogical introduction into the main concepts of the theory of structure formation in the universe. The text is suited for graduate students of astronomy with a moderate background in general relativity. A special focus is laid on deriving the results formally from first principles.

During my PhD studies on high redshift galaxy groups I had the endeavor to understand the theoretical framework on which my research was based. I did not only want to be able to reproduce statements that were made in the books, but to understand where they came from and what their underlying assumptions were. Therefore, I read several books on cosmology and worked out the basic theory the way I found it the most accessible. In the course of time, I filled a couple of notebooks this way, until I finally started to convert part of them into digital form as an introduction for my PhD thesis. But, as the introduction turned out to be way too long, I eventually did not include it into my thesis. However, since I had already spent much work compiling this text and since I was encouraged by the positive feedback of students who had read it, I decided to assemble it into a self-contained introduction on cosmological structure formation in order to make it also available to other graduate students with the same interest as me.

I attempted to make a common theme obvious, which is the question of how structures in the universe were created and grew to the present time large-scale structure of galaxies that is observable today. The selected material is supposed to be self-contained, but nevertheless concise. I put a special focus on clarifying how the results are formally derived from underlying fundamental principles and which assumptions were made. This still does not imply every single calculation to be included in detail. For instance, the derivation of the Robertson-Walker metric is not performed specifically, as ``maximally symmetric spaces'' are a special part of general relativity rather than astrophysical cosmology. However, I show certain conditions to be satisfied so that I can refer to a common theorem of general relativity in the literature that uniquely leads to the Robertson-Walker metric. Besides it is generally only the simplest possible case worked out in a given context, since these cases often allow elegant, rigorous derivations. This approach is motivated by the fact that the simplest cases already allow a sufficient qualitative understanding and that realistic quantitative results in the context of structure formation usually require large numerical computations.

Unfortunately, technical texts with a high aspiration for completeness and rigorousness are prone to become longish, while concise texts have the tendency to be incomplete, inaccurate, or ambiguous. For this reason this introduction contains unusually many footnotes compared to other astronomical texts. In order to keep the central theme as straight and concise as possible, I included many minor comments and sometimes also short derivations in the form of footnotes. That is, the basic text (without footnotes) is self-contained, while the footnotes provide additional comments and assistance. It is exactly these footnotes, which can be helpful for students to understand certain subtleties. Those readers who are not interested in these details can just omit them whithout losing the thread.

There are still many topics that would fit into this introduction, which were omitted, such as the derivation of the cosmic microwave background fluctuations or a more detailed discussion of the nonlinear regime of the large-scale structure (e.g., the Zel'dovich approximation, numerical $N$-body simulations). So naturally the material selected represents only a tiny fraction of interesting and important topics in the context of structure formation. Also the literature cited is not comprehensive. I mainly included the material which the text is based on and which I find useful for further reading. If I have accidentally omitted an important reference that should definitely be included, please let me know.

Since I aim to maintain this text, all sorts of comments are welcome. If you find typos or if you have suggestions on how to improve the content, I would be grateful, if you sent me a message to \href{mailto:christian.knobel@phys.ethz.ch}{christian.knobel@phys.ethz.ch}, so that I can update it.

\subsection*{Structure of the introduction}

This introduction is divided into four chapters. The first two are easier to understand than the last two and need only little input from general relativity (only for the derivation of the Robertson-Walker metric and the Friedmann equation). The second chapter is even fully based on Newtonian physics and yet contains most of the results that are presented. On the other hand, the last two chapters and the appendix are entirely based on general relativity and are much more technical than the first two chapters. They require a moderate background in general relativity, although we still derive all results starting with the field equations. Readers who are not interested in the theory of general relativistic structure formation can just omit the latter two chapters and focus on the former ones, which are mostly self-contained.

In \textbf{Chapter \ref{sec:homogeneous_and_isotropic_universe}}, we discuss the universe in the homogeneous and isotropic limit. Starting with the Robertson-Walker metric, we derive the Friedmann equation and describe the dynamics of the universe for given energy contents. We introduce many basic concepts that will be needed in further chapters, such as redshift, comoving distance, and horizons. Then we give an overview of the current concordance cosmology (i.e., the $\Lambda$CDM model) and briefly summarize the history of the universe. Finally, we provide an introduction into the phenomenology of the simplest models of inflation.

In \textbf{Chapter \ref{sec:structure_formation}}, we present the theory of structure formation based on Newtonian physics which is valid well inside the horizon. Interestingly, using Newtonian physics it is possible to derive basically all of the main results for the formation of structures in the universe (except of the form of the primordial power spectrum) and this is the reason why many books entirely omit a treatment of the general relativistic case. We first derive the equations governing the growth of fluctuations at first (linear) order within an expanding universe from the basic hydrodynamical equations and discuss the different possible fluctuation modes. We afterwards introduce the correlation function and the power spectrum to describe the perturbations in statistical terms. In a further step, we leave the linear regime and describe the formation of nonlinear bound structures (``halos'') by means of the (simplistic) ``spherical top hat model''. We motivate approximative formulas for the number density and spatial correlation of dark matter halos in the universe. Finally, we introduce the ``halo model'', which is the current scheme for analyzing the clustering of galaxies in the universe.

In \textbf{Chapter \ref{sec:general_relativistic_treatment}}, we give an introduction into the linear theory of hydrodynamic perturbations in the general relativistic regime, which is needed to understand the evolution of structures outside the horizon. Without this theory, it is impossible to understand how structures that were created during inflation evolved until present time. Compared with the Newtonian linear theory, the general relativistic case is much more complicated and also allows perturbations which have no Newtonian counterpart (e.g., gravitational waves). We first introduce the Scalar-Vector-Tensor decomposition to simplify the perturbed field equations, since they decouple into independent scalar, vector, and tensor equations. Next, we further simplify the field equations by introducing ``gauge transformations'' and choosing particular gauges. We compare our results to the Newtonian ones from the previous chapter for the limiting case well inside the horizon and give a sketch for the general treatment which treats the fluctuations by means of the general relativistic Boltzmann equation.

\textbf{Chapter \ref{sec:generation_of_initial_perturbations}}, is the most technical chapter of all and basically consists of one single big calculation. The aim is to compute the form of the primordial dark matter power spectrum that results from the generation of fluctuations by the simplest models of inflation. The chapter is divided up into three parts: First, we quantize the perturbations of a scalar field inside the horizon during inflation and compute the power spectrum of the fluctuations in the ground state, second we show that under certain conditions the perturbations remain constant outside the horizon, and finally we compute the spectrum of the perturbations after they have reentered the horizon during the matter dominated era and compute the deviations from scale invariance that are expected from slow-roll inflation.

The \textbf{appendix} gives an introduction into the theory of a classical scalar field in the context of general relativity. We derive the equations of motion for the scalar field in a smooth Friedmann-Robertson-Walker universe and then for the general relativistic linear theory of perturbations.

Some key terms that are frequently used are abbreviated: dark matter (DM), large-scale structure (LSS), Friedmann-Lema\^itre-Robertson-Walker (FLRW), cosmic microwave background (CMB), and $\Lambda$ cold dark matter ($\Lambda$CDM).

Finally, we want to briefly introduce some conventions on the notation we are going to adopt. Throughout the text, $c$ denotes the speed of light, $\hbar$ the reduced Planck constant, $G$ the gravitational constant, and $h_{\rm B}$ the Boltzmann constant. Greek indices $\mu$, $\nu$, etc.~generally run over the four spacetime coordinates, while latin indices $i$, $j$, etc.~run only over the three spatial coordinates. Repeated indices are automatically summed over. For spacetime coordinates $x = (x^0,x^1,x^2,x^3)$, the derivatives are abbreviated by $\partial_\mu = \partial /\partial x^\mu$.  We will often adopt a 1+3 formalism $x = (x^0,\bv x)$ with $x^0$ the timelike coordinate and $\bv x = (x^1,x^2,x^3)$ the spacelike coordinates. In the Chapters 1 and 2 we use $x^0 = ct$ with $t$ the cosmic time, and in the Chapters 3 and 4 we use $x^0 = \tau$ with $\tau$ the conformal time and natural units, i.e., $c = \hbar = 1$. Correspondingly, in the Chapters 1 and 2 a dot denotes the derivative with respect to cosmic time $t$ and in the Chapters 3 and 4 the derivative with respect to conformal time $\tau$. Spatial hypersurfaces of constant time $x^0$ are called ``slices''. Comoving Fourier modes on spatially flat slices are denoted by the vector $\bv k$ with $k = |\bv k|$. Using `$\simeq$' we indicate approximations and using `$\sim$' we indicate orders of magnitude. To introduce new symbols and to emphasize equalities we sometimes use `$\equiv$' instead of `$=$'.

\subsection*{Acknowledgments}

I want to thank Simon J.~Lilly, Norbert Straumann, Cristiano Porciani, and Uros Seljak for useful discussions. Their unpublished lecture notes were a well of inspiration for me. I give special thanks to Norbert Straumann, who read through the entire manuscript and provided valuable comments. Furthermore, I thank Damaris and Tobias Holder, who also read through the manuscript. Tobias was a great help for issues regarding the formatting and the layout of the manuscript, and Damaris patiently pondered with me for many hours on how to improve the text and the appearance of the figures. Last but not least, I want to thank Neven \v Caplar for reporting many typos in the equations and for detailed comments. He was a great help to make the text cleaner and clearer.

\cleardoublepage
\phantomsection

\tableofcontents


\mainmatter

\settocdepth{2}


\chapter{Homogeneous and isotropic universe}\label{sec:homogeneous_and_isotropic_universe}

Astrophysical cosmology needs a theoretical framework that allows the interpretation of observational data. Without such a framework not even the most basic observational properties of galaxies, such as redshift, apparent luminosity or apparent size, could be interpreted properly. The current theoretical framework accepted by most astronomers is the ``concordance model'', which is a special case of a Friedmann-Lema\^itre-Robertson-Walker (FLRW) world model. These models are based on the assumption that the universe is governed by general relativity and are essentially homogeneous and isotropic, if smoothed over large enough scales. 

In this chapter, we introduce the FLRW models and how observational data is interpreted within them. It builds the basis for all other chapters. In Section \ref{sec:cosmological_principle}, we briefly discuss the philosophical assumptions behind the FLRW models and in the Sections \ref{sec:robertson_walker_metric} and \ref{sec:friedmann_equation} we develop the mathematical formulation of the FLRW models. In Section \ref{sec:observational_cosmology}, we introduce redshift, peculiar velocities and discuss the structure of causality within the FLRW world models. Then in Section \ref{sec:concordance_cosmology}, we restrict the FLRW world models to the current concordance cosmology and in Section \ref{sec:inflation} we discuss some particularities of the concordance model and what they might tell us about the very early universe.

\section{Cosmological principle}\label{sec:cosmological_principle}

Modern cosmology is based on two fundamental assumptions: First, the dominant interaction on cosmological scales is gravity, and second, the cosmological principle is a good approximation to the universe. The \textbf{cosmological principle} states that the universe, smoothed over large enough scales, is essentially homogeneous and isotropic. ``Homogeneity'' has the intuitive meaning that at a given time the universe looks the same everywhere, and ``isotropy'' refers to the fact that for any observer moving with the local matter the universe looks (locally) the same in all directions. The precise formulation and the consequences of these two concepts in the context of general relativity will be discussed in the next section. But first we want to explore a bit more the philosophical issues of the cosmological principle\footnote{\nocite{ellis2006} Ellis (2006) provides a systematic discussion of philosophical issues for cosmology, which can be warmly recommended.}.

How can the cosmological principle be justified? Obviously, the universe is not homogeneous and isotropic on scales as big as our Solar System, our Galaxy or even our Local Group of galaxies. Nevertheless the cosmological principle has been invoked from the beginning of modern cosmology in the first half of the 20th century, when almost nothing about the large-scale structure in the universe was known. The main reasons for its acceptance were simplicity and the ``Copernican principle''. Applying the cosmological principle to general relativity yields rather strong constraints and leads to the simplest category of realistic cosmological models.\footnote{Weinberg (1972, p.~408)\nocite{weinberg1972} expresses this spirit by writing:
\begin{quotation}
The real reason, though, for our adherence here to the Cosmological Principle is not that it is surely correct, but rather, that it allows us to make use of the extremely limited data provided to cosmology by observational astronomy. If we make any weaker assumptions, as in the anisotropic or hierarchical models, then the metric would contain so many undetermined functions (whether or not we use the field equations) that the data would be hopelessly inadequate to determine the metric. On the other hand, by adopting the rather restrictive mathematical framework described in this chapter, we have a real chance of confronting theory with observation. If the data will not fit into this framework, we shall be able to conclude that either the Cosmological Principle or the Principle of Equivalence is wrong. Nothing could be more interesting.
\end{quotation}
For a historical account of the beginning of modern cosmology we refer to \cite{nussbaumer2009}.} On the other hand, the \textbf{Copernican principle} according to which we do not occupy any special place in the universe fits the cosmological principle perfectly \citep[][Sect.~4.2.2]{ellis2006}. If we perceive the universe around us isotropically, the Copernican principle asserts that also other observers should see the universe isotropically, since otherwise we would occupy a special place in the universe. Since a universe that is isotropic everywhere is also homogeneous (in fact, isotropy around three distinct observers suffices), the cosmological principle is a relatively straightforward conclusion from an observed isotropy and the Copernican principle.

\begin{figure}[tp]\centering
 \includegraphics[scale = 0.39]{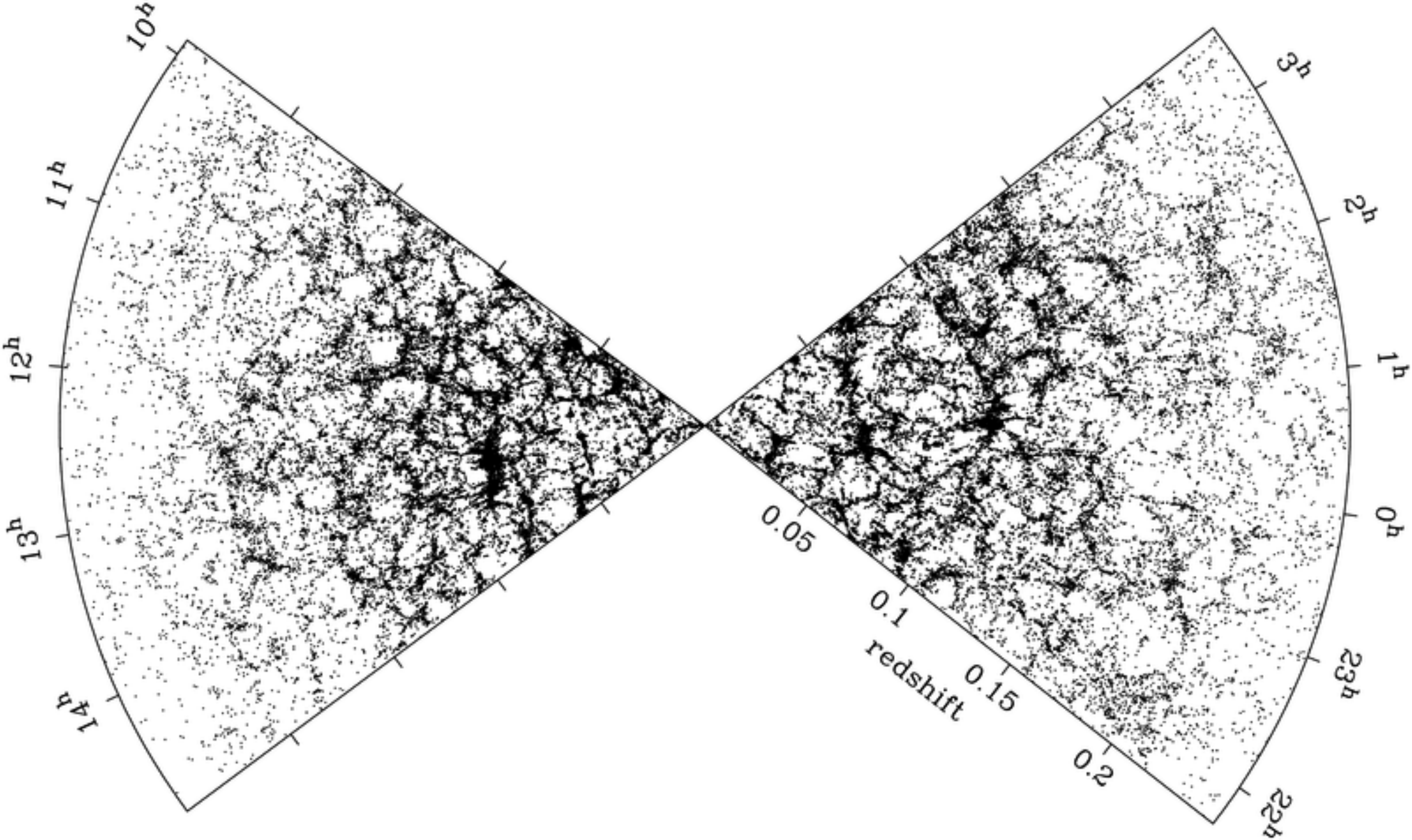}
   \includegraphics[scale = 0.5]{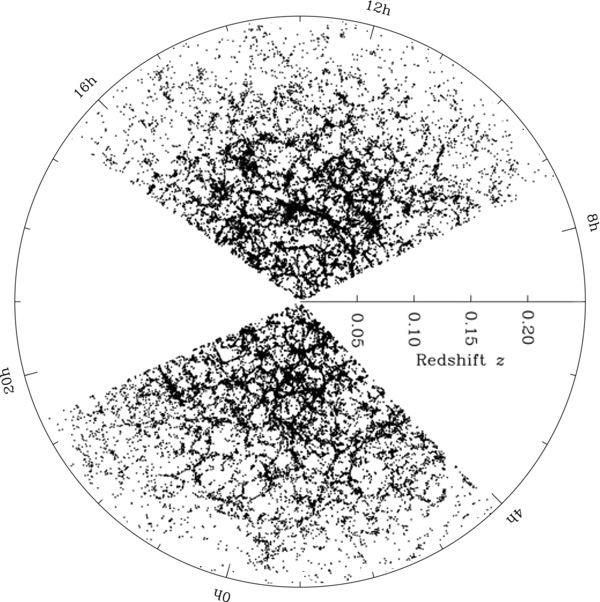}
 \caption{The large-scale structure (LSS) as observed with the largest current, spectroscopic galaxy surveys. The upper panel displays two cones of the 2dfGRS and the lower panel two cones of SDSS. Every point in the figure corresponds to a galaxy. For the SDSS cones, the galaxies are colored according to the ages of their stars, where red corresponds to older stellar populations. The structures are shown out to redshift $z \simeq 0.2$ which corresponds to a light travel time of about 2.6 Gyr and a comoving distance of about 800 Mpc for the concordance cosmology (see Sects.~\ref{sec:observational_cosmology} and \ref{sec:concordance_cosmology}). For both surveys the flux limit becomes apparent at this redshift. Obviously, the LSS is made up of sheets and filaments of galaxies, which can be as big as 100 Mpc. It should be noted that only the LSS of the luminous matter is visible on such diagrams. (Credits: 2dfGRS team, and M.~R.~Blanton and the SDSS team)}\label{fig:LSS}
\end{figure}

Over the last two decades, the amount of data in astronomy has grown immensely so that today the cosmological principle can be discussed in the context of a wealth of different and detailed observations, even though only consistency statements are possible.\footnote{Consistency is generally very important in cosmology as it is often the only way to ``test'' a paradigm. Since the conversion of astronomical observations such as redshifts, apparent luminosities and apparent sizes into distances, absolute luminosities and physical sizes depend on the adopted cosmological framework, also our reconstruction of the universe depends on cosmology. This is why only consistency statements within a given framework are possible. In principle there could be several cosmological frameworks based on different assumptions leading to consistent interpretations of the observations.} For instance, the isotropy of the universe with respect to the Milky Way has been strongly confirmed by the remarkable isotropy of the cosmic microwave background (CMB, see Sect.~\ref{sec:timeline}) as observed by the satellites COBE and WMAP. If the dipole\footnote{The dipole leads to a relative motion of the center of the Milky Way with respect to the rest frame of the CMB of about 552 km/s \citep{kogut1993}. This is comparable to the peculiar velocities of other galaxies which are typically in the range of a few hundred km/s depending on the cosmic environment, thus interpreting the dipole as due to the relative motion of the Milky Way is consistent within the concordance model.} in the CMB is interpreted as relative motion of the earth with respect to the CMB rest frame, the degree of isotropy is as high as $10^{-5}$ \citep{smoot1992}. On the other hand, huge low-redshift galaxy surveys such as the 2-degree field galaxy redshift survey (2dfGRS, \citealt{colless2001}) and the Sloan digital sky survey (SDSS, \citealt{york2000}) have convinced most cosmologists that not only isotropy but also homogeneity is in fact a reasonable assumption for the universe. Taking a glance at light cones produced by these surveys (see Figure \ref{fig:LSS})
one can see (even without any statistical tools) that the fractal nature of the universe stops at a certain scale and builds a net of clusters and filaments which is called \textbf{large scale structure} (LSS). However, it is still difficult to exactly estimate the scale on which the universe becomes homogeneous. \cite{hogg2005} investigated the spatial distribution of a big sample of luminous red galaxies from SDSS and found that after applying a smoothing scale of about 100 Mpc, the spatial distribution approaches homogeneity within a few percent. In a more recent study, \cite{scrimgeour2012} demonstrated by selecting 200,000 blue galaxies within an unprecedented huge volume that a fractal distribution of galaxies on scales from about 100 Mpc to 400 Mpc can be excluded with high confidence. Also the analysis performed by \cite{hoyle2013} did not find any evidence for inhomogeneities on large scales.

Despite these observational confirmations, the cosmological principle remains a fundamental assumption \citep[][Sect.~4.2.2]{ellis2006}, and moreover we can hardly make any reasonable statement about the state of the universe far beyond our current horizon of causality (cf.~Sect.~\ref{sec:revisited}, where the cosmological principle is revisited in the context of inflation). But even regarding the universe within our horizon, there are still some cosmologists sharing doubts about the validity of the cosmological principle or at least exploring other possibilities. Doubts are mainly raised by the apparent acceleration of the universe as observed by type Ia supernovae, which is accounted for by invoking dark energy (cf.~Sect.~\ref{sec:concordance_cosmology}). It has been claimed that this acceleration could just be an artefact caused by inhomogeneities in the universe due to the nonlinearity of Einstein's field equations without any actual acceleration taking place. This effect is called \textbf{backreaction} (see \citealt{clarkson2011} for a review). Although it was not yet possible to rule out that the interpretations of our observations are to some extent affected by backreaction, there are good arguments why it should be negligible at least on cosmological scales.
However, the best argument for the validity of the cosmological principle is presumably the remarkable consistency of several independent observables, such as CMB anisotropies, galaxy power spectra, type Ia supernovae, cluster abundances and others (see, e.g., \citealt{dunkley2009}, Sect.~4.2), within the framework of the concordance model.

Throughout this introduction we will assume that the metric and the dynamics of the universe are well described to zeroth order by the smoothed homogeneous and isotropic universe, and that the observed inhomogeneities in the universe can be treated as perturbations within the homogeneous and isotropic background.

\section{Robertson-Walker metric}\label{sec:robertson_walker_metric}

In this section, we discuss the metric of the universe as required by the cosmological principle. This metric determines all geometrical properties of the universe, such as the distance between two points or the apparent extension of an object with known diameter if seen at a given distance. In a first step, we will present the mathematical formulation and in a second step its physical interpretation.

\subsection{Mathematical formulation}\label{sec:mathematical_formulation}

According to the first fundamental assumption of the previous section, cosmology is based on general relativity being the best theory of gravity so far. So we regard spacetime as a pseudo-Riemannian manifold $\mathcal{M}$ with metric $g_{\mu\nu}$, where the latter is determined by the Einstein field equations. Then, the second fundamental assumption (i.e., the cosmological principle) states that the universe is essentially homogeneous and isotropic. In order to apply these properties to the manifold $\mathcal{M}$, we have to carefully paraphrase the intuitive notion of these terms in the context of the general relativistic spacetime. Hereto we mainly follow the outline of \nocite{wald1984}Wald (1984, Ch.~5) and \nocite{misner1973}Misner (1973, Sect.~27.3).
\begin{figure}[tp]\centering
  \includegraphics[scale = 0.45]{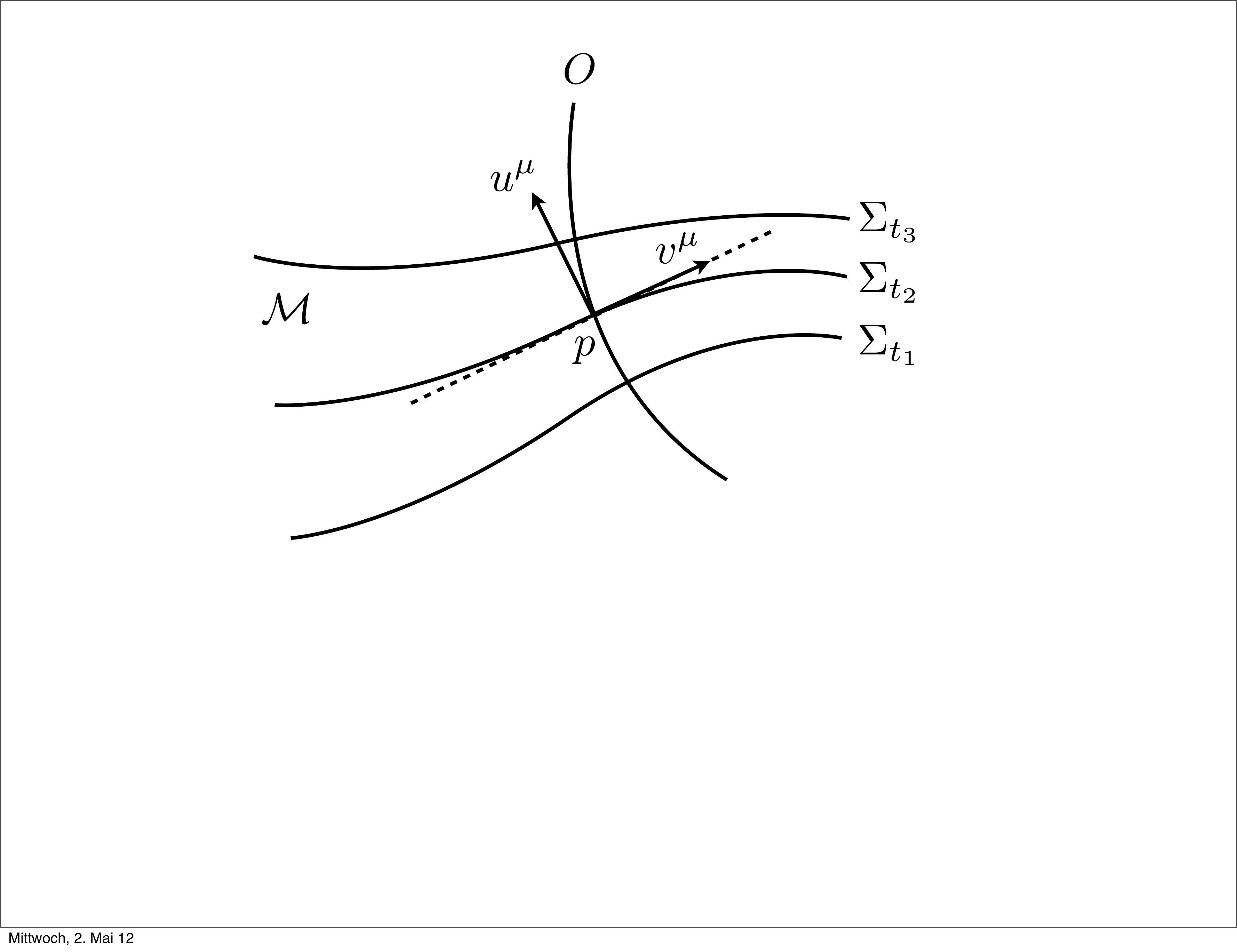}
  \caption{A schematic illustration of the manifold $\mathcal{M}$. Shown is the world line of a fundamental observer $O$ that pierces through the spatial slices $\Sigma_{t_1}$, $\Sigma_{t_2}$, and $\Sigma_{t_3}$ of constant time $t_1$, $t_2$, and $t_3$, respectively. The tangent vector of the line $O$ in the point $p$ is denoted by $u^\mu$, whereas $v^\mu$ is a spatial vector perpendicular to it. Since a fundamental observer is free falling, the metric in the restframe of the observer is locally Minkowskian (i.e., the observer does not feel any gravity) and thus diagonal. That is, the vector $u^\mu$ must be perpendicular to its local slice of simultaneity (dashed line). If the slices are homogeneous, the fundamental observer will intersect them perpendicularly.}\label{fig:manifold}
\end{figure}

``Homogeneity'' refers to the intuitive notion that at a given time $t$ spacetime looks the same at any place. However, in general relativity there is no ``absolute simultaneity'', since whether two events happen at the same time depends not only on the chosen reference frame, but also on the metric $g_{\mu\nu}$. ``At a given time'' means in general relativity ``on a given spacelike hypersurface''. So \textbf{homogeneity} is interpreted such that the whole manifold $\mathcal{M}$ can be sliced up into a one-parameter family of spacelike hypersurfaces (slices) $\Sigma_t$ (see Fig.~\ref{fig:manifold}), which are homogeneous. That is, for any time $t$ and for any two points $p,q \in \Sigma_t$, there exists a diffeomorphism (i.e., a coordinate transformation) of spacetime that carries $p$ into $q$ and leaves the metric $g_{\mu\nu}$ invariant.

To introduce the concept of ``isotropy'' we first note that an isotropic universe will not appear isotropic to \textit{any} observer. For instance, if the universe appears isotropic to an observer at rest within the Milky way, then it would not appear so to an observer moving away from the Milky way at half the speed of light. Such an observer would detect light coming toward him with much higher intensity than from behind. So, to study isotropy we have to consider the ``world lines'' of observers. Let $u^\mu$ be the tangent vector along the world line in a point $p$ and $v^\mu_1$ and $v^\mu_2$ any two unit vectors perpendicular to it (see Fig.~\ref{fig:manifold}). \textbf{Isotropy} then means that there exists a diffeomorphism of spacetime with fixed $p$ and $u^\mu$ that carries $v^\mu_1$ into $v^\mu_2$ and leaves the metric $g_{\mu\nu}$ invariant. An observer that sees the universe (locally)  isotropic at any time is called a ``fundamental observer''.

As mentioned in the previous section, a fundamental observer must be moving with his local matter (otherwise the local flow of matter would indicate a preferred direction). Moreover, such an observer must also be free falling (otherwise the gravitational force acting on the observer would introduce a preferred direction) and thus the metric in its restframe is locally Minkowskian (i.e., the observer does not feel any gravity). That is, we can choose coordinates in the restframe of the observer such that the metric takes the familiar form $g_{\mu\nu} = {\rm diag}(-1,1,1,1)$ at any point along the world line. Since for an observer at rest the tangent vector $u^\mu$ along the world line points in the direction of the time coordinate and since the metric is diagonal, the tangent vector $u^\mu$ is always perpendicular to the observer's local slice of simultaneity (see Fig.~\ref{fig:manifold}). 

Invoking both, homogeneity and isotropy, is already more than needed as mentioned in the previous section. It is sufficient to assume the existence of a single fundamental observer in addition to homogeneity to fully determine the metric $g_{\mu\nu}$ as long as this observer crosses every slice $\Sigma_t$. Then it is easy to see that the world line of the fundamental observer always pierces perpendicular through the homogeneous slices $\Sigma_t$, i.e., $u^\mu$ is perpendicular to $\Sigma_t$ for any $t$. (Geometrically speaking, the slice of local simultaneity perpendicular to $u^\mu$ in the point $p$ is tangential in $p$ to the slice $\Sigma_t$ going through $p$.) If $u^\mu$ was not perpendicular to $\Sigma_t$, the observer would be in motion relative to the homogeneous slice and would therefore observe a preferred direction in the universe.\footnote{Here, we assumed that the homogeneous slices $\Sigma_t$ are unique, i.e., for a given point $p$ there is exactly one homogeneous slice that passes through $p$. There are, however, cases (e.g., Minkowski spacetime, de Sitter spacetime) for which there is not a unique way how spacetime can be split into homogeneous slices $\Sigma_t$. Nevertheless, for these cases we can always find a corresponding family of homogeneous slices which are perpendicular to the fundamental observer \citep[see][Ch.~5]{wald1984}.} Due to this perpendicularity of the fundamental observer to the slice $\Sigma_t$, the isotropy condition fully applies to the slice $\Sigma_t$.
That is, if we regard the slice $\Sigma_t$, which is a submanifold of $\mathcal{M}$, as an independent manifold, the isotropy around our fundamental observer implies the following: For any two tangential vectors $v^\mu_1$ and $v^\mu_2$ of $\Sigma_t$ in $p$ there exists a diffeomorphism on $\Sigma_t$ which carries $v^\mu_1$ into $v^\mu_2$, while leaving $p$ and $g_{\mu\nu}$ (restricted on $\Sigma_t$) invariant. This is exactly the definition of an arbitrary manifold being isotropic around a point $p$. Thus, any slice $\Sigma_t$ is not only homogeneous, but also isotropic around the point where the fundamental observer intersects it. Since for any manifold homogeneity and isotropy around a point entail maximal symmetry \citep[e.g.,][Sect.~13.1]{weinberg1972}, any slice $\Sigma_t$ constitutes a three-dimensional maximally symmetric space, i.e., a three-dimensional space with constant curvature.

A spacetimes that is made up of maximally symmetric spatial slices has almost no remaining degrees of freedom. It can be shown \citep[e.g.,][Sect.~13.5]{weinberg1972} that for such a spacetime there always exist coordinates $x = (x^0,x^1,x^2,x^3) = (ct,\chi,\theta,\varphi) = (ct,\bv{x}) $ such that the metric $g_{\mu\nu}$ takes the form of the \textbf{Robertson-Walker metric} whose line element is
\begin{equation}\label{eq:robertson_walker_metric1}
	ds^2 = g_{\mu\nu}(x)\: dx^\mu dx^\nu = -c^2 dt^2 + R^2(t)\: \gamma_{ij}(\chi,\theta,\varphi)\: dx^i dx^j
\end{equation}
with $\gamma_{ij}$ being the metric of a three-dimensional space of constant curvature, which is generally described by\footnote{The generality of this expression is guaranteed by the uniqueness theorem for maximally symmetric spaces. Two maximally symmetric spaces with the same curvature and the same metric signature are always isometric to each other \citep[see][Sect.~13.2]{weinberg1972}. Since $K$ is the curvature of $\gamma_{ij}$ and can take any value, we can for a given maximally symetric space (with the right metric signature) choose coordinates, so that the metric takes the form of $\gamma_{ij}$.}
\begin{equation}\label{eq:gamma1}
	\gamma_{ij}(\chi,\theta,\varphi)\: dx^i dx^j = \frac{d\chi^2}{1 - K \chi^2}+ \chi^2 \left(d\theta^2 + \sin^2(\theta)d\varphi^2 \right)\:.
\end{equation}
By adjusting the coordinate $\chi$, the constant $K$ can always be normalized to one of the three discrete values $1$, $0$, or $-1$ specifying the geometry of the slice, where $K=1$ corresponds to positvely curved, $K = 0$ to flat, and $K = -1$ to negatively curved space.
In Eq.~(\ref{eq:robertson_walker_metric1}), $t$ is called \textbf{cosmic time} (or \textbf{epoch}), $R(t)$ is the \textbf{cosmological world radius}, and $(\chi,\theta,\varphi)$ are spatial spherical \textbf{comoving coordinates} for reasons that will become clear in the next section. While $R(t)$ takes units of length, the comoving coordinates $(\chi,\theta,\varphi)$ are dimensionless. The ranges of values for the coordinates are\footnote{We have not yet said anything about the mathematical topology of our spacetime $\mathcal{M}$. With the ranges of coordinates in Eq.~(\ref{eq:coordinate_range}) the spatial slices are topologically homeomorph to the 3-sphere $S^3$ in the case $K = 1$, to the Euclidian space $\mathbb{R}^3$ in the case $K=0$, and to the hyperbolic 3-space $\mathbb{H}^3$ in the case $K = -1$. These topological spaces are all simply connected, that is any closed line on these slices can be continuously contracted to a point, and the volumes of the universe in the cases $K = 0, -1$ are infinite. However, general relativity is a local theory, i.e., the metric $g_{\mu\nu}$ determines the local properties of spacetime, but not its global structure. There are also multi-connected topological spaces consistent with our maximal symmetric slices and for these topologies the allowed parameter range is smaller than the one admitted in Eq.~(\ref{eq:coordinate_range}). Some of these topologies even allow finite volumes for the universe in the cases $K = 0, -1$. As different topologies can lead to different observational results and can also affect the growth of structure in the universe, they have to be considered as possible models for the universe. A systematic and pedagogical introduction into this topic is given by \cite{lachieze-rey1995}. Since the standard simply-connected topologies are so far consistent with all measurements, we will stick to this case for simplicity.}
\begin{equation}\label{eq:coordinate_range}
	0 \leq \chi < \left\{
\begin{array}{ll}
 \infty, &  K = 0,-1\\
 1, &  K = 1\:,
\end{array} \right.\quad\quad 0 \leq \theta < \pi\:, \quad\quad 0 \leq \varphi < 2 \pi\:,
\end{equation}
where $\theta$ and $\varphi$ are the standard angle coordinates on the sphere, and $\chi$ is a sort of radial coordinate. A detailed discussion of the physical interpretation of the Robertson-Walker-metric will be given in the next section.

Although Eq.~(\ref{eq:robertson_walker_metric1}) is already the general Robertson-Walker metric, it is often convenient to use slightly different forms of it. If we perform the substitution $\chi = f_K(\tilde r)$, where $f_K(\tilde r)$ is defined by
\begin{equation}
	f_K(\tilde r) = \left\{
\begin{array}{ll}
\sin(\tilde r) & \text{if}\ K = 1\\
\tilde r &  \text{if}\  K = 0\\
\sinh(\tilde r) &  \text{if}\  K = -1\:,
\end{array} \right.
\end{equation}
the Eqs.~(\ref{eq:robertson_walker_metric1}) and (\ref{eq:gamma1}) together become
\begin{equation}\label{eq:robertson_walker_metric2}
	ds^2 = -c^2 dt^2 + R^2(t) \Big[d\tilde r^2 + f_K^2(\tilde r)\left(d\theta^2 + \sin^2(\theta)d\varphi^2\right)\Big]\:.
\end{equation}
The new coordinate $\tilde r$ is still dimensionless, but it is now proportional to the physical distance from the coordinate origin as we shall see in the next section. Note that for $K=1$, the spatial part of Eq.~(\ref{eq:robertson_walker_metric2}) takes the standard form of the 3-sphere with radius $R(t)$, where $\tilde r$ only takes values in the range $0 \leq \tilde r < \pi$ and just plays the role of another angle coordinate in addition to $\theta$ and $\varphi$. The Robertson-Walker metric in the form of Eq.~(\ref{eq:robertson_walker_metric2}) is very compact and allows a straightforward interpretation of the comoving coordinates $(\tilde r, \theta, \phi)$. However, there is still another form which is more common among cosmologists even if slightly less compact. To derive it, we introduce the dimensionless \textbf{scale factor}
\begin{equation}
	a(t) \equiv \frac{R(t)}{R(t_0)}\:,
\end{equation}
where $t_0$ denotes an arbitrary reference epoch that is usually chosen to be the present time. With the coordinate transformation $r = R(t_0)\tilde r$, Eq.~(\ref{eq:robertson_walker_metric2}) becomes
\begin{equation}\label{eq:robertson_walker_metric3}
	\boxed{ds^2 = -c^2 dt^2 + a^2(t) \Big[dr^2 + R_0^2\: f_K^2(r/R_0)\left(d\theta^2 + \sin^2(\theta)d\varphi^2\right)\Big]\:,}
\end{equation}
where $R_0= R(t_0)$. Here $a(t)$, $\theta$, and $\varphi$ are dimensionless, and $r$ takes units of length. This version of the Robertson-Walker metric has the advantage that it holds $a(t_0) = 1$, and the comoving coordinates $(r,\theta,\phi)$ are the usual spherical coordinates taking physical units. Eq.~(\ref{eq:robertson_walker_metric3}) is the form of the metric we will mainly work with.

Sometimes it is convenient to work with another time coordinate. So we introduce the \textbf{conformal time} $\tau$ by setting
\begin{equation}\label{eq:conformal_time}
	d\tau = \frac{c}{a(t)}\:dt\:.
\end{equation}
Using conformal time $\tau$ instead of cosmic time $t$, the scale factor moves in front of the total metric
\begin{equation}\label{eq:robertson_walker_metric4}
	ds^2 = a^2(\tau)\Big[-d\tau^2 + dr^2 + R_0^2\: f_K^2(r/R_0)\left(d\theta^2 + \sin^2(\theta)d\varphi^2\right)\Big]\:.
\end{equation}
The main advantage of using conformal time is that it allows analytic solutions to the time evolution of open and closed universes (cf.~Sect.~\ref{sec:spherical_top_hat_collapse}), and the metric undergoes just a conformal transformation as $\tau$ changes.

\subsection{Physical interpretation}

It is relatively easy to see that the fundamental observers are the observers at fixed comoving coordinates, i.e., those at rest to the homogeneous slices. For instance, the observer at $\bv x = 0$ is a fundamental observer, since he sees an entirely isotropic universe at any time due to the rotational symmetry of $\gamma_{ij}$. But this also holds for any other point with fixed comoving coordinates, since $\gamma_{ij}$ is maximally symmetric and so any point with fixed comoving coordinates could have been chosen as the spatial origin. We will term the fundamental observers also \textbf{comoving observers}.

How is the proper time $\tau_{\rm pr}$ of a comoving observer related to the cosmic time $t$? Since for comoving observers the line element (\ref{eq:robertson_walker_metric3}) reduces to $ds^2 = -c^2 dt^2$ and since comoving observers follow timelike world lines (like every physical observer), their lapse of proper time $d\tau_{\rm pr}$ is related to the line element by $c^2\:d\tau_{\rm pr}^2 = -ds^2$. Hence, we get the simple relation
\begin{equation}\label{eq:dt_dtau_pr}
	dt = d\tau_{\rm pr}\:.
\end{equation}
This means that for comoving observers the cosmic time $t$ in the Robertson-Walker metric is just their proper time $\tau_{\rm pr}$ as given by a standard clock in their rest frame. This also means that if we synchronize a set of comoving observers on a slice of constant time, they will stay synchronized on every subsequent slice as time goes on.

How are the comoving observers related to the matter (e.g., galaxies) in the universe? As already mentioned in the previous section, the fundamental observers must be at rest relative to local flow of matter. So, as long as the symmetry of the universe is perfect, all matter will stay at fixed comoving coordinates. All matter is, of course, also free falling. This can be formally shown by means of the geodesic equation. Let $x(\tau_{\rm pr}) = (ct(\tau_{\rm pr}),\bv{x}(\tau_{\rm pr}))$ be the world line of a comoving observer, i.e., $\bv{x}(\tau_{\rm pr}) \equiv \bv{x}$, with $\tau_{\rm pr}$ its proper time. With Eq.~(\ref{eq:dt_dtau_pr}) and since for the Robertson-Walker metric $\Gamma_{00}^\mu = 0$, comoving observers satisfy the geodesic equation
\begin{equation}\label{eq:geodesic}
	\frac{d^2x^\nu}{d\tau_{\rm pr}^2} + \Gamma_{\mu\sigma}^\nu \frac{dx^\mu}{d\tau_{\rm pr}}\frac{dx^\sigma}{d\tau_{\rm pr}} = 0
\end{equation}
and are thus free falling.

The \textbf{physical distance} (or \textbf{proper distance}) $D_{\rm pr}(t)$ between two points $\bv x_1$ and $\bv x_2$ on the slice $t$ is defined by the physical length of the shortest connection on the slice between them. While it might be complicated to find and parametrize this connection for two arbitrary points $\bv x_1$ and $\bv x_2$, we can simplify this problem substantially by taking advantage of the underlying symmetry of the Robertson-Walker metric and choosing coordinates such that $\bv x_1 = 0$ (homogeneity) and $\bv x_2 = (r,0,0)$ (isotropy). With the parametrization $\bv x(\lambda) = (\lambda,0,0)$\footnote{Note that for $K=1$ there are always (at least) two possibilities being the great circle segments between the two points. In this case we take the shorter one.},
the physical distance $D_{\rm pr}(t)$ is obtained by
\begin{equation}\label{eq:spatial distance}
	D_{\rm pr}(t) = a(t)\int_{0}^{r} \sqrt{\gamma_{ij}(\lambda,0,0)\frac{dx^i}{d\lambda}\frac{d x^j}{d\lambda}}\;d\lambda = a(t) \int_{0}^{r} d\lambda = a(t)\: r\:.
\end{equation}
This confirms the interpretation of the coordinate $r$ as the standard radial coordinate (up to the scale factor). Moreover, Eq.~(\ref{eq:spatial distance}) tells us that the physical distance of any two comoving observer scales with the time dependent scale factor $a(t)$. Assuming $a(t)$ was a monotonically increasing function of $t$ as indicated by the redshift of galaxies (see Section \ref{sec:cosmological_redshift}), it follows that any two galaxies in the universe are receding from each other and all separations between galaxies increase by the same factor with time. This global, coherent motion is called \textbf{Hubble flow} and describes the expansion of the universe.\footnote{It is important to note that the expansion of the universe not only means that galaxies are receding from each other, but rather that the universe as a whole is growing. For instance, in the case of $K=1$ the proper volume of the universe is given by $V = 2 \pi^3 R^3(t_0)a^3(t)$, thus it is finite and grows proportionally to $a^3(t)$.} It is often convenient to measure distances irrespective of the expansion of the universe. So we define the \textbf{comoving distance} $D(t)$ as
\begin{equation}\label{eq:comoving_distance}
\boxed{	D(t) = \frac{D_{\rm pr}(t)}{a(t)}\:.}
\end{equation}
Note that the comoving distance between comoving observers is constant and equal to the proper distance at the time $t_0$.

\section{Friedmann equations}\label{sec:friedmann_equation}

In a homogeneous and isotropic universe, the dynamics of spacetime and matter are determined solely by the scale factor $a(t)$. In order to determine the scale factor, we have to know the content of the universe in form of the energy-momentum tensor $T_{\mu\nu}$ and solve the Einstein field equations.

\subsection{Field equations and equation of motion}\label{sec:Field equations and equation of motion}

Fortunately, the symmetries of the universe not only set strong constraints on the metric $g_{\mu\nu}$, but also on the energy-momentum tensor. Since spatial coordinate transformations affect only the $i = 1,2,3$ components of $T_{\mu\nu}$, it follows immediately that $T_{00}$ transforms like a 3-scalar, $T_{i0}$ like a 3-vector, and $T_{ij}$ like a 3-tensor under such transformations.\footnote{The general transformation behavior
\begin{equation}
 \tilde{T}_{\alpha\beta}(\tilde{x}) = T_{\mu\nu}(x)\frac{\partial x^\mu}{\partial {\tilde{x}}^\alpha}\frac{\partial x^\nu}{\partial {\tilde{x}}^\beta}
\end{equation}
under a spacetime coordinate transformation $x \rightarrow \tilde{x}$ reduces for purely spatial coordinate transformations $(ct,\bv{x})\rightarrow (ct,\tilde{\bv{x}})$ to
\begin{equation}\label{eq:T_transform}
\tilde{T}_{00}(t,\tilde{\bv{x}}) = T_{00}(t,\bv{x})\:,\qquad \tilde{T}_{a0}(t,\tilde{\bv x}) = T_{i0}(t,\bv{x})\frac{\partial x^i}{\partial \tilde{x}^a}\:,\qquad  \tilde{T}_{ab}(t,\tilde{\bv{x}}) = T_{ij}(t,\bv{x})\frac{\partial x^i}{\partial {\tilde{x}}^a}\frac{\partial x^j}{\partial {\tilde{x}}^b}\:.
\end{equation}
This is the transformation behavior of a 3-scalar, a 3-vector, and a 3-tensor respectively.}

Moreover, since $T_{\mu\nu}$ has the same transformation behavior as the metric tensor $g_{\mu\nu}$ and the latter is substantially restricted by the symmetry of the spatial slice as shown in the previous section, the same restrictions hold for $T_{\mu\nu}$. It can be proven \citep[see][Sect.~13.4]{weinberg1972} that $T_{00}$, $T_{i0}$, and $T_{ij}$ must take the form
\begin{equation}\label{eq:form_of_tmunu}
	T_{00} = \rho(t)c^2\:, \quad\quad\quad T_{i0} = 0\:, \quad\quad\quad T_{ij} = p(t)\:g_{ij}\:,
\end{equation}
where the functions $\rho(t)$ and $p(t)$ can depend only on $t$. However, this means nothing else than that the energy-momentum tensor takes automatically the form of an \textbf{ideal fluid}
\begin{equation}\label{eq:energy_momentum_tensor1}
\boxed{	T_{\mu\nu} = \left(\rho + \frac{p}{c^2} \right)u_{\mu}u_{\nu} + p\:g_{\mu\nu}\:,}
\end{equation}
where the function $\rho(t)$ gets the interpretation of the matter density, $p(t)$ that of the pressure\footnote{This term should not connote that an energy component taking the form of an ideal fluid always features a pressure that could accomplish mechanical work (like moving a wall). For instance, a gas of weakly interacting relativistic particles such as neutrinos exhibits a pressure $p = \rho c^2/3$ and yet could hardly move a wall. Therefore the term ``pressure'' here is rather a property of the system regarding its momentum distribution.}, and where $u^{\mu} = -u_{\mu} = (c,0,0,0)$ is the 4-velocity of the fluid in comoving coordinates (which is vanishing for a comoving fluid). (Raising and lowering of indices is defined in the usual way, i.e., $u^\mu = u_{\nu} g^{\mu\nu}$ and $u_\mu = u^{\nu} g_{\mu\nu}$ with $g^{\mu\nu}$ being the inverse of $g_{\mu\nu}$. The component $u_0$ is constrained by the normalization condition $u^\mu u^{\nu} g_{\mu\nu} = -c^2$, which holds for every 4-velocity.) The energy-momentum tensor with mixed indices takes the simple form $T_{\mu}^{\phantom{\mu} \nu} = {\rm diag}(\rho c^2,p,p,p)$. It should also be noted that the universe may consist of several ideal fluids $T_{\mu\nu} = \sum_I [T_I]_{\mu\nu}$ for $I = 1,\ldots,N$, since the sum of an ideal fluid is also an ideal fluid.

The Einstein field equations are
\begin{equation}
	G_{\mu\nu} = \frac{8\pi G}{c^4}\: T_{\mu\nu}\:,
\end{equation}
where $G_{\mu\nu}$ is the Einstein tensor for the Robertson-Walker metric (\ref{eq:robertson_walker_metric3}) and $T_{\mu\nu}$ is the energy-momentum tensor (\ref{eq:energy_momentum_tensor1}). The computation of the Einstein tensor is straightforward but tedious. Appendix 2.3 of \cite{durrer2008} provides a table of all geometrical quantities of interest for the Robertson-Walker metric. Making use of them, we immediately find the \textbf{Friedmann-(Lema\^itre) equations}\footnote{The first equation is the $G^0_{\phantom 0 0} = 8\pi G\: T^0_{\phantom 0 0}$ component and the second equation corresponds to the trace $G^i_{\phantom i i}= 8\pi G\: T^i_{\phantom i i}$.}
\begin{empheq}[box=\fbox]{align}
	H^2 = \left(\frac{\dot a}{a}\right)^2 &= \frac{8 \pi G}{3}\rho-\frac{Kc^2}{R_0^2 a^2}\label{eq:friedmann_equation}\\
	\dot H-H^2 = \frac{\ddot a}{a} &= -\frac{4\pi G}{3}\left(\rho  + 3\frac{p}{c^2}\right)\:,\label{eq:second_friedmann_equation}
\end{empheq}
where a dot denotes the derivative with respect to cosmic time $t$. In these equations we also introduced the \textbf{Hubble parameter}
\begin{equation}\label{eq:hubble}
	H(t) \equiv \frac{\dot a(t)}{a(t)}\:.
\end{equation}

The \textbf{equation of motion} for the ideal fluid (\ref{eq:energy_momentum_tensor1}) is given by the general relativistic energy-momentum conservation
\begin{equation}
		\nabla_\nu T^{\mu\nu} = \partial_\nu T^{\mu\nu} + \Gamma_{\beta\nu}^\mu T^{\beta\nu}+\Gamma_{\beta\nu}^\nu T^{\mu\beta} = 0\:.
\end{equation}
Again using the table in the Appendix 2.3 of \cite{durrer2008} we obtain\footnote{This equation is obtained from the $\nabla_\mu T^{0\mu}=0$ component, while the $\nabla_\mu T^{i\mu}=0$ components just yield $dp/dx^i = 0$, i.e., homogeneity.}
\begin{equation}\label{eq:equation_of_motion_total}
\dot \rho = -3 H \left(\rho  +  \frac{p}{c^2}\right)\:.
\end{equation}
Note that this equation is not independent from the Friedmann equations, but could be derived from them. However, if the energy-momentum tensor consists of many separate fluids $T_{\mu\nu} = \sum_I [T_I]_{\mu\nu}$ for $I = 1,\ldots,N$, which are non-interacting (except for gravity), then the equation of motion holds for each fluid separately, i.e.
\begin{equation}\label{eq:equation_of_motion1}
\boxed{\dot \rho_I = -3 H \left(\rho_I +  \frac{p_I}{c^2}\right), \quad I = 1,\ldots, N\:.}
\end{equation}
This is information which could not be obtained from the Friedmann equations. We will always assume that the fluids considered are non-interacting. The total density and total pressure are obviously just the sums of the different components, i.e., $\rho(t) = \sum_I \rho_I(t)$ and $p(t) = \sum_I p_I(t)$, respectively.

Throughout this introduction, we will denote every time dependent quantity evaluated at the time $t_0$ for which $a(t_0) = 1$ by a subscript 0. For instance, $H_0 = H(t_0)$ is the \textbf{Hubble constant} and $R_0 = R(t_0)$ is the curvature radius of the universe at time $t_0$ in the case of $K\neq 0$. A model of the universe that is described in the framework of the Robertson-Walker metric and whose dynamics are determined by the Friedmann equations is called a \textbf{Friedmann-Lema\^itre-Robertson-Walker (FLRW) universe} or \textbf{FLRW (world) model}.\footnote{We include Georges Lema\^itre in this acronym for his substantial contributions to the early development of these cosmological models. For a historical review we refer to \cite{nussbaumer2009}.}

\subsection{Equation of state}\label{sec:equation_of_state}

In order to solve the Eqs.~(\ref{eq:friedmann_equation}) and (\ref{eq:equation_of_motion1}), we have to know how $\rho_I(t)$ and $p_I(t)$ are related for each separate fluid component. These relations are usually expressed by the \textbf{equation of state}
\begin{equation}\label{eq:equation_of_state}
	w_I(t) \equiv \frac{p_I(t)}{\rho_I(t)c^2}
\end{equation}
for each component.
If $w_I(t)$ is known for every fluid component and if the fluids are non-interacting, then we can solve the Friedmann equation for given initial conditions $\rho_{I0}$ and for a given $K$. That is, the Friedmann equation (\ref{eq:friedmann_equation}) together with the Eqs.~(\ref{eq:equation_of_motion1}) and (\ref{eq:equation_of_state}) form a closed system of equations. Note that for given initial densities $\rho_{I0}$ there is in general a (locally) expanding and a (locally) contracting solution due to the square on the left hand side of the Friedmann equation (\ref{eq:friedmann_equation}).

If the fluid $I$ with $w_I(t)$ is non-interacting with the other fluid components, $\rho_I(t)$ is determined by Eq.~(\ref{eq:equation_of_motion1}) and has the general solution
\begin{equation}
	\rho_I(t) = \rho_{I0} \:a(t)^{-3[1+w_{{\rm eff} I}(t)]}\:, \qquad w_{{\rm eff}I}(t) = \frac{1}{\ln(a)}\int_0^{\ln(a)} \frac{w_I(a)}{a}\: da\:.
\end{equation}
For a constant equation of state, i.e., $w_I(t)\equiv w_I$, this reduces to
\begin{equation}\label{eq:rho_of_t}
	\boxed{\rho_I(t) = \rho_{I0}\: a(t)^{-3(1+w_I)}\:.}
\end{equation}
For a fluid of weakly interacting non-relativistic ``particles'' (e.g., DM, galaxies) holds $w = 0$, while for a fluid of radiation or relativistic particles holds $w = 1/3$.\footnote{This can be easily seen by representing the fluid as a set of $N$ point particles (e.g., DM particles, galaxies, photons). In the special relativistic limit, the energy-momentum tensor of the fluid is then given by \citep[Sect.~2.10]{weinberg1972}
\begin{equation}\label{eq:gas}
	T^{\mu\nu}(x) = c \sum_{i=1}^N \frac{[p_i]^\mu [p_i]^\nu}{[p_i]^0}\:\delta^3\big(\bv x - \bv x_i(t)\big)\:,
\end{equation}
where $\bv x_i(t)$ is the world line of the $i$th particle and $[p_i]^\mu = ([p_i]^0,\bv p_i)$ its 4-momentum. Interpreting this expression as an ideal fluid (see Eq.~(\ref{eq:energy_momentum_tensor1})), it follows
\begin{equation}
	p = \frac{1}{3} T^i_{\phantom i i} = \frac{c}{3}\sum_{i=1}^N \frac{\bv p_i^2}{[p_i]^0}\:\delta^3(\bv x - \bv x_i)\:,\quad\quad \rho c^2 = T^{00} = c \sum_{i=1}^N [p_i]^0\: \delta^3(\bv x - \bv x_i)\:.
\end{equation}
With $[p_i]^\mu {[p_i]}_\mu = -[p_i]^0[p_i]^0 + \bv p_i^2 = -m c^2$ it follows for such a fluid in general $0 \leq p \leq \rho c^2/3$. Moreover, for non-relativistic particles, i.e., $\bv p_i^2 \ll mc^2$, it holds $p \ll \rho c^2/3$, and for relativistic particles, i.e., $\bv p_i^2 \gg mc^2$, it holds $p \simeq \rho c^2/3$.} A fluid with $w_I = -1$ is special in the sense that it has constant energy density with time, i.e., $\rho_I(t) \equiv {\rho_I}_0$, thus such a fluid can be interpreted as a property of space itself. Formally, it is equivalent to the inclusion of a cosmological constant term in the field equations:
\begin{equation}
	G_{\mu\nu} \rightarrow G_{\mu\nu} + g_{\mu\nu} \Lambda\:.
\end{equation}
The \textbf{cosmological constant} $\Lambda$ is then related to ${\rho_I}_0$ and ${p_I}_0$ by
\begin{equation}
	{\rho_I}_0 = \frac{\Lambda c^2}{8 \pi G}\:,\quad\quad\quad {p_I}_0 = - {\rho_I}_0 c^2 = -\frac{\Lambda c^4}{8 \pi G}\:,
\end{equation}
and the Friedmann equations correspondingly become 
\begin{equation}
		\left(\frac{\dot a}{a}\right)^2 = \frac{8 \pi G}{3}\rho-\frac{Kc^2}{R_0^2 a^2} + \frac{\Lambda c^2}{3}\:,\quad\quad\quad \frac{\ddot a}{a} = -\frac{4\pi G}{3}\left(\rho + 3\frac{p}{c^2}\right)+ \frac{\Lambda c^2}{3}\:,
\end{equation}
where $\rho$ and $p$ do not include the $I$th component anymore.

For a flat universe, i.e., $K=0$, that is governed by a single energy component $T_{\mu\nu}$ with a constant equation of state $w$, the time evolution of the scale factor can be explicitly given.\footnote{In Section \ref{sec:spherical_top_hat_collapse} we discuss the solution of an overcritical, i.e., $K = 1$, universe for $w = 0$.} Let $t_\ast$ be an arbitrary epoch. For $w>-1$ and with Eq.~(\ref{eq:rho_of_t}), the function
\begin{equation}\label{eq:a_of_t}
	\boxed{a(t) = a(t_\ast)\:\left(\frac{t}{t_\ast}\right)^{2/[3(1+w)]}}
\end{equation}
is an expanding solution of the Friedmann equation (\ref{eq:friedmann_equation}), and yields the Hubble parameter (\ref{eq:hubble})
\begin{equation}\label{eq:meaning_of_H0}
H(t) = \frac{\dot a(t)}{a(t)} = \frac{2}{3 (1+w) t}\:.
\end{equation}
The origin of the time coordinate $t=0$ has been chosen such that the scale factor vanishes at that epoch, which in our simple model marks the beginning of the universe (``big bang''). Thus, a flat, expanding universe with a single energy component ($w > -1$) has a beginning and will expand at any time, where it follows from Eq.~(\ref{eq:meaning_of_H0}) that its age is given by the inverse of the corresponding Hubble parameter (up to a constant of order unity). On the other hand, for $w = -1$ the energy density is constant (see  Eq.~(\ref{eq:rho_of_t})) and so the Friedman equation (\ref{eq:friedmann_equation}) has the solution
\begin{equation}\label{eq:a_of_t_for_w=-1}
	\boxed{a(t) = a(t_\ast)\: e^{H_0 \left(t-t_\ast\right)}}
\end{equation}
with $H_0 = H(t_\ast) = \sqrt{8\pi G\rho(t_\ast)/3} \equiv$ const. In this case the scale factor never vanishes and the universe is formally infinitely old. While $H_0$ is a free parameter for the solution (\ref{eq:a_of_t_for_w=-1}), it is entirely specified for the solution (\ref{eq:a_of_t}) by Eq.~(\ref{eq:meaning_of_H0}), since we have fixed $a(0) = 0$. So the only remaining free parameter in the latter case is $R_0$, which however for a flat universe is just an arbitrary scaling with no observational consequences. Hence the evolution of a flat, expanding universe with a single energy component ($w>-1$) has effectively no degree of freedom.

For a flat universe that is governed by several fluids $T^{\mu\nu}_I$ with different, but still constant equations of state $w_I$, the expansion history is slightly more complicated and can in general be only computed numerically. However, for a certain time interval (``era'') between $t_{\rm i}$ and $t_{\rm f}$, when the universe is \textbf{dominated} by the $I$th energy component, i.e.,
\begin{equation}\label{eq:domination}
	T_{\mu\nu}(t) \simeq [T_I]_{\mu\nu}(t)\:,\quad t_{\rm i} \leq t \leq t_{\rm f}\:,
\end{equation}
we can find approximate solutions. For $w_I > -1$ , the scale factor $a(t)$ for an expanding universe is approximately described by
\begin{equation}\label{eq:a_of_t_zero_offset}
	a(t) \simeq a(t_{\rm m})\left(\frac{t-\tilde t}{t_m-\tilde t}\right)^{2/[3(1+w_I)]}\:,
\end{equation}
where $\tilde t$ is a time shift that is determined by the Friedmann equation (\ref{eq:friedmann_equation}) for $t= t_{\rm m}$ yielding
\begin{equation}
	\left(\frac{2}{3} \frac{1}{(t_{\rm m}-\tilde t)(1+w)}\right)^2 = H^2(t_{\rm m}) = \frac{8\pi G}{3}\rho (t_{\rm m})
\end{equation}
and $t_{\rm_m}$ is a fixed epoch between $t_{\rm i}$ and $t_{\rm f}$. For $w_I = -1$ we have instead the approximation
\begin{equation}\label{eq:a_t_ext_shift}
	a(t) \simeq a(t_{\rm i}) e^{H(t_{\rm i}) \left(t-t_{\rm i}\right)}\:.
\end{equation}
Thus a comparison of the approximations for the multi-component system to the solutions of the single-component systems (see Eqs.~(\ref{eq:a_of_t}) and (\ref{eq:a_of_t_for_w=-1})) shows that the difference is just a time shift $\tilde t$. For a given era that is dominated by the component $I$, this shift is usually so small relative to the corresponding age of the universe $t_{\rm m}$ that it can be neglected. Hence for many applications it is sufficient to just use the Eqs.~(\ref{eq:a_of_t}) and (\ref{eq:a_of_t_for_w=-1}) even in the case of a multi-component system.

What can we say about the acceleration $\ddot a(t)$ of the universe? The second Friedmann equation (\ref{eq:second_friedmann_equation}) tells us that, irrespective of the curvature, the expansion of the universe is decelerating, i.e., $\ddot a(t) < 0$, if the universe is dominated by an equation of state $w_I > -1/3$, and accelerating, i.e., $\ddot a(t) > 0$, if it is dominated by an equation of state $w_I < -1/3$. Thus, a fluid with $w_I < -1/3$ has the remarkable property to act repulsively by gravitation. This means it violates the strong energy condition which requires physical fluids to satisfy $\rho_I + 3 p_I > 0$.

\subsection{Density parameters}

In order to study and compare different cosmological models, it is convenient to introduce the dimensionless \textbf{density parameters} $\Omega_I(t)$ defined as
\begin{equation}\label{eq:density_parameter}
	\boxed{\Omega_I(t) = \frac{\rho_I(t)}{\rho_{\rm c}(t)}\:,\quad\quad\rho_{\rm c}(t) = \frac{3H^2(t)}{8\pi G}\:,}
\end{equation}
where $\rho_{\rm c}(t)$ is the \textbf{critical density}. The first Friedmann equation (\ref{eq:friedmann_equation}) then simplifies to
\begin{equation}\label{eq:omega_k}
	 1-\sum_{I=1}^{N} \Omega_I(t) = -\frac{Kc^2}{H^2(t) R_0^2 a^2(t)} \equiv \Omega_{K}(t)\:,
\end{equation}
where $\Omega_K(t)$ is the \textbf{curvature density} acting phenomenologically like a fluid with an equation of state $w_K = -1/3$ (cf.~Eq.~(\ref{eq:rho_of_t})). Note that such a fluid does not contribute to the second Friedmann equation (\ref{eq:second_friedmann_equation}) due to $\rho_K(t) + 3p_K(t)/c^2 = 0$. Moreover, it follows from the definition that
\begin{equation}
	\Omega_K(t) < 0\ \Leftrightarrow\ K = 1\:, \quad\quad \Omega_K(t) = 0\ \Leftrightarrow\ K = 0\:, \quad\quad \Omega_K(t) > 0\ \Leftrightarrow\ K = -1\:,
\end{equation}
and $\Omega_K(t)$ cannot change the sign during the evolution of the universe. The first Friedmann equation then takes by construction the very compact form
\begin{equation}\label{eq:friedmann_equation3}
	\boxed{\sum_{I=1}^{N} \Omega_I(t) + \Omega_K(t)  = 1\:.}
\end{equation}

The cosmological models are usually characterized by the present day values of the density parameters. For ease of notation we will omit the subscript 0 for the values of the density parameters at $t_0$, i.e., if no particular epoch is indicated, it holds $\Omega_I = \Omega_I(t_0)$ and $\Omega_K = \Omega_K(t_0)$. Assuming that the different fluid components $I$ are not interacting with each other, the evolution of each density parameter can be expressed using Eq.~(\ref{eq:rho_of_t}) as
\begin{equation}\label{eq:density_parameter_evolution}
	\Omega_I(t) = \Omega_I \:a(t)^{-3(1+w_I)} \left(\frac{H_0}{H(t)}\right)^2\:,\quad\quad\quad \Omega_K(t) = \Omega_K \:a^{-2}(t) \left(\frac{H_0}{H(t)}\right)^2\:,
\end{equation}
so that the Friedmann equation (\ref{eq:friedmann_equation3}) becomes
\begin{equation}\label{eq:friedmann_equation4}
	\boxed{H(t) = H_0 \sqrt{\sum_{I=1}^N \Omega_I \:a(t)^{-3(1+w_I)} + \Omega_K \:a^{-2}(t)}\:.}
\end{equation}

\section{Observational cosmology}\label{sec:observational_cosmology}

To connect the theoretical framework developed in the previous two sections with astronomical observables, we have to understand how photons behave in this framework. This directly leads to the redshift of galaxies, which is probably the most important observable of extragalactic astronomy.

\subsection{Cosmological redshift}\label{sec:cosmological_redshift}

Consider a photon being emitted by a distant galaxy at the spatial coordinate $\bv x_1$ that arrives at the earth being at $\bv x_0$ at the present epoch. Without loss of generality, we can assume that the Milky Way lies at the origin of the spatial comoving coordinate system, i.e., $\bv x_0 = 0$, and that the distant galaxy has the comoving coordinate $\bv x_1 = (r,0,0)$. Like in special relativity, the world line of a photon in general relativity is characterized by $ds = 0$ due to the principle of equivalence, so with the Robertson-Walker metric (\ref{eq:robertson_walker_metric3}) it holds for a photon coming toward us
\begin{equation}\label{eq:photon}
	dr = -\frac{c}{a(t)}\:dt\:.
\end{equation}
Now consider two wave crests of the photon leaving the distant galaxy at $t$ and $t+\delta t$, respectively, and arriving the earth at $t_0$ and $t_0+\delta t_0$, respectively. Since the two galaxies are at fixed comoving coordinates $\bv x_1$ and $\bv x_0$, the comoving distance traveled by the crests of the photon is the same for the two crests, i.e.,
\begin{equation}\label{eq:comoving_distance_for_photon}
	r = -\int_{t}^{t_0} \frac{c}{a(t)}dt = -\int_{t+\delta t}^{t_0+\delta t_0} \frac{c}{a(t)}dt\:.
\end{equation}
This leads to
\begin{equation}
	0 = \int_{t_0}^{t_0+\delta t_0} \frac{c}{a(t)}\:dt - \int_{t}^{t+\delta t} \frac{c}{a(t)}\:dt \simeq \frac{c\:\delta t_0}{a(t_0)} - \frac{c\:\delta t}{a(t)}\:,
\end{equation}
where the last approximation is very accurate since $dt \sim dt_0 \sim 10^{-14}\ {\rm s}$ for visible light. Since $\delta t$ and $\delta t_0$ are just the periods of the wave of the photon at the epochs $t$ and $t_0$ respectively, it holds for the frequencies of the emitted photon, i.e., $\nu_{\rm em} = 1/\delta t$, and the observed photon, i.e., $\nu_{\rm obs} = 1 / \delta t_0$,
\begin{equation}\label{eq:redshift}
	\boxed{\frac{\nu_{\rm em}}{\nu_{\rm obs}} = \frac{\delta t_0}{\delta t} = \frac{a(t_0)}{a(t)} = \frac{1}{a(t)} \equiv 1+z\:.}
\end{equation}
This means that a photon experiences a frequency shift inversely proportional to the expansion of the universe during its journey.\footnote{Our derivation was slightly heuristic. For a more formal derivation by means of the collisionless relativistic Boltzmann equation see, e.g., \nocite{durrer2008}Durrer (2008, Sect.~1.3.3).} The new introduced quantity $z$ is called \textbf{cosmological redshift}, if the frequency shift is towards smaller frequencies, and \textbf{cosmological blueshift}, if the frequency shift is towards larger frequencies. Since essentially all galaxies exhibit a redshift, we will call $z$ just (cosmological) redshift. Thus, assuming that a galaxy is a comoving observer and that there is no other contribution to its redshift, the ratio $a(t_0)/a(t)$ between the emission and arrival of the photon can be measured by studying the spectrum of the galaxy. Unfortunately, as we will see in the next section, for instance the fact that galaxies are not perfect comoving observers produces an additional redshift contribution, which cannot be disentangled from the cosmological one. But if not mentioned otherwise, we assume that the redshift is purely cosmological.

Eq.~(\ref{eq:comoving_distance_for_photon}) shows that for a photon arriving at the earth at the present epoch there exists a one-to-one correspondence between the comoving distance $D = r$ (see Eq.~(\ref{eq:comoving_distance})) and the emission epoch $t$ of the photon. Then there is a one-to-one correspondence between the scale factor $a(t)$ and the emission epoch $t$, if $a(t)$ is a monotonically increasing function, and there is a one-to-one correspondence between the scale factor $a$ and the redshift $z$ given by Eq.~(\ref{eq:redshift}). So finally, if the universe is monotonically increasing, there exist one-to-one correspondences between any pair of the four quantities $D$, $t$, $a$, and $z$, and we can always express any of these quantities as a function of any other. To derive the relation $D(z)$ explicitly, we again consider the world line of a photon
\begin{equation}\label{eq:dr}
	a(t)\:dr \supeq{(\ref{eq:photon})}{=} -c\: dt = -\frac{c}{\dot a(t)}\:da =  -\frac{c}{a(t)H(t)}\:\frac{d a}{d z}\: dz = \frac{c}{H(t)}\:a(t)\: dz\:,
\end{equation}
where $t$ is the emission epoch for a photon and where we have used $d a(z)/d z = -1/(1+z)^2 = -a^2(t)$. So it holds
\begin{equation}\label{eq:hz}
dr = \frac{c}{H(z)}dz\:,\quad H(z) = H_0 \sqrt{\sum_{I=1}^N \Omega_I \:(1+z)^{3(1+w_I)} + \Omega_K \:(1+z)^{2}}\:,
\end{equation}
where the explicit expression for $H(z) = H(t(z))$ is immediately obtained by using Eq.~(\ref{eq:friedmann_equation4}) and the definition of the redshift. The comoving distance to a galaxy with redshift $z$ is then
\begin{equation}\label{eq:comoving_distance_z}
	\boxed{D(z) = \frac{c}{H_0} \int_0^z \frac{1}{\sqrt{\sum_{I=1}^N \Omega_I \:(1+z)^{3(1+w_I)} + \Omega_K \:(1+z)^{2}}}\:dz\:.}
\end{equation}

Surveys encompassing thousands or even millions of galaxies have shown that essentially all galaxies exhibit a redshift meaning $z > 0$. In the light of Eq.~(\ref{eq:redshift}), this is direct confirmation that $a(t)$ was indeed smaller when the observed galaxy photons were emitted. Today, it is well established that there exists a one-to-one correspondence between the distance and the redshift of a galaxy. On the one hand, this is directly demonstrated with the aid of supernovae Ia acting as ``standard candles'' up to redshifts of $z \sim 1$, and, on the other hand, it is a consequence of the concordance cosmology to be introduced in the next section. This justifies a posteriori our assumption of $a(t)$ being a monotonically increasing function of time.

\subsection{Peculiar velocities}\label{sec:peculiar_velocities}

As mentioned in Section \ref{sec:cosmological_principle}, our universe is only homogeneous for length scales $\gtrsim 100$ Mpc. On smaller scales, however, it is strongly inhomogeneous, which leads to deviations from the overall Hubble flow of the order of a few 100 km/s. These deviations are termed \textbf{peculiar velocities}. The radial components of the peculiar velocities of galaxies add a contribution to their total redshift by means of the Doppler effect at the position of the galaxies which observationally cannot be disentangled from their cosmological contribution defined by Eq.~(\ref{eq:redshift}).

Consider a galaxy which resides at the position corresponding to the cosmological redshift $z_{\rm cos}$ and which has a peculiar velocity $\delta v$ in radial direction. Due to the local Doppler effect a photon emitted from that galaxy in our direction as observed by a comoving observer at the position of the galaxy is redshifted by
\begin{equation}
	1+z_{\rm p} \equiv \frac{\nu}{\nu'}= \sqrt{\frac{1+\delta v / c}{1-\delta v / c}} \simeq 1 + \frac{\delta v}{c}
\end{equation}
for $|\delta v /c| \ll 1$, where $\nu$ is the frequency of the photon in the restframe of the galaxy and $\nu'$ the frequency observed by the comoving observer at the position of the galaxy.\footnote{Note that only the radial components of the peculiar velocities lead to redshift perturbations, as the transverse Doppler effect caused by the velocity components perpendicular to the line of sight can be neglected due to the non-relativistic motions of galaxies.} After its emission the redshifted or blueshifted photon travels all the way to the earth and is further redshifted due to the expansion of the universe, i.e., $\nu' / \nu'' = 1 + z_{\rm cos}$, where $\nu''$ is the frequency of the photon arriving at the earth. Thus, the total, observable redshift $z$ of the galaxy is
\begin{equation}\label{eq:peculiar_redshift}
	1 + z = \frac{\nu}{\nu''} = \frac{\nu}{\nu'}\frac{\nu'}{\nu''} = \left(1+z_{\rm cos}\right)\left(1+z_{\rm p}\right) \simeq \left(1+z_{\rm cos}\right)\left(1+\frac{\delta v}{c}\right)\:,
\end{equation}
leading to the redshift perturbation
\begin{equation}\label{eq:redshift_perturbation}
\delta z = z - z_{\rm cos} \simeq \left(1+z_{\rm cos}\right)\frac{\delta v}{c}\:.
\end{equation}
If the observed redshift is interpreted as purely cosmological, this redshift perturbation produces a spurious displacement $\delta D$ of the galaxy along the line of sight of the order
\begin{equation}\label{eq:delta_l}
	\delta D = D(z)-D(z_{\rm cos}) \simeq \frac{d D}{d z}(z_{\rm cos})\:\delta z \supeq{(\ref{eq:comoving_distance_z})}{=} \frac{c}{H(z_{\rm cos})}\:\delta z \supeq{(\ref{eq:redshift_perturbation})}{\simeq} \frac{1+z_{\rm cos}}{H(z_{\rm cos})}\:\delta v\:,
\end{equation}
where we have used the first order expansion $D(z) = D(z_{\rm cos}+\delta z) \simeq D(z_{\rm cos}) + d D/d z\:\delta z$. Therefore, in observational cosmology one has to distinguish between the ideal \textbf{real space}, where the true distances of galaxies are known, and the observable \textbf{redshift space}, in which distances are inferred from their observed redshift.

Peculiar velocities are particularly prominent in groups (and clusters) of galaxies. A galaxy group is a gravitationally bound system (typically associated with a DM halo, see Sect.~\ref{sec:dm_halos}) containing several galaxies and other forms of matter, which are moving in the gravitational potential of the group. Due to the gravitational boundedness, the system is decoupled from the Hubble flow and hence photons moving through the group are not affected by the expansion of the universe until they leave the group. Thus, the observable redshifts of galaxies within groups consist of the redshift $z_{\rm gr}$ of the group as a whole\footnote{This redshift can be purely cosmological or it can itself feature a peculiar velocity, if the whole group is moving with respect to the Hubble flow.} and their line of sight peculiar velocities within the group. As a first consequence, the redshift of the galaxies does not contain any information about the line of sight position of the galaxy within the group. As a second consequence, the peculiar velocities of galaxies in groups lead to an elongated shape of groups along the line of sight in redshift space \citep{jackson1972,tadros1999}, as only the components of the peculiar velocities parallel to the line of sight contribute to the redshift perturbations. Since these elongations are always pointing toward us, they are termed \textbf{fingers-of-god}. If a group with redshift $z_{\rm gr}$ has a line of sight velocity dispersion $\sigma_{\rm v}$, i.e., the standard deviation of the line of sight components of the peculiar velocities of its galaxies, its finger-of-god has the comoving radial length (see Eq.~(\ref{eq:delta_l}))
\begin{equation}
	\boxed{D_{\rm FOG} = \frac{c}{H(z_{\rm gr})}\:\sigma_{\rm z} = \frac{1+z_{\rm gr}}{H(z_{\rm gr})}\:\sigma_{\rm v}\:,}
\end{equation}
where $\sigma_{\rm z}$ denotes the redshift dispersion of the galaxies. This is a convenient formula for measuring the velocity dispersions in groups by means of their redshift distribution. This example illustrates that peculiar velocities of galaxies can be both a blessing and a curse; they obscure the real positions of galaxies, but provide us valuable information on the dynamics of galaxies.

\subsection{Horizons}\label{sec:horizons}

The distance that photons can travel in the universe during a given time interval defines the radius of causality, within which information can propagate during the time interval. This radius is called horizon. In cosmology, there are two kinds of horizons of interest.

The \textbf{particle horizon} at time $t$ is the distance that a photon can travel from the beginning of the universe up to this time. This means that a point in space is causally connected only to the region within its particle horizon. This region with us being a the center is called the \textbf{observable universe}. With Eq.~(\ref{eq:photon}), the comoving particle horizon $D_{\rm p}(t)$ at time $t$ is given by
\begin{equation}\label{eq:com_dp}
\boxed{D_{\rm p}(t) = \int_{t_{\rm i}}^{t} \frac{c}{a(t')}\: dt'\:,}
\end{equation}
where $t_{\rm i}$ is the beginning of the universe. If the universe is flat with a single ideal fluid component, we can set $t_{\rm i} = 0$ and use Eq.~(\ref{eq:a_of_t}). It follows immediately that for an equation of state $w>-1/3$ the particle horizon is finite and takes the value
\begin{equation}\label{eq:particle_horizon}
	D_{\rm p}(t) = \frac{t}{a(t)}\:\frac{c}{1-\frac{2}{3(1+w)}} \supeq{(\ref{eq:meaning_of_H0})}{=} \frac{c}{H(t)a(t)}\: \frac{1}{\frac{3}{2} (1+w)-1}\:.
\end{equation}
On the other hand, for $w\leq -1/3$ there is no particle horizon (i.e., it is infinite) even though the age of the universe is finite. Note that formally the particle horizon corresponds to redshift $z = \infty$ (see Eq.~(\ref{eq:redshift})).

The \textbf{event horizon} $D_{\rm e}$ is the distance that a photon can travel from now until the end of the universe $t_{\rm f}$. This means that a photon emitted at the present epoch from a galaxy outside our event horizon will never reach us even if we wait infinitely long. Formally the event horizon $D_{\rm e}(t)$ is
\begin{equation}\label{eq:event_horizon}
\boxed{D_{\rm e}(t) = \int_{t}^{t_{\rm f}} \frac{c}{a(t')}\: dt'\:.}
\end{equation}
By means of a similar argument like for the particle horizon, we find that the event horizon for a flat universe with a single component with an equation of state $-1 \leq w < -1/3$ is finite and given by
\begin{equation}\label{eq:event_horizon2}
	D_{\rm e}(t) = \frac{t}{a(t)}\:\frac{c}{\frac{2}{3(1+w)}-1} \supeq{(\ref{eq:meaning_of_H0})}{=}  \frac{c}{H(t)a(t)}\: \frac{1}{1-\frac{3}{2} (1+w)}\:,
\end{equation}
where for such models the end of the universe is $t_{\rm f} = \infty$. (For the case $w=-1$ we have used Eq.~(\ref{eq:a_of_t_for_w=-1}).) For $w \geq -1/3$ there is no event horizon, i.e., any photon emitted at the present epoch will reach us at some time in the future.

Note that both the comoving particle horizon as well as the comoving event horizon are essentially $D_{\rm H}(t) = c/(Ha)$ times a numerical constant of order unity. This shows that $D_{\rm H}$ is the typical length scale in a FLRW universe at time $t$ and we call $D_{\rm H}$ the comoving \textbf{Hubble length}. We will often approximate either horizon by this quantity. The \emph{proper} Hubble length is then just $D^{\rm pr}_{\rm H}(t) = D_{\rm H}a = c/H$. Note that for $w = -1$ the proper Hubble length is constant, since in this case $H(t) \equiv H_0$ is constant, and identical to the proper event horizon.

\section{Concordance model}\label{sec:concordance_cosmology}

In the last three sections, we have developed the FLRW framework for a general expanding universe. Fortunately, the growing amount of observational data in astronomy, particularly the CMB and huge galaxy surveys, have allowed the determination of the constituents of the universe $T^{\mu\nu}_I$ and the present day values of the cosmological parameters (e.g., $H_0$, $\Omega_I$, $\Omega_K$) to the impressive precision of a few percent. This led to the currently favored \textbf{concordance model}, which is a flat universe whose energy budget at the present epoch is dominated by some sort of exotic ``cold dark matter'' (CDM) and exotic ``dark energy'' in the form of a cosmological constant $\Lambda$, where exotic refers to the fact that these constituents must represent physics beyond the standard model of particle physics and could not yet be observed in human made experiments. Due to these two main contributions, the concordance model ist also called \textbf{$\bv \Lambda$CDM model}.

A summary of the current values of the present day cosmological parameters as obtained by a combination of the CMB WMAP 7-year and complementary data sets \citep{komatsu2011} is given in Table \ref{tab:cosmological_parameters}, where the Hubble constant is parametrized by means of $h$ as $H_0 = 100\: h\ {\rm km}\: {\rm s}^{-1}\: {\rm Mpc}^{-1}$. The density parameters $\Omega_{\rm b}$, $\Omega_{\rm m}$, $\Omega_\Lambda$, $\Omega_\gamma$, and $\Omega_\nu$ will be discussed in detail in the following section, and the cosmological parameters $n_{\rm s}$ and $\sigma_8$, which describe the clustering in the universe, will be introduced in the Sections \ref{sec:Initial_conditions} and \ref{eq:filtering} respectively.
\begin{table}[tp]\centering
\caption{Present day cosmological parameters as obtained by the WMAP 7-year data set combined with supernovae Ia data, and acoustic baryonic oscillations \citep{komatsu2011}.\\[-2mm]}
\begin{tabular}{lcl}
\toprule
Parameter & Present day value & Equation of state\\
\midrule
Hubble constant & $h = 0.702 \pm 0.014$  & --\\
baryonic matter & $\Omega_{\rm b} = 0.0458 \pm 0.0016$ & $w_{\rm b} = 0$ \\		
dark matter & $\Omega_{\rm d} = 0.229 \pm 0.015$ & $w_{\rm d} = 0$\\
dark energy & $\Omega_\Lambda = 0.725 \pm 0.016$ & $w_\Lambda = -0.980 \pm  0.053$\\
curvature density & $ -0.0133 < \Omega_K < 0.0084$ &  $w_K = -1/3$  \\
photons & $\Omega_\gamma = 2.47 \times 10^{-5}\:h^{-2}$ & $w_\gamma = 1/3$ \\
neutrinos & $\Omega_\nu = 1.71 \times 10^{-5}\:h^{-2}$ & $w_\nu = 1/3$ \\
spectral index  & $n_{\rm s} = 0.968 \pm 0.012$ & -- \\
linear fluctuation amplitude  & $\sigma_8 = 0.816 \pm 0.024$ & --\\
\bottomrule\\[-2mm]
\end{tabular}
\begin{minipage}{.85\textwidth}
Note: The values and errorbars of the parameters $h$, $\Omega_{\rm b}$, $\Omega_{\rm d}$, $\Omega_\Lambda$, $n_{\rm s}$, and $\sigma_8$ correspond to the mean and $68 \%$ confidence limits (CL), respectively, of the marginalized distributions after fitting the data to a flat 6-parameter $\Lambda$CDM model. To estimate $w_\Lambda$ the cosmology was kept flat and only WMAP and supernovae data were used, and to estmate $\Omega_K$ an equation of state $w_\Lambda = -1$ was assumed. The errorbar of $w_\Lambda$ corresponds to the 68\% CL and the one of $\Omega_K$ to the 95\% CL. The estimation of $\Omega_\gamma$ and $\Omega_\nu$ is described in Sect.~\ref{sec:content}.
\end{minipage}
\label{tab:cosmological_parameters}
\end{table}

One of the most striking properties of our universe is that its geometry is essentially flat, i.e., (see Tab.~\ref{tab:cosmological_parameters})
\begin{equation}
	\boxed{\Omega_K \simeq 0\:.}
\end{equation}
On the one hand, within a flat universe the formalism developed so far (and the one to be developed in the other chapters) simplifies a lot. From now on we will stick to the case of a precisely flat universe. On the other hand, without a reason at hand why the universe should be flat this result is somewhat surprising and became what is known as the ``flatness problem'', which will be discussed in Section \ref{sec:connection_to_the_standard_model}.

\subsection{Content of the present day universe}\label{sec:content}

We see from Table \ref{tab:cosmological_parameters} that the energy content of the present day universe mainly consists of
\begin{equation}\label{eq:energy_component}
	\boxed{\Omega_\Lambda \simeq 0.73\:,\quad\quad \Omega_{\rm d} \simeq 0.23\:, \quad\quad \Omega_{\rm b} \simeq 0.04\:,}
\end{equation}
where $\Omega_\Lambda$ corresponds to \textbf{dark energy}, $\Omega_{\rm d}$ to cold \textbf{dark matter} (DM), and $\Omega_{\rm b}$ to \textbf{baryons}\footnote{Unlike in particle physics, in cosmology ``baryons'' also comprise electrons, i.e., the term just refers to ``normal matter'', of which gas, stars and planets etc.~are made in contrast to more exotic matter like neutrinos and DM.}. Thus, dark energy is the dominant energy contribution of the present day universe and acts repulsively due to its equation of state being smaller than $-1/3$, i.e., dark energy is responsible for the observed acceleration of the universe, which was directly observed for the first time by means of supernovae Ia (see \citealt{weinberg2012} for a review on observational probes of the cosmic acceleration). Moreover, all measurements are so far consistent with $w_\Lambda = -1$ (for this reason it has been given the subscript of the cosmological constant $\Lambda$), and yet there is no clue from fundamental physics what dark energy could be. Hence dark energy is not only the dominant, but also the most mysterious\footnote{The ``mystery'' of dark energy has been extensively discussed in the literature \citep[for reviews see, e.g.,][]{weinberg1989,carroll2001,padmanabhan2003,sahni2005}. The most prominent issue is the so-called ``cosmological constant problem'' stating that, if the cosmological constant is interpreted as the zero-point energy of the vacuum, attempts to estimate its value from quantum field theory yield a value that is typically about 120 orders of magnitude too large compared with cosmological measurements. In this context, we also refer to the discussion of \cite{bianchi2010}.} component of the universe. DM being the second ranked dominant energy contribution at the present epoch must constitute some sort of non-baryonic, cold (i.e., non-relativistic), very weakly interacting massive particle (WIMP), which has not been detected in laboratories yet. Its non-relativistic nature can be inferred from the observed cosmic structures at small scales, and its non-baryonic nature is a consequence of the theory of nucleosynthesis in the early universe (see the next section for a brief history of the universe). Since baryons and DM have the same equation of state, it is meaningful for many applications to add them together yielding the total \textbf{matter density}
\begin{equation}
	\Omega_{\rm m} = \Omega_{\rm d} + \Omega_{\rm b} \simeq 0.27\:.
\end{equation}

The universe also contains energy in the form of relativistic particles such as photons and neutrinos. By far the biggest contribution to the energy density of such particles comes from the equilibrium distribution produced in the early universe. The energy density for a relativistic particle species of a given temperature $T$ is
\begin{equation}
	u(T) = g \frac{\pi^2}{30}k_{\rm B}T \left(\frac{kT}{\hbar c}\right)^3\:,
\end{equation}
where $g$ denotes the effective number of degrees of freedom of the particle.\footnote{For a derivation of this formula we refer, for instance, to \nocite{mukhanov2005}Mukhanov (2005, Sect.~3.3). The effective number of degrees of freedom for a photon is $g_\gamma = 2$ due to the two polarization states and $g_\nu = 2 \times 7/8$ for each neutrino species. In the latter case the factor 2 is due to existence of neutrinos and anti-neutrinos and the factor 7/8 must be included for all fermionic particles.} With $g_\gamma = 2$ for photons and the current CMB temperature of $T_\gamma = 2.725 \pm 0.002$ K ($95 \%$ confidence, \citealt{mather1999}) we obtain the photon density
\begin{equation}
	\Omega_\gamma = \frac{u(T_\gamma)/c^2}{\rho_{\rm c}} \simeq 2.47 \times 10^{-5}\:h^{-2}\:.
\end{equation}
Since for the neutrinos it holds $g_\nu = 2\times 7/8$ and since the temperature of the neutrino background is predicted to be $T_\nu = (4/11)^{1/3} T_\gamma = 1.95$ K by the theory of the early universe, the neutrino density is
\begin{equation}
	\Omega_\nu = \frac{7}{8}\left(\frac{4}{11}\right)^{4/3} N_\nu\: \Omega_\gamma \simeq 1.71 \times 10^{-5}\:h^{-2}
\end{equation}
with $N_\nu = 3.04$ the standard value for the effective number of neutrino species \citep{komatsu2011}. Note that in contrast to the CMB and its temperature, the expected cosmic neutrino background could not yet be detected, but is predicted by the cosmological model of the early universe, which will not be discussed in this introduction. The sum of the photon and neutrino densities yields a total \textbf{radiation density} of
\begin{equation}
	\boxed{\Omega_{\rm r} = \Omega_\gamma + \Omega_\nu \simeq 4.2 \times 10^{-5}\: h^{-2}\:.}
\end{equation}
Comparing this value to those in Eq.~(\ref{eq:energy_component}) shows that the radiation density is negligible in the present universe, and so the relation (\ref{eq:comoving_distance_z}) between comoving distance $D$ and redshift $z$ simplifies for the redshift range accessible by optical astronomy, i.e., $z \lesssim 10$, to
\begin{equation}\label{eq:H_z_concordance_model}
	\boxed{
	D(z) = \int_0^z \frac{c}{H(z)}\:dz\:,\quad\quad H(z) = H_0\sqrt{\Omega_\Lambda + \Omega_{\rm m} \:(1+z)^{3}}\:.
	}
\end{equation}

\subsection{History of the universe}\label{sec:timeline}

With the cosmological parameters specified in the previous section we are able to give an outline of the history of the universe. Details can be found in almost any textbook of cosmology. We will follow the summary in \nocite{mukhanov2005}Mukhanov (2005, Sect.~3.2). At the present time the universe is mainly driven by $\Omega_\Lambda$, while radiation $\Omega_{\rm r}$ is entirely negligible. However, due to the different equation of states the relative importance of the different energy constituents change with time according to Eq.~(\ref{eq:rho_of_t}). The larger $w_I$ the more important is the corresponding energy constituent at earlier times. So there was a time when the universe was radiation dominated, then there was a time when it was matter dominated, and now it is about to become $\Lambda$ dominated.

The model of the history of the universe is obtained by extrapolating the current state back in time and feeding it with the inputs from observations. By doing this we find a remarkably consistent model back to the time when the universe was about $10^{-5}$ seconds old. The basic idea is that the universe becomes smaller and smaller as we go back in time and so the matter density higher and higher. At an early enough point in time the universe was so dense that matter and radiation were in the state of a plasma and the different energy contributions cannot be treated as non-interacting anymore (e.g., Eq.~(\ref{eq:equation_of_motion1})). The further we go back in time the higher the temperature and the more particle species are being created and interact with the plasma. At about $10^{-5}$ seconds or equivalently at a temperature of $T = 200$ MeV$/k_{\rm B}$ the quark-gluon transition takes place, which is not fully understood yet. Going much further back in time leads to the problem that we cannot probe the physics anymore in our accelerators, since the involved particle energies become too large. In the following, we will outline the main stages in the history of the universe:

\begin{description}
	\item[Very early universe $(\lesssim\! 10^{-14}\ {\rm s})$] For temperatures $\gtrsim\! 10$ TeV/$k_{\rm B}$ we have no clue about the physical interactions from accelerator experiments and so any model for this stage of the universe will necessarily be very speculative. This is the era where hypothetic processes, such as the origin of baryon asymmetry or inflation, might have taken place.
	
	\item[Early universe $(\sim\! 10^{-5} \textmd{--} 1\ {\rm s})$] After the temperature dropped to 200 MeV/$k_{\rm B}$, the quark-gluon transition takes places: free quarks and gluons become confined within baryons and mesons. This is the starting point from when we understand the history of the universe in great detail. In this era, the universe is a hot plasma where many particle species (e.g., electrons, neutrinos, photons, baryons) are in thermal equilibrium with each other. As soon as the inverse of the interaction rate for a given particle species with all the other species exceeds the characteristic timescale $1/H(t)$ of the universe, the corresponding species ``freezes out'' and remains as a thermal relict from the early universe. As the temperature reaches $\sim 0.5$ MeV/$k_{\rm B}$ only electrons, photons, protons and neutrons remain in the plasma. All other particles froze out.
	
	\item[Nucleosynthesis $(\sim\! 3  \textmd{--} 5\ {\rm min})$] At temperatures of $\sim\! 0.05$ MeV/$k_{\rm B}$ nuclear reactions become efficient, so that free protons and neutrons form helium and other light elements.
	
	\item[Matter-Radiation-Equality $(t_{\rm eq}\sim 60,000\ {\rm yr})$] This is the epoch when the energy density of matter and radiation was equal, i.e., $\Omega_{\rm r}(t_{\rm eq}) = \Omega_{\rm m}(t_{\rm eq})$. Before this epoch the universe was radiation dominated and afterwards matter dominated.
	
	\item[Recombination $(t_{\rm dec} \sim 380,000\ {\rm yr})$] Electrons and protons recombine to form neutral atoms. The universe becomes transparent and the cosmic microwave background (CMB) is released as a cosmic relict. This process is also called \textbf{decoupling} and corresponds to a redshift $z_{\rm dec} \simeq 1089$ with a thickness of $\Delta z \simeq 200$ \citep{bennett2003}.
	
	\item[Structure formation $(\sim\! 0.1 - 13.7\ {\rm Gyr})$] As time goes on, tiny fluctuations in the distribution of matter start to grow under the action of gravity leading to the LSS at the present day $t_0 \sim 13.7$ Gyr. This is the topic which will be discussed in detail in the next three chapters.
	
\end{description}

These different stages are, of course, not totally separated from each other, but blend and interact leading to a complicated history of the universe, which can in detail only be modeled by numerical simulations. As we will see in the next chapter, DM fluctuations can efficiently start to grow as soon as the universe becomes matter dominated, while the baryonic fluctuations are prevented from growing due to the interaction with the photons (and remain at the temperature of the photons even for some time after recombination).

What is the age of our universe? To answer this question we first need an event that we can interprete as the beginning of our universe and then we need the full knowledge about the expansion history of the universe since that beginning. But as discussed before, we have no firm knowledge about the physics in the very early universe and so any model of the universe at that time is unavoidably very speculative. Since we can only count back as long as we understand the universe, it is reasonable to define the beginning of the universe as the epoch, when the scale factor formally becomes zero at very early time during the radiation dominated era. This is the time coordinate that we used in the outline of the history of the universe above and according to this definition of the beginning, the universe basically starts with inflation, which is a meaningful starting point as will be discussed in the next section. However, the questions whether inflation really took place and, if yes, what was before inflation, cannot be answered today, and it is open whether we will ever be able to answer them.

With the definition of the origin of the time coordinate $t$ at hand, the present age of the universe can be computed as follows. It holds
\begin{equation}
dt = \frac{1}{\dot a(t)} \: da = \frac{1}{a(t) H(t)} \: da
\end{equation}
and so the age of the universe is
\begin{equation}
t_0 = \int_0^{a(t_0)=1} \frac{1}{aH} \: da \supeq{(\ref{eq:friedmann_equation4})}{=} \frac{1}{H_0}\:\int_0^{1} \frac{1}{a} \:\frac{1}{\sqrt{\Omega_\Lambda + \Omega_{\rm m}a^{-3} + \Omega_r a^{-4}}} \: da \simeq 13.7\ \rm Gyr\:,
\end{equation}
where we applied the values in Table~\ref{tab:cosmological_parameters}. Note that for this precision it does not matter whether or not we consider the contribution from radiation, since $\Omega_{\rm r}$ is very small and can change the age only about 6 Myr. In fact, the age of the universe at the epoch of matter-radiation-equality $t_{\rm eq} \simeq 56,000$ years is so small compared to the typical age of the universe during the matter dominated era that we can safely neglect the radiation and approximate the scale factor during matter domination by means of Eq.~(\ref{eq:hubble}) instead of Eq.~(\ref{eq:a_of_t_zero_offset}), i.e.,
\begin{equation}\label{eq:a_t_matter}
a(t) \propto t^{2/3}\:, \qquad H(t) = \frac{\dot a(t)}{a(t)} =\frac{2}{3t}\:.   
\end{equation}
Similarly, to describe the evolution of the scale factor during the radiation dominated era, we can neglect the first second, when complicated processes might have taken place, and just write
\begin{equation}\label{eq:a_t_radiation}
a(t) \propto t^{1/2}\:, \qquad H(t) = \frac{\dot a(t)}{a(t)} = \frac{1}{2t}\:.   
\end{equation}

\section{Inflation}\label{sec:inflation}

In the previous section, it became clear that from the time when the temperature was about 200 MeV/$k_{\rm B}$ until the present time, we have a mostly consistent story of the universe, and the physics of the universe is well understood and tested in our laboratories (except DM and dark energy, but we nevertheless understand their phenomenological behaviors quite well today). However, as we go further back in time, the story of the universe becomes much more fuzzy until we do not know anything safe about the involved fundamental physics. Nonetheless, it was possible to propose a consistent scenario for the universe at very early times called \textbf{inflation}, which has the potential to produce the known radiation dominated early universe from a preexisting chaotic state and to solve a couple of independent shortcomings of the concordance model. The basic idea of inflation is that the very early universe underwent a short stage of accelerated expansion (i.e., $\ddot a(t) > 0$) driven by a scalar field $\phi$.

In Section \ref{sec:connection_to_the_standard_model}, we briefly discuss the shortcomings of the concordance model and how inflation is able to solve them. Then in Section \ref{sec:slow-roll-inflation}, we outline the phenomenology of the simplest class of inflation models (``slow-roll'' inflation). The rather technical formalism of the theory of scalar fields is provided in Appendix \ref{sec:classical_scalar_field} to focus on conceptual issues here. The most important aspect of inflation in the context of this introduction is that it can produce density perturbations which act as the starting point for structure formation in the early universe. The main idea of this process is summarized in Section \ref{sec:Generation of the primordial perturbations}, whereas the technical details are presented in Chapter \ref{sec:generation_of_initial_perturbations}. We finally conclude this section with some remarks on the plausibility of inflation and on the light inflation shed on the cosmological principle.

\subsection{Connection to the concordance model}\label{sec:connection_to_the_standard_model}

There are certain features associated with the concordance model that seem weird. The most famous of these features are the ``flatness problem'' and the ``horizon problem''.

\paragraph{Flatness problem}

The curvature density $\Omega_K$ defined in Eq.~(\ref{eq:omega_k}) can be expressed in terms of the comoving Hubble length $D_{\rm H}(t)$ (see Sect.~\ref{sec:horizons})
\begin{equation}\label{eq:omea_k}
	\Omega_{\rm K}(t) = -K \left(\frac{D_{\rm H}(t)}{R_0}\right)^2,
\end{equation}
where $R_0$ is the curvature radius at the present epoch and thus equal to the comoving curvature radius of the universe. That is, $\Omega_k(t)$ basically measures the ratio between the Hubble length being roughly the radius of the observable universe and the radius of curvature of the universe at time $t$. The smaller $|\Omega_{\rm K}(t)|$ the less curved space appears within our particle horizon. The value of $\Omega_{\rm K}$ at the present epoch is consistent with a flat universe to a high precision (see Table \ref{tab:cosmological_parameters}). However, since our universe was dominated by either radiation or matter during most time of its history, it was essentially decelerating and so $D_{\rm H}(t) = c/(Ha) = c / \dot a$ was always increasing with time. This means that $|\Omega_{\rm K}(t)|$ was even much closer to unity in the past. So the question naturally arises why the universe was so close to flat in the past or why it is still so flat at present.

\newpage
\paragraph{Horizon problem}

At the epoch of decoupling $z_{\rm dec} \sim 10^4$, the comoving horizon $D_{\rm H}(t)$ was so small that observed today perpendicular to the line of sight at corresponding distance it would subtend only about 1 degree on the sky. Yet the CMB exhibits the same temperature in any direction to a precision of $10^{-5}$. How is it possible that two causally absolutely disconnected parts in the universe can exhibit the same temperature to such a high degree although they never were in causal contact and thus never in thermodynamic equilibrium?\\

Both problems do not constitute inconsistencies in the framework of the concordance model, but the concordance model does not give any clue why the initial conditions should be as described in either problem. They give the impression of some sort of fine tuning related to the initial conditions of the universe. Yet there are a couple of further similar questions:

\begin{itemize}
	\item Why are there no ``topological defects'' (e.g., magnetic monopoles) in the universe, although they are expected by extensions to the standard model of particle physics to be created in the very early universe?
	\item Why is the universe expanding at all, although it was decelerating during most of its history?
	\item How were the density seeds generated in the early universe (along with their characteristic power spectrum), which led to the LSS observed in the universe?
\end{itemize}

\begin{figure}[tp]\centering
  \includegraphics[scale = 0.6]{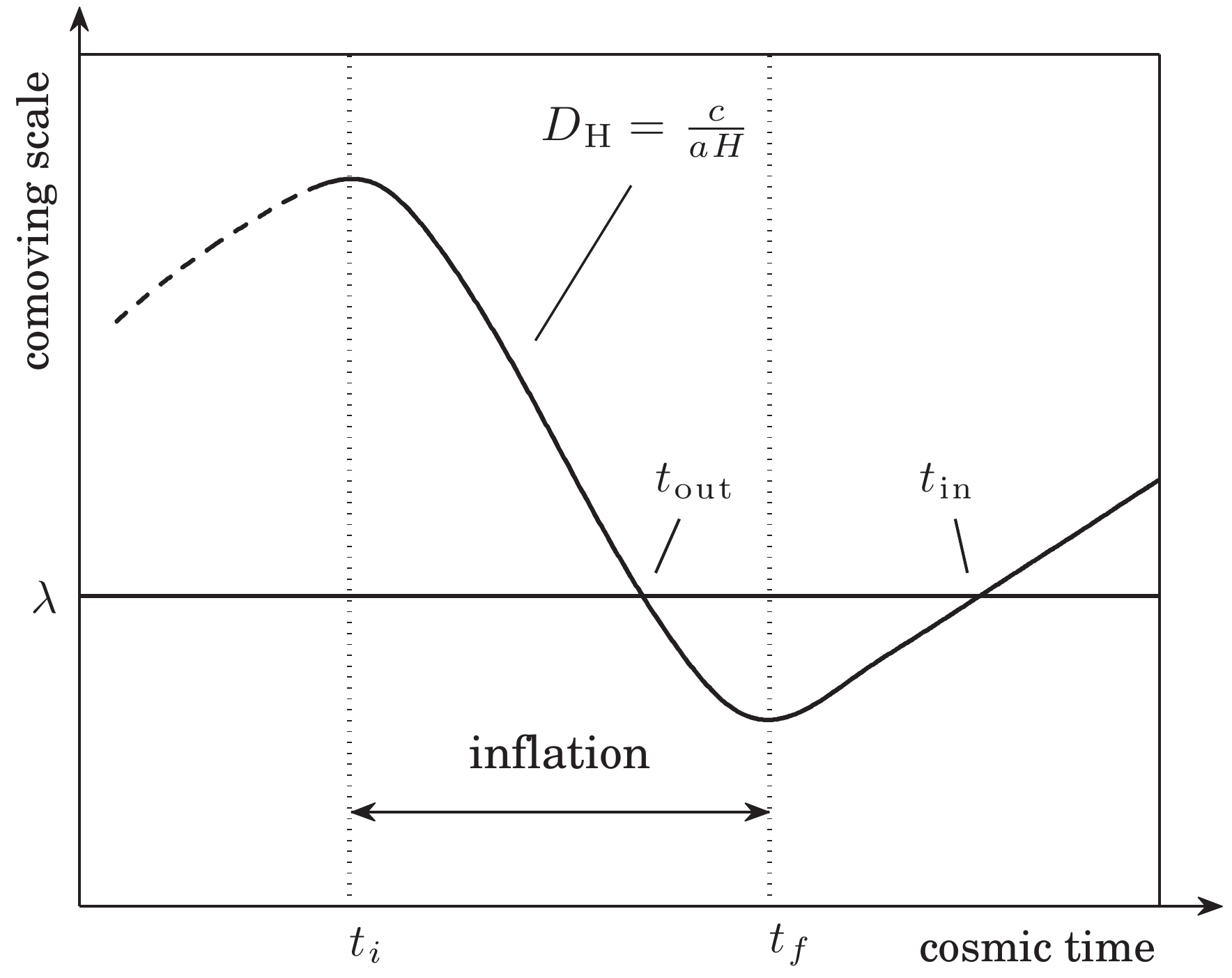}
  \caption{Schematic sketch for the evolution of the comoving Hubble length $D_{\rm H}$ during and after inlation. Since $D_{\rm H}$ is decreasing during inflation, a comoving scale $\lambda$ can exit the horizon and reenter it after inflation finished and $D_{\rm H}$ started to increase again.}\label{fig:inflation}
\end{figure}

It would be tempting to solve all these problems by a single additional ingredient of the concordance model and yet this is exactly what inflation aims to achieve. Inflation being a stage of accelerated expansion in the early universe is characterized by the condition $\ddot a(t) > 0$. With 
\begin{equation}
	\dot D_{\rm H}(t) = \frac{d}{d t}\left(\frac{c}{Ha}\right) = c\: \frac{d}{d t} \left(\frac{1}{\dot a}\right) = - c\:\frac{\ddot a}{\dot a^2}
\end{equation}
and Eq.~(\ref{eq:omea_k}) follows the relation
\begin{equation}
\boxed{\ddot a(t) > 0 \quad \Leftrightarrow \quad \dot D_{\rm H}(t) < 0 \quad \Leftrightarrow \quad \frac{d}{d t}\left|\Omega_K(t) \right| < 0\:.}
\end{equation}
From the last expression it is clear that inflation can solve the flatness problem by decreasing $|\Omega_K(t)|$ to an arbitrary small value. Since the comoving Hubble length $D_{\rm H}$ is decreasing during inflation, also the horizon problem can be solved. To see this, suppose the comoving particle horizon at the beginning of inflation was $D_{\rm p}(t_{\rm i})$ and consider a comoving scale $\lambda$ within this horizon (see Figure \ref{fig:inflation}).
Now as the expansion of the universe starts accelerating there starts to exist an event horizon $D_{\rm e}(t) \simeq D_{\rm H}(t)$ in the universe.\footnote{We assume here that we can approximate $t_{\rm f}$ in Eq.~(\ref{eq:event_horizon}) by $\infty$ and that with Eq.~(\ref{eq:event_horizon2}) follows $D_{\rm e}(t) \simeq D_{\rm H}(t)$. The longer inflation lasts, the better is this approximation. Note that the event horizon is independent from what happened before inflation.} If $l < D_{\rm H}(t_{\rm i})$ at the beginning of inflation, there will be a time $t_{\rm out}$ when the comoving scale $\lambda$ crosses the horizon, i.e., $\lambda = D_{\rm H}(t_{\rm out})$. This means that a scale being in causal contact at the beginning of inflation will not be in causal contact anymore at $t_{\rm out}$, i.e., any signal emitted at one end of a scale $\lambda$ at time $t_{\rm out}$ will not be able to reach the other end as long as inflation is going on. What happens after the end $t_{\rm f}$ of inflation? The expansion of the universe starts decelerating again and thus the event horizon vanishes. The region defined by (cf.~Eq.~(\ref{eq:com_dp}))
\begin{equation}
	D_{\rm H}(t_{\rm f}) + \int_{t_{\rm f}}^t \frac{c}{a(t')}\: dt' \simeq D_{\rm H}(t)
\end{equation}
is the region of causal contact for events that happen after inflation and thus defines some kind of ``apparent particle horizon'' for such events. For instance, a photon that is emitted at some time after inflation cannot reach us from distances larger than $D_{\rm H}(t)$ at time $t$. Since $D_{\rm H}(t)$ is an increasing function of $t$ after inflation, each scale $\lambda$ that exited the event horizon during inflation reenters the horizon at time a $t_{\rm in}$ when $L = D_{\rm H}(t_{\rm in})$. That is, regions that lost causal contact during inflation start interacting again. This is why it is possible for the CMB to have essentially the same temperature in all directions, although the $D_{\rm H}(t_{\rm dec})$ at time of decoupling is much smaller than the scales of the CMB observed today. The reason is that the actual particle horizon $D_{\rm p}(t_{\rm dec})$ is much larger than $D_{\rm H}(t_{\rm dec})$, i.e., at some time in the past there already were interactions on scales much larger than $D_{\rm H}(t_{\rm dec})$.

In a similar manner, inflation can also solve the other raised questions. It is not yet clear what initiated the inflationary era. Theorists argue that inflation might happen under very general conditions and that our universe might be only a homogeneous and isotropic ``patch'' within a huge chaotic and inhomogeneous universe (``chaotic inflation'', \citealt{mukhanov2005}, Sect.~5.6). If this was the case, inflation would not only explain why our observable universe is remarkably flat, it would also explain why we live in a FLRW universe at all \citep{weinberg2008}. So, for the sake of simplicity, we will assume that for the description of inflation we can start with a flat FLRW universe (if it was not, inflation would make it so) and try a proper assessment of the whole inflationary paradigm in Section \ref{sec:revisited}. 

\subsection{Slow-roll inflation}\label{sec:slow-roll-inflation}

For this section we choose natural units, i.e., $c = \hbar = 1$. As discussed in Section \ref{sec:equation_of_state}, the universe is only accelerating if it is dominated by one or several energy components with equations of state $w_I < -1/3$. So this condition must be satisfied during inflation. The simplest way to achieve this is by invoking a real scalar field $\phi(x)$. The associated theory is rather technical and therefore presented in Appendix \ref{sec:classical_scalar_field}. So far, there has been no definitive experimental detection of a scalar field in nature, but the Higgs boson of the standard model of particle physics would constitute such a field. On 4 July 2012, two separate experimental teams at the Large Hadron Collider at CERN announced that they had each independently confirmed the existence of a previously unknown particle with a mass between 125 and 127 GeV. Although the identification of this particle is not definitely resolved yet, it is not unlikely that it may in fact be the higgs boson.

In the FLRW framework, the scalar field $\phi(t)$ is homogeneous and thus only a function of time. As is shown in the Appendix \ref{sec:scalar_field_in_cosmology}, a scalar field $\phi$ moving in a potential $V(\phi)$ behaves like an ideal fluid with an effective matter density and pressure
\begin{equation}\label{eq:scalar_field_energy_pressure}
	\rho_{\phi} = \frac{1}{2} \dot \phi^2 + V(\phi)\:, \quad\quad p_{\phi} = \frac{1}{2} \dot \phi^2 - V(\phi)\:,
\end{equation}
respectively, and obeys the equation of motion
\begin{equation}\label{eq:equation_of_motion_phi}
	\ddot \phi + 3 H\dot \phi + V'(\phi) = 0\:,
\end{equation}
where $V'(\phi) = dV/d\phi$. The term $\phi^2/2$ is like the ``kinetic energy'' of the field, and the term $3 H\dot \phi$ in the equation of motion comes from the expansion of the universe and acts like a friction. With the expressions in Eq.~(\ref{eq:scalar_field_energy_pressure}) the equation of state of the scalar field is
\begin{equation}
	w_\phi(t) = \frac{\frac{1}{2}\dot \phi^2 - V(\phi)}{\frac{1}{2}\dot \phi^2 + V(\phi)}
\end{equation}
with the bounds $-1 \leq w_\phi(t) \leq 1$. This equation of state is generally time dependent. However, if $\dot \phi^2 < 4 V(\phi)$ then $w_{\phi}<-1/3$ and the condition for inflation is satisfied. Moreover, if $\dot \phi^2 \ll V(\phi)$, then the equation of state even becomes $w \simeq -1$ and is constant. The associated expansion of the universe is then exponential (see Eq.~(\ref{eq:a_of_t})) with $H$ and $\phi$ being roughly constant.

So far, the form of the potential $V(\phi)$ is undetermined and there are many possible inflationary scenarios proposed in the literature (see, e.g., \citealt{liddle2000}). However, the predictions from the simplest models of inflation are rather robust and so it is not necessary to specify all the details as long as certain general features are satisfied. The simplest class of inflationary models is called \textbf{slow-roll inflation}. These models contain a single scalar field $\phi$ called the \textbf{inflaton} and are characterized by the conditions
\begin{equation}\label{eq:slow_roll_conditions}
	\dot \phi^2 \ll V(\phi)\:, \quad\quad \big|\ddot \phi \big| \ll \big|3 H \dot \phi \big| \sim \big|V'(\phi)\big|\:.
\end{equation}
That is the ``kinetic energy'' $\dot \phi^2/2$ of the field is small compared to the potential $V(\phi)$ and thus the $\phi$ is rolling slowly down the potential $V(\phi)$. In this approximation, the equation of state is in fact $w \simeq -1$ and the expansion of the universe is roughly exponential. The Friedmann equation (\ref{eq:friedmann_equation}) for $K = 0$ and the equation of motion (\ref{eq:equation_of_motion_phi}) simplify to
\begin{equation}\label{eq:equations_slow_roll}
	H^2 = \frac{8\pi G}{3} V(\phi)\:, \quad\quad 3 H \dot \phi + V'(\phi) = 0\:,
\end{equation}
respectively. Slow-roll inflation is usually quantified by the dimensionless \textbf{slow-roll parameters}
\begin{equation}\label{eq:slow_roll_parameters}
\boxed{\epsilon \equiv \frac{1}{16 \pi G} \:\left(\frac{V'}{V}\right)^2 = - \frac{\dot H}{H^2} = \frac{3}{2} \frac{\dot \phi^2}{V} = \frac{3}{2}\left(1 + w_\phi\right)\:, \quad\quad \eta \equiv \frac{1}{8\pi G} \:\frac{V''}{V} = \epsilon - \frac{\ddot \phi}{H \dot \phi}\:,}
\end{equation}
where the different expressions are obtained by using the Eqs.~(\ref{eq:scalar_field_energy_pressure}) and (\ref{eq:equations_slow_roll}). Comparing the Eqs.~(\ref{eq:slow_roll_conditions}) and (\ref{eq:slow_roll_parameters}) shows that the slow-roll conditions are equivalent to
\begin{equation}
\boxed{\epsilon \ll 1\:, \quad\quad \eta \ll 1\:.}
\end{equation}
Since $\epsilon \ll 1$ and $\eta \ll 1$ are equivalent to $|V'/V| \ll 1$ and $|V''/V|\ll 1$, respectively, the slow roll approximation is automatically satisfied if $V(\phi)$ is sufficiently flat.

\subsection{Generation of the primordial perturbations}\label{sec:Generation of the primordial perturbations}

Probably the most important aspect of inflation in the context of structure formation is the possible associated production of tiny perturbations in the early universe that act as density seeds at the beginning of structure formation. The mechanism employs quantum mechanical processes. \nocite{mukhanov2005}Mukhanov (2005, Sect.~8.2.3) gives an intuitively very clear description of how this works:

Actually, inflation smooths classical inhomogeneities by stretching them to very large scales. However, it cannot remove quantum fluctuations because in place of the stretched quantum fluctuations, new ones are generated by means of the Heisenberg uncertainty relation. But why is inflation needed for this process? The reason is that in Minkowski space, typical amplitudes of vacuum metric fluctuations are very small. They are only large near the Planckian scale. On galactic scales they are smaller than $10^{-58}$ and thus could never produce the perturbations of $10^{-5}$ as measured in the CMB. The only way of producing such fluctuations on large scales is by stretching the very short wavelength fluctuations without decreasing their amplitude. As long as the fluctuations are within the horizon, they in fact continuously decrease when they are stretched, but as soon as they cross the horizon, the quantum mechanical fluctuations become classical and are indeed ``frozen''. That is, they are stretched to galactic scales with almost no change of amplitudes. So inflation is necessary for the generation of perturbations, since only during an inflationary era the comoving Hubble length $D_{\rm H}(t)$ decreases such that comoving scales can exit the horizon. That perturbations are frozen outside the horizon is also the only reason why we are able to use inflationary theories to make any predictions at all about observational perturbations \citep{weinberg2008}. Remember that we basically know nothing about fundamental physics at the time when inflation happens and nobody knows exactly how the inflationary era turned into the radiation dominated universe of the concordance model. So all these unknown processes happened when the perturbations that are observable today were well outside the horizon and thus unaffected by any unknown physics.

Also the statistical properties of the spatial perturbations created by inflation coincide very well with observations. A measure for the distribution of perturbations is the power spectrum $P(k)$ (see Sect.~\ref{sec:Correlation function and power spectrum}). Slow-roll inflation predicts a functional form of $P(k)$ as (see Sect.~\ref{sec:primordial_power_spectrum})
\begin{equation}
	P(k) \propto k^{n_{\rm s}}\:, \quad\quad\quad n_{\rm s} = 1 - 6 \epsilon + 2 \eta\:,
\end{equation}
where $n_{\rm s}$ is the spectral index and $\epsilon$ and $\eta$ are the slow-roll parameters (see Eq.~(\ref{eq:slow_roll_parameters})). The current observed value is in fact $n_{\rm s} \simeq 0.97$ (see Tab.~\ref{tab:cosmological_parameters}) in excellent agreement with the expectations from slow-roll inflation. Other predictions from inflation about the nature of perturbation are that perturbations should be Gaussian and adiabatic (see Ch.~\ref{sec:generation_of_initial_perturbations}).

\subsection{Final remarks}\label{sec:revisited}

In this section, we have outlined the simplest scenario of inflation and we have shown how it can solve a couple of problems arising in the concordance model and how it can set the framework for a FLRW universe. How sure can we be that such a scenario really took place in the very early universe? A conclusive answer to this question cannot be given. On the one hand, there are a couple of robust predictions of the simplest class of inflationary models, but on the other hand, by introducing extra parameters and by fine-tuning one can alter these robust predictions and, for instance, also produce FLRW universes that are open. While so far the robust predictions of the simplest inflationary models seem to be confirmed by observations, the whole underlying physics of inflation is very speculative and there might be other reasons for why our universe is very close to flat etc. Mukhanov (2005, Sect.~8.6)\nocite{mukhanov2005} argues for a proper consideration of the ``price-to-performance'' ratio of inflationary theories in the sense that by an increase of the complexity of the models their predictive power is decreasing. The most attractive feature of inflation is certainly its simplicity and so for a proper assessment of its usefulness one has to presumably separate its phenomenology (i.e., the basic processes and robust predictions) from the underlying physical models (i.e., what fields are involved, how did it start, how did it end) which are extremely speculative and which can possibly never be confirmed by observations in the future. 

The concept of chaotic inflation -- whether true or not -- also puts a new complexion to the cosmological principle (see Sect.~\ref{sec:cosmological_principle}). It constitutes sort of a mechanism to produce a homogeneous and isotropic ``patch'' within a much larger inhomogeneous and anisotropic chaotic universe. As long as this homogeneous and isotropic patch is much larger than the Hubble length (i.e., our effective observable universe) we are entirely unaffected by the universe outside this patch and our observable universe behaves like a proper FLRW universe. This shows how idealistic and far reaching the cosmological principle is if interpreted in a strict sense. In the end we are, in fact, unable to make any firm statements about the universe beyond our observable universe and if chaotic inflation is taken at face value, our initial assumption of the homogeneous and isotropic universe leads us via inflation to the conclusion that (globally) the universe is not homogeneous and isotropic at all.


\chapter{Newtonian theory of structure formation}\label{sec:structure_formation}

The Universe is homogeneous and isotropic on scales larger than 100 Mpc, but on smaller scales we observe huge deviations from the mean density in the form of galaxies, galaxy clusters, and the cosmic web being made of sheets and filaments of galaxies. How do structures grow in the universe and how can we describe them? In this chapter, we develop the Newtonian theory of structure formation and introduce the basic statistical equipment for quantifying them. Since we will entirely work within the concordance model outlined in Section \ref{sec:concordance_cosmology}, we will stick to a flat geometry. This simplifies our treatment enormously insofar as it allows the usage of Fourier transformation to decompose the structures of the universe into single independent modes. We prefer to first present the theory of Newtonian structure formation before going to the general relativistic theory in the next chapter, since the Newtonian approach is not only technically much simpler, but also sufficient to understand most of the processes which are well within the horizon.\footnote{The relation between Newtonian physics and general relativity in the context of structure formation is discussed in \cite{lima1997} and \cite{noh2006}. It is interesting that not only the homogeneous and isotropic world models of Chapter \ref{sec:homogeneous_and_isotropic_universe}, but also their linear structures were first studied in the context of general relativity and not Newtonian physics.} Nevertheless a full analysis of structure formation in the universe starting with small perturbations generated during inflation is not possible in terms of Newtonian physics and therefore requires a general relativistic treatment. This will be done in the next two chapters, where we will use the Newtonian results developed in this chapter to interpret the general relativistic results.

In Section \ref{sec:linear_theory}, we develop the basic equations governing the growth of structures by solving the corresponding hydrodynamical equations at linear order. The domain where these equations are valid defines the linear regime. In Section \ref{sec:statistics_overdensity_field}, we introduce the correlation function and the power spectrum for a precise and quantitative description of cosmological structures and we explore how these statistics evolve with time as the structure grows. Using these statistics we will define the two cosmological parameters $\sigma_8$ and $n_{\rm s}$, which we have already encountered in Section \ref{sec:inflation}. Finally, in Section \ref{sec:dm_halos} we introduce approximations to treat the nonlinear growth of cosmic structures in an analytic approximative way. This will lead to the concept of the ``halo'' and the corresponding theories about their abundance and statistical distribution in space. We finish this section by presenting the ``halo model'', which is a very simple, but powerful theory for describing the galaxy correlation function in the linear and nonlinear regimes.

\section{Linear perturbation theory}\label{sec:linear_theory}

In the framework which was outlined in Section \ref{sec:timeline}, the LSS that we can see today started with very small initial deviations from the homogeneous FLRW model and grew by gravitational instability. At epochs when these deviations are very small, they can be treated as perturbations around the smooth background, while we keep only terms of first order in perturbation quantities. This is called ``linear theory'' and the regime where this approach is valid is called ``linear regime''. The corresponding Newtonian theory was initially formulated by \cite{bonnor1957} for an expanding universe. We follow mainly the introductions given in \nocite{coles2002}Coles \& Lucchin (2002, Ch.~10) and  \nocite{mukhanov2005}Mukhanov (2005, Ch.~6).

\subsection{Newtonian hydrodynamics in an expanding universe}

Suppose the universe is filled with an inhomogeneous, dissipationless, ideal fluid with matter density $\rho(t,\bv{r})$, velocity field $\bv{v}(t,\bv{r})$, pressure $p(t,\bv{r})$, gravitational potential $\Phi(t,\bv{r})$, and entropy per unit mass $S(t,\bv{r})$, where $t$ denotes the cosmic time and $\bv{r}$ cartesian physical coordinates. The fluid is governed by the basic hydrodynamical equations of Newtonian physics:
\begin{IEEEeqnarray}{ll}\label{eq:hydrodynamical1}
\text{continuity equation:} &\frac{\partial \rho}{\partial t}+\bv{\nabla}\cdot(\rho \:\bv{v})=0\\
\text{Euler equation:} &\frac{\partial \bv{v}}{\partial t}+\left(\bv{v}\cdot\bv{\nabla}\right)\bv{v}+\frac{\bv{\nabla} p}{\rho}+\bv{\nabla}\Phi = 0\\
\text{Poisson equation:} & \nabla^2 \Phi = 4 \pi G \rho\\
\text{conservation of entropy:}\qquad  &\frac{\partial S}{\partial t} + \left(\bv{v}\cdot \bv{\nabla}\right)S = 0\:.\label{eq:hydrodynamical4}
\end{IEEEeqnarray}
These equations taken together with the equation of state $p = p(\rho,S)$ form a closed system of equations and determine the seven unknown functions $\rho$, $\bv{v}$, $p$, $\Phi$, and $S$. Note that the Eqs.~(\ref{eq:hydrodynamical1})-(\ref{eq:hydrodynamical4}) are only valid for non-relativistic matter, i.e., it must hold $|\bv{v}| \ll c$ and $p \ll \rho c^2$. Since for a fluid a unique velocity vector is associated to every point in space, the motion of matter must be in the single stream regime, i.e., adjacent particles  move in approximately parallel trajectories and do not cross. This is a reasonable assumption in the linear regime. Mixing of different streams at the same point in space (``shell crossing'') does not occur until the structures have grown nonlinearly. Moreover, our simple ideal fluid model does not account for dissipative processes (e.g., ``free-streaming'' of relativistic particles) which erase small scale perturbations. However, since DM is cold and does hardly interact with other particles, we can neglect such effects for the study cold DM. In order to get accurate models for structure formation, one would have to solve the general relativistic Boltzmann equation taking into account all sorts of energy contributions in the universe and their interactions. The corresponding procedure is outlined in Section \ref{sec:complete_treatement}.

Unfortunately, the hydrodynamical equations (\ref{eq:hydrodynamical1}) are nonlinear and it is very difficult to find their general solution. So, assuming that the universe is close to a FLRW universe, we can perturb the fluid around its Hubble flow and solve the hydrodynamical equations at first order (or ``linear order'') in the perturbed quantities. Thus, we split each quantity into a homogeneous background (indicated by a bar) and an inhomogeneous perturbation (indicated by a $\delta$), where the perturbations are small compared to their background:
\begin{equation}\label{eq:hydrodynamical_split}
\begin{array}{ll}
\rho(t,\bv{r}) = \bar{\rho}(t)+\delta \rho(t,\bv{r})\:,\quad\quad & \bv{v}(t,\bv{r}) = \bar{\bv{v}}(t,\bv{r})+\delta \bv{v}(t,\bv{r})\:, \\
p(t,\bv{r}) = \bar{p}(t)+\delta p(t,\bv{r})\:, &  \Phi(t,\bv{r}) = \bar{\Phi}(t,\bv{r})+\delta \Phi(t,\bv{r})\:,\\
S(t,\bv{r}) = \bar{S}(t)+\delta S(t,\bv{r})\:.
\end{array}
\end{equation}
Note that since $\bv{r}$ are physical coordinates rather than comoving, the homogeneous velocity field $\bar{\bv{v}}(t,\bv{r})$ and the homogeneous gravitational potential $\bar{\Phi}$ are nonzero and even depend on the physical position. The homogeneous velocity field is given by the Hubble Law
\begin{equation}
\bar{\bv{v}}(t,\bv{r}) = H(t)\bv{r}\:.
\end{equation}
This is easily seen using Eq.~(\ref{eq:comoving_distance}), since the homogeneous matter field is at rest with respect to comoving coordinates. The perturbation velocity field $\delta \bv{v}$ is the field of \textbf{peculiar velocities} (cf.~Sect.~\ref{sec:cosmological_redshift}).

How do the background quantities in Eq.~(\ref{eq:hydrodynamical_split}) relate to the general relativistic approach of the homogeneous universe in Chapter \ref{sec:homogeneous_and_isotropic_universe}? The hydrodynamical equations for the homogeneous background yield
\begin{equation}\label{eq:newtonian_friedmann}
\dot{\bar{\rho}}+3 H \bar{\rho} =0\:, \quad\quad\quad \dot H+H^2=-\frac{4\pi G \bar{\rho}}{3}\:,
\end{equation}
the first being the continuity equation and the second the divergence of the Euler equation and substituting the Poisson equation. The Eqs.~(\ref{eq:newtonian_friedmann}) are the equation of motion (\ref{eq:equation_of_motion_total}) and the second Friedmann equation (\ref{eq:second_friedmann_equation}) for pressureless matter. Note that the pressure $p$ of the fluid does not enter the Friedmann equation in the Newtonian approach and we have already assumed that it is small. Thus, we are consistent with general relativity and can assume that the hydrodynamical equations (\ref{eq:hydrodynamical1})-(\ref{eq:hydrodynamical4}) for the homogeneous quantities are satisfied. If one wants to treat fluids with considerable amount of pressure in a Newtonian approach, one has to be very careful to avoid inconsistencies (see \citealt{lima1997}) as discussed in the following paragraph.

If there are further energy contributions in the universe which do not interact with our fluid except by gravitation and if these contributions are homogeneous, e.g.,  a cosmological constant or a homogeneous radiation background not interacting with the matter, these fluids enter our formalism only through a homogeneous (i.e., unperturbed) term in the Poisson equation and so do not alter our first order perturbation equations. They alter, however, the expansion of the universe. Note that such additional fluids could introduce (relativistic) pressure. To reconcile the Newtonian and general relativistic approach for our calculation, we just \emph{assume} that the hydrodynamical equations (\ref{eq:hydrodynamical1})-(\ref{eq:hydrodynamical4}) for the homogeneous quantities are satisfied (although we are aware that they may not be due to relativistic effects) and that the expansion is governed by the relativistic Friedmann equations (\ref{eq:friedmann_equation}).

Substituting the Eqs.~(\ref{eq:hydrodynamical_split}) into the hydrodynamical equations (\ref{eq:hydrodynamical1})-(\ref{eq:hydrodynamical4}) and keeping only terms at first order in perturbation quantities yields
\begin{equation}\label{eq:hydrodynamical2}
\begin{aligned}
&\text{continuity equation:} \quad\frac{\partial \delta \rho}{\partial t}+\bar{\rho}\:\bv{\nabla}\cdot\delta\bv{v}+\bv{\nabla}\cdot(\delta \rho \:\bar{\bv{v}})=0\\
&\text{Euler equation:}\quad\frac{\partial \delta \bv{v}}{\partial t} +\left(\delta\bv{v}\cdot\bv{\nabla}\right)\bar{\bv{v}} +\left(\bar{\bv{v}}\cdot\bv{\nabla}\right)\delta\bv{v}+\frac{\bv{\nabla}}{\bar{\rho}}\left(c_{\rm s}^2 \: \delta \rho + \sigma\: \delta S\right)+\bv{\nabla}\delta\Phi = 0\\
&\text{Poisson equation:}\quad \nabla^2 \delta \Phi = 4 \pi G \:\delta \rho\\
&\text{conservation of entropy:}\quad \frac{\partial \delta S}{\partial t} + \left(\bar{\bv{v}}\cdot \bv{\nabla}\right)\delta S = 0\:.
\end{aligned}
\end{equation}
For the Euler equation we have used the expansion $1/(\bar{\rho}+\delta \rho) \simeq 1/\bar{\rho} + \delta \rho/\bar{\rho}^2 + \ldots$ and substituted the equation of state $p(\bar{\rho}+\delta \rho,\bar{S}+\delta S)$ at first order
\begin{equation}\label{eq:adiabatic_dp}
	\delta p = c_{\rm s}^2 \: \delta \rho + \sigma\: \delta S
\end{equation}
with $c_{\rm s}^2 = (\partial \bar{p}/\partial \bar{\rho})_{\bar{S}}$ the square of the speed of sound and $\sigma = (\partial \bar{p}/\partial \bar{S})_{\bar{\rho}}$.

We can further simplify the equations (\ref{eq:hydrodynamical2}) by transforming them into the comoving frame. This is done by the following transformation from physical coordinates $(t,\bv{r})$ to comoving coordinates $(t,\bv{x})$:
\begin{equation}
\bv{r} = a\: \bv{x}\:,\quad\quad\quad
\bv{\nabla}_{\bv r} = \frac{1}{a}\bv{\nabla}_{\bv x}\:,\quad\quad\quad
\left.\frac{\partial}{\partial t}\right|_{\bv r} = \left.\frac{\partial}{\partial t}\right|_{\bv x} - \frac{1}{a} \:\bar{\bv v} \cdot \bv{\nabla}_{\bv x}\:.
\end{equation}
Substituting these expressions in the Eqs.~(\ref{eq:hydrodynamical2}) and taking the Fourier transform with respect to comoving coordinates, i.e., for any perturbation quantity $dq(t,\bv x)$ holds\footnote{For these integrals to converge, we formally assume that $dq(t,\bv x)=0$ for $|\bv x|>L$ with the cut off scale $L$ being much larger than any other scale of interest so that it does not play any role.}
\begin{equation}
	 \delta q (t, \bv k) = \int \delta q(t,\bv x) e^{-i \bv k \bv x} dx^3\:, \quad\quad\quad \delta q (t, \bv x) = \frac{1}{\left(2 \pi \right)^3} \int \delta q(t,\bv k) e^{i \bv k \bv x} dk^3
\end{equation}
with $\bv k$ the comoving Fourier modes, we obtain the first order hydrodynamical equations for a given mode $\bv k$
\begin{equation}\label{eq:hydrodynamical3}
	 \boxed{
\begin{aligned}
&\text{continuity equation:} \quad \delta \dot{\rho} + 3 H \delta \rho + \frac{i \bar \rho \bv{k}}{a} \cdot \delta \bv{v} = 0 \\
&\text{Euler equation:}\quad \delta \dot{\bv{v}} + H \delta \bv{v} + \frac{i\bv{k}}{a \bar{\rho}}  \left(c_{\rm s}^2\delta \rho + \sigma\delta S\right) + \frac{i\bv{k}}{a}\:  \delta \Phi = 0\\
&\text{Poisson equation:}\quad k^2 \delta \Phi + 4 \pi G a^2 \delta \rho = 0\\
&\text{conservation of entropy:}\quad \delta \dot{S}=0\:,
\end{aligned}}
\end{equation}
where $k = |\bv{k}|$ and a dot denotes the derivative with respect to $t$ at fixed $\bv{k}$. For the first equation we have used $\bv{\nabla}\cdot \bar{\bv{v}} = 3 H$ and for the second equation $(\delta\bv{v}\cdot\bv{\nabla})\:\bar{\bv{v}}/a = \delta\bv{v} H$, since $\bar{\bv{v}} = H \bv{r} = \dot{a} \bv{x}$.

\subsection{Perturbation modes}

The Eqs.~(\ref{eq:hydrodynamical3}) are five coupled linear first order differential equations and one algebraic relation, so we expect the general solution to be a superposition of five linear independent modes. These can be characterized as follows:

\paragraph{Entropy mode}

The conservation of entropy allows a static entropy perturbation
\begin{equation}\label{eq:entropc_mode}
\delta S(t,\bv{k}) = \delta S(\bv{k})
\end{equation}
with appropriate $\delta \rho$, $\delta\bv{v}$ and $\delta \Phi$, so that the other equations are satisfied. Note that entropy perturbations can only occur in a multi-component fluid (e.g., photon-baryon plasma before recombination). Since the matter is dominated by cold DM and this matter does hardly interact with any other energy contribution, we will neglect entropy perturbations in our following discussion. Furthermore, entropy perturbations are rather ``unnatural'' insofar as the simplest models of inflation do not predict any entropy perturbations (see Sect.~\ref{sec:adiabaticity}). If there are no entropy perturbations at the beginning there will be none created due to Eq.~(\ref{eq:entropc_mode}), and preexisting entropy perturbation might even become erased (see discussion in \citealt[Sect.~5.4]{weinberg2008}). To gain insight into the behavior of entropy modes, we assume that the universe is static\footnote{Note that a static homogeneous universe with $\bar{\rho} \neq 0$ is no solution of the Eqs.~(\ref{eq:hydrodynamical1})-(\ref{eq:hydrodynamical4}), because in a static universe we have $\bar{\bv{v}}\equiv 0$, so that the Euler equation becomes $\bv{\nabla} \delta \bar{\Phi} = 0$. Inserting this into the Poisson equation yields $\bar{\rho} \equiv 0$ contradicting our assumption. So a homogeneous universe in Newtonian physics must either expand or contract. This problem is solved by arbitrarily assuming that the Poisson equation holds only for the perturbation quantities, which is called the ``Jeans swindle'' (see, e.g., \citealt[Sect.~5.2.2]{binney2008}).}, i.e., $H \equiv 0$ and $\bar{\rho}\equiv \text{const}$. The entropy mode is solved by $\delta \bv{v} = 0$ and $\delta \rho = \text{const}$, so density fluctuations do not grow in this mode. Thus, entropy fluctuations are not interesting in the context of structure formation.

\paragraph{Vortical modes}

Setting $\delta \rho = \delta \Phi = \delta S = 0$ and $\bv{k}\cdot \delta\bv{v} = 0$, the Euler equation becomes $\delta \dot{\bv{v}} + H\delta \bv{v} = 0$ with the  solution
\begin{equation}
\delta \bv{v} \propto \frac{1}{a}\:.
\end{equation}
In an expanding universe these modes can only decay and so we can neglect them.

\paragraph{Adiabatic modes}

We set $\delta S = 0$ and $\bv{k}\parallel \delta\bv{v}$. Expressing the continuity equation in terms of the \textbf{matter overdensity}
\begin{equation}\label{eq:overdensity}
\delta(\bv{k},t) \equiv \frac{\delta \rho(\bv{k},t) }{ \bar{\rho}(t)} = \frac{\rho(\bv{k},t)-\bar{\rho}(t)}{\bar{\rho}(t)}
\end{equation}
and using the first equation in Eq.~(\ref{eq:newtonian_friedmann}) yields
\begin{equation}\label{eq:peculiar_velocity_field}
\dot{\delta}+\frac{ik}{a}\:\delta v = 0\;.
\end{equation}
If we differentiate this equation, we can first eliminate $\delta\dot{v}$ from the Euler equation and then $\delta v$ with Eq.~(\ref{eq:peculiar_velocity_field}). Then eliminating $\delta \Phi$ with the aid of the Poisson equation we obtain
\begin{equation}\label{eq:linear_perturbation_equation}
	 \boxed{
\ddot \delta + 2H\: \dot \delta + \left(\frac{c_{\rm s}^2k^2}{a^2}-4 \pi G \bar{\rho}\right)\delta  = 0\:.}
\end{equation}
This is a linear second order differential equation and allows two independent solutions, which can grow under certain conditions. Note that with these solutions at hand we can immediately compute the peculiar velocity field that is generated by the perturbations $\delta$ by means of Eq.~(\ref{eq:peculiar_velocity_field}).

So we have found all modes. The most general solution of the Eqs.~(\ref{eq:hydrodynamical3}) is a superposition of two vortical modes, two adiabatic modes, and one entropy mode. Among these, only the adiabatic modes are interesting in the context of structure formation and we will merely focus on these in the following.

\subsection{Linear growth function}\label{eq:lin_growth}

To gain insight into the dynamics of adiabatic perturbations, we again consider the case of a static universe. Eq.~(\ref{eq:linear_perturbation_equation}) has then simple analytic solutions for each $\bv{k}$-mode. If the third term in Eq.~(\ref{eq:linear_perturbation_equation}) is negative, there are exponential growing and decaying solutions, but if it is positive, both solutions are oscillating. This is called the Jeans criterion. Only modes whose physical wavelength $\lambda$ is larger than the \textbf{Jeans length} $\lambda_{\rm J}$ can grow, i.e.,
\begin{equation}
\lambda  = \frac{2 \pi a}{k} > \lambda_{\rm J} \equiv c_{\rm s}\:\sqrt{\frac{\pi}{G \rho}}\:.
\end{equation}
Similarly, we can define the \textbf{Jeans mass} as
\begin{equation}
M_{\rm J} = \frac{4 \pi}{3} \left(\frac{\lambda_{\rm J}}{2}\right)^3 \bar{\rho} = \frac{\pi}{6}\: \lambda_{\rm J}^3 \: \bar{\rho}\:.
\end{equation}
Only perturbations which are more massive than the Jeans mass are able to grow. Note that the Jeans mass is proportional to the speed of sound $c_{\rm s}$ of the fluid. For baryons the speed of sound is a strong function of cosmic time. As long as the baryons are coupled to the photons in the early universe, the speed of sound is huge due to the photon pressure, while during recombination it decreases dramatically and becomes practically negligible (however see the discussion in Sect.~\ref{sec:dark_matter_and_baryons}). On the other hand, the speed of sound of DM is basically negligible at all times. So baryonic perturbations can grow only after recombination, whereas DM is not constrained to this condition.

In an expanding universe the behavior of the perturbations is qualitatively similar to that in the static universe. Perturbations which are larger than the Jeans length exhibit a growing and the decaying mode, while smaller perturbations constitute oscillating sound waves. However, the term  in Eq.~(\ref{eq:linear_perturbation_equation}) which is proportional to $H$ is now nonzero and acts like a ``friction'' (``Hubble drag'') counteracting the dynamics of the perturbations. Let us concentrate on Fourier modes which are much larger than the Jeans length so that equation (\ref{eq:linear_perturbation_equation}) reduces to
\begin{equation}\label{eq:linear_perturbation_equation_simplified}
\ddot \delta + 2H\: \dot \delta -4 \pi G \bar{\rho}\:\delta  = 0\:,
\end{equation}
which has the general solution
\begin{equation}
\delta(t,\bv{k}) = \delta_+(\bv{k})\:D_+(t) + \delta_-(\bv{k})\:D_-(t)\:,
\end{equation}
where the subscript $+$ denotes the growing and $-$ the decaying mode. The functions $D_+(t)$ and $D_-(t)$ are independent real valued ($\bv{k}$-independent) solutions of Eq.~(\ref{eq:linear_perturbation_equation_simplified}) and $\delta_+(\bv{k})$ and $\delta_+(\bv{k})$ are complex valued initial conditions. The growing solution $D_+(t)$ is called \textbf{linear growth function} and is normalized such that $D_+(t_0) = 1$.

How fast do perturbations grow in an expanding universe? When the universe is matter dominated, the expansion rate is $H(t) = 2/(3t)$ (see Eq.~(\ref{eq:a_t_matter})), and the time evolution of perturbations is given by $D_+ \propto  t^{2/3}$ and $D_- \propto t^{-1}$. Thus, matter perturbations can only grow as a power law in the linear regime during matter domination and so structure formation is much more inefficient in an expanding universe than in a static one. During radiation domination matter perturbations grow even slower. Assuming that the universe is filled with DM and radiation, where radiation is assumed to be homogeneous distributed\footnote{This is, of course, an approximation, since the radiation perturbations are not zero and, if the adiabaticity condition holds (see Sect.~\ref{sec:adiabaticity}), they even have the same amplitude outside the horizon as the DM perturbations. However, Weinberg (2008, p.~296)\nocite{weinberg2008} showed that well inside the horizon the radiation perturbations are smaller than the DM perturbations and so can be neglected for studying the growth of DM perturbations.}, it can be shown that the growing mode during the whole radiation dominated era grows maximally about a factor of 2.5 \citep[section 10.11]{coles2002}. This is called the \textbf{Meszaros effect} and says that DM perturbations during radiation domination are basically frozen even for perturbations much larger than the Jeans length. During the era dominated by the cosmological constant, the perturbations are even entirely frozen \cite[Sect.~6.3.4]{mukhanov2005}. So DM perturbations can basically grow only during matter domination, where they grow proportional to the scale factor $a$, i.e.,
\begin{equation}\label{eq:dat}
\boxed{
D_+(t) \propto a(t) \propto t^{2/3}\:.}
\end{equation}

Unfortunately, Eq.~(\ref{eq:linear_perturbation_equation_simplified}) does not allow a closed, analytic solution for the concordance cosmology. However, the following fitting formula provides a sufficiently accurate approximation for all practical purposes at low redshift \citep{carroll1992}\footnote{For a flat universe and realistic values of $\Omega_{\rm m}$ and $\Omega_\Lambda$ it is more accurate than one percent at any redshift, for which the radiation density is negligible. An exact solution for a universe with $\Omega_{\rm m} + \Omega_\Lambda = 1$ is provided in Mukhanov (2005, Eq.~(6.67)) in terms of an integral expression.}
\begin{equation}\label{eq:growth_function_lambda}
D_+(z) = \frac{1}{(1+z)}\frac{g(z)}{g(0)}\:,\quad\quad g(z) = \frac{\Omega_m(z)}{\Omega_m^{4/7}(z)-\Omega_{\Lambda} (z)+\left(1+\Omega_m(z)/2\right)\left(1+\Omega_{\Lambda}(z)/70\right)}\:,
\end{equation}
where
\begin{equation}
\Omega_m(z) \supeq{(\ref{eq:density_parameter_evolution})}{=} \Omega_m(1+z)^3\left[ \frac{H_0}{H(z)}\right]^2\:,\quad\quad \Omega_{\Lambda}(z) \supeq{(\ref{eq:density_parameter_evolution})}{=} \Omega_{\Lambda}\left[ \frac{H_0}{H(z)}\right]^2\:,
\end{equation}
and $H(z)$ is given by Eq.~(\ref{eq:H_z_concordance_model}).

\subsection{Transfer function}\label{sec:transferfunction}

The considerations in the last section concern only perturbations which are well within the horizon. For the dynamics of the perturbations outside the horizon, a general relativistic treatment is required. It can be shown (see Sect.~\ref{sec:primordial_power_spectrum}) that outside the horizon a given mode $\bv k$ effectively grows like\footnote{It should be noted that in Eq.~(\ref{eq:power_spectrum_relation2}) only the term $(k/\mathcal H)^4$ depends on time, since the second last term is evaluated at the fixed time $\tau_{\rm out}(\bv k)$ for a given Fourier mode $\bv k$. So Eq.~(\ref{eq:power_spectrum_relation2}) yields $\delta^2 \propto 1/\mathcal H^4$ and hence $\delta \propto 1/\mathcal H^2 = 1/(Ha)^2$.}
\begin{equation}\label{eq:effective_growth_outside}
\delta \left(t,\bv{k}\right) \supeq{(\ref{eq:power_spectrum_relation2})}{\propto} \frac{1}{H^2(t) a^2(t)} \propto \left\{
\begin{array}{ll}
a^2(t) & \text{(radiation domination)}\\
a(t) & \text{(matter domination)}\:,
\end{array} \right.
\end{equation}
where for the last step we have used Eq.~(\ref{eq:a_t_radiation}) and Eq.~(\ref{eq:a_t_matter}) respectively. Since during matter domination, the perturbations outside and inside the horizon grow both proportional to $a$, perturbations that enter during matter domination are considered as ``benchmark'' to which other perturbations are related. The amplitude of the perturbation at horizon entry is called ``primordial''. The \textbf{transfer function} $T(k)$ is then introduced by 
\begin{equation}\label{eq:lin_overdensity}
	\delta(t,\bv{k}) = \delta(t_{\rm in},\bv{k})\: T(k)\:\frac{D_+(t)}{D_+(t_{\rm in})}\:,
\end{equation}
where $D_+$ is the linear growing function that includes matter and a cosmological constant, but no radiation. This means that $T(k)$ considers all deviations in the evolution of the primordial perturbation which happen at early times during radiation domination and due to physical processes well within the horizon (e.g., matter-radiation plasma). For a mode $\bv{k}$ which enters well within the matter dominated regime and after decoupling, we have by definition $T(k) \simeq 1$.

How does the transfer function look like for a DM mode that enters during radiation domination? Since DM perturbations cannot grow inside the horizon during radiation domination due to the Meszaros effect, modes at smaller scales are suppressed compared to those of larger scales. For simplicity, we assume that during radiation domination a mode entering the horizon freezes immediately and starts to grow like $D_+ \propto a$ at matter-radiation equality $t_{\rm eq}$. Since during radiation domination, perturbations outside the horizon effectively grow proportionally to $a^2$ (see (Eq.~\ref{eq:effective_growth_outside})) relative to those inside the horizon (which remain constant), the amplitude of a mode entering at $t_{\rm in} < t_{\rm eq}$ is suppressed by the factor $(a(t_{\rm in})/a(t_{\rm eq}))^2$. Since for a mode $\bv k$ that enters the horizon at $t_{\rm in}$ it holds $k \propto H(t_{\rm in})a(t_{\rm in})$ (see Sect.~\ref{sec:horizons}), we obtain
\begin{equation}
\frac{a^2(t_{\rm in})}{a^2(t_{\rm eq})} = \frac{k^2}{k_{\rm eq}^2} \frac{H^2(t_{t_{\rm eq}})}{H^2(t_{\rm in})} \supeq{(\ref{eq:a_t_radiation})}{=} \frac{k^2}{k_{\rm eq}^2} \frac{t_{\rm in}^2}{t_{\rm eq}^2} \supeq{(\ref{eq:a_t_radiation})}{=} \frac{k^2}{k_{\rm eq}^2} \frac{a^4(t_{\rm in})}{a^4(t_{t_{\rm eq}})}\:,
\end{equation}
where $k_{\rm eq}$ is the mode that enters the horizon at matter-radiation equality $t_{\rm eq}$. So in total the transfer function for DM has the asymptotical behavior
\begin{equation}\label{eq:transfer_function_asymptotic}
\boxed{T(k) \simeq \left\{
\begin{array}{ll}
1 & \text{if } k_{\rm eq}/k \gg 1\\
(k_{\rm eq}/k)^2 & \text{if } k_{\rm eq}/k \ll 1\:.
\end{array} \right.}
\end{equation}

To obtain the precise form for the transfer function at all scales, we would have to solve the general relativistic Boltzmann equation taking into account all sorts of energy contributions in the universe and their interactions. The corresponding procedure is outlined in Section \ref{sec:complete_treatement}. For instance, the corresponding transfer function $T_{\rm b}(k)$ for baryons does also contain effects such as acoustic oscillations leading to a strong oscillation pattern in $T_{\rm b}(k)$. These oscillations are then transferred by gravitational interactions to the spatial distribution of DM particles producing a weak oscillation pattern in $T(k)$ which is called ``baryonic acoustic oscillations''. This feature, which has been convincingly detected \citep[e.g.,][]{percival2010}, is a powerful confirmation of our model of the history of the universe and is useful for estimating cosmological parameters. A discussion of common fitting formulas for the transfer function is given in Section 6.5 of \cite{weinberg2008}.

\subsection{Nonlinear regime}

So far, we have assumed that $\delta \ll 1$ on all scales within the horizon. This condition is satisfied to high accuracy in the early matter dominated regime (e.g., $\delta \sim 10^{-5}$ for a mode entering the horizon around recombination). At these times, Eq.~(\ref{eq:lin_overdensity}) is a good approximation. However, as time goes by, $\delta$ grows and there will be a point, where $\delta \lesssim 1$ and the linear approximation fails to be a good approximation. Around this time, nonlinear effects become important. Thus, it will prove to be useful to divide up the overdensity $\delta$ into a linear and a nonlinear part at every time, i.e.,
\begin{equation}\label{eq:nl_definition}
	\delta(t,\bv{k}) = \delta_{\rm lin}(t,\bv{k}) + \delta_{\rm nl}(t,\bv{k})\:,
\end{equation}
where the linear part $\delta_{\rm lin}$ is defined by Eq.~$(\ref{eq:lin_overdensity})$ and the nonlinear part $\delta_{\rm nl}$ is defined by Eq.~(\ref{eq:nl_definition}). At early times, $\delta_{\rm nl}$ is basically zero and $\delta \simeq \delta_{\rm lin}$ holds at all scales of interest (linear regime).

What can we say about nonlinear structure formation? Unfortunately, not surprisingly the nonlinear evolution of the density field is very complicated and there is no analytical formula describing the general case. However, for certain special cases, analytic solutions can be found (see, e.g., Sect.~\ref{sec:spherical_top_hat_collapse}) and there are several approximation methods (see, e.g., \citealt{sahni1995} for a review). To deal with the general case, one has to make use of large numerical simulations. This has become an extensive field within the branch of astronomy and a review would go beyond the scope of this introduction.\footnote{The reader is referred to \cite{frenk2012} for an easily readable review of the development and the results from numerical DM simulations.} In general, nonlinear effects mix different $\bv k$-modes and lead to a cosmic web which is made of sheets and filaments and in whose nodes are big galaxy clusters (see Fig.~\ref{fig:LSS}). The collectivity of this cosmic web along with clusters and groups of galaxies is called the ``large-scale structure'' (LSS).

\section{Statistics of the overdensity field}\label{sec:statistics_overdensity_field}

In principal, the overdensity $\delta(t,\bv{x})$ contains all information about the LSS in the universe at any time. However, in order to characterize the structure in the universe and to compare observations of $\delta$ with theory, it is meaningful to think of $\delta$ as a realization of a stochastic process. We can think of it like the initial inhomogeneities in the universe were created by a stochastic process and that this process was the same at every position. This lays the theoretical foundation for the cosmological principle.

A possible candidate for such a stochastic process is discussed in Chapter \ref{sec:generation_of_initial_perturbations}. Since such a stochastic process is not only constrained to one point but to all space, the mathematical theory needed here is the theory of random fields.\footnote{For a general introduction into the theory of random fields see \cite{miller1975}, \cite{adler1981}, or \cite{adler2007}.} In the following, we regard $\delta$ as a realization of a homogeneous and isotropic random field with zero mean. The random field itself will also be denoted by $\delta$.

\subsection{Correlation function and power spectrum}\label{sec:Correlation function and power spectrum}

The simplest nontrivial statistics of an inhomogeneous universe is the 2-point correlation function
\begin{equation}\label{correlationfunction}
\xi(\bv{x},\bv{x}') = \left\langle \delta(\bv{x}) \delta(\bv{x}')\right\rangle,
\end{equation}
where $\left\langle \ldots\right\rangle$ denotes the ensemble average (expectation value) of the stochastic process underlying the random field $\delta$. Since our random field is homogeneous and isotropic, the correlation function can only depend on $|\bv{x}-\bv{x}'|$, i.e., we write $\xi(r) \equiv \xi(\bv{x},\bv{x}')$ for $r = |\bv{x}-\bv{x}'|$. If $\xi(r)$ is additionally continuous at $r = 0$, there exists a spectral representation of the field, i.e., we can decompose it in Fourier modes $\delta(\bv k)$ and it holds (spectral representation theorem)\footnote{See, e.g., \cite{adler1981} Theorem 2.4.1 together with Theorem 2.2.1.
}
\begin{equation}\label{eq:power_spectrum_definition}
	\left\langle \delta(\bv{k}) \delta^\ast(\bv{k}')\right\rangle = \left(2\pi\right)^{3}\:\delta_{\rm D}\!\!\left(\bv{k}-\bv{k}'\right)\: \mathcal P(\bv k)\:,\quad\quad
	\mathcal P(\bv k) = \int \xi(x) e^{-i\bv{k}\bv{x}}\:dx^3\:.
\end{equation}
The function $\mathcal P(k)$ is called \textbf{power spectrum}. Due to the reality of $\xi(r)$ it holds $\mathcal P(\bv k)=\mathcal P^\ast(-\bv k)$ and due to the rotational invariance of $\xi(r)$ it holds $\mathcal P(k) \equiv \mathcal P(\bv k)$, so $\mathcal P(k)$ is a real function. Moreover, it is also nonnegative for all $k$. The power spectrum contains exactly the same amount of information as the correlation function, but depending on the application it may, however, be useful to prefer one or the other \citep[e.g.,][Sect.~1]{feldman1994}.

A widely used kind of random fields are the Gaussian random fields being the simplest and most natural. A homogeneous real Gaussian random field $\delta$ with zero mean is fully characterized by its finite-dimensional distributions. That is, for $N$ points in space $\bv{x}_1 \ldots \bv{x}_N$ the probability density function is a multivariate Gaussian
\begin{equation}
f\big(\delta(\bv{x}_1), \ldots, \delta(\bv{x}_N)\big) = \frac{1}{(2\pi)^{N/2} \sqrt{\det(V)}} \:\exp\left(- \frac{1}{2} \sum_{i,j=1}^N \delta(\bv{x}_i)\: (V^{-1})_{ij}\: \delta(\bv{x}_j) \right)
\end{equation}
with the covariance matrix given by $V_{ij} = \xi(|\bv{x}_{i}-\bv{x}_{j}|)$. Hence the correlation function $\xi(r)$ determines the random field entirely.

The Fourier transform $\delta(\bv{k})$ of a Gaussian random field has for each $k$-mode a real and imaginary part, which are independent and Gaussian distributed with zero mean and variance $\mathcal P(k)/2$. This is equivalent for $\delta(\bv k)$ having a uniformly distributed random phase and a modulus $|\delta(\bv k)|$ which is Rayleigh distributed with variance $\mathcal P(k)$. Additionally, each $\bv k$-mode is independent from the others.

By introducing the ensemble average $\left\langle \ldots\right\rangle$ we referred to a stochastic process taking place in the early universe. However, the observable LSS constitutes a single realisation of this process, so the question arises how these ensemble averages might be measured in practice. To be able to infer anything about the underlying stochastic process one has to postulate some sort of ``ergodicity'' or ``fair sample hypothesis''. \textbf{Ergodicity} refers to the mathematical property of random fields that volume averages converge to ensemble averages as the survey volume goes to infinity. In general it is hard to prove that a random field has this property. However, it can be shown that a zero mean, homogeneous, real Gaussian random field is ergodic if $\xi(r) \rightarrow 0$ for $r \rightarrow \infty$ \citep[Thm.~6.5.4]{adler1981}. For a general random field, ergodicity is a valid assumption if the length scale over which the average is computed is large enough, so that the spatial correlation become negligibly small (see \citealt{weinberg2008}, App.~D). On the other hand, the \textbf{fair sample hypothesis} \citep{peebles1980} states that well separated parts of the universe can be regarded as independent realizations of the underlying stochastic process and that the observable universe contains many such realizations.

While ergodicity is a precise mathematical statement which may or may not apply to a given random field, the fair sample hypothesis is more vague. However, \cite{watts2003} pointed out that the fair sample hypothesis is stronger than ergodicity and probably more useful for studying the LSS, because to obtain a fair sample it is not necessary to average over an infinite volume, which is practically impossible. Whichever hypothesis finally applies, most present day galaxy surveys are way too small to constitute a fair sample (especially at high redshift) and thus averages over the volumes of such surveys are subjected to statistical fluctuations. This phenomenon is called \textbf{sample variance} or \textbf{cosmic variance} if the sample is constrained by the size of the observable universe (e.g., CMB). The two terms are, however, often used interchangeably.

\subsection{Initial conditions and linear power spectrum}\label{sec:Initial_conditions}

Even in the absence of any mechanism producing perturbations in the early universe and well before the idea of inflation, there was a preferred ansatz for the initial power spectrum of the form \citep{harrison1970,zeldovich1970,peebles1970}
\begin{equation}\label{eq:primordial_power_spectrum}
	\mathcal P(k) \propto k^{n_{\rm s}}
\end{equation}
with $n_{\rm s}$ the \textbf{spectral index}. The virtue of this ansatz is that it does not introduce any particular length scale. If $n_{\rm s} = 1$ then this spectrum is called the \textbf{Harrison-Peebles-Zel'dovich spectrum} and has the preference of being scale invariant (see Sect.~\ref{sec:primordial_power_spectrum}), which is a natural expectations. In any case, there are fairly general reasons to constrain the spectral index within $-3 < n_{\rm s} < 4$ (see, e.g., \citealt{peacock1999}, Sect.~16.2). A further assumption about the initial perturbation is that it constitutes a realization of a Gaussian random field which, again, is probably the simplest and most natural choice.

Today, we are in the favorable situation that both of these assumptions could be verified to a high degree by observations of the CMB \citep[e.g.,][]{komatsu2011} and that we also have a theory at hand which explains how this initial state could have been produced. As we will show in Chapter \ref{sec:generation_of_initial_perturbations}, the simplest models of inflation being governed by a single inflaton field produce an initial density field which can be regarded as a realization of a Gaussian random field whose power spectrum for modes entering the horizon during matter domination is (see Sect.~\ref{sec:primordial_power_spectrum})
\begin{equation}
	\mathcal P(t,k) \propto k^{n_{\rm s}}\:,\quad\quad\quad n_{\rm s} = 1 - 6 \epsilon + 2 \eta\:,
\end{equation}
where $\epsilon$ and $\eta$ are the slow-roll parameters (see Eq.~(\ref{eq:slow_roll_parameters})) evaluated at the time when the mode $\bv k$ exits the horizon.

Similar to Eq.~(\ref{eq:nl_definition}) it is reasonable to split up the power spectrum (and likewise the correlation function) into a linear and a nonlinear part
\begin{equation}
\mathcal P(t,k) = \mathcal P_{\rm lin}(t,k)+\mathcal P_{\rm nl}(t,k)\:,
\end{equation}
where the linear power spectrum is just the power spectrum of the linear overdensity field $\delta_{\rm lin}$ and is given for any time after $t_{\rm eq}$ by (see Eqs.~(\ref{eq:lin_overdensity}) and (\ref{eq:primordial_power_spectrum}))
\begin{equation}\label{eq:lin_power_spectrum}
\boxed{\mathcal P_{\rm lin}(t,k) = A_0\:k^{n_{\rm s}}\: T^2(k)\:D_+^2(t)\:}
\end{equation}
with $A_0$\footnote{Since $k$ is not dimensionless, we have to be careful in interpreting Eq.~(\ref{eq:lin_power_spectrum}). The expression $k^{n_{\rm s}} = e^{n_{\rm s}\ln(k)}$ is not well defined in general, as we cannot build the logarithm for a dimensioned quantity. To make sense of $k^{n_{\rm s}}$ we have to introduce a reference mode $k_0$ (e.g., $k_0 = 1$ Mpc$^{-1}$) to scale out the unit of $k$ as $(k/k_0)^{n_{\rm s}}$. Therefore the normalization $A_0$ carries the full dimension of the power spectrum and depends on the adopted reference mode $k_0$. For this reason, a numerical value of $A_0$ always has to be stated for a corresponding reference mode $k_0$.} its normalization at the present time $t_0$ (recall that we set $D_+(t_0)=1$). The nonlinear part $\mathcal P_{\rm nl}$ is then defined by the difference between the total and the linear power spectrum.

\begin{figure}[t]\centering
  \includegraphics[scale = 0.6]{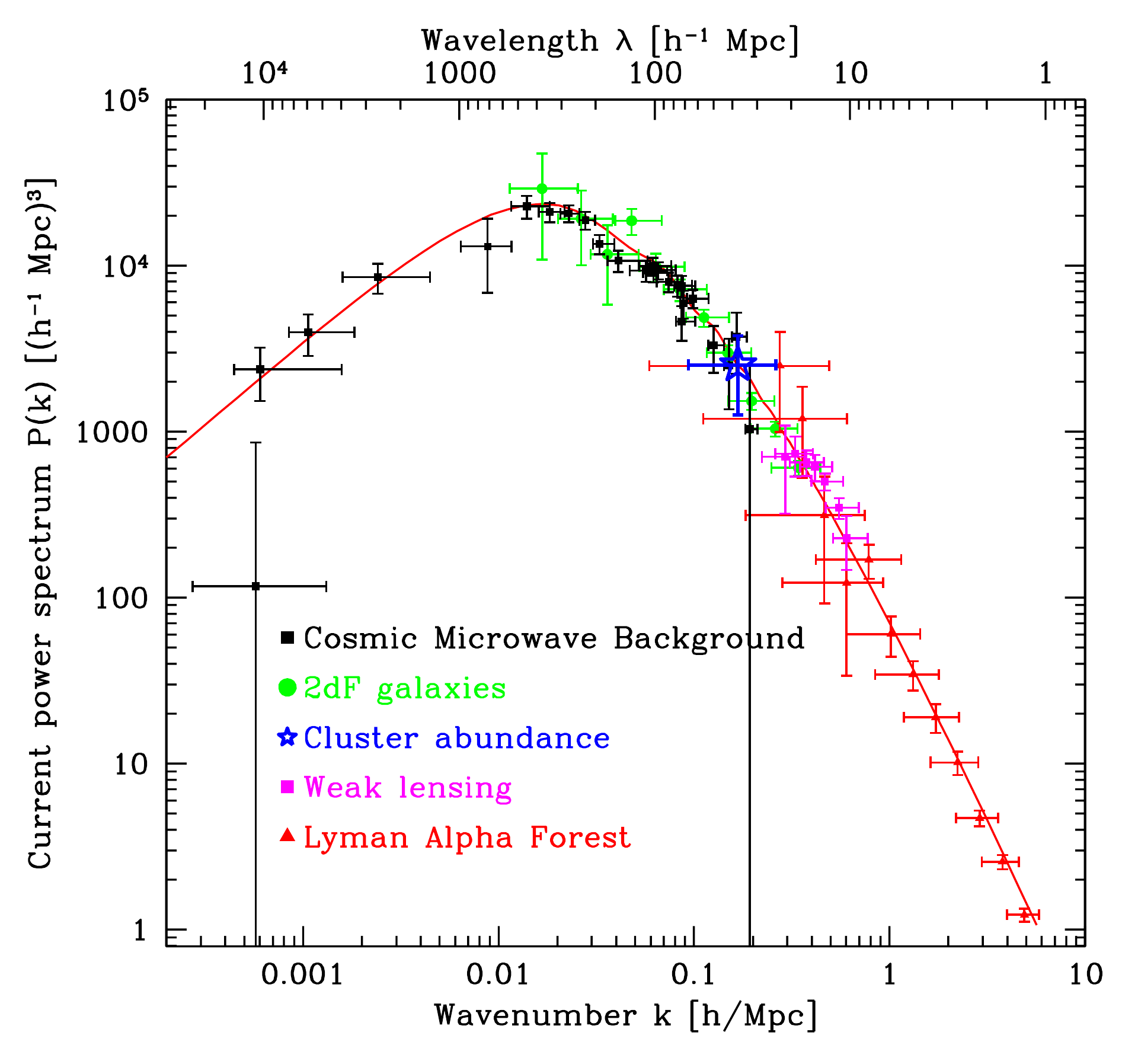}
  \caption{Linear power spectrum at the present epoch. The solid line is the model of the linear power spectrum for the concordance model and the points with errorbars show the different measurements as indicated in the legend. The methods by which these measurements were maped onto $k$-space are explained in \cite{tegmark2002}. It is obvious that it holds on large scales $P(k) \propto k$ and on small scales $P(k) \propto k^{-3}$. The turnaround roughly marks the scale of the horizon at the epoch of radiation-matter equality. \citep[Taken from][Copyright (2002) by The American Physical Society.]{tegmark2002}}\label{fig:linear_power_spectrum}
\end{figure}

The linear power spectrum for the concordance model together with a compilation of measurements is shown in Figure \ref{fig:linear_power_spectrum}.
At large scales the power spectrum is basically given by the primordial power spectrum $P(k) \propto k$ and on small scales it is affected by the physical processes inside the horizon (e.g., Meszaros effect) so that $P(k) \propto k^{-3}$ as expected by the asymptotical behavior of the transfer function (\ref{eq:transfer_function_asymptotic}). Obviously, the stagnation of the growth of perturbation during radiation domination introduces a distinct length scale into the linear power spectrum, which separates these two regimes and which is of the size of the horizon at radiation-matter equality. The overlap of the different measurements shown in Figure \ref{fig:linear_power_spectrum} impressively demonstrates the success and the consistency of the paradigm of structure formation in the linear regime.

The total power spectrum is equal to the linear power spectrum if $\delta_{\rm lin} \ll 1$. The contribution due to $\mathcal P_{\rm nl}$ is practically negligible at early times and becomes more important as perturbations grow. However, since the primordial power spectrum is a power law on large scales, there is for any time $t$ a threshold $k_{\rm th}(t)$ so that for modes with $k \lesssim k_{\rm th}(t)$ the overdensity $\delta$ is so small that we can assume $\mathcal P(t,k) \simeq \mathcal P_{\rm lin}(t,k)$ (linear regime). The regime with $k \gtrsim k_{\rm th}(t)$, where $\mathcal P(t,k) \simeq \mathcal P_{\rm nl}$, is called the \textbf{nonlinear regime}. Note that nonlinear effects not only mix different $\bv k$-modes, but also spoil Gaussianity\footnote{A simple way to see this is by noting that by definition $\delta \geq -1$, while on the positive side there is no such constraint, so the distribution function of $\delta$ becomes asymmetric and cannot be Gaussian anymore.}.

Sometimes it is convenient to express the linear power spectrum by means of the \textbf{effective spectral index} $n_{\rm eff}$ defined as
\begin{equation}\label{eq:effective_spectral_index}
n_{\rm eff}(k) = \frac{d \ln \left(k^{n_{\rm s}} T^2(k)\right)}{d \ln(k)} = n_{\rm s}+ 2\frac{d \ln \left(T(k)\right)}{d \ln(k)}
\end{equation}
such that
\begin{equation}
k^{n_{\rm eff}(k)} = k^{n_{\rm s}} T^2(k) \:.
\end{equation}
With Eq.~(\ref{eq:transfer_function_asymptotic}) we obtain immediately $n_{\rm eff}(k) \simeq n_{\rm s}$ for $k \gg k_{\rm eq}$ and $n_{\rm eff}(k) \simeq n_{\rm s}-4$ for $k \ll k_{\rm eq}$.

\subsection{Filtering and moments}\label{eq:filtering}

In the real universe, the overdensity field $\delta$ has powers on all scales. Particularly on very small scales, deviations of the matter density from the mean density can become huge. For instance, if we apply $\delta = (\rho-\bar{\rho})/\bar{\rho}$ naively to the earth, we obtain an overdensity of about $10^{30}$ for every position inside the earth, but this huge overdensity has no meaning for cosmology and makes it meaningless to search the overdensity field for peaks with cosmological relevance. Hence, an important concept in cosmology is filtering, where contributions to the density field below a given length scale are filtered out. Mathematically, this is obtained by convolving the overdensity with some window function $W(r)$, i.e.,
\begin{equation}
\delta_{R_{\rm f}}(t,\bv{x}) = (\delta \ast W)(t,\bv{x}) = \int \delta (t,\bv{x}-\bv{x'})W(|\bv{x'}|,R_{\rm f})\:{dx'}^3\:,
\end{equation}
where the window function $W$ is associated with a comoving length scale $R_{\rm f}$ beyond which it is essentially zero and is normalized for all $R_{\rm f}$ such that $\int W(|\bv{x}|,R_{\rm f})dx^3 = 4 \pi \int W(r,R_{\rm f})r^2 dr = 1$. So the filtered overdensity $\delta_{R_{\rm f}}(t,\bv{x})$ is the overdensity smoothed at every position over a scale of $R_{\rm f}$ and features that are smaller than this length scale are washed out. The most common window function in cosmology is the top-hat filter with radius $R_{\rm f}$
\begin{equation} \label{eq:tophat}
W_{\rm TH}(r,R_{\rm f}) = \left\{ \begin{array}{ll}
3/(4\pi R_{\rm f}^3) &  \text{if}\ r \leq R_{\rm f}\\
0 & \text{if}\   r > R_{\rm f}
\end{array} \right.
\end{equation}
with its Fourier transform
\begin{equation}
W_{\rm TH}(k,R_{\rm f}) = \frac{3}{(kR_{\rm f})^3}\Big(\sin (kR_{\rm f}) - kR_{\rm f}\:\cos (kR_{\rm f})\Big)\:.
\end{equation}
We will stick to this case throughout this introduction.

In the following, we are interested in the statistical moments of the filtered \emph{linear} overdensity field. Since the linear overdensity $\delta_{\rm lin}$ has zero mean, we obtain for the first moment immediately
\begin{equation}
\big\langle \delta_{R_{\rm f}}(t,\bv{x})\big\rangle = \int \left\langle \delta_{\rm lin}(t,\bv{x}-\bv{x'})\right\rangle W(\bv{x'},{R_{\rm f}})\:{dx}'^3 \equiv 0\:.
\end{equation}
The second moment, however, is nontrivial and can be expressed by the linear power spectrum as
\begin{align}
	\sigma_{R_{\rm f}}^2(t) &\equiv \left\langle \delta_{R_{\rm f}}^2(t,\bv{x}) \right\rangle = \left\langle \delta_{R_{\rm f}}(t,\bv{x})\delta_{R_{\rm f}}^\ast(t,\bv{x}) \right\rangle
	 =\left\langle \left(\mathcal F^{-1}\mathcal F \delta_{R_{\rm f}}\right)  \left(\mathcal F^{-1}\mathcal F\delta_{R_{\rm f}}\right)^\ast \right\rangle\\
	&= \frac{1}{(2\pi)^6}\int \int \left\langle \delta_{\rm lin}(t,\bv{k}) \delta_{\rm lin}^\ast(t,\bv{k}') \right\rangle \:W(k,{R_{\rm f}}) W^\ast(k',{R_{\rm f}}) \:e^{i(\bv{k}-\bv{k}')\bv{x}} \:dk^3 {dk'}^3\\
	&= \frac{1}{(2\pi)^3} \int \mathcal P_{\rm lin}(k)\:|W(k,{R_{\rm f}})|^2\: dk^3\:,
\end{align}
where we have denoted the Fourier transformation operator by $\mathcal F$ and we have applied the definition of the power spectrum (\ref{eq:power_spectrum_definition}) to the linear overdensity field. If the $k$-dependence of $W(k,R_{\rm f})$ appears only in the combination $kR_{\rm f}$ (cf.~Eq.~\ref{eq:tophat}), we can define $\widetilde{W}(kR_{\rm f}) = W(k,R_{\rm f})$ so that
\begin{equation}\label{eq:sigma_R_particular}
\begin{aligned}
\sigma_{R_{\rm f}}^2(t) &= \frac{A_0 D_+^2(t)}{2 \pi^2} \int k^{n_{\rm eff}(k)+2}\: |\widetilde{W}(kR_{\rm f})|^2\: dk\\
&= \frac{A_0 D_+^2(t)}{2 \pi^2} \int \frac{1}{R_{\rm f}} \left(\frac{y}{R_{\rm f}}\right)^{n_{\rm eff}(y/R_{\rm f})+2}\: |\widetilde{W}(y)|^2\: dy\:,
\end{aligned}
\end{equation}
where we used the explicit form of the linear power spectrum (\ref{eq:lin_power_spectrum}) with the effective spectral index (\ref{eq:effective_spectral_index}). If we scale out $R_{\rm f}^{-(n_{\rm eff}+3)}$ with $n_{\rm eff}$ evaluated at $2\pi/R_{\rm f}$, the integral becomes approximately independent of $R_{\rm f}$ in the range $0.8\ {\rm Mpc}\lesssim R_{\rm f} \lesssim 40\ {\rm Mpc}$, so that it holds for this range\footnote{For a top-hat filter the integral changes about a factor of 2 over the indicated range. For simplicity we will assume the integral to be roughly constant and use it only for order of magnitude calculations.}
\begin{equation}\label{eq:sigma_W_rf}
\sigma_{R_{\rm f}}(t) \propto D_+(t)\:R_{\rm f}^{-(n_{\rm eff}+3)/2}\:.
\end{equation}

For certain applications it is convenient to express the filtered density field in terms of the mass involved. We define the typical mass associated to each point of $\delta_{R_{\rm f}}(t,\bv{x})$ by $M(R_{\rm f}) = \bar{\rho}_0 V(R_{\rm f})$ with the comoving volume $V(R_{\rm f}) =4 \pi R_{\rm f}^3/3$. Since $R_{\rm f}$ and $M$ are uniquely related to each other, we will use the notations ($\delta_{R_{\rm f}}$, $\sigma_{R_{\rm f}}$) and ($\delta_M$, $\sigma_M$) interchangeably. With Eq.~(\ref{eq:sigma_W_rf}) it follows the approximation
\begin{equation}\label{eq:scaling_relation_sigma_M}
\boxed{\sigma_M(t) \propto D_+(t)\:M^{-(n_{\rm eff}+3)/6}}
\end{equation}
for masses in the range $10^{11}\ M_\odot \lesssim M \lesssim 10^{16}\ M_\odot$ and $n_{\rm eff}$ evaluated at the corresponding scale $2\pi/R_{\rm f}$.

For a top hat filter $W_{\rm TH}$ with radius $R_{\rm f} = 8\: h^{-1}$ Mpc, $\sigma_{R_{\rm f}}(t_0)$ for the present time $t_0$ is denoted by $\sigma_8$ being a cosmological parameter. This parameter fixes the normalization $A_0$ of the linear power spectrum (\ref{eq:lin_power_spectrum}) as
\begin{equation}
A_0 = \frac{2\pi^2\:\sigma_8^2}{9}\left(\int k^{n_{\rm s}+2}\: T^2(k) \left[\frac{\sin (kR_{\rm f}) - kR_{\rm f}\:\cos (kR_{\rm f})}{(kR_{\rm f})^3}\right]^2 dk\right)^{-1},
\end{equation}
with $R_{\rm f}=8\: h^{-1}$ Mpc and $D_+(t_0) = 1$. The physical meaning of $\sigma_8$ is how much the mass within boxes of $R_{\rm f}=8\: h^{-1}$ fluctuates from one place to another in the present day universe. Since at the present time the scale $8 \: h^{-1}$ Mpc is already mildly nonlinear and since $\sigma_8$ considers only the linear overdensity field, $\sigma_8$ is a rather abstract quantity from an observational point of view and cannot be estimated, for instance, by counting galaxies in boxes of $8\:h^{-1}$ Mpc. As we shall see in Section \ref{sec:press_schechter_theory}, $\sigma_8$ is very sensitive to the number density of clusters in the universe. The current estimated value is (see Tab.~\ref{tab:cosmological_parameters})
\begin{equation}
\sigma_8 \simeq 0.8\:,
\end{equation}
although different methods such as measuring number density of galaxy clusters and weak lensing surveys typically yield slightly different values for $\sigma_8$ probably owing to systematic errors inherent in these methods.

\section{Dark matter halos}\label{sec:dm_halos}

In this section, we discuss how the linear theory that has been developed in the last two sections can be related to the complex nonlinear evolution of the LSS. First, we present the spherical top hat model as a simple model for the formation of DM halos, then we try to gain insight into the statistics of the these halos by computing approximative expressions for their number density and spatial correlations. Finally, we introduce the halo model which is a simple approach to understanding the behavior of the galaxy correlation function in the linear and nonlinear regime.

\subsection{Spherical top hat collapse}\label{sec:spherical_top_hat_collapse}

Suppose a small spherical, homogeneous perturbation that is imbedded in a homogeneous background universe. Since the present day structures grew from tiny inhomogeneities produced in the very early universe, we require that the perturbation at early times approaches the density of the background and expands in accordance with it. In this section we investigate how the overdensity of such a perturbation evolves with time.

For simplicity we assume that the background universe with density $\bar \rho(t)$ is flat and matter dominated and that our perturbation with density $\rho(t)>\bar{\rho}(t)$ and radius $R_{\rm p}(t)$ is symetrically placed within a spherical cavity that expands with the background. The radius $\bar R(t)$ of the cavity is chosen such that $\rho(t) R_{\rm p}^3(t) = \bar{\rho}(t) \bar{R}^3(t)$, i.e., if the mass of the overdensity was uniformly distributed within the cavity, it would just approach the density of the background. With these (simplistic) assumptions the evolution of the overdensity $\delta(t) = (\rho(t)-\bar \rho(t))/\bar \rho(t)$ can be easily computed analytically. Our requirements on the initial conditions translate to $\rho(t) \simeq \bar{\rho} (t)$ and $R_{\rm p}(t) \simeq \bar{R}(t)$ for small $t$.

The condition $\rho(t) R_{\rm p}^3(t) = \bar{\rho}(t) \bar{R}^3(t)$ guarantees that the background expands undisturbed by the perturbation due to the Newtonian theorem for a spherical mass distribution.\footnote{This condition is, of course, only adopted for simplicity to obtain exact mathematical results. The case of a more general perturbation is treated, for instance, in \nocite{mukhanov2005}Mukhanov (2005, Sect.~6.4.1). The corresponding behavior of the background is identical with our case for large distances from the perturbation.} Moreover, since the background is distributed spherically symmetric around the perturbation, the perturbation also evolves independently from the background. Thus, the perturbation and the background are entirely decoupled.\footnote{This remains valid even in a general relativistic treatment due to the Birkhoff theorem and the fact that the background and the perturbation are not overlapping (see, e.g., \citealt{weinberg2008}, Sect.~8.2).} The dynamics of the background are then simply determined by the familiar Friedmann equation (\ref{eq:friedmann_equation}) for $K = 0$
\begin{equation}\label{eq:friedmann_5}
	\left(\frac{d a}{dt}\right)^2 = \frac{8 \pi G}{3} \bar{\rho}a^2
\end{equation}
with the solution $a(t) \propto t^{2/3}$ (see Eq.~(\ref{eq:a_of_t})). Inserting this solution into the Friedmann equation we obtain the explicit time dependence of the background density as
\begin{equation}\label{eq:mean_density}
	\bar{\rho}(t) = \frac{1}{6 \pi G t^2}\:.
\end{equation}

How can we describe the dynamics of the perturbation? If the radius of the perturbation was continuously increased until the whole universe was covered by the perturbation, we would simply have an overcritical FLRW world model, i.e., a universe with $K = 1$ that is also described by the Friedmann equation (\ref{eq:friedmann_equation}). However, due to the Newtonian theorem for spherical symmetric mass distributions, our perturbation does not know whether it is imbedded within a critical or overcritical background universe as long as the matter is distributed spherically around it. So the perturbation must obey a scaled down version of the overcritical Friedmann equation. In the following, we will study the expanding solutions for such a universe and then apply it to our perturbation.

Expressed in terms of the cosmological world radius $R(t) = R_0 a(t)$ (see Sect.~\ref{sec:mathematical_formulation}) the Friedmann equation for a closed universe is
\begin{equation}\label{eq:closed_friedmann}
\left(\frac{dR}{dt}\right)^2 = \frac{8 \pi G}{3}\rho R^2-c^2\:.
\end{equation}
To find an analytic solution, we express this equation in terms of conformal time\footnote{In this formulation, the conformal time is dimensionless in contrast to Eq.~(\ref{eq:conformal_time}). The difference between these two formulations is, however, just the constant factor $R_0$.}
\begin{equation}
	d\tau = \frac{c}{R}\:dt\:.
\end{equation}
Note that there is a one-to-one correspondence between $t$ and $\tau$, whereby these two time coordinates have the same zero point. The Friedmann equation (\ref{eq:closed_friedmann}) then becomes
\begin{equation}\label{eq:closed_friedmann_conformal_time}
	\left(\frac{d}{d\tau}\frac{R}{R_\ast}\right)^2 = 2 \frac{R}{R_\ast}- \left(\frac{R}{R_\ast}\right)^2
\end{equation}
with the constant
\begin{equation}\label{eq:R_ast}
	R_\ast = \frac{4 \pi G \rho_0 R_0^3}{3 c^2} = \frac{1}{2}\:\frac{c}{H_0}\frac{\Omega_{\rm m}}{(\Omega_{\rm m}-1)^{3/2}}\:.
\end{equation}
To derive the expressions for $R_\ast$ we used $\rho(t) R^3(t) = \rho_0 R_0^3$ and 
\begin{equation}\label{eq:r_0_Omega_m}
R_0 = \frac{c}{H_0} \frac{1}{\sqrt{\Omega_{\rm m}-1}}\:,
\end{equation}
which is obtained by means of Eqs.~(\ref{eq:density_parameter}) and (\ref{eq:closed_friedmann}) for an arbitrary reference time $\tau_0$ (corresponding to $t_0$). Equation (\ref{eq:closed_friedmann_conformal_time}) has the simple solution
\begin{equation}\label{eq:closed_solution_R}
	R(\tau) = R_\ast \left(1 - \cos(\tau)\right)\:,\quad\quad t(\tau) = \int_0^\tau \frac{R(\tau')}{c}\: d\tau'=\frac{R_\ast}{c}\left(\tau-\sin(\tau)\right)\:.
\end{equation}

What can we say about the initial conditions? Obviously, the evolution of the closed universe is entirely determined by $R_\ast$, so we have one degree of freedom. Similar to the flat universe (see Sect.~\ref{sec:equation_of_state}), the Hubble parameter is determined by the choice of the zero point $R(0) = 0$, i.e., $H(\tau) = \dot R(\tau)/R(\tau) = \sin(\tau)/(1-\cos(\tau))$ is merely a function of $\tau$. The free parameter $R_\ast$ can be fixed, for instance, by specifying $R_0$, $\rho_0$, or $\Omega_{\rm m}$ at an arbitrary reference time $t_0$. Moreover, for small $\tau$, we can expand the Eqs.~(\ref{eq:closed_solution_R}) to the first nonvanishing order in $\tau$ yielding
\begin{equation}\label{eq:tau_first_order}
R(\tau) \simeq R_\ast \frac{\tau^2}{2}\:,\qquad t(\tau) \simeq \frac{R_\ast}{c} \frac{\tau^3}{6}\:.
\end{equation}
Inserting $t(\tau)$ into $R(\tau)$ leads to
\begin{equation}
R(t) \simeq R_0 \:\Omega_{\rm m}^{1/3} \left(\frac{3}{2}H_0 t \right)^{2/3}\:,\qquad H(t) = \frac{\dot R(t)}{R(t)} = \frac{2}{3t}\:,
\end{equation}
where we used the Eqs.~(\ref{eq:R_ast}) and (\ref{eq:r_0_Omega_m}). After dividing $R(t)$ by $R_0$, these expressions are identical to those of a flat, matter dominated universe (see Eqs.~(\ref{eq:a_of_t}) and (\ref{eq:meaning_of_H0}) for $w=0$) up to the factor $\Omega_{\rm m}^{1/3}$.
However, since $\Omega_{\rm m} (t) \rightarrow 1$ for $t \rightarrow 0$,\footnote{This can be seen as follows: For a universe that started from a big bang and that contains at least one energy contribution $I$ with an equation of state $w_I>-1/3$, it holds for early enough times that the second term on the right hand side of the Friedmann equation (\ref{eq:friedmann_equation}) always becomes negligible relative to the first term due to Eq.~(\ref{eq:rho_of_t}). With the definition of the density parameters (\ref{eq:density_parameter}) it follows immediately $\sum_{I} \Omega_I(t) \simeq 1$ at these times.} we have $\Omega_{\rm m} \equiv \Omega_{\rm m}(t_0) \simeq 1$ for small $t_0$ so that every closed, matter dominated universe is indistinguishable from a flat, matter dominated universe at early times. That is, for any choice of $\Omega_{\rm m}$, our perturbation approaches the density of the background at early times and expands in accordance with it as required.

\begin{figure}[t]\centering
    \includegraphics[scale = 0.6]{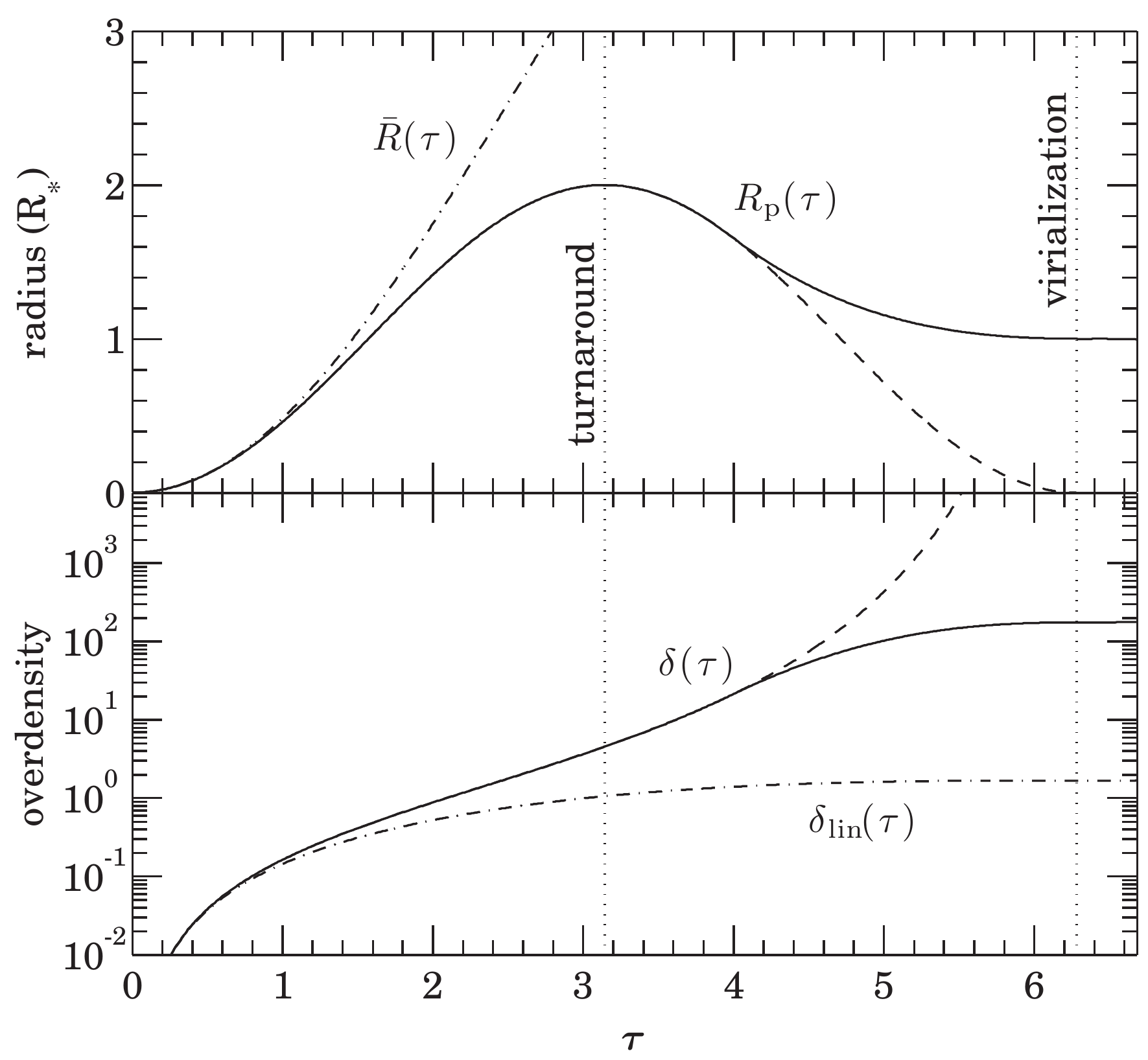}
  \caption{Evolution of the spherical top hat perturbation as a function of the (dimensionless) conformal time $\tau$. The growth of the radius $R_{\rm p}(t)$ (in units of $R_\ast$) and the corresponding overdensity $\delta(t)$ of the spherical perturbation are shown by the solid line in the upper and lower panels, respectively. At the beginning, $R_{\rm p}(t)$ grows in accordance with the expansion of the background universe $\bar R(t)$ (dashed-dotted line), decouples from the background at $\tau = \pi$ (turnaround), and finally collapses. If perfect sphercial symmetry was established, the overdensity would actually collapse into a singularity at $\tau = 2\pi$ (dashed line). However, since perfectly symmetric overdensities do not exist in reality, the overdensity virializes at $\tau = 2 \pi$ with a radius that is about half of its maximal extension. In the lower panel the linear overdensity $\delta_{\rm lin}(t)$ (dashed-dotted line) is shown for comparison.}\label{fig:top_hat}
\end{figure}

Our perturbation, however, has another degree of freedom being its mass $M$. For a given $\rho_0$ and mass $M$, the scale factor $R(\tau)$ is uniquely determined and the radius of the perturbation is given by $R_{\rm p}(\tau_0) = (3M/(4\pi \rho_0 ))^{1/3}$. Since $R_{\rm p}(\tau)$ must evolve in proportion to $R(\tau)$, we introduce a constant $C$ such that $R_{\rm p}(\tau) = C R(\tau)$. The only constraint on $M$ is that it must be smaller than the total mass within our overcritical universe, which formally leads to $C < \pi$, since $\pi R(\tau)$ is half the circumference of such an overcritical universe with world radius $R(\tau)$.\footnote{To make this point clearer we recall the meaning of $R(\tau)$ and $R_{\rm p}(\tau)$. $R(\tau)$ is the world radius of an abstract, inexistent universe that we would obtain, if we expanded the physical radius of our perturbation $R_{\rm p}(\tau)$ until it covered the whole universe. This (abstract) universe constitutes a closed matter dominated universe, whose dynamics is determined by Eq.~(\ref{eq:closed_friedmann}). Since for a closed universe the world radius $R(\tau)$ just corresponds to the physical radius of the spatial slice at time $\tau$ being the 3-sphere with radius $R(\tau)$, the circumference of this slice at time $\tau$ is $2 \pi R(\tau)$ and so the radius of our perturbation would completely fill this universe for $R_{\rm p}(\tau) = \pi R(\tau)$. This is, however, a rather formal issue.} With the Eqs.~(\ref{eq:closed_solution_R}) and (\ref{eq:R_ast}) the dynamics of the perturbation are then given by
\begin{equation}\label{eq:closed_solution}
	R_{\rm p}(\tau) = \tilde{R}_{\ast} \left(1 - \cos(\tau)\right)\:,\qquad \tilde{R}_\ast = C R_\ast = \frac{1}{C^2}\:\frac{MG}{c^2}\:.
\end{equation}
Using the explicit expressions (\ref{eq:closed_solution_R}),  (\ref{eq:closed_solution}), and (\ref{eq:mean_density}) for the radius of the perturbation, the conformal time, and the density of the background, respectively, we are able to compute the overdensity of a perturbation with mass $M$ as
\begin{equation}\label{eq:non_linear_delta}
\boxed{\delta(\tau)+1 = \frac{\rho(\tau)}{\bar{\rho}(\tau)} = \left(\dfrac{M}{4\pi R_{\rm p}^3(\tau)/3}\right) / \left(\dfrac{1}{6 \pi G t^2(\tau)}\right)= \frac{9}{2} \frac{\left(\tau-\sin(\tau)\right)^2}{\left(1-\cos(\tau)\right)^3}\:.}
\end{equation}
This equation is exact within our simplistic picture and describes the full nonlinear growth of our spherical overdensity. Moreover, it even remains exact in the general relativistic framework and is valid inside and outside the horizon. Note that $\delta(\tau)$ is independent of the mass of the perturbation and becomes zero for small conformal times as expected (cf.~Eq.~(\ref{eq:delta_lin})). In the following, we will take a closer look at several states of the evolution of $\delta(\tau)$. The results are summarized in Figure \ref{fig:top_hat} and Table \ref{tab:top_hat_model}.
\begin{table}[tp]\centering
\caption{Summary of the different stages of the spherical top hat model.\\[3mm]}
\begin{tabular}{lcccc}
\toprule
Stage & $\tau$ & $t$ & $\delta_{\rm lin}$ & $\delta+1$\\
\midrule
\vspace*{13pt}Turnaround & $\pi$ & $\pi \dfrac{R_\ast}{c}$ & $\dfrac{3}{20}(6 \pi)^{2/3} \simeq 1.1$ & $\dfrac{9 \pi^2}{16} \simeq 5.55$\\
		\vspace*{5pt}Virialization & $2\pi$ & $2 \pi \dfrac{R_\ast}{c}$ & $\dfrac{3}{20}(12\pi)^{2/3} \simeq 1.69$ & $\frac{9}{2}(2\pi)^2 \simeq 178$\\
\bottomrule
\end{tabular}
\label{tab:top_hat_model}
\end{table}

\paragraph{Linear regime}

At early times, $\tau \ll 1$, we can expand Eq.~(\ref{eq:non_linear_delta}) to the first nonvanishing order in $\tau$ yielding
\begin{equation}\label{eq:delta_lin}
	\delta(\tau) \simeq \frac{3}{20}\tau^2\:,
\end{equation}
so that by eliminating $\tau$ using Eq.~(\ref{eq:tau_first_order}) we obtain
\begin{equation}\label{eq:d_lin}
	  \delta(t) \simeq \frac{3}{20}\left( \frac{6ct}{R_\ast} \right)^{2/3} \equiv \delta_{\rm lin}(t)\:.
\end{equation}
Not surprisingly, we recover the relation $\delta(t) \propto t^{2/3}$ from the linear perturbation theory for a matter dominated universe since the overdensity $\delta(t)$ is small at early times. We denote the linear density by $\delta_{\rm lin}(t)$.

\paragraph{Turnaround}

As time passes, the perturbation grows and leaves the linear regime. Eq.~(\ref{eq:closed_solution}) shows that for $\tau_{\rm max} = \pi$ the radius $R_{\rm max} = R_{\rm p}(\tau_{\rm max})$ finally becomes maximal and the perturbation stops expanding. This state is called ``turnaround'' and marks the epoch when the perturbation decouples entirely from the Hubble flow of the homogeneous background. The overdensity at this stage is $\delta(\tau_{\rm \max}) \simeq 4.55$.

\paragraph{Virialization}

After the turnaround the perturbation starts contracting. For a perfect spherical symmetry and perfect pressureless matter, the pertubation would collapse to a single point for $\tau_{\rm coll} = 2 \pi$ becoming infintely dense. However, there is hardly any perfect spherical symmetric overdensity in the universe, so the perturbation does not collapse to a single point, but rather extreme shell crossing occurs and finally a virialized object of a certain finite size is formed which is called \textbf{halo}.\footnote{For a DM perturbation there is no interaction between the DM particles (except of gravity) and so there are no collisions between them, in contrast to a baryonic gas. Thus, the exchange of momentum between particles can only occur through gravitation as each DM particle moves in the fluctuating gravitational field of all other particles. This process is called ``violent relaxation'' and it can be shown that it leads to a Maxwellian distribution of velocities \citep[Sect.~17.1]{peacock1999}.} To find the size of the halo, we search for the radius $R_{\rm vir}$ for which the virial condition $2 E_{\rm kin}(R_{\rm vir})+E_{\rm pot}(R_{\rm vir}) = 0$ is satisfied. Since at the turnaround the kinetic energy is zero, it holds $E_{\rm pot}(R_{\rm max}) = E_{\rm tot}$, and since the potential energy of a homogeneous sphere of mass $M$ is
\begin{equation}\label{eq:potential_energy}
	E_{\rm pot}(R) = -\frac{3}{5}\frac{G M^2}{R}\:,
\end{equation}
we obtain at the radius $R_{\rm max}/2$
\begin{equation}
\begin{aligned}
		E_{\rm kin}\!\left(\frac{R_{\rm max}}{2}\right) &= E_{\rm tot} - E_{\rm pot}\! \left(\frac{R_{\rm max}}{2}\right) = E_{\rm pot}\!\left(R_{\rm max}\right) - E_{\rm pot}\!\left(\frac{R_{\rm max}}{2}\right)\\
		&\supeq{(\ref{eq:potential_energy})}{=} -\frac{1}{2} E_{\rm pot}\left(\frac{R_{\rm max}}{2}\right)\:,
\end{aligned}
\end{equation}
since $2 E_{\rm pot}(R) = E_{\rm pot}(R/2)$. This is exactly the virial relation and so we can set $R_{\rm vir} = R_{\rm max}/2$.

What is the epoch of virialization? Following Eq.~(\ref{eq:closed_solution}), the conformal time is $3 \pi/2$ when $R_{\rm max}/2$ is reached, however, Eq.~(\ref{eq:closed_solution}) considers only the single stream limit without any crossing of trajectories. So virialization takes some additional time and usually the time $\tau_{\rm vir} = 2 \pi$ is assumed, i.e., the epoch when the perfect symmetric perturbation would have collapsed to a point. So we estimate the overdensity $\delta_{\rm vir}$ of a virialized halo by evaluating the nominator of Eq.~(\ref{eq:non_linear_delta}) at $\tau = 2 \pi$ and its denominator at $\tau = 3\pi/2$, i.e.,
\begin{equation}\label{eq:delta_vir}
	\Delta_{\rm vir} \equiv \delta_{\rm vir}+1 \supeq{(\ref{eq:non_linear_delta})}{\equiv} \frac{9}{2}\frac{\left.\left(\tau-\sin(\tau)\right)^2\right|_{\tau = 2 \pi}}{\left.\left(1-\cos(\tau)\right)^3\right|_{\tau = 3 \pi/2}} = \frac{9}{2}(2\pi)^2 \simeq 178\:.
\end{equation}
This overdensity does not depend on the mass of the perturbation, on the initial overdensity, nor on the epoch of virialization $t_{\rm vir}$. Thus, whenever we observe an overdensity of the order of $\delta_{\rm vir}$, we can assume that the corresponding structure is virialized (or close to virialization) irrespective of its mass or formation history. Moreover, since virialization always happens at $\tau_{\rm vir} = 2\pi$, the corresponding epoch is $t_{\rm vir} = 2\pi R_\ast/c$ (see Eq.~(\ref{eq:closed_solution_R})) being fully specified by $R_\ast$, which in turn is given by the (initial) overdensity of the perturbation (see Eq.~(\ref{eq:R_ast})). That is,  the formation epoch is independent of the mass $M$, but depends only on the initial overdensity. Numerical simulations of collapsing halos show that the choice of $\tau_{\rm vir} = 2 \tau_{\rm max}$ (or equivalently $t_{\rm vir} = 2 t_{\rm max}$) is in fact of the right order of magnitude. One typically finds $t_{\rm vir} \simeq 3 t_{\rm max}$ \citep[Sect.~14.1]{coles2002}. However, we can assume that for $t_{\rm vir}= 2 t_{\rm max}$ the density of the halo is already of the right order.

In the same time, in which the perturbation grows to an overdensity of $\delta_{\rm vir}$, the linear overdensity grows to
\begin{equation}\label{eq:delta_c_lin}
	\delta_{\rm c} \equiv \delta_{\rm lin}(t_{\rm vir}) \supeq{(\ref{eq:d_lin})}{=} \dfrac{3}{20}(12\pi)^{2/3} \simeq 1.69
\end{equation}
being also independent of the mass $M$, the initial overdensity, and the epoch of virialization. Hence, we may postulate that whenever the linear overdensity of a perturbation exceeds the threshold $\delta_{\rm c}$, a halo of overdensity $\delta_{\rm vir}$ has formed at its place. This is in fact a very simplistic approach, but will prove to be astonishingly successful.

\paragraph{After virialization} Just after its formation, a halo has a mean density of about $\Delta_{\rm vir} \simeq 178$ times the mean matter density $\bar \rho(t_{\rm vir})$ (or the critical density $\rho_{\rm c}(t_{\rm vir})$ being the same in our case) irrespective of its mass $M$ or formation history. As the halo decoupled from the homogeneous background and virialized, we may expect that not only its mass, but, due to the dissipationless nature of DM, also its radius stays constant with time. However, in reality the halo is not isolated from the background and there will take place a constant inflow of material into the halo or the halo might even merge with another halo. Thus, the evolution of a halo after its formation is complicated and cannot be analyzed within our simplistic model. However, if we assume that the continuous inflow of material and the merging with other halos produce new halos which are still characterized by the same density ratio $\Delta_{\rm vir} \simeq 178$, then we would expect that the mean density of a typical halo with a given mass $M$ roughly scales with redshift like $\rho_{\rm vir}(z) = \Delta_{\rm vir}\bar\rho(z) \propto (1+z)^3$ and its physical radius as $r_{\rm vir}(z) \propto \rho_{\rm vir}^{-3}(z) \propto (1+z)^{-1}$.

In practice, there are several operational definitions for halos, which in one way or another are all based on our simplistic analysis within our flat, matter dominated universe.\footnote{\label{footnote:delta_c}This analysis can also be conducted for other cosmologies. For example, \cite{eke1996} derived an analytical prescription for the case $\Omega_{\rm m}+\Omega_\Lambda = 1$. While $\delta_{\rm c}$ defined as in Eq.~(\ref{eq:delta_c_lin}) turns out to be rather insensitive to the presence of a cosmological constant, the density contrast $\Delta_{\rm c}$ with respect to the critical density $\rho_{\rm c}$ significantly decreases for increasing $\Omega_\Lambda$. \cite{bryan1998} provide a fitting formula for the $\Delta_{\rm c}$ of \cite{eke1996} as
\begin{equation}
\Delta_{\rm c}(\Omega_\Lambda) = 18\pi-82\:\Omega_\Lambda+39\:\Omega_\Lambda^2\:.
\end{equation}
For $\Omega_\Lambda = 0$ we recover the value $\Delta_{\rm c} = \Delta_{\rm vir} \simeq 178$ from Eq.~(\ref{eq:delta_vir}) and for $\Omega_\Lambda = 0.7$ we find $\Delta_{\rm c} \simeq 101$. It should be noted that in the presence of a cosmological constant $\Lambda$ very small initial perturbations do not collapse at all due the repelling force of $\Lambda$ (see also \citealt{weinberg2008}, Sect.~8.2). For a discussion of the case $-1 < w < -1/3$ we refer to \cite{horellou2005}.} In numerical simulations, halos are often defined by a density contrast $\Delta_{\rm m} = 200$ with respect to the mean density $\bar \rho(t)$ or by a density contrast $\Delta_{\rm c} = 200$ with respect to the critical density $\rho_{\rm c}(t)$.\footnote{Halos defined by a density constant $\Delta_{\rm m}$ are not necessarily equivalent to halos defined by $\Delta_{\rm c}$. If the constants $\Delta_{\rm m}$ and $\Delta_{\rm c}$ are related by $\Delta_{\rm c} = \Delta_{\rm m}\Omega_{\rm m}(t)$ at a given epoch $t$, the definitions are completely equivalent at this epoch. However, for an extended period of time, over which $\Omega(t)$ is not constant, these definitions are not equivalent.} Another common approach in simulations is to identify DM halos as ``friends-of-friends'' (percolation) groups for a linking length of $b = 0.2$ times the mean interparticle separation within the simulation. For a general discussion and comparisons between these definitions we refer to \cite{white2001}, \cite{cuesta2008}, and \cite{more2011}.

\subsection{Press-Schechter theory}\label{sec:press_schechter_theory}

In the previous section, we have encountered the simplistic concept, in which a halo forms at a given position in space, whenever the linear overdensity field reaches a threshold of $\delta_{\rm c} \simeq 1.69$.\footnote{The numerical value of $\delta_{\rm c}$ is often taken to be the one in Eq.~(\ref{eq:delta_c_lin}) even if the universe is not matter dominated (e.g., in the case of the concordance cosmology), because the value of $\delta_{\rm c}$ is rather insensitive to the cosmology (see the footnote \ref{footnote:delta_c} in this chapter).} We will apply this concept to estimate the mean number density of halos in the universe at a given time. In the following, the number density of halos of mass $M$ at position $\bv x$ and time $t$ is denoted by $n_{\rm h}({\bv x},M,t)$ and the corresponding mean density by $\bar n_{\rm h}(M,t) = \langle n_{\rm h}({\bv x},M,t)\rangle$. For ease of notation, we will often suppress the time dependence. We mainly follow the presentation of \nocite{longair2008}Longair (2008, Sect~16.3).

Suppose the filtered density field smoothed over a given mass $M$ is $\delta_M$ (see Sect.~\ref{eq:filtering}). The main postulate of the Press-Schechter approach is the assumption that if $\delta_M(t,\bv x)$ for a given point $\bv x$ is larger than a certain threshold $\delta_{\rm c}$ this point is contained within a halo of mass $>\! M$. As it holds $\delta_M(t,\bv x) \rightarrow 0$  for $M \rightarrow \infty$, we will always find a mass $\widetilde{M} > M$ such that $\delta_{\widetilde{M}}(t,\bv x) = \delta_{\rm c}$ and this is the mass associated to the halo at the position $\bv x$. Since $\delta_M(t,\bv x)$ is a zero mean Gaussian random field with standard deviation $\sigma_M$, the probability that at a random point $\delta_M$ exceeds this threshold $\delta_{\rm c}$ is
\begin{equation}
p(\sigma_M) = \frac{1}{\sqrt{2 \pi}\sigma_M} \int_{\delta_{\rm c}}^\infty \exp\left(-\frac{x^2}{2 \sigma_M^2}\right)\:dx
= \frac{1}{2}\left[1-\rm erf \left(\frac{\nu}{\sqrt{2}}\right)\right]\:,
\end{equation}
where $\nu = \delta_{\rm c}/\sigma_M$ and the error function ${\rm erf}(x) = 2/\sqrt{\pi} \int_0^x e^{-y^2}dy$. Since a halo of mass $M$ has effectively swept up the mass within a comoving volume $V(M) = M/\bar{\rho}_0$, we will consider $\delta_{\rm c}$ at random positions $x_i$, $i = 1,2,\ldots$ such that their associated volumes $V(M)$ do not overlap. The fraction of such points with $\delta_M(t,\bv x_i) \geq \delta_{\rm c}$ for masses within $[M,M+dM]$ is then simply $(dp/dM)(\sigma_M) dM$. The \textbf{mass function} (or ``multiplicity function''), which is the mean number of halos of mass $M$ per unit comoving volume and unit mass, is then given by
\begin{equation}\label{eq:press-schechter}
\boxed{\frac{d\bar n_{\rm h}}{dM}(M) = -\frac{2}{V(M)}\frac{d p}{d M} = -2\frac{\bar{\rho}_0}{M}\frac{d p}{d \sigma_M}\frac{d \sigma_M}{d M} = - \sqrt{\frac{2}{\pi}}\frac{\bar \rho_0}{M \sigma_M} \frac{d \sigma_M}{d M}\:\nu\: e^{-\nu^2/2}\:,}
\end{equation}
where $\bar n_{\rm h}(M) = \langle n_{\rm h}({\bv x},M)\rangle$ is the mean comoving number density of halos of mass $M$. Note that we have multiplied the whole expression by an additional factor of 2 in order to be consistent with simulations (see the discussion below). The formula (\ref{eq:press-schechter}) was first derived by \cite{press1974} and is therefore called the ``Press-Schechter mass function''. To roughly obtain its explicit mass dependence, we use 
\begin{equation}
\frac{\nu(t)}{\sqrt{2}} = \frac{\delta_{\rm c}}{\sqrt{2}\sigma_M(t)} \supeq{(\ref{eq:scaling_relation_sigma_M})}{\simeq} A_\ast(t)  M^{\tfrac{3+n_{\rm eff}}{6}}\:,
\end{equation}
where $A_\ast(t) \propto D_+^{-1}(t)$ is a time dependent normalization factor and $n_{\rm eff}(M)$ the effective spectral index evaluated at the corresponding mass $M$.  Setting $\gamma(M) = 1 + n_{\rm eff}(M)/3$, it follows approximately
\begin{equation}
\frac{1}{\sigma_M}\frac{d \sigma_M}{d M} \simeq - \frac{\gamma}{2} \frac{1}{M}\:,
\end{equation}
since $n_{\rm eff}(M)$ is only a slowly varying function of mass. The Press-Schechter mass function (\ref{eq:press-schechter}) can then be expressed as
\begin{equation}\label{eq:press_schechter}
\frac{d\bar n_{\rm h}}{dM}(M) \simeq  \gamma \: \frac{  A_\ast(t)}{\sqrt{\pi}} \: \frac{\bar{\rho}_0}{M^2}\: M^{\gamma/2} \:\exp\left(-A_\ast^2(t) M^\gamma \right)\:.
\end{equation}
Thus, at a given time, it is proportional to $\rho_0 \propto \Omega_{\rm m} h^2$ and approximately scales like a power law in mass with an exponential cut off at the high mass end.

\begin{figure}[tp]\centering
  \includegraphics[scale = 0.55]{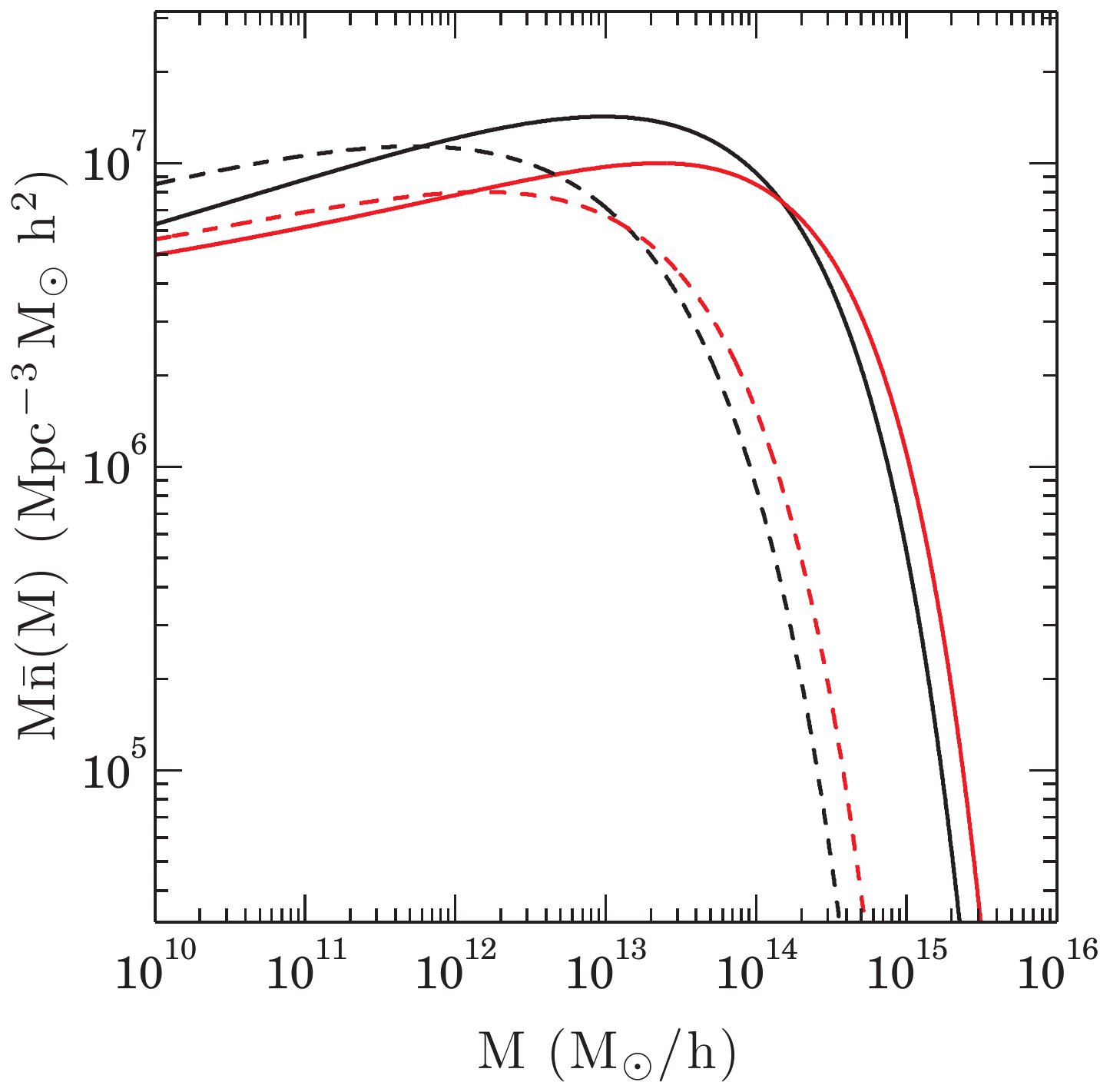}
  \caption{Mean mass density $\bar \rho_{\rm h}(M) = M\bar n_{\rm h}(M)$ of halos for two different redshifts. The mass density $\bar \rho_{\rm h}(M)$ instead of the mass function $d\bar n /dM$ is shown for clarity, as it allows to squeeze the $y$-axis. The black lines correspond to the Press-Schechter mass function (\ref{eq:press-schechter}) and the red lines to the fitting formula of \cite{pillepich2010}.  The solid lines refer to $z = 0$ and the dashed lines to $z = 1$. For the relation $\nu(M)$ a concordance cosmology was assumed and the transfer function $T(k)$ was taken from \cite{eisenstein1999}.}\label{fig:mass_function}
\end{figure}

What can we learn from this analysis? It became clear that the Press-Schechter approach cannot be regarded as a ``rigorous derivation''. It is not only based on very simplistic assumptions, but also needs certain ad-hoc modifications, such as the multiplication by an additional factor of 2. The initial assumptions do not only neglect to large part the nonlinear evolution of the density field, they also suffer from at least two major drawbacks: First, being based on the spherical top hat model we assumed that the collapse of the DM halos is a spherical symmetric process, while in reality the halos can have complicated three-axial shapes. Second, there is the ``peaks-within-peaks problem'' asserting that the Press-Schechter approach does not take into account whether a halo of a certain mass is included in a halo of some larger mass. Despite these caveats, the comparison between the Press-Schechter approach and numerical simulations showed that the former gives roughly the right shape of the mass function and is correct up to an order of magnitude (see Fig.~\ref{fig:mass_function}). In particular, it shows the exponential dependence on $\sigma_M$ (and thus on $\sigma_8$) at the high mass end.

Since the publication of the Press-Schechter mass function there has been great effort to deal with the previously mentioned caveats. For reviews of the most important improvements we refer, for example, to \cite{zentner2007} and \nocite{mo2010}Mo et al.~(2010, Sect.~7.2). Today, one typically uses simulation calibrated formulas or fitting formulas derived from simulations (e.g., \citealt{sheth1999,jenkins2001,sheth2001,reed2003,warren2006,tinker2008,pillepich2010}). In earlier work it was suggested that there might be a ``universal mass function'' (i.e., same functional form and numerical parameters) for different cosmologies and over a broad range of redshift. However, most recent studies have shown that if one aims at a precision of $\lesssim 5\%$ such a universal mass function cannot be found, neither for different cosmologies nor for a broad redshift range. For a $\Lambda$CDM-cosmolgy and for the redshift range $z \lesssim 1$, fitting formulas at a precision of a few percent in the mass range relevant for cosmological studies of the LSS are provided by \cite{tinker2008} and \cite{pillepich2010}. These accuracies should, however, be taken with with caution. Uncertainties in the halo mass function are not only introduced by the definition of halos in simulations (see, e.g., \citealt{white2001}, \citealt{cuesta2008}, and \citealt{more2011}), but also by effects due to the baryonic physics, which may cause larger deviations from the mass functions in pure DM simulations than the uncertainty stated above \citep[e.g.,][]{stanek2009,cui2012,balaguera2012}.

\subsection{Linear bias}

The Press-Schechter approach does not only open the door for an analytic calculation of the mean number density of DM halos, it also allows insight into how these DM halos are correlated in space. This leads to the concept of ``bias'' \citep{kaiser1984}.\footnote{There are different kinds of bias. Here we focus on the ``linear bias'' being the oldest and the most obvious of them. For a discussion of other kinds of bias, such as the ``assembly bias'' or the ``nonlinear bias'', we refer to Sect.~7.4 of \cite{mo2010}.}

In a first step, we express the auto-correlation function of halos of mass $M$ in terms of their comoving number density $n_{\rm h}(\bv x,M)$.
With the overdensity of halos of mass $M$ (cf.~Eq.~(\ref{eq:overdensity}))
\begin{equation}
\delta_{\rm h}(\bv x,M) = \frac{n_{\rm h}(\bv x,M) -\bar n_{\rm h}(M)}{\bar n_{\rm h}(M)}\:,
\end{equation}
where $\bar n_{\rm h}(M) = \langle n_{\rm h}(\bv x,M) \rangle$, the corresponding halo auto-correlation function is (cf.~Eq.~(\ref{correlationfunction}))
\begin{equation}\label{eq:correlatinfunction2}
\xi_{\rm hh}(r,M) = \langle \delta_{\rm h}(\bv x,M) \delta_{\rm h}(\bv x',M) \rangle = \frac{\langle n_{\rm h}(\bv x,M) n_{\rm h}(\bv x',M)\rangle}{\bar n_{\rm h}(M) ^2}-1\:,
\end{equation}
where $r = |\bv x'-\bv x|$.
This formulation has the straightforward interpretation of the correlation function as a measure of the excess of halo-pairs at separations $r$ compared to the mean number density of the halos. In the following, we try to relate the halo correlation function $\xi_{\rm hh}$ to the linear correlation function $\xi_{\rm lin}$, which is the Fourier transform of the linear power spectrum (\ref{eq:lin_power_spectrum}).

The Press-Schechter mass function gives us the mean number density $\bar n_{\rm h}$ of halos, but it does not tell us how it varies from place to place or how it depends on its cosmic environment. The probably simplest way to show how the local number of halos depends on the environment is the \textbf{Peak-Background split} \citep{bardeen1986,cole1989,mo1996}. Suppose we have an overdensity field $\delta(\bv x)$ that can be decomposed into a short wavelength part $\delta_{\rm h}$ and into a long wavelength background part $\delta_{\rm b}$ such that
\begin{equation}
\delta = \delta_{\rm h} + \delta_{\rm b}\:.
\end{equation}
The short wavelength perturbation $\delta_{\rm h}$ is the progenitor of the halos we want to study and the long wavelength perturbation $\delta_{\rm b}$ plays the role of a smooth background density being in the linear regime, i.e., it holds $\delta_{\rm b} \ll 1$. We will assume that $\delta_{\rm b}$ is essentially constant over the region where $\delta_{\rm h}$ collapses into the halos. The effect of $\delta_{\rm b}$ is to perturb the critical threshold that the linear part of $\delta_{\rm h}$ has to reach for a collapse. If the linear part of $\delta_{\rm h}$ reaches the effective threshold
\begin{equation}
\tilde \delta_{\rm c} = \delta_{\rm c}-\delta_{\rm b}\:,
\end{equation}
the linear part of the total perturbation $\delta$ reaches the actual threshold $\delta_{\rm c}$ that is needed for a structure to collapse. The effective threshold $\tilde \delta_{\rm c}$ depends on the linear background field $\delta_{\rm b}$ and causes the fluctuation of a given strength to collapse at different places at slightly different times. This causes the local number density $n_{\rm h}$ to vary from place to place depending on $\delta_{\rm b}$.

A quantitative estimation of this effect can be gained by using the explicit form of the Press-Schechter mass function (\ref{eq:press-schechter})
\begin{equation}
\frac{d\bar n_{\rm h}}{dM}(M,\tilde \delta_{\rm c}) \propto \nu(\tilde \delta_{\rm c}) e^{-\nu^2(\tilde \delta_{\rm c})/2}\:.
\end{equation}
Since $\delta_{\rm b} \ll 1$, we can expand $\bar n_{\rm h}$ at first order after $\delta_{\rm b}$ yielding
\begin{equation}\label{eq3:expansion_of_n}
\frac{dn_{\rm h}}{dM}(M,\delta_{\rm b}) = \frac{d\bar n_{\rm h}}{dM}(M)+\frac{d^2\bar n_{\rm h}}{dMd \tilde \delta_{\rm c}}\frac{d \tilde \delta_{\rm c}}{d \delta_{\rm b}}\delta_{\rm b} = \frac{d\bar n_{\rm h}}{dM}(M)\left[1+\frac{\nu^2-1}{\nu \sigma}\:\delta_{\rm b}\right]\:.
\end{equation}
Thus we find the result
\begin{equation}
\delta_{\rm h}(M) = \frac{n_{\rm h}(M,\delta_{\rm b})-\bar n_{\rm h}(M)}{\bar n_{\rm h}(M)} = \kfrac{\frac{dn_{\rm h}}{dM}(M,\delta_{\rm b})-\frac{d\bar n_{\rm h}}{dM}(M)}{\frac{d\bar n_{\rm h}}{dM}(M)} = \frac{\nu^2-1}{\nu \sigma}\: \delta_{\rm b}\:.
\end{equation}

In the analysis so far, we have assumed that the positions of the halos that form during the growth of $\delta$ remain unchanged relative to each other during the evolution of $\delta$. This is, however, a very crude approximation. To obtain a more realistic picture, we have to account for the fact that the region of the background density $\delta_{\rm b}$ shrinks during its linear growth and thus moves the halos that have been created closer together as time goes by. This means that the halos which collapse at the time when the background field reaches the strength $\delta_{\rm b}$ were farther apart at earlier times. Since the background fluctuation $\delta_{\rm b}$ was initially a very small perturbation $\delta_{\rm i} \ll \delta_{\rm b}$, the current region of the background density was once smaller by a factor $\rho_{\rm b}/\bar \rho(1+\delta_{\rm i}) \simeq \rho_{\rm b}/\bar \rho = 1+\delta_{\rm b}$. Taking this factor into account we obtain a more accurate relation between the overdensity of halos and the background
\begin{equation}\label{eq3:relation_delta_h_delta_b}
\delta_{\rm h}(M)=\left(1+\delta_{\rm b}\right)\frac{\nu^2-1}{\nu \sigma}\: \delta_{\rm b} \simeq \left(1+\frac{\nu^2-1}{\nu \sigma}\right) \delta_{\rm b}\:,
\end{equation}
where we have neglected terms of second order in $\delta_{\rm b}$.

Since we know the correlation of the linear density field $\delta_{\rm lin}$, we are finally able to compute the correlation of halos. Reckoning the definition of the correlation function and the fact that $\delta_{\rm b}$ played the part of the linear theory, it follows immediately from Eq.~(\ref{eq3:relation_delta_h_delta_b}) that
\begin{equation}\label{eq3:xi_hh}
	\boxed{\xi_{\rm hh}(r,M) = b^2(M)\: \xi_{\rm lin}(r)\:,\qquad b(M) = 1+\frac{\nu^2-1}{\nu \sigma}\:,}
\end{equation}
where $b$ is called \textbf{linear bias}. This result shows that DM halos are biased tracers of the underlying mass field with a bias depending on the mass of the halo. The higher the mass the stronger the bias. Note that this result is only valid for scales $r$ within the linear regime \citep{seljak2000,cooray2002}. The bias for the nonlinear regime would be scale dependent.

\begin{figure}[tp]\centering
  \includegraphics[scale = 0.53]{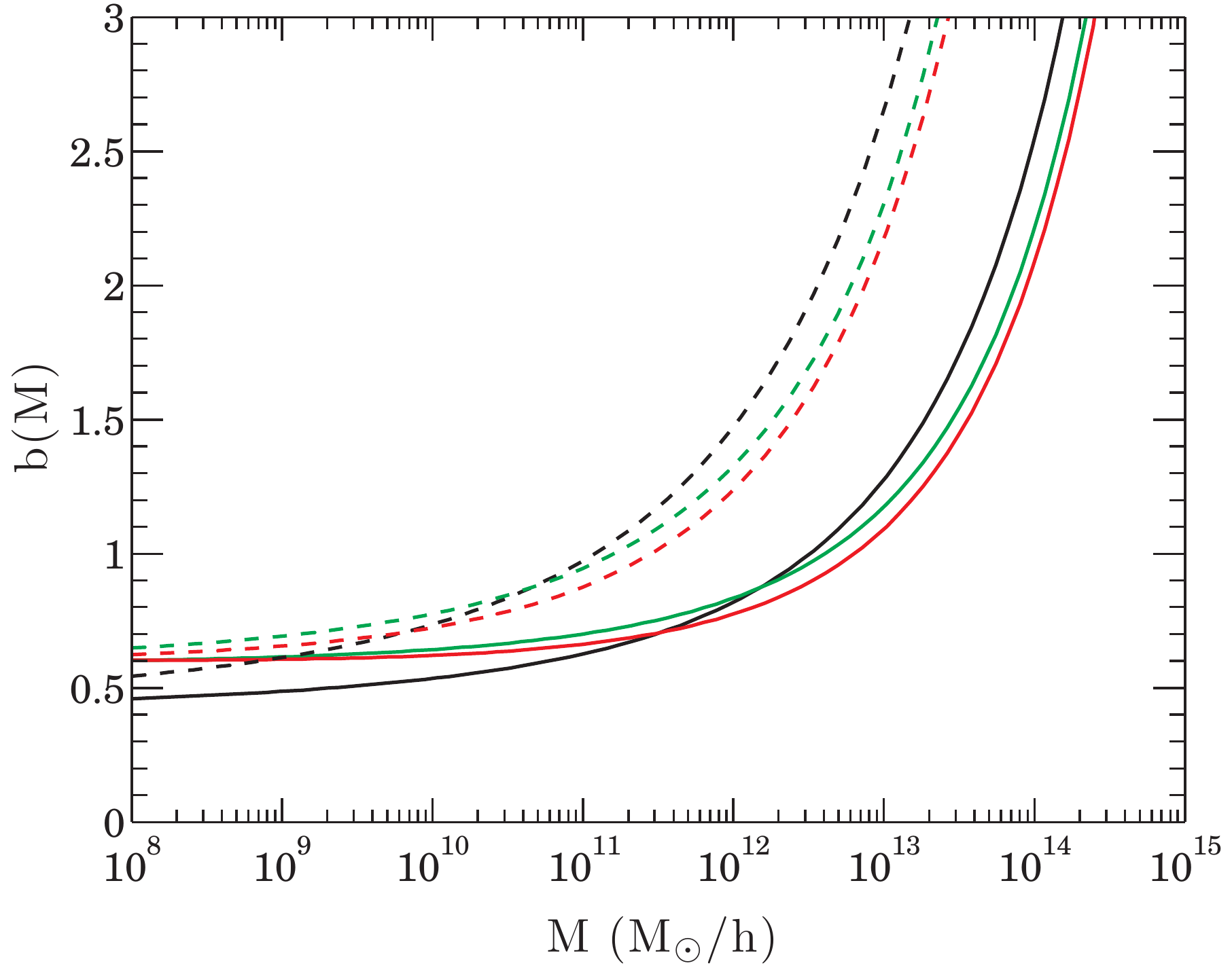}
  \caption{Bias of DM halos as a function of mass $M$ for two different redshifts. The black lines correspond to the linear bias given by Eq.~(\ref{eq3:xi_hh}), the green lines to the fitting formula of \cite{tinker2010}, and the red lines to the fitting formula of \cite{pillepich2010}. The solid lines refer to $z = 0$ and the dashed lines to $z = 1$. For the displayed mass, the accuracy of the linear bias is $\lesssim \! 20\%$ and the relative difference between the two fitting formulas is $\lesssim \!  5\%$. For the relation $\nu(M)$ a concordance cosmology was assumed and the transfer function $T(k)$ was taken from \cite{eisenstein1999}.}\label{fig:bias}
\end{figure}

The relevance of this simple bias model is similar to that of the Press-Schechter mass function. It allows an understanding of the general behavior of the bias within our cosmological framework, but is less suited as a tool for a precision cosmology. A comparison of our simple model with the fitting formuals of \cite{tinker2010} and \cite{pillepich2010} are shown in Figure \ref{fig:bias} for two different redshifts.
The deviations of our simple model from the results of numerical simulations are $\lesssim \! 20\%$. Observationally, the linear bias $b$ was first detected by \cite{bahcall1983} using Abell clusters. The growth of $b(M)$ with halo mass $M$ was recently measured in SDSS in the low-redshift universe including a detailed comparison to the $\Lambda$CDM model \citep[e.g.,][]{berlind2006b,wang2008} and the effect was also confirmed at high redshift \citep{knobel2012b}. These analyses were performed by interpreting galaxy clusters and galaxy groups as DM halos, which directly leads us to the next section.

\subsection{Halo model}

So far, we have only considered the DM part of the universe. Unfortunately, it is hardly possible to measure DM halos directly. So it is an important question how to connect the theory that has been developed in this chapter to the ``bright part'' of the universe which can be easily observed. How do the galaxies fit into this framework?

To answer this question we would need a theory on how galaxies form within the DM framework discussed so far. By now, many details of this process are not well understood.\footnote{The reader is, for example, referred to \cite{silk2012}, who provide a recent review on galaxy evolution with a special focus on open problems.} There is, however, the general agreement that galaxies form at the centers of DM halos as baryonic matter falls into the halo and cools \citep[][cf.~the scenario discussed in Sect.~\ref{sec:dark_matter_and_baryons}]{white1978}. These are called ``central galaxies''. During the evolution of the LSS, some of the DM halos merge to build larger halos containing several galaxies. If a big halo merges with a small one, the galaxies of the small halo become ``satellite galaxies'' within the resulting halo, while the central galaxy of the big halo usually becomes the central galaxy of the resulting halo. Central galaxies and satellite galaxies may evolve differently over cosmic time owing to their different places within the halo. In the course of time more and more galaxies get assembled in DM halos. A DM halo containing several galaxies is called a ``galaxy group'' or, in the case of a very huge halo containing hundreds of galaxies, a ``galaxy cluster''.

This is the theoretical foundation of the \textbf{halo model} \citep{peacock2000,seljak2000}, which is an intriguingly simple attempt to describe correlation function of galaxies and galaxy groups in the linear and nonlinear regime. It is based on a few straightforward assumptions in the light of the framework that we have sketched above (see, e.g., \citealt{cooray2002} Ch.~4 and 5 for a review):
First, all galaxies reside within halos according to a certain spherical density profile, where there is always a galaxy at the center of the halo. Second, the distribution of the number of galaxies in a halo $p(N|M)$ and their spatial distribution depend for a given galaxy sample only on the mass $M$ of the halo. The distribution $p(N|M)$ is called \textbf{halo occupation distribution} (HOD) and is the main ingredient to the halo model.\footnote{Alternatively to the HOD, some authors use the conditional luminosity function (CLF) $\Phi(L|M)dL$ instead being the average number of galaxies with luminosity $L$ residing in halos of mass $M$. The two approaches are equivalent to each other if the galaxy sample in the case of the HOD is selected by luminosity (see, e.g., the comments in \citealt{skibba2009}, Sect.~1, and \citealt{zehavi2011}, Sect.~2.3).} The dependence of the galaxy populations from the halo mass is the reason why different galaxy population cluster differently. Further common assumptions are that the galaxy density profile within halos follows that of the DM or that central and satellite galaxies constitute different galaxy populations.

All of these assumptions are reasonable to some extent or are warranted by observations (e.g., \citealt{cooray2002}), but they also show the limitations of the halo model. The assumption, for instance, that the spherical distribution of galaxies within halos depends only on the Mass of the halo is certainly a mere approximation, since it is well known that even the DM profile of a halo of a certain mass varies from halo to halo within some range. There are also discussions in the literature about the dependency of the HOD on the cosmic environment in addition to the mass of the halo \citep[e.g.,][]{Gil-Marin2011,croft2012}.

In the following, we develop the formalism of the halo model keeping it as simple as possible. Suppose we have two samples of galaxies $g$ and $g'$, respectively, with no intersections between the samples. From the definition of the correlation function in the form of Eq.~(\ref{eq:correlatinfunction2}) and the assumption that all galaxies reside within DM halos, it follows immediately that the the cross-correlation function $\xi_{\rm gg'}$ between these samples divides into two terms, i.e.,
\begin{equation}
\xi_{\rm gg'}(r) = \xi_{\rm gg'}^{\rm (h1)}(r)+\xi_{\rm gg'}^{\rm (h2)}(r)\:,
\end{equation}
where the \textbf{one-halo term} $\xi_{\rm gg'}^{\rm (h1)}$ contains the contribution from galaxy pairs within the same halo and the \textbf{two-halo term} $\xi_{\rm gg'}^{\rm (h2)}$ from pairs within different halos. We can derive approximate expressions for these two terms from the basic assumptions.

For the one-halo term, we get
\begin{equation}\label{eq3:1h-term}
\xi_{\rm gg'}^{\rm (h1)}(r) \simeq \int \frac{d\bar n_{\rm h}}{dM}(M) \frac{\langle N_{\rm g} N_{\rm g'}|M \rangle}{\bar n_{\rm g} \bar n_{\rm g'}} \int u_{\rm g}(|\bv r'|,M)u_{\rm g'}(|\bv r'-\bv r|,M)\: {dr'}^3\:dM\:,
\end{equation}
where $d\bar n_{\rm h}/dM$ is the mass function of halos, $\bar n_{\rm g}$ and $\bar n_{\rm g'}$ are the mean number densities of the galaxies, $u_{\rm g}(r,M)$ and $u_{\rm g'}(r,M)$ are for either galaxy sample the normalized mean radial galaxy density profiles within halos of mass $M$, and $\langle N_{\rm g} N_{\rm g'}|M \rangle$ is the mean number of pairs determined by the HODs of $g$ and $g'$. It is a common practice to set $u_{\rm g} \simeq u_{\rm g'} \simeq u_{\rm h}$ with $u_{\rm h}(r,M)$ the normalized mean DM density profile for the halos of mass $M$. In Eq.~(\ref{eq3:1h-term}) we have not accounted for the assumption that there is always a galaxy at the center of the halo. An analytic approximation to deal with this complication is given in \cite{cooray2002}, whereby the integral over the convolution of the density profiles reduces to $u_{\rm g}$ if $\langle N_{\rm g} N_{\rm g'}|M \rangle \lesssim 1$. Since the dependence of $\xi_{\rm gg'}^{\rm (h1)}$ on $r$ comes only from the density profiles it is clear that it contributes only on scales comparable to the extension of the halos ($\lesssim \! 1$ Mpc). On larger scales it can be neglected. Since convolutions become simple multiplications in Fourier space, the one-halo term becomes much simpler if expressed by means of the power spectrum. This is why the formalism is usually developed in Fourier space.

The two-halo term can be approximated by
\begin{equation}\label{eq:two-halo-term}
\xi_{\rm gg'}^{\rm (h2)}(r) \simeq b_{\rm g}b_{\rm g'} \:\xi_{\rm lin}(r)\:,
\end{equation}
where we have introduced the linear bias $b_{\rm g}$ and $b_{\rm g'}$ for the two galaxy species, respectively, as
\begin{equation}
b_{\rm g} = \int \frac{d\bar n_{\rm h}}{dM}(M) b(M) \frac{\langle N_{\rm g}|M\rangle}{\bar n_{\rm g}} \: dM\:,\qquad b_{\rm g'} = \int \frac{d\bar n_{\rm h}}{dM}(M) b(M) \frac{\langle N_{\rm g'}|M\rangle }{\bar n_{\rm g'}} \: dM\:
\end{equation}
with $\langle N_i |M \rangle$, $i = \{g, g'\}$, the mean numbers of galaxies in halos of mass $M$ determined by the HODs. In Eq.~(\ref{eq:two-halo-term}) we have neglected the extensions of the halos and formally just placed all galaxies at the centers of the halos which is a good approximation for scales much larger than the typical extension of a halo.

As for the case of the cross-correlation function we can similarly write down the correlation function for the following three special cases:
\begin{itemize}
	\item For the auto-correlation function $\xi_{\rm gg}$ of the species $g$ the one- and two-halo terms reduce to
\begin{equation}
	\xi_{\rm gg}^{\rm (h1)}(r) \simeq \int \frac{d\bar n_{\rm h}}{dM}(M) \frac{\langle N_{\rm g} (N_{\rm g}-1)|M\rangle}{\bar n_{\rm g}^2} \int u_{\rm g}(|\bv r'|,M)u_{\rm g}(|\bv r'-\bv r|,M)\: {dr'}^3\:dM\:
\end{equation}
and
\begin{equation}
	\xi_{\rm gg}^{\rm (h2)}(r) \simeq b_{\rm g}^2 \:\xi_{\rm lin}(r)\:, \qquad b_{\rm g} = \int \frac{d\bar n_{\rm h}}{dM}(M) b(M) \frac{\langle N_{\rm g}|M\rangle}{\bar n_{\rm g}}\: dM\:.
\end{equation}
Note that the modification in the one-halo term $\langle N_{\rm g} \left(N_{\rm g'}-1\right) \rangle$ is due two the fact that the autocorrelation function is a cross-correlation between two samples with intersecting points.

\item For the cross-correlation between the galaxy species $g$ and halos $h$ in the range $M_{\rm min} < M < M_{\rm max}$, the one- and two-halo terms become
\begin{equation}
	\xi_{\rm gh}^{\rm (h1)}(r) \simeq \kfrac{\int_{M_{\rm min}}^{M_{\rm max}} \frac{d\bar n_{\rm h}}{dM}(M) \frac{\langle N_{\rm g} |M\rangle}{\bar n_{\rm g}} u_{\rm g}(r,M)\:dM}{\int_{M_{\rm min}}^{M_{\rm max}} \frac{d\bar n_{\rm h}}{dM}(M)\:dM}\:,\qquad \xi_{\rm gh}^{\rm (h2)}(r) \simeq b_{\rm g}\: b_{\rm h} \:\xi_{\rm lin}(r)
\end{equation}
with
\begin{equation}
	b_{\rm g} = \int \frac{d\bar n_{\rm h}}{dM}(M) b(M) \frac{\langle N_{\rm g}|M\rangle}{\bar n_{\rm g}}\: dM\:,\qquad b_{\rm h} = \kfrac{\int_{M_{\rm min}}^{M_{\rm max}} \frac{d\bar n_{\rm h}}{dM}(M) b(M) \: dM}{  \int_{M_{\rm min}}^{M_{\rm max}} \frac{d\bar n_{\rm h}}{dM}(M)\:dM}\:.
\end{equation}

\item The DM halo auto-correlation function $\xi_{\rm hh}$(r) consists for $r\neq 0$ only of the two-halo term, i.e.,
\begin{equation}
\xi_{\rm hh}(r) = \xi_{\rm hh}^{\rm (h2)}(r) \simeq b_{\rm h}^2\:\xi_{\rm lin}(r)\:,\qquad b_{\rm h} = \kfrac{\int  \frac{d\bar n_{\rm h}}{dM}(M) b(M)\:dM}{\int  \frac{d\bar n_{\rm h}}{dM}(M)\:dM}\:.
\end{equation}

\item For the autocorrelation function of DM particles $\xi_{\rm dd}$ we have
\begin{equation}
	\xi_{\rm dd}^{\rm (h1)}(r) \simeq \int \frac{d\bar n_{\rm h}}{dM}(M)\frac{M}{\bar \rho_0} \int u_{\rm h}(|\bv r'|,M)u_{\rm h}(|\bv r'-\bv r|,M)\: {dr'}^3\:dM\:
\end{equation}
and
\begin{equation}
	\xi_{\rm dd}^{\rm (h2)}(r) \simeq b_{\rm d}^2 \:\xi_{\rm lin}(r)\:,\qquad b_{\rm d} = \int \frac{d\bar n_{\rm h}}{dM}(M) \frac{M}{\bar \rho} b(M)\:dM\:.
\end{equation}
\end{itemize}

In total, we find the neat result that in the linear regime any correlation function is proportional to $\xi_{\rm lin}$ with a constant bias. This is, however, only true for the linear regime, where the one-halo term is negligible. As soon as the one-halo term becomes important, we enter the nonlinear regime and the bias becomes scale dependent \citep{seljak2000,cooray2002}. This means, for instance, that the galaxy-galaxy correlation function $\xi_{\rm gg}$ is not just a scaled version of the DM correlation function $\xi_{\rm dd}$.

\begin{figure}[tp]\centering

\begin{minipage}[c]{0.55\textwidth}
\centering\includegraphics[scale = 0.82]{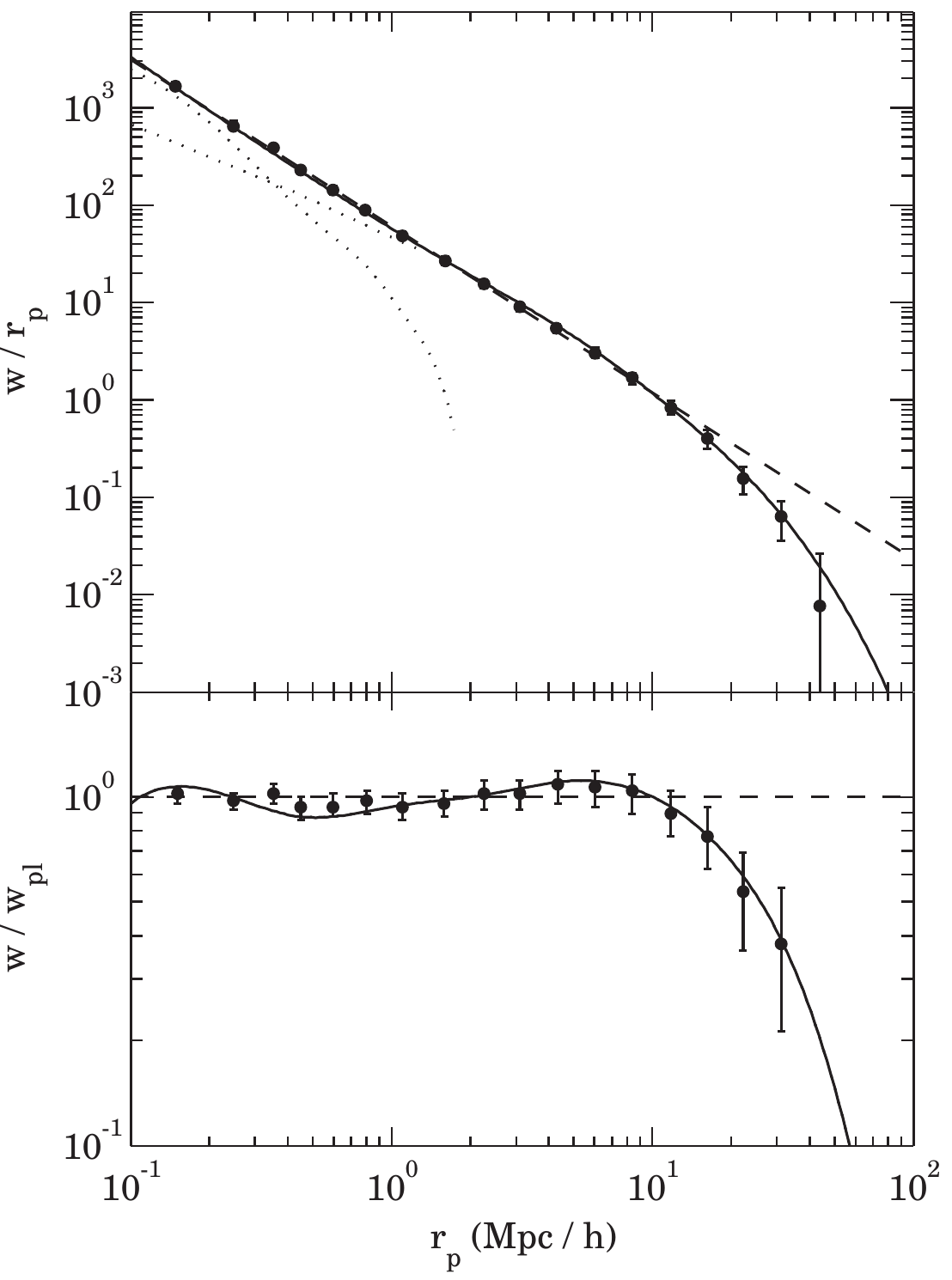}
\end{minipage}
\hfill
\begin{minipage}[c]{0.44\textwidth}
\centering\includegraphics[scale = 0.94]{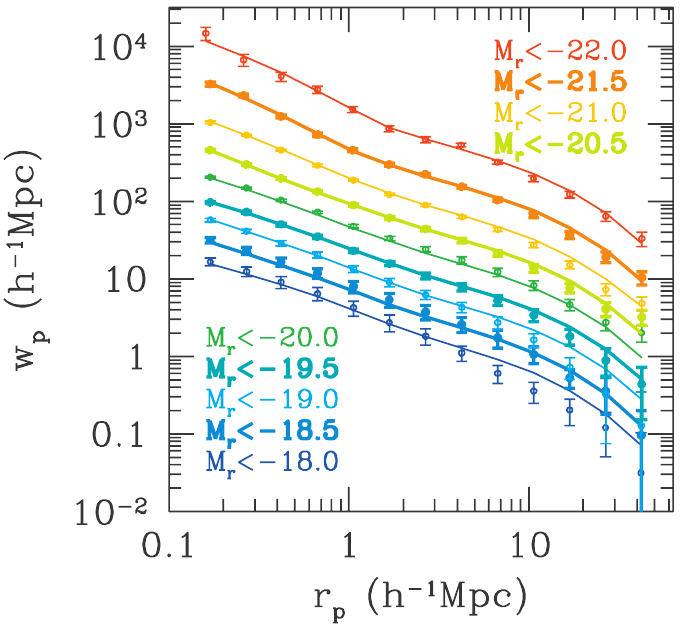}
\centering\includegraphics[scale = 0.94]{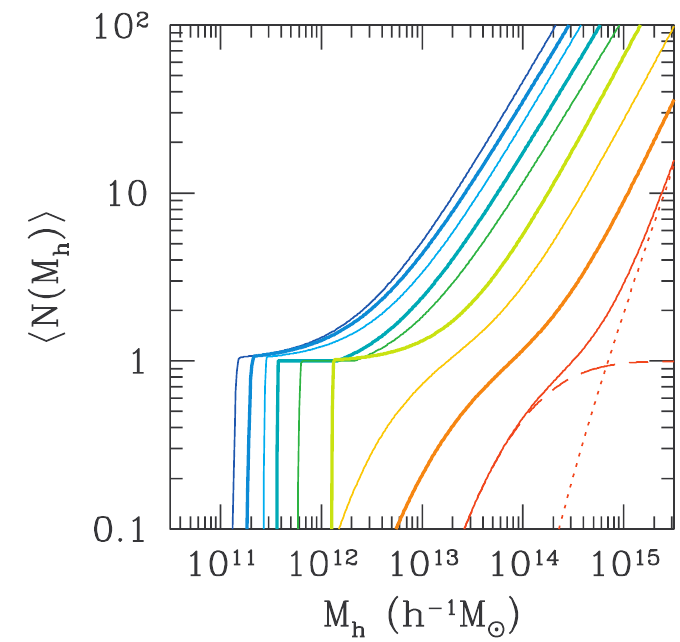}
\end{minipage}

\caption{Comparison of the halo model to data from 2dfGRS and SDSS in terms of the projected correlation function $w(r_{\rm p})$.\\
\emph{Left panels} (adapted from \citealt{collister2005}): In the upper panel the data points represent the observed projected correlation function (divided by $r_{\rm p}$)
for galaxies within 2dFGRS, the dot-dashed line is a power-law fit to the data, the solid line is the prediction from the halo model, and the dotted lines show the corresponding one- and two-halo terms. It can be clearly seen how the one- and two-halo terms approximately sum up to a power law in the range from $0.1\: h^{-1}$ to $10\: h^{-1}$ Mpc. Note that the solid line is not a fit to the data, but shows the prediction of the halo model for the HOD that was measured by means of galaxy groups within 2dfGRS. Thus, the curve constitutes a nice self-consistence test of the halo model. The lower panel shows the ratio of the data points and the solid curve to the power law fit. 
\\
\emph{Right panels} (taken from \citealt{zehavi2011}, reproduced by permission of the AAS): The upper panel shows the observed projected correlation functions (data points) and the corresponding halo model fits (solid lines) for different volume-limited galaxy samples from SDSS as indicated within the panel. For the brightest samples (red curves) the transition from the one-halo to the two-halo term around $r_{\rm p} \simeq 1.5 \: h^{-1}$ Mpc is clearly visible. Note that the correlation functions are each staggered by 0.25 dex for clarity. The lower panel shows the corresponding HODs as a function of halo mass $M_{\rm h}$ for the different galaxy samples. It is shown for the brightest sample how the HOD combines from the HOD of central (dashed line) and of satellite (dotted line) galaxies.}
\label{fig:halo_model}
\end{figure}

Unfortunately, the halo model cannot be tested by measuring the correlation function $\xi(r)$ for galaxies (or other objects) directly. Since the distance to galaxies is measured using their redshift $z$, we can only determine the positions of the galaxies in comvoving redshift space, but not in comoving real space (see Sect.~\ref{sec:peculiar_velocities}). That is, the positions of galaxies include a small random component along the line of sight and hence the corresponding correlation function appears distorted. To deal with this difficulty, we can estimate the correlation function for galaxy separations parallel and perpendicular to the line of sight. If $\bv s$ is the separation vector of galaxy pairs in redshift space, we can estimate $\xi(|\bv s_\parallel|,|\bv s_\perp|)$ for $\bv s = \bv s_\parallel + \bv s_\perp$, where $\bv s_\parallel$ is the component parallel and $\bv s_\perp$ perpendicular to the line of sight.
By integrating this correlation function along the line of sight, we obtain the \textbf{projected correlation function}
\begin{equation}
w(r_p) = \int_{-\infty}^{\infty} \xi(|\bv s_\parallel|,|\bv s_\perp|)\:ds_\parallel = 2 \int_{0}^{\infty} \xi(\pi,r_p)\:d\pi\:,
\end{equation}
where we denoted $\pi = |\bv s_\parallel|$ and $r_p = |\bv s_\perp|$, which is independent of the redshift space distortions. The projected correlation function can then either be converted to the real space correlation function $\xi(r)$ by means of an inverse Abel transform \citep[see, e.g.,][Sect.~16.5]{peacock1999} or can be directly compared to the corresponding project correlation function from the halo model. The latter is shown in Figure \ref{fig:halo_model} for the projected correlation functions from the big low-redshift galaxy surveys 2dfRGS and SDSS. It is evident that the halo model is very successful in reproducing the correlation function in both the linear and nonlinear regime for different galaxy samples.


\chapter{General relativistic treatment of linear structure formation}\label{sec:general_relativistic_treatment}

In the previous chapter, we presented the theory of structure formation in the context of Newtonian physics. However, this theory is only valid well inside the horizon and only in a spatially flat universe. If we want to understand how initial perturbations were created during inflation and how they evolved to the present time, we also have to study perturbations outside the horizon which is only possible by a purley general relativistic treatement.

In this chapter, we give an introduction to the full general relativistic treatment of linear structure formation. Our goal is to complement and confirm the results from our study of perturbations in the Newtonian regime and to understand the ingredients of modern cosmological codes like CMBfast, which produce the most accurate transfer functions $T(k)$ for cosmology. We will see how the Newtonian treatment in Section \ref{sec:linear_theory} naturally arises as a limiting case for perturbations well inside the horizon, and in the next chapter we will apply the general relativistic framework developed in this chapter to derive the power spectrum of perturbations that is created by the simplest models of inflation.

The term ``linear theory'' formally means that we perform our calculations with a set of perturbation quantities which are very small (of the order of $10^{-5}$ in the early universe such as the relative amplitude of the CMB temperature fluctuations) and always keep only terms that are linear in perturbation quantities. Due to the smallness of the perturbations, this approach is very accurate in the early universe, and an immediate consequence of this procedure is that the resulting field equations and equations of motion are linear differential equations. This simplifies the analysis a great deal and enables us to solve these equations independently of the (stochastic) initial conditions that were created during inflation. The corresponding theory was first developed by \cite{lifshitz1946} in a remarkable paper treating the problem of relativistic structure formation with impressive generality.\footnote{\cite{bertschinger1995} commented on Lifshitz's paper: 
\begin{quotation}
This classic paper was remarkably complete, including a full treatment of the scalar, vector, and tensor decomposition in open and closed universes and a concise solution to the gauge mode problem; it presented solutions for perfect fluids in matter- and radiation-dominated universes; and it contrasted isentropic (adiabatic) and entropy fluctuations.
\end{quotation}
} In the following, we will be mainly guided by the discussion in Seljak (unpublished lecture notes), \cite{durrer2008}, and \cite{weinberg2008}.

We will entirely stick to the case of a spatially flat universe (i.e., $K = 0$). This condition further simplifies our calculations enormously, since it allows us to decompose the perturbations into the familiar Fourier modes. This restriction is justified by the measured 95\% confidence limit on the present day curvature being $-0.0133 < \Omega_K < 0.0084$ (see Table \ref{tab:cosmological_parameters}) and by the fact that according to the discussion in Section \ref{sec:connection_to_the_standard_model} the early universe was even much flatter than the present day universe. For a treatment of linear perturbation theory in non-flat universes we refer, for instance, to \cite{kodama1984}, \cite{hu1998}, and \nocite{durrer2008}Durrer (2008, Ch.~2 and App.~9).

\section{Perturbations}

Regarding the universe as perfectly homogeneous and isotropic leads to the FLRW framework that was discussed in Chapter \ref{sec:homogeneous_and_isotropic_universe}. In this case the universe is approximated by a manifold $\bar{\mathcal{M}}$ with a high degree of symmetry whose metric is the Robertson-Walker metric. We denote this manifold by a bar to indicate that it constitutes the smooth background universe and any quantity associated to this background will also be indicated by a bar. For our general relativistic discussion it is convenient to use conformal time $\tau$ (see Eq.~(\ref{eq:conformal_time})) instead of cosmic time $t$ and natural units, i.e., $c = \hbar = 1$. In this case, the Robertson-Walker metric takes the form (\ref{eq:robertson_walker_metric4}) and reduces for $K=0$ to the very simple expression
\begin{equation}\label{eq:ds_2}
	ds^2 = \bar g_{\mu\nu} dx^\mu dx^\nu = a^2(\tau) \left(-d\tau^2 + \bar \gamma_{ij}\:dx^i dx^j \right),
\end{equation}
where $\bar \gamma_{ij} = \delta_{ij}$ for spatial cartesian comoving coordinates $\bv x$. The energy-momentum tensor $T_{\mu\nu}$ then takes automatically the form of an ideal fluid, which reduces in these coordinates to 
\begin{equation}\label{eq:tmunu_bar}
	\bar T_{\mu\nu} = \left(\bar \rho + \bar p \right)\bar u_\mu \bar u_\nu + \bar p \:\bar g_{\mu\nu} = \frac{1}{a^2}{\rm diag} \left(\bar \rho, \bar p,\bar p,\bar p  \right),
\end{equation}
where $\bar \rho(\tau)$, $\bar p(\tau)$, and $\bar u^\mu(\tau) = -a^{-2}(\tau)\:\bar u_\mu(\tau) = a^{-1}(\tau)(1,0,0,0)$ are the energy density, the pressure, and the 4-velocity of the fluid, respectively. The dynamics of the background universe are conveniently expressed in terms of the \textbf{conformal Hubble parameter} $\mathcal H = \dot a/a = Ha$, where the derivative is taken with respect to conformal time $\tau$. The Friedmann equations (\ref{eq:friedmann_equation}) and (\ref{eq:second_friedmann_equation}) then take the form
\begin{equation}\label{eq:friedmann_equations_conformal}
	\mathcal{H}^2 = \frac{8 \pi G}{3}  \bar \rho \:a^2\:, \quad\quad\quad \dot{\mathcal{H}} = - \frac{4 \pi G}{3} a^2 \left(\bar \rho + 3 \bar p \right)\:,
\end{equation} 
and the equation of motion (\ref{eq:equation_of_motion_total}) becomes
\begin{equation}\label{eq:equation_of_motion_total_2}
\dot{\bar\rho} = -3 \mathcal{H} \left(\bar \rho  +  \bar p\right)\:.
\end{equation}
Using Eq.~(\ref{eq:rho_of_t}), it is easy to see that for a constant equation of state $w$
\begin{equation}\label{eq:a_tau}
a(\tau) \propto \tau^{2/(1+3w)}
\end{equation}
is a solution to the Friedmann equation (\ref{eq:friedmann_equations_conformal}).\footnote{Similarly to the derivation of the Eqs.~(\ref{eq:a_t_matter}) and (\ref{eq:a_t_radiation}), we assume here that the timespan, when the universe was \textit{not} dominated by the current fluid with the constant equation of state $w$, is negligible compared to the time $\tau$. This approximation is acceptable for the radiation dominated epoch as well as for the matter dominated epoch (see the discussion in Sect.~\ref{sec:timeline}).}

\subsection{Definition of the perturbations}\label{sec:perturbations_definion}

If we want to consider deviations from the homogeneous and isotropic background, we can no longer choose coordinates such that the metric $g_{\mu\nu}$ takes the form (\ref{eq:ds_2}). That is, our universe will correspond to a different manifold $\mathcal{M}$ whose metric is, in general, a complicated function of the coordinates. We will assume that the deviations from our smooth background are small enough so that we can choose coordinates such that our metric $g_{\mu\nu}$ is ``close'' to the metric $\bar g_{\mu\nu}$ of the smooth background.\footnote{Being given observational data within a ``lumpy universe'' one first has to identify the corresponding ``optimal'' FLRW universe that constitutes the background. This is known as the so-called ``fitting problem'' and is related to the issue of averaging  and backreaction (see, e.g., \citealt{clarkson2011} for a general discussion).} However, when comparing these two quantities we encounter the problem that these two metrics correspond to different manifolds and hence mathematical operations like addition, subtraction or transformation laws are not well defined. There are different ways to solve this problem mathematically (see, e.g., \citealt{malik2008} for a discussion). We will adopt the following simple approach:\footnote{This approach is, for example, also pursued in \cite{weinberg2008} and \cite{mukhanov2005}, but without making it very explicit.} if $x$ are any coordinates on $\mathcal{M}$, we will require $\bar g_{\mu\nu}(x)$ to have the fixed functional form (\ref{eq:ds_2}). That is, $\bar g_{\mu\nu}(x)$ is \emph{no} geometrical object on $\mathcal{M}$. If we perform a coordinate transformation $x \rightarrow \tilde x$ on $\mathcal{M}$, $\bar g_{\mu\nu}(\tilde x)$ will still have the functional form (\ref{eq:ds_2}) on the new coordinates $\tilde x$.

Adopting this approach, we can define the metric perturbation
\begin{equation}\label{eq:g_munu_split}
\delta g_{\mu \nu}(x) \equiv g_{\mu\nu}(x) - \bar g_{\mu\nu}(x)
\end{equation}
and we assume that there are coordinates $x = (\tau,\bv x)$ on $\mathcal{M}$ such that $|\delta g_{\mu\nu}(x)| \ll \max_{\mu\nu}|\bar g_{\mu\nu}(x)| = a^2(\tau)$. It should be stressed that this splitting of $g_{\mu\nu}$ into a ``background'' $\bar g_{\mu\nu}$ and a ``perturbation'' $\delta g_{\mu\nu}$ is a \emph{non-covariant} procedure, since $\bar g_{\mu\nu}$ is not a geometrical object on $\mathcal{M}$. As a consequence, the perturbation $\delta g_{\mu\nu}$ is no geometrical quantity either, i.e., it will not transform like a 4-tensor, which will lead to the gauge transformations discussed in Section \ref{eq:gauge_transformations}. In a similar way, we can define for any geometrical obejct $q$ on $\mathcal{M}$ its perturbation
\begin{equation}
\delta q(x) \equiv q(x)- \bar q(x)\:, 
\end{equation}
where $\bar q$ is the corresponding quantity on the smooth background $\bar{\mathcal{M}}$ having a fixed functional form of the coordinates. For the same reason as for the metric perturbation, the perturbation $\delta q$ is no geometrical quantity and it will not transform, in general, like a 4-scalar, 4-vector or 4-tensor. We make the basic assumption that the deviations from a homogeneous and isotropic universe are so small that all perturbation quantities $\delta q$ can be treated at ``first'' (or ``linear'') order, i.e., any term containing a product of perturbation quantities is set to zero.

Our approach can be simply extended to include raising and lowering indices for a perturbation $\delta q_\mu$. Defining $\bar q^\mu \equiv \bar q_\nu \bar g^{\mu\nu}$ with $\bar g^{\mu\nu}$ being the inverse of $\bar g_{\mu\nu}$, we obtain at first order
\begin{equation}
q^\mu = q_\nu  g^{\mu\nu} = \big[ \bar q_\nu + \delta q_\nu \big]  \big[ \bar g^{\mu\nu} + \delta g^{\mu\nu} \big] =  \bar q^\mu + \delta q_\nu \bar g^{\mu \nu} + \bar q_\nu \delta g^{\mu \nu}\:.
\end{equation}
 Thus, we can define
\begin{equation}\label{eq:raising}
\delta q^{\mu} \equiv q^\mu - \bar q^\mu = \delta q_\nu \bar g^{\mu \nu} + \bar q_\nu \delta g^{\mu \nu} 
\end{equation}
and similarly for lowering indices. That is, raising and lowering indices for perturbations quantities is nontrivial in general. However, if a perturbation $\delta q_\mu$ does not have counterpart on the smooth background, i.e., $\bar  q_\mu = 0$, then raising and lowering indices simply reduces to applying $\bar g^{\mu\nu}$ directly on the perturbation $\delta q_\mu$.

It should be noted that in the limiting case, in which the deviations from the smooth background vanish, i.e., $g_{\mu\nu}$ reduces to $\bar g_{\mu\nu}$, the interpretation of the coordinates $x = (\tau,x)$ will automatically conform to the familiar interpretation of $x = (\tau,\bv x)$ in Eq.~(\ref{eq:ds_2}), i.e., $\tau$ being the conformal time and $\bv x$ spatial cartesian comoving coordinates. For this reason and since the deviations from $\bar g_{\mu\nu}$ are small, we will nevertheless interprete $\tau$ as conformal time and $\bv x$ as comoving coordinates, even if $\delta g_{\mu \nu}$ does not vanish.

\subsection{Metric and energy-momentum tensor}
\label{sec:energy_momentum_tensor}

Performing the split (\ref{eq:g_munu_split}), we can write the perturbation $\delta g_{\mu\nu}$ in general as
\begin{equation}\label{eq:perturbed_metric_2}
	\delta g_{\mu\nu}dx^\mu dx^\nu = a^2(\tau) \left[- 2 A\: d\tau^2 + B_i\: d\tau dx^i + 2 \left(H_{\rm L} \delta_{ij} + H_{ij}\right) dx^i dx^j  \right]\:,
\end{equation}
where $A(\tau,\bv x)$, $H_{\rm L}(\tau,\bv x)$, $B_i(\tau,\bv x)$, and $H_{ij}(\tau,\bv x)$ are perturbation quantities and where $H_{ij}$ is symmetric and traceless, i.e., $H_{ij}g^{ij} = 0$. (Note that $3 H_{\rm L}$ basically plays the role of the trace of $H_{ij}$, which was taken out from $H_{ij}$ for later convenience.) The function $a(\tau) = \bar a(\tau)$ is just the background scale factor (\ref{eq:a_tau}), since perturbations to $a(\tau)$ can be neglected at first order or absorbed by the other perturbation quantities. We will write $a(\tau)$ instead of $\bar a(\tau)$ for the ease of notation. The full perturbed metric in the form of a 1+3 block matrix is then given by
\begin{equation}\label{eq:metric}
\boxed{
  g_{\mu\nu}=  a^2\left(
      \begin{array}{cc}
        -(1+2A) & B_i  \\
        B_i & (1+2H_{\rm L})\delta_{ij} + 2H_{ij} 
      \end{array} \right)}
  \end{equation}
and its inverse at first order
 \begin{equation}\label{eq:g_inverse}
  g^{\mu\nu}=  a^{-2}\left(
      \begin{array}{cc}
        -(1-2A) & B_i  \\
        B_i & (1-2H_{\rm L})\delta_{ij} - 2H_{ij} 
      \end{array} \right)\:.
  \end{equation}
We have denoted $B_i$ and $H_{ij}$ with lowered indices in $g^{\mu\nu}$ to indicate that they are the same functions as in $g_{\mu\nu}$. None of the perturbations $A$, $H_{\rm L}$, $B_i$, and $H_{ij}$ are geometrical quantities except under special transformations that will be discussed in the following.

Since the background metric $\bar g_{\mu\nu}$ is invariant under spatial translations and rotations on a slice of constant conformal time $\tau$, the metric perturbation $\delta g_{\mu\nu}$ transforms like a 4-tensor under these transformation, i.e., $\delta g_{00}$ transforms like a 3-scalar, $\delta g_{0i} = \delta g_{i0}$ like a 3-vector, and $\delta g_{ij}$ like a 3-tensor (see the discussion in Sect.~\ref{sec:Field equations and equation of motion}). Formally, this can be seen as follows. For a spatial translation or rotation $x = (\tau,\bv x) \rightarrow \tilde{x} = (\tau,\tilde{\bv x})$ the transformation of $\delta g_{\mu\nu}$ is as given in Eq.~(\ref{eq:T_transform}). Since  $\bar g_{\mu\nu}(\tau,\bv x) \equiv \bar g_{\mu\nu}(\tau)$ is independent of $\bv x$ and $\bar g_{i0} = 0$, it follows immediately
\begin{align}
\delta \tilde{g}_{00}(\tilde{x}) &=  \tilde{g}_{00}(\tilde{x})-\bar g_{00}(\tau) =  g_{00}(x)-\bar g_{00}(\tau) = \delta g_{00}(x)\\
\delta \tilde{g}_{a0}(\tilde{x}) &=  \tilde{g}_{a0}(\tilde{x}) = \frac{\partial x^i}{\partial {\tilde{x}}^a}  g_{i0}(x) = \frac{\partial x^i}{\partial {\tilde{x}}^a}  \delta g_{i0}(x)\:.
\end{align}
For the $ij$-components we use the fact that 
\begin{equation}
\frac{\partial x^i}{\partial {\tilde{x}}^a}\frac{\partial x^j}{\partial {\tilde{x}}^b} \bar g_{ij}(\tau)  = \bar g_{ab}(\tau) \:,
\end{equation}
since $g_{ij} = a^2(\tau) \delta_{ij}$ is diagonal and $\partial x^i/\partial {\tilde{x}}^a$ is a (constant) orthogonal matrix. Then we have
\begin{equation}\label{eq:delta_g_transform}
\delta \tilde{g}_{ab}(\tilde{x}) = \tilde{g}_{ab}(\tilde{x}) - \bar g_{ab}(\tau)  = \frac{\partial x^i}{\partial {\tilde{x}}^a}\frac{\partial x^j}{\partial {\tilde{x}}^b} \big( g_{ij}(x)  - \bar g_{ij}(\tau) \big) =  \frac{\partial x^i}{\partial {\tilde{x}}^a}\frac{\partial x^j}{\partial {\tilde{x}}^b} \delta g_{ij}(x)\:.
\end{equation}
It should be stressed that to derive the transformation behavior of $\delta g_{\mu\nu}$, we have not treated $\bar g_{\mu\nu}$ as a geometrical object, but have just exploited the symmetry of its functional form. As a consequence, under spatial translations and rotations, $A$ and $H_{\rm L}$ transform like 3-scalars, $B_i$ like a 3-vector, and $H_{ij}$ like a 3-tensor. From now on, whenever we talk about geometrical quantities on a slice of constant $\tau$, we refer to their transformation behavior under spatial translations and rotations.\footnote{This analysis remains valid in the case of a nonflat universe, i.e., $K \neq 0$, since the Robertson-Walker metric is generally invariant under spatial translations and rotations.}

Similarly, we decompose the perturbed energy-momentum tensor $T_{\mu\nu}$ into its background $\bar{T}_{\mu\nu}$ and a small perturbation
\begin{equation}
	\delta T_{\mu\nu} \equiv T_{\mu\nu}  - \bar{T}_{\mu\nu} \:,
\end{equation}
where $\bar T_{\mu\nu}$ has the fixed functional form (\ref{eq:tmunu_bar}). In a perturbed universe, the form of $T_{\mu\nu}$ is in general not restricted to the form of an ideal fluid anymore. In general, it will take the form of a \textbf{real fluid}
\begin{equation}
	T_{\mu\nu} = \left( \rho + p \right)u_\mu u_\nu + p\: g_{\mu \nu} + \Pi_{\mu \nu}\:,
\end{equation}
where $\rho(\tau,\bv x) = \bar\rho(\tau)+\delta \rho (\tau,\bv x)$ is the energy density, $p(\tau,\bv x) = \bar p(\tau) + \delta p (\tau,\bv x)$ the pressure, $u_\mu(\tau,\bv x) = \bar u_\mu(\tau)+\delta u_\mu(\tau,\bv x)$ the 4-velocity, and $\Pi_{\mu \nu}(\tau,\bv x)$ the \textbf{anisotropic stress} (or anisotropic inertia) of the fluid, which accounts for the deviations from an ideal fluid (e.g., dissipative corrections like ``free-streaming''). The quantities $\delta \rho$, $\delta p$, $\delta u_\mu$ are treated as small perturbations. Since the anisotropic stress $\Pi_{\mu \nu}$ has no counterpart in the ideal fluid of the FLRW background, it is treated as a perturbation as well. Furthermore, the anisotropic stress obeys the restrictions
\begin{equation}
	\Pi_{\mu\nu} = \Pi_{\nu\mu}\:, \quad\quad\quad \Pi_{\mu\nu} u^\mu = 0\:, \quad\quad\quad \Pi^\mu_{\phantom{\mu} \mu} = 0\:,
\end{equation}
i.e., it is a traceless, symmetric 4-tensor perpendicular to the 4-velocity $u^\mu$. Making the general ansatz $\delta u^\mu = (\delta u^0, v_i)$, the normalization constraint for the 4-velocity, i.e.,
\begin{equation}
	g_{\mu \nu} u^\mu u^\nu = -1\:,
\end{equation}
yields $\delta u^0 = -A$ at first order, while $v_i$ remains unconstrained. We call $v_i$ the (peculiar) \textbf{velocity} of the fluid. Thus, we obtain at first order
\begin{equation}
	\delta u^\mu = \frac{1}{a}\left(-A, v_i \right)\:, \quad\quad\quad \delta u_\mu = g_{\mu\nu} u^\nu - \bar g_{\mu\nu}\bar u^\nu = a (-A, B_i + v_i)\:.
\end{equation}
With this form of $\delta u^\mu$ the condition $\Pi_{\mu\nu}u^\mu$ yields at first order $\Pi_{00} = \Pi_{i0} = 0$, and accordingly $\Pi^i_{\phantom i i} = \Pi^\mu_{\phantom{\mu} \mu} = 0$. So $\Pi_{\mu\nu}$ can be restricted to its spatial part $\Pi_{ij}$. Taking everything together, the general form of the energy-momentum $T_{\mu \nu}$ at first order is
\begin{equation}\label{eq:energy_momentum_tensor}
\boxed{
  T_{\mu\nu}=  a^2\left(
      \begin{array}{cc}
        \bar \rho\left(1+2A \right)+\delta \rho\ \ \ & -\bar \rho \:B_i - \left(\bar \rho + \bar p \right) v_i \\
         -\bar \rho \:B_i - \left(\bar \rho + \bar p \right) v_i \ \ \ \ & \bar p \:\left[ \left(1 + 2 H_{\rm L} \right)\delta_{ij} + 2 H_{ij} \right] +\delta p \:\delta_{ij} + \Pi_{ij} 
      \end{array} \right)\:.}
  \end{equation}
Analog to the perturbation part of the metric $\delta g_{\mu\nu}$, the perturbation part $\delta T_{\mu\nu}$ of the energy-momentum tensor is a 4-tensor under spatial translations and rotations, since $\bar T_{\mu\nu}$ is invariant under these transformations. So $\delta \rho$ and $\delta p$ are 3-scalars, $v_i$ a 3-vector, and $\Pi_{ij}$ a 3-tensor.

If the fluid $T_{\mu\nu}$ is made up of several non-interacting fluids $[T_I]_{\mu\nu}$, $I=1,\ldots,N$ such that
\begin{equation}
	T_{\mu\nu} = \sum_{I=1}^N [T_I]_{\mu\nu}\:,
\end{equation}
we immediately see using Eq.~(\ref{eq:energy_momentum_tensor}) that the perturbation quantities add as follows:
\begin{equation}
	\delta \rho = \sum_{I=1}^N \delta \rho_I\:, \quad \delta p = \sum_{I=1}^N \delta p_I\:, \quad \left( \bar \rho+\bar p \right)v_i = \sum_{I=1}^N \left(\bar \rho_I+\bar p_I\right)[v_I]_i\:, \quad \Pi_{ij} = \sum_{I=1}^N [\Pi_I]_{ij}\:.
\end{equation}

\section{Scalar-Vector-Tensor (SVT) decomposition}\label{sec:SVT}

Our goal is to solve the field equations and equations of motion at first order. However, with the general first order metric (\ref{eq:metric}) and energy-momentum tensor  (\ref{eq:energy_momentum_tensor}) this would lead to horribly complicated equations.\footnote{See, e.g., Weinberg (2008, p.~219-224) who derives the field equations and equations of motion in real space by ``brute force''. He then denotes the obtained system of equations as ``repulsively complicated''.} Fortunately, the rotational symmetry of the underlying homogeneous FLRW universe allows us to decompose the perturbations into 3-scalars, divergenceless 3-vectors, and divergenceless, traceless, symmetric 3-tensors. This will simplify our analysis considerably, inasmuch as these different contributions are not coupled to each other by the field equations or equations of motion. We will first describe this decomposition in real space and then move into Fourier space for a more detailed analysis. Finally, we give a proof of the decomposition theorem.

Before we start with the decomposition of perturbation quantities, we briefly want to comment on the geometry on the tangent bundle of a slice of constant time $\tau$ in a first order calculation.  First, we will fully stay in comoving space. That is, we scale out the expansion of the universe, i.e., the 3-metric is $\gamma_{ij} = a^{-2}g_{ij}$. Second, having a perturbation quantity $\delta q$ on the tangent bundle of a manifold with the metric $\gamma_{ij}$, we can essentially treat $\delta q$ as if our slice was the three-dimensional Euclidean space. This arises from the fact that in a first order calculation, the product $\delta q \: \gamma_{ij} = \delta q \: \bar \gamma_{ij}$ always reduces to the corresponding expression involving the background metric $\bar \gamma_{ij} = \delta_{ij}$, which is trivial. For instance, the covariant derivative $\nabla_i$ applied to $\delta q$ reduces to the ordinary derivative $\partial_i$. Since 3-vectors $\delta q_i$ and symmetric, traceless 3-tensors $\delta q_{ij}$ do not have an unperturbed counterpart\footnote{Following the discussion of Section \ref{sec:Field equations and equation of motion}, any 3-vector $\bar q_i$ must be zero, and any 3-tensor must have the form $\bar q_{ij} = f \bar g_{ij}$ with a function $f(\tau)$. The trace of $\bar q_{ij}$ is $\bar q_{ij} \bar g^{ij} = f \delta_i^{\phantom{i} i} = 3f$. If $\bar q_{ij}$ is traceless, $f$ must be zero and so $\bar q_{ij}$ must be zero too.}, i.e., $\bar q_i=0$ and $\bar q_{ij} = 0$, also raising and lowering indices becomes trivial, i.e., using Eq.~(\ref{eq:raising}) we have $\delta q_i =\delta q_i \: \bar \gamma^{ij} = \delta q^i$. Morevoer, we can fully decompose the tangent bundle into Fourier modes. On a manifold with metric $\gamma_{ij}$ the corresponding integral reads as
\begin{equation}\label{eq:integral_g}
	\delta q (\bv k) = \int \delta q(\bv x) e^{-i \bv k \bv x} \sqrt{\gamma(\bv x)}\: dx^3\:,
\end{equation}
where $\gamma = |\det(\gamma_{ij})|$ and $\bv k$ is a comoving Fourier mode. However, with $|\delta \gamma_{ij}| \ll 1$ the determinant is at first order
\begin{equation}
\det(\gamma_{ij}) =\det(\bar \gamma_{ij}+\delta \gamma_{ij}) \simeq \det(\bar \gamma_{ij}) +{\rm tr}(\delta \gamma \cdot \bar \gamma^{-1}) =  1 + {\rm tr}(\delta \gamma) \:,\label{eq:order_g_term}
\end{equation}
where here ${\rm tr}(\delta \gamma) = \sum_i \delta \gamma_{ii}$ denotes the simple trace in the sense of matrices, and thus the integral (\ref{eq:integral_g}) reduces in a first order calculation to the usual integral in an Euclidean space. In the following, we consider the  3-scalar $S(\tau,\bv x)$, the 3-vector $V_i(\tau,\bv x)$, and the traceless, symmetric 3-tensor $D_{ij}(\tau,\bv x)$ being arbitrary perturbation quantities in our comoving frame.

\subsection{SVT decomposition in real space}\label{sec:SVT_decomposition_in_real_space}

The analysis of the 3-scalar $S$ is trivial, since $S$ cannot be decomposed any further and so has just a scalar part, i.e., $S = S^{(S)}$. However, it is always possible to decompose the 3-vector $V_i$ into a 3-scalar part $V^{(S)}$ (``gradient'') and a divergenceless 3-vector part $V^{(V)}$ (``curl''), i.e.,
\begin{equation}\label{eq:vector_decomposition_real_space}
	V_i = \partial_i V^{(S)} + V^{(V)}_i\:,
\end{equation}
where
\begin{equation}\label{eq:divergencelessness1}
	\partial_i V^i = 0\:.
\end{equation}
Similarly, a spatial traceless, symmetric 3-tensor field can always be decomposed as
\begin{equation}\label{eq:tensor_decomposition_real_space}
	D_{ij} = \left(\partial_i \partial_j - \frac{1}{3}\delta_{ij}\Delta \right)D^{(S)} + \frac{1}{2}\left(\partial_j D_i^{(V)} + \partial_i D_j^{(V)} \right) + D_{ij}^{(T)}\:,
\end{equation}
where
\begin{equation}\label{eq:divergencelessness2}
	\partial^i D^{(V)}_i = {D_i^{(T)}}^i = \partial_j {D_i^{(T)}}^j = 0
\end{equation}
and $\Delta = \partial^i \partial_i$. Here $D^{(S)}$ transforms like a 3-scalar, $D_{i}^{(V)}$ like a 3-vector, and $D_{ij}^{(T)}$ like a 3-tensor. This is the \textbf{scalar-vector-tensor (SVT) decomposition} in real space.

To see how this decomposition arises, we briefly describe how it can be constructed. To obtain Eq.~(\ref{eq:vector_decomposition_real_space}), we define $V^{(S)}$ as the solution of
\begin{equation}
	\Delta V^{(S)} = \partial_i V^i
\end{equation}
and then we define $V^{(V)}_i$ simply by
\begin{equation}
V^{(V)}_i	= V_i - \partial_i V^{(S)}\:.
\end{equation}
Similarly, to obtain Eq.~(\ref{eq:tensor_decomposition_real_space}), we can define $D^{(S)}$ as the solution of
\begin{equation}
	\partial^i \partial^j \left(\partial_i \partial_j -\delta_{ij} \Delta \right)D^{(S)} = \partial^i \partial^j D_{ij}\:,
\end{equation}
$D_i^{(V)}$ as the solution of
\begin{equation}
\Delta D_i^{(V)}	= \partial^j D_{ji}-\partial_i \left[\partial^a \partial^b \left(\partial_a \partial_b -\delta_{ab} \Delta \right)D^{(S)}\right]
\end{equation}
and we finally set $D_{ij}^{(T)}$ to
\begin{equation}
	D_{ij}^{(T)} = D_{ij} - \left(\partial_i \partial_j - \frac{1}{3}\delta_{ij}\Delta \right)D^{(S)} - \frac{1}{2}\left(\partial_j D_i^{(V)} + \partial_i D_j^{(V)} \right)\:.
\end{equation}
According to this construction it is obvious that the conditions in the Eqs.~(\ref{eq:divergencelessness1}) and (\ref{eq:divergencelessness2}) are satisfied automatically.

\subsection{SVT decomposition in Fourier space}\label{sec:SVT_decomposition_in_fourier_space}

The meaning of the SVT decomposition is much easier grasped in Fourier space. We can decompose each perturbation quantity $\delta q(\tau,\bv x)$ into comoving Fourier modes $\bv k$ on each slice of constant $\tau$, i.e.,
\begin{equation}
\delta q (\tau, \bv k) = \int \delta q(\tau,\bv x) e^{-i \bv k \bv x} dx^3\:, \quad\quad\quad \delta q (\tau, \bv x) = \frac{1}{\left(2 \pi \right)^3} \int \delta q(\tau,\bv k) e^{i \bv k \bv x} dk^3\:.
\end{equation}
Now consider an arbitrary Fourier mode $\bv k$ on a given slice and choose two normalized vectors $\bv e_1$ and $\bv e_2$ perpendicular to $\bv k$, so that the set $\{ \bv e_1, \bv e_2, \hat{\bv k} \}$ constitutes an orthonormal basis for our comoving spaces.\footnote{Recall that our slice is essentially flat, because we work with perturbations in a first order calculation, and thus parallel transport of vectors and tensors becomes trivial. For this reason, we can define the basis $\{ \bv e_1, \bv e_2, \hat{\bv k} \}$ globally without explicitely specifying a tetrad. Also note that the vectors and tensors in our tangential spaces are complex, because we are in Fourier space. So any vector can be represented by the basis $\{ \bv e_1, \bv e_2, \hat{\bv k} \}$ with complex coefficients.} Then Fourier transforming the real space SVT decomposition of the 3-Vector $V_i$ (see the Eqs.~(\ref{eq:vector_decomposition_real_space}) and (\ref{eq:divergencelessness1})) we immediately see that the scalar part must be parallel to $\hat{\bv k}$ in Fourier space, while the vector part must be perpendicular to it. Similarly, Fourier transformation of the SVT decomposition of the 3-tensor $D_{ij}$ (see the Eqs.~(\ref{eq:tensor_decomposition_real_space}) and (\ref{eq:divergencelessness2})) shows that the scalar part has two components along $\hat{\bv k}$, the vector part has one component along and one perpendicular to $\hat{\bv k}$, and the tensor part has two components perpendicular to $\hat{\bv k}$. With this information we are able to construct a basis for 3-scalars, 3-vectors and traceless, symmetric 3-tensors in Fourier space such that each basis element is associated to either a scalar, vector, or tensor part. Expressed by means of the helicity basis
\begin{equation}
	\bv e^\pm = \frac{1}{\sqrt{2}}\left[\bv e_1 \pm i\bv e_2 \right]
\end{equation}
the three sets of basis elements are:\footnote{These basis elements are a representations of the so-called \textbf{harmonic functions} $Y$, $Y_i$, and $Y_{ij}$ for $K=0$ (up to the factor $e^{i\bv k \bv x}$ which is omitted in our basis). In general, the harmonic functions are also defined for non-flat FLRW universes and are eigenfunctions of the generalized Laplace operator $\nabla_j \nabla_i \bar{\gamma}^{ij}$, where $\bar{\gamma}_{ij}$ is the spatial metric of the background FLRW universe and $\nabla^i$ the covariant derivative. For each $K$ the set of these solutions constitutes a complete set for decomposing 3-scalars, 3-vectors, and 3-tensors, respectively. For a summary of the properties of the harmonic functions see the Appendix C of \cite{kodama1984}.}
\begin{itemize}
	\item \textit{3-scalar:}
\begin{equation}
	\mathcal{S}^{(0)} = 1\:.\label{eq:scalar_basis}
\end{equation}	\item \emph{3-vector:}
	\begin{equation}
	\mathcal{V}_i^{(0)} = - i \hat{k}_i\:, \ \ \ \ \mathcal{V}_i^{(\pm 1)} = e^\pm_i\:.
	\end{equation}
	\item \emph{traceless, symmetric 3-tensor:}
		\begin{equation}
	\mathcal{D}_{ij}^{(0)} = -\hat{k}_i\hat{k}_j + \frac{1}{3}\delta_{ij}\:,\ \ \ \ \mathcal{D}_{ij}^{(\pm 1)} = -\frac{i}{2}\left[\hat{k}_j\: e^\pm_i +\hat{k}_i \: e^\pm_j \right]\:,\ \ \ \ \mathcal{D}_{ij}^{(\pm 2)} =  e^\pm_i e^\pm_j\:.\label{eq:tensor_basis3}
	\end{equation}
\end{itemize}
It is easy to see that basis elements with $m = 0$ are associated to the scalar parts, those with $m = \pm 1$ to the vector parts, and those with $m = \pm 2$ to the tensor part. This is the \textbf{SVT-decomposition} in Fourier space. The meaning of the index $m$ will become clear, when we study the behavior of these basis elements under rotations. The sets of bases contain 1 element for 3-scalars, 3 elements for 3-vectors, and 5 elements for 3-tensors according to the degrees of freedom of 3-scalars, 3-vectors, and traceless, symmetric 3-tensors, respectively. Thus, to proof that these sets indeed are bases, we just have to show that their elements are linear independent.
This is easily done by introducing the inner product $\langle \cdot \:,\cdot\rangle$ defined by the hermetian contraction of vectors or tensors as follows
\begin{eqnarray}
	\left\langle \mathcal{V}_i^{(m)},\mathcal{V}_i^{(m')}\right\rangle &=& {\mathcal{V}^{(m)}}^i\:{\mathcal{V}^{(m')}_i}^\ast = \delta_{mm'}\:,\ \ \ \ m,m' = 0,\pm 1\\
	\left\langle \mathcal{D}_{ij}^{(m)},\mathcal{D}_{ij}^{(m')}\right\rangle &=& {\mathcal{D}^{(m)}}^{ij}\:{\mathcal{D}^{(m')}_{ij}}^\ast = \delta_{mm'}\:,\ \ \ \ m,m' = 0,\pm 1,\pm 2\:.
\end{eqnarray}
Thus the 3-vectors $\mathcal{V}_i^{(m)}$, for $m=0,\pm 1$, and the 3-tensors $\mathcal{D}_{ij}^{(m)}$, for $m=0,\pm 1,\pm 2$, are all orthogonal to each other and hence automatically linearly independent.

The meaning of the tacitly introduced index $m$ describes the transformation property of the basis elements under spatial rotations. If we rotate $\bv e_1$ and $\bv e_2$ counterclockwise around $\hat{\bv k}$ by an angle $\varphi$, we find immediately
\begin{equation}
	\tilde{\bv e}^\pm = e^{\mp i \varphi}\: \bv e^\pm\:.
\end{equation}
So the basis elements transform according to their definition
\begin{equation}
	\tilde{\mathcal{S}}^{(m)} = e^{-i m\varphi}\mathcal{S}^{(m)}\:,\ \ \ \ \tilde{\mathcal{V}}^{(m)} = e^{-i m\varphi}\mathcal{V}^{(m)}\:,\ \ \ \ \tilde{\mathcal{D}}^{(m)} = e^{-i m\varphi}\mathcal{D}^{(m)}\:.
\end{equation}
The quantities $\tilde{\mathcal{S}}^{(m)}$, $\tilde{\mathcal{V}}^{(m)}$, and $\tilde{\mathcal{D}}^{(m)}$ each constitute a new basis according to the choice of new unit vectors $\tilde{\bv e}_1$ and $\tilde{\bv e}_2$. Since under spatial rotations $S$ transforms like a 3-scalar , $V$ like a 3-vector, and $D$ like a 3-tensor, we can expand $S$, $V$, and $D$ into these sets of basis elements, i.e.,
\begin{equation}\label{eq:D_ij_expanded}
\begin{aligned}
		S &= S^{(0)} \mathcal{S}^{(0)} = \tilde{S}^{(0)} \tilde{\mathcal{S}}^{(0)}\:,\\
		V &= \sum_{m=-1}^1 V^{(m)}\mathcal{V}^{(m)} = \sum_{m=-1}^1 \tilde V^{(m)}\tilde{\mathcal{V}}^{(m)}\:,\\
		D &= \sum_{m=-2}^2 D^{(m)}\mathcal{D}^{(m)}= \sum_{m=-2}^2 \tilde D^{(m)}\tilde{\mathcal{D}}^{(m)}\:,
		\end{aligned}
\end{equation}
and the corresponding coefficients must transform the opposite way, i.e.,
\begin{equation}
	\tilde S^{(m)} = e^{i m\varphi}S^{(m)}\:,\ \ \ \ \tilde V^{(m)} = e^{i m\varphi}V^{(m)}\:,\ \ \ \ \tilde D^{(m)} = e^{i m\varphi}D^{(m)}\:.
\end{equation}
This means, under rotations around $\hat{\bv k}$, the coefficients transform like helicity states of \textbf{helicity} (or spin) $m$. Helicity states with $m=0$ are \textbf{scalar perturbations}, helicity states with $m = \pm 1$ \textbf{vector perturbations}, and helicity states with $m = \pm 2$ \textbf{tensor perturbations}. Scalar perturbations correspond to the usual energy overdensities of Newtonian physics and vector perturbations correspond to velocity perturbations in Newtonian physics. Tensor perturbations are also called \textbf{gravitational waves} and have no Newtonian analogon. We are mostly interested in scalar perturbations, since these are the perturbations that can undergo gravitational instability and can lead to structure formation in the universe.

By means of the Eqs.~(\ref{eq:D_ij_expanded}), $S$, $V$, and $D$ are now entirely expressed in terms of helicity states. In the basis $\{ \bv e_1, \bv e_2, \hat{\bv k} \}$ for a given Fourier mode $\bv k$ they have the explicit representation
\begin{equation}\label{eq:helicity_representation}
\boxed{\begin{aligned}
S &= S^{(0)}\\\\
V_i &= \left(\frac{1}{\sqrt{2}} \left[V^{(+1)}+V^{(-1)}\right],\frac{i}{\sqrt{2}} \left[V^{(+1)}-V^{(-1)}\right],-i\:V^{(0)}\right)\\\\
D_{ij} &=\left(
\begin{array}{c c c}
\frac{D^{(0)}}{3} + \frac{D^{(+2)} + D^{(-2)}}{2} &
i\frac{D^{(+2)} - D^{(-2)}}{2} & 
-i\frac{D^{(+1)} + D^{(-1)}}{2\sqrt{2}} \\
i\frac{D^{(+2)} - D^{(-2)}}{2} &
\frac{D^{(0)}}{3} -\frac{D^{(+2)} + D^{(-2)}}{2} &
\frac{D^{(+1)} - D^{(-1)}}{2\sqrt{2}} \\
-i\frac{D^{(+1)} + D^{(-1)}}{2\sqrt{2}} &
\frac{D^{(+1)} - D^{(-1)}}{2\sqrt{2}}  &
-\frac{2D^{(0)}}{3}
\end{array}
\right)\:.
\end{aligned}}
\end{equation}

We can summarize the results of the last two sections as follows: In Fourier space, 3-scalars are functions of helicity $m = 0$, 3-vectors are superpositions of functions of helicity $m = 0,\pm 1$, and traceless, symmetric 3-tensors are superpositions of functions of helicity $m = 0,\pm 1, \pm 2$, while in real space, the helicity $m=0$ states correspond to 3-scalars, the helicity $m=\pm 1$ states to divergenceless 3-vectors, and the helicity $m=\pm 2$ states to transverse, traceless, symmetric 3-tensors. Regarding the metric perturbation $\delta g_{\mu\nu}$ (see Eq.~(\ref{eq:metric})), it follows that it can be decomposed into the helicity states $A^{(0)}$, $H_{\rm L}^{(0)}$, $B_i^{(m)}$, $m = 0,\pm 1$, and $H_{ij}^{(m)}$, $m=0,\pm 1, \pm 2$. These are exactly 10 degrees of freedom as expected from the 10 independent components of the general perturbed metric. This confirms the generality of our analysis so far.

\subsection{Independence of different Fourier modes}

Before proving the decomposition theorem, we first have to show that at linear order the field equations and equations of motion for different $\bv k$ modes decouple. This follows from the translational invariance of the background FLRW universe.

Suppose the field equations and equations of motion contain $N$ perturbation quantities $\delta_A$, $A=1,\ldots,N$. At linear order, the evolution of any perturbation $\delta_A$ can generally be expressed as
\begin{equation}\label{eq:transfer_function}
	\delta_A(\tau,\bv k) = \sum_{B=1}^N \int T_{AB}(\tau,\tau_{\rm i};\bv k, \bv k')\: \delta_B(\tau_{\rm i},\bv k') \:{dk'}^3\:,
\end{equation}
where $T_{AB}(\tau,\tau_{\rm i};\bv k, \bv k')$ is the transfer function for the perturbation $\delta_A$ and $\tau_{\rm i}<\tau$ is an arbitrary initial time. Note that $T_{AB}(\tau,\tau_{\rm i};\bv k, \bv k')$ can only depend on the background FLRW universe, since if it was dependent on a perturbation quantity, the term $T_{AB}(\tau,\tau_{\rm i};\bv k, \bv k') \delta_B(\tau_{\rm i},\bv k')$ would be of second order and thus would be neglected in our linear treatment. Now we perform a coordinate transformation $\tilde{\bv x} = \bv x + \bv{\Delta x}$ with $\bv{\Delta x}$ a constant translation. Since $\delta_A(\tau,\bv x)$ transforms like a geometrical object under spatial translations (cf.~Eq.~(\ref{eq:T_transform})) and since $\partial x_i/\tilde \partial x_j = \delta_{ij}$, it holds in real space $\tilde \delta_A(\tau,\tilde{\bv x}) = \delta_A(\tau,\bv x)$ for any perturbation quantity. The transformation behavior in Fourier space is then
\begin{eqnarray}
	\tilde\delta_A(\tau,\bv k) &=& \int \tilde \delta_A (\tau,\tilde{\bv x}) e^{-i \bv k \tilde{\bv x}} d\tilde{x}^3  = \int \delta_A (\tau,\bv x) e^{-i \bv k\left( \bv x + \bv{\Delta x}\right)} dx^3 = e^{-i \bv k \bv{\Delta x}} \delta_A(\tau,\bv k)\:.
\end{eqnarray}
Thus, transforming Eq.~(\ref{eq:transfer_function}) yields
\begin{equation}\label{eq:translational_invariance}
\begin{split}
	\tilde \delta_A(\tau,\bv k) &= e^{-i \bv k \bv{\Delta x}}\:\delta_A(\tau,\bv k) = e^{-i \bv k \bv{\Delta x}} \sum_{B=1}^N \int T_{AB}(\tau,\tau_{\rm i};\bv k, \bv k')\: \delta_B(\tau_{\rm i},\bv k') \:{dk'}^3\\
	&= \sum_{B=1}^N  \int e^{-i \bv k \bv{\Delta x}}\:T_{AB}(\tau,\tau_{\rm i};\bv k, \bv k')\: e^{i \bv k' \bv{\Delta x}}\:\tilde\delta_B(\tau_{\rm i},\bv k') \:{dk'}^3\\
	&= \sum_{B=1}^N \int \tilde T_{AB}(\tau,\tau_{\rm i};\bv k, \bv k')\: \tilde\delta_B(\tau_{\rm i},\bv k') \:{dk'}^3\:.
\end{split}
\end{equation}
Since the background FLRW universe is translational invariant, the transfer function must be translational invariant, i.e., $\tilde T_{AB}(\tau,\tau_{\rm i};\bv k, \bv k') = T_{AB}(\tau,\tau_{\rm i};\bv k, \bv k')$, and so we obtain from the last two equalities in Eq.~(\ref{eq:translational_invariance})
\begin{equation}
	e^{i \left(\bv k'-\bv k\right) \bv{\Delta x}}\: T_{AB}(\tau,\tau_{\rm i};\bv k, \bv k')  = T_{AB}(\tau,\tau_{\rm i};\bv k, \bv k')
\end{equation}
for any $\bv{\Delta x}$. This means that for $\bv k \neq \bv k'$ the transfer function must vanish. Hence different Fourier modes are not coupled to each other and, for the further analysis, we can focus on a single arbitrary Fourier mode $\bv k$.

\subsection{Decomposition Theorem}\label{sec:decomp}

Now we are able to proof the decomposition theorem that will simplify the subsequent analysis of the field equations and equations of motion immensely. The theorem states that due to the rotational symmetry of the FLRW background universe, perturbations of different helicity $m$ evolve independently from each other at first order. The proof is very similar to the one given in the last section.\footnote{A proof of the decomposition theorem for a general FLRW background universe is, for instance, given in Kodama \& Sasaki (1984, App.~B)\nocite{kodama1984} or \cite{straumann2008}.}

Again, suppose the field equations and equations of motion contain a set of $N$ perturbation quantities $\delta_A$, $A = 1,\ldots,N$, where $m_A$ is the helicity of the perturbation $\delta_A$. Since different Fourier modes are decoupled, we can now express the evolution of a given perturbation $\delta_A$ as
\begin{equation}\label{eq:transfer_function2}
	\delta_A(\tau,\bv k) = \sum_{B=1}^N T_{AB}(\tau,\tau_{\rm i};\bv k) \delta_B(\tau_{\rm i}, \bv k)\:,
\end{equation}
where $T_{AB}(\tau,\tau_{\rm i};\bv k)$ is the transfer function associated to the perturbation $\delta_A$ for a given Fourier mode $\bv k$ and $\tau_{\rm i}$ is again some arbitrary initial time. If we perform a spatial rotation around $\hat{\bv k}$ by some angle $\varphi$, Eq.~(\ref{eq:transfer_function2}) becomes
\begin{equation}\label{eq:rotational_invariance}
\begin{split}
	\tilde{\delta}_A(\tau,\bv k) &= e^{i m_A \varphi}\: \delta_A(\tau,\bv k) = e^{i m_A \varphi}\sum_{B=1}^N  T_{AB}(\tau,\tau_{\rm i};\bv k)\: \delta_B(\tau_{\rm i},\bv k) \\
 &= \sum_{B=1}^N e^{i m_A \varphi} \:T_{AB}(\tau,\tau_{\rm i};\bv k) \:e^{-i m_B \varphi}\:\tilde{\delta}_B(\tau_{\rm i},\bv k) \\
 &= \sum_{B=1}^N \tilde T_{AB}(\tau,\tau_{\rm i};\bv k) \: \tilde\delta_B(\tau_{\rm i},\bv k)\:.
\end{split}
\end{equation}
Due to the rotational symmetry of the background FLRW universe, the transfer function must be rotationally invariant, i.e., $\tilde T_{AB}(\tau,\tau_{\rm i};k) = T_{AB}(\tau,\tau_{\rm i};k)$ depending only on $k$. So we end up with 
\begin{equation}
	e^{i \left(m_A-m_B\right)\varphi }\: T_{AB}(\tau,\tau_{\rm i};k)  = T_{AB}(\tau,\tau_{\rm i};k)
\end{equation}
for any angle $\varphi$. This means that for every index pair $(A,B)$ such that $m_A \neq m_B$, the transfer function must vanish. Thus, different helicity states are indeed decoupled from each other.

\section{Field equations}

With the preliminaries of the previous section we are now able to compute the field equations and equations of motion in a sensible way. We compute them in Fourier space, where the independence of different Fourier modes allows us to focus on an arbitrary mode $\bv k$, and split them in equations of different helicity according to the decomposition theorem (see Sect.~\ref{sec:decomp}).

The field equations are
\begin{equation}
	G_{\mu\nu} = 8 \pi G \:T_{\mu\nu}\:,
\end{equation}
where $G_{\mu\nu}$ is the perturbed first order Einstein tensor. Since the unperturbed quantities satisfy the field equations, i.e., $\bar G_{\mu \nu} = 8 \pi G \bar T_{\mu\nu}$, we have to compute
\begin{equation}
	\delta G_{\mu\nu} = 8 \pi G\: \delta T_{\mu\nu}\:,
\end{equation}
where $\delta G_{\mu\nu} \equiv G_{\mu\nu}-\bar G_{\mu\nu}$ can be computed at first order from $g_{\mu\nu}$. To compute the field equations in Fourier space, we choose the basis $\{ \bv e_1, \bv e_2, \hat{\bv k} \}$ and represent the 3-scalars, 3-vectors, and 3-tensors of the metric perturbation $\delta g_{\mu\nu}$ and the energy-momentum perturbation $\delta T_{\mu\nu}$ in terms of helicity states so that the particular expressions are given by Eq.~(\ref{eq:helicity_representation}). The computation of $\delta G_{\mu\nu}$ is straightforward, but lengthy and tedious. We will not go into the details of this calculation, but refer to the Appendix D of \cite{kodama1984}, who provide explicit expressions for the perturbations of many geometrical quantities (e.g., Christoffel symbols $\delta \Gamma^\alpha_{\mu\nu}$, scalar curvature $\delta \mathcal{R}$, Einstein tensor $\delta G^\mu_{\phantom \mu \nu}$) in Fourier space in terms of helicity states using the same basis as Eqs.~(\ref{eq:scalar_basis})-(\ref{eq:tensor_basis3}). For a given (comoving) Fourier mode $\bv k$, the field equations then become\footnote{In Appendix D of \cite{kodama1984} the perturbations of the Einstein Tensor are given in the form $\delta G^\mu_{\phantom \nu \nu}$ for all helicity states. So we have to compute $8\pi G \delta T^\mu_{\phantom \nu \nu}$ and equate it with the corresponding expression for $\delta G^\mu_{\phantom \nu \nu}$ (for $K=0$ and $n=3$) for all helicity states separately. (It should be noted that $B_i$ is defined with the opposite sign in \cite{kodama1984}.) With Eqs.~(\ref{eq:tmunu_bar}) and (\ref{eq:g_inverse}) the energy-momentum perturbation with mixed indices is at first order given by
\begin{equation}
  \delta T^{\mu}_{\phantom \nu \nu}  = T_{\alpha\nu} g^{\alpha \mu} =  \left(
      \begin{array}{cc}
        -\delta \rho & (\bar \rho + \bar p) (v_i +  B_i)  \\
        -(\bar \rho + \bar p) v_i & \delta p\: \delta_{ij} + \Pi_{ij} 
      \end{array} \right)\:.
  \end{equation}
The field equations are then obtained from the following components (up to a constant): For the scalar perturbations, the density equation corresponds to the $3 \mathcal{H}/k\:\: \: \delta G^0_{\phantom 0 j} - \delta G^0_{\phantom 0 0}$ component, the momentum equation to $\delta G^0_{\phantom 0 j}$, the pressure equation to $\delta G^i_{\phantom i i} - \delta G^0_{\phantom 0 0}$, and the anisotropic stress equation to $\delta G^1_{\phantom 1 1} - 1/3\: \delta G^i_{\phantom i i}$. For the vector perturbations, the momentum equation corresponds to the $\delta G^0_{\phantom 0 j}$ component and the anisotropic stress equation to $\delta G^i_{\phantom i j}$. The tensor perturbations correspond to the $\delta G^i_{\phantom i j}$ component.}
\begin{equation}\label{eq:scalar_field_equations}
\boxed{\begin{aligned}
&\text{\textbf{Scalar field equations:}}\\\\
&\text{density:}\\
&k^2\left(H_{\rm L}^{(0)} +\frac{1}{3}H^{(0)} \right)-\mathcal{H} \left(kB^{(0)}+\dot H^{(0)}\right) = 4 \pi G a^2 \left[\delta \rho^{(0)} + 3 \frac{\mathcal{H}}{k}\left(\bar \rho + \bar p\right)\left(v^{(0)}+B^{(0)}\right)\right]\\
	&\text{momentum:}\\
	&\mathcal{H}A^{(0)}-\dot H_{\rm L}^{(0)} -\frac{1}{3}\dot H^{(0)} = 4 \pi G a^2 \frac{1}{k}\left(\bar \rho + \bar p\right)\left(v^{(0)}+B^{(0)}\right)\\
	&\text{pressure:}\\
	&\left(2\dot{\mathcal{H}}+\mathcal{H}\partial_\tau -\frac{1}{3}k^2\right)A^{(0)} - \left(\partial_\tau+\mathcal{H}\right)\left(\dot H_{\rm L}^{(0)} - \frac{k}{3}B^{(0)}\right) = 4 \pi G a^2\left(\delta p^{(0)} + \frac{1}{3}\delta \rho^{(0)} \right)\\
	&\text{anisotropic stress:}\\
	&k^2\left(A^{(0)}+H_{\rm L}^{(0)}+\frac{1}{3}H^{(0)} \right)-\left(\partial_\tau + 2\mathcal{H} \right)\left(k B^{(0)}+\dot H^{(0)}\right) = -8\pi G a^2 \Pi^{(0)}
\end{aligned}}
\end{equation}
\begin{equation}
\boxed{\begin{aligned}
&\text{\textbf{Vector field equations:}}\\\\
&\text{momentum:}\quad k B^{(\pm 1)} + \dot H^{(\pm 1)} = - 16 \pi G a^2\frac{1}{k} \left(\bar \rho + \bar p \right)\left(v^{(\pm 1)}+B^{(\pm 1)}\right)\\
&\text{anisotropic stress:}\quad \left(\partial_\tau + 2\mathcal{H}\right)\left(k B^{(\pm 1)}+\dot H^{(\pm 1)}\right) = 8 \pi G a^2 \Pi^{(\pm 1)}
\end{aligned}}
\end{equation}
\begin{equation}\label{eq:tensor_field_equation}
\boxed{\begin{aligned}
&\text{\textbf{Tensor field equation:}}\\\\
&\text{anisotropic stress:}\quad \ddot H^{(\pm 2)} + 2 \mathcal{H} \dot H^{(\pm 2)} + k^2 H^{(\pm 2)} = 8 \pi G a^2 \Pi^{(\pm 2)}\:.
\end{aligned}}
\end{equation}

The field equations are complemented by the equations of motion given by the general relativistic energy-momentum conservation
\begin{equation}\label{eq:energy_momentum_conservation}
	\nabla_\nu T^{\mu\nu} = \partial_\nu T^{\mu\nu} + \Gamma_{\beta\nu}^\mu T^{\beta\nu}+\Gamma_{\beta\nu}^\nu T^{\mu\beta} = 0\:.
\end{equation}
Again, Fourier transforming and using Appendix D of \cite{kodama1984} (and additionally Appendix A for the unperturbed expressions), we find for a given (comoving) Fourier mode $\bv k$
\begin{equation}\label{eq:scalar_equations_of_motion}
\boxed{\begin{aligned}
&\text{\textbf{Scalar equations of motion:}}\\\\
&\text{continuity:}\quad \left(\partial_\tau + 3\mathcal{H} \right)\delta \rho^{(0)}+3\mathcal{H} \delta p^{(0)} = -\left(\bar \rho + \bar p\right)\left(k v^{(0)}+3 \dot H_{\rm L}^{(0)}\right)   \\
	&\text{Euler:}\quad \frac{1}{k}\left(\partial_\tau + 4 \mathcal{H} \right)\left[\left(\bar \rho + \bar p\right)\left(v^{(0)}+B^{(0)}\right)\right] = \delta p^{(0)}-\frac{2}{3}\Pi^{(0)}+\left(\bar \rho + \bar p\right)A^{(0)} 
\end{aligned}}
\end{equation}
\begin{equation}
\boxed{\begin{aligned}
&\text{\textbf{Vector equation of motion:}}\\\\
&\text{Euler:}\quad \frac{2}{k}\left(\partial_\tau + 4 \mathcal{H} \right)\left[\left(\bar \rho + \bar p\right)\left(v^{(\pm 1)}+B^{(\pm 1)}\right)\right] = -\Pi^{(\pm 1)}\:. 
\end{aligned}}
\end{equation}

The equations of motion are not independent from the field equations. However, since the field equations are second order differential equations and the equations of motion only first order differential equations, it might be quite useful to use the latter in place of two of the field equations. Furthermore, if the universe is made up of $N$ non-interacting fluids $[T_I]^{\mu\nu}$, $I = 1,\ldots,N$, the energy-momentum conservation is satisfied separately by each fluid, i.e., $\nabla_\nu [T_I]^{\mu\nu} = 0$ for $I = 1,\ldots,N$. This information could not be derived from the field equations.

Similar to the unperturbed case discussed in Chapter \ref{sec:homogeneous_and_isotropic_universe}, the field equations together with the equations of motion do not constitute a closed set of equations, since there are more variables than equations. Like we needed the equation of state (\ref{eq:equation_of_state}) in the unperturbed case, we also need additional relations to close the set of equations here. This is usually done in two different ways: First, DM and baryons can often be approximated to high accuracy by an ideal fluid, i.e., $\Pi = 0$ and an equation of state between $\rho$ and $p$. Second, for photons and neutrinos the ideal fluid approximation is not sufficient and so the energy-momentum tensor has to be modeled using the general relativistic Boltzmann equation. The former case is applied in Section {\ref{sec:dark_matter_and_baryons}} and the latter is outlined in Section \ref{sec:complete_treatement}. Once the energy momentum tensor is fully specified, the scalar and vector perturbations of the metric tensor (including their initial conditions) are fully specified too by means of the field equations and equations of motion. Hence, to solve the scalar and vector equations, only initial conditions for the energy-momentum tensor and its derivative are needed. However, the situation is different for tensor perturbations. It is obvious that even if $\Pi^{(\pm 2)} = 0$, Eq.~(\ref{eq:tensor_field_equation}) allows a nonzero solution for $H^{(\pm 2)}$, and so to specify this solution initial conditions for $H^{(\pm 2)}$ and $\dot H^{(\pm 2)}$ are required. That is, the tensor perturbations of the metric tensor have their own degrees of freedom additionally to the energy-momentum tensor. This has consequences for the generation of perturbations during inflation (see the discussion in the Footnote \ref{foot:deg} of Ch.~\ref{sec:generation_of_initial_perturbations}).

The field equations and equations of motion became much simpler by decoupling them into different helicity states, but they are still very complicated. Fortunately, there is still a way to simplify them without committing any further approximations within our linear treatment. This is choosing the coordinates $x$ in a sophisticated way, which will be discussed in the next section.

\section{Gauge transformations}\label{eq:gauge_transformations}

So far we have described the perturbations in a given coordinate system $x = (\tau,\bv x)$, but we have not said much about the coordinate system itself. The only thing we know about this coordinate system is that it was chosen such that the metric $g_{\mu\nu}$ approximates very closely the FLRW metric $\bar{g}_{\mu\nu}$ allowing us to treat the difference of the two, i.e., $\delta g_{\mu\nu}$, as a small perturbation. However, while in the limiting case of a homogeneous and isotropic universe there is a preferred coordinate system with the metric becoming particularly simple, i.e., it takes the form $\bar{g}_{\mu\nu}$, there is no such preferred choice in the presence of perturbations. Therefore it might be useful to change the coordinate system by a small coordinate transformation, so that $\delta g_{\mu\nu}$ in the new coordinate system is still very small. Such coordinate transformations are called \textbf{gauge transformations}.\footnote{\cite{malik2008} provide a review about different perspectives on the mathematical interpretation of gauge transformations.} While we have performed all calculations keeping the full generality of the metric $g_{\mu\nu}$, we are, in principle, free to choose a specific coordinate system (or \textbf{gauge}) such that it takes a suitable form.

From a fundamental point of view, no gauge is better than the other, but there may be gauges which are particularly appropriate for certain applications. More important, in the general gauge used so far, not all of the perturbations $A$, $H_{\rm L}$, $B_i$, and $H_{ij}$ correspond to ``physical perturbations'' in the universe. Some of them are spurious and exist only in a particular gauge choice. This ``unphysical gauge modes'' impeded the interpretation of the perturbations and were responsible for some confusion in the past (see \citealt{kodama1984} Sect.~1 for a historical review). Therefore, it is important to eliminate these unphysical gauge modes. In the literature there exist two different ``schools'' how this can be achieved. One way is by introducing ``gauge independent perturbations'' (e.g., \citealt{bardeen1980,kodama1984,mukhanov1992,mukhanov2005,straumann2006,durrer2008}), the other way is to ``fix the gauge'' by setting constraints on $A$, $H_{\rm L}$, $B_i$, and $H_{ij}$ such that the coordinate system is completely specified (e.g., \citealt{ma1995,seljak1996,hu1998,liddle2000,weinberg2008}). None of these two approaches is superior to the other. The most important thing is to know how to correctly interpret the perturbation quantities. This is probably equally difficult in both approaches. Here, we will follow the second school and treat the gauge issue by fixing the gauge and changing between different gauges.

\subsection{Transformation laws}\label{sec:gauge_transformations}

A general gauge transformation has the form
\begin{equation}
	\tilde x^\mu = x^\mu + \xi^\mu(x)\:,
\end{equation}
where $\xi^\mu(x)$ is treated as a small perturbation. In the following, we will denote all transformed quantities by a tilde (e.g., $\tilde{q}$). How do perturbations transform under gauge transformations? It was already mentioned that in general perturbations are no geometrical objectes, since splitting variables into a background and a perturbation is a non-covariant procedure. To derive the transformation laws for perturbations, we need the following first order relation
\begin{equation}
\xi^\mu(x) = \xi^\mu(\tilde x-\xi) = \xi^\mu(\tilde x) - \partial_\alpha \xi^\mu(\tilde x)\:\xi^\alpha = \xi^\mu(\tilde x)\:.
\end{equation}
Let $q(x)$ be a 4-scalar, i.e., $\tilde q(\tilde x) = q(x)$. It holds
\begin{equation}
\tilde q(\tilde x) = q(x) = q(\tilde x-\xi) = q(\tilde x) - \partial_\mu q(\tilde x)\:\xi^\mu(\tilde x) = q(\tilde x)-\partial_\mu \bar q(\tilde x)\: \xi^\mu(\tilde x)\:.
\end{equation}
Thus, with the definitions of the perturbations $\delta\tilde q(\tilde x) \equiv\tilde q(\tilde x) - \bar q(\tilde x)$ and $\delta q(\tilde x) \equiv q(\tilde x) - \bar q(\tilde x)$ we obtain the transformation behavior for perturbations of 4-scalars: 
\begin{equation}\label{eq:scalar_transformation}
	\delta \tilde q(\tilde x) = \delta q (\tilde x) - \partial_\mu \bar q (\tilde x)\:\xi^\mu(\tilde x)\:.
\end{equation}
(Recall that $\bar q(x)$ is not a geometrical object on $\mathcal{M}$, but has a fixed functional form irrespective of the choice of coordinates, i.e., it holds $\bar q(x) = \bar q(\tilde x)$ for $x = \tilde x$.)  For a 4-vector $q_\mu$, i.e., $\tilde q_\mu(\tilde x) = \partial x^\alpha/\partial \tilde x^\mu\: q_\alpha(x)$, it works quite similar. With $\partial x^\alpha / \partial \tilde x^\mu = \delta_\mu^\alpha - \partial^\alpha \xi_\mu(\tilde x)$ we have
\begin{equation}
\begin{aligned}
	\tilde q_\mu(\tilde x) &= \frac{\partial x^\alpha}{\partial \tilde x^\mu}\: q_\alpha(x) = \frac{\partial x^\alpha}{\partial \tilde x^\mu}\: q_\alpha\left(\tilde x - \xi \right) = \Big[ \delta_\mu^\alpha - \partial^\alpha \xi_\mu (\tilde x)\Big]\Big[q_\alpha (\tilde x)-\partial_\beta q_\alpha(\tilde x)\:\xi^\beta (\tilde x)\Big]\\
	&= q_\mu(\tilde x)-\partial_\beta q_\mu (\tilde x)\:\xi^\beta(\tilde x)-q_\alpha(\tilde x)\:\partial_\mu \xi^\alpha(\tilde x)\:.
\end{aligned}
\end{equation}
Thus again, with the definitions $\delta \tilde q_\mu(\tilde x) \equiv \tilde q_\mu(\tilde x) -  \bar q_\mu(\tilde x)$ and $\delta q_\mu(\tilde x) \equiv q_\mu(\tilde x) -  \bar q_\mu(\tilde x)$ we find
\begin{equation}\label{eq:vector_transformation}
\delta \tilde q_\mu(\tilde x) = \delta q_\mu(\tilde x) -  \partial_\beta \bar q_\mu(\tilde x)\: \xi^\beta(\tilde x) - \bar q_\alpha (\tilde x) \:\partial_\mu \xi^\alpha (\tilde x)\:.
\end{equation}
Analog we have for 4-tensors $q_{\mu\nu}$
\begin{equation}
	\tilde q_{\mu\nu}(\tilde x) = \frac{\partial x^\alpha}{\partial \tilde x^\mu} \frac{ \partial x^\beta}{\partial \tilde x^\nu}\: q_{\alpha\beta}(x)
	= \Big[\delta^\alpha_\mu-\partial_\mu \xi^\alpha(\tilde x)\Big]\Big[\delta^\beta_\nu- \partial_\nu\xi^\beta(\tilde x)\Big]\Big[q_{\alpha\beta}(\tilde x)-\partial_\gamma q_{\alpha\beta}(\tilde x)\:\xi^\gamma(\tilde x)\Big]\:,
\end{equation}
and obtain by the same argument 
\begin{equation}\label{eq:tensor_transformation}
	\delta \tilde q_{\mu\nu}(\tilde x) = \delta q_{\mu\nu}(\tilde x)- \partial_\beta \bar q_{\mu\nu}(\tilde x)\:\xi^\beta (\tilde x) - \bar q_{\mu\beta}(\tilde x)\:\partial_\nu\xi^\beta(\tilde x)-\bar q_{\alpha\nu}(\tilde x)\:\partial_\mu \xi^\alpha(\tilde x)\:.
\end{equation}
So we can summarize the transformation behavior for scalar, vector, and tensor perturbations as
\begin{empheq}[box=\fbox]{align}
\delta \tilde q(\tilde x) &= \delta  q(\tilde x) - \partial_\mu \bar q (\tilde x)\:\xi^\mu(\tilde x)\label{eq:gauge_transformation_scalar}\\
\delta \tilde q_\mu(\tilde x) &= \delta q_\mu(\tilde x) -  \partial_\beta \bar q_\mu(\tilde x)\: \xi^\beta(\tilde x) - \bar q_\alpha (\tilde x) \:\partial_\mu \xi^\alpha (\tilde x)\\
\delta \tilde q_{\mu\nu}(\tilde x) &= \delta q_{\mu\nu}(\tilde x)- \partial_\beta \bar q_{\mu\nu}(\tilde x)\:\xi^\beta (\tilde x) - \bar q_{\mu\beta}(\tilde x)\:\partial_\nu\xi^\beta(\tilde x)-\bar q_{\alpha\nu}(\tilde x)\:\partial_\mu \xi^\alpha(\tilde x)\:.\label{eq:gauge_transformation_tensor}
\end{empheq}
Note that the argument is always the same on both sides and is thus \emph{not} referring to the same point in spacetime unlike the usual transformation, e.g., $\tilde q(\tilde x) = q(x)$, where $\tilde{x}$ and $x$ are different arguments but refer to the same point in spacetime.

Knowing how $\delta g_{\mu\nu}$ transforms we are able to derive the transformation behavior  in Fourier space for the single helicity states that are contained in $\delta g_{\mu\nu}$. That is, we apply Eq.~(\ref{eq:gauge_transformation_tensor}) to $\delta g_{\mu\nu}$, Fourier transform it, and simply relate the left and the right-hand-side seperately for different helicity states in the basis $\{ \bv e_1, \bv e_2, \hat{\bv k} \}$ for a given Fourier mode $\bv k$. To do this, we decompose $\xi^\mu$ itself into helicity states in the basis $\{ \bv e_1, \bv e_2, \hat{\bv k} \}$
\begin{equation}
	\xi^\mu = \left(T, L_i \right) \supeq{(\ref{eq:helicity_representation})}{=} \left(T^{(0)},\frac{L^{(+1)}+L^{(-1)}}{\sqrt{2}},i \frac{L^{(+1)}-L^{(-1)}}{\sqrt{2}},-iL^{(0)}  \right)\:.
\end{equation}
Since there are no tensor modes involved, it follows immediately from Eq.~(\ref{eq:gauge_transformation_tensor}) that tensor perturbations are invariant under gauge transformations. Scalar and vector modes, however, transform nontrivially. Using the relation
\begin{equation}
	\partial_\gamma \bar g_{\mu\nu} \xi^\gamma= \partial_0 \bar g_{\mu\nu} \xi^0 + \partial_i \bar g_{\mu\nu} \xi^i = 2 \mathcal{H}\bar g_{\mu \nu}\xi^0
\end{equation}
we obtain for a given Fourier mode $\bv k$ after a straightforward calculation
\begin{empheq}[box=\fbox]{align}
\tilde A^{(0)} &= A^{(0)}-\dot T^{(0)} -\mathcal{H}T^{(0)}\label{eq:gauge_transformation_1} \\
\tilde B^{(0)} &= B^{(0)}-\dot L^{(0)}  - k T^{(0)}\label{eq:gauge_transformation_2} \\
\tilde H_{\rm L}^{(0)} &= H_{\rm L}^{(0)}-\frac{k}{3}L^{(0)} - \mathcal{H} T^{(0)}\label{eq:gauge_transformation_3}\\
\tilde H^{(0)} &= H^{(0)}+k L^{(0)}\label{eq:gauge_transformation_4}\\\nonumber\\
\tilde B^{(\pm 1)} &= B^{(\pm 1)}-\dot L^{(\pm 1)}\label{eq:gauge_transformation_5}\\
\tilde E^{(\pm 1)} &= E^{(\pm 1)} + k L^{(\pm 1)}\label{eq:gauge_transformation_6}\\\nonumber\\
\tilde E^{(\pm 2)} &= E^{(\pm 2)}\label{eq:gauge_transformation_7}
\end{empheq}
and similarly for the helicity states of the energy-momentum perturbation $\delta T_{\mu\nu}$
\begin{empheq}[box=\fbox]{align}
\delta \tilde \rho^{(0)} &= \delta \rho^{(0)} - \dot{\bar \rho} (x) T^{(0)}\label{eq:gauge_transformations_energy_1}\\
\delta \tilde p^{(0)} &= \delta p^{(0)} - \dot {\bar p} T^{(0)}\label{eq:gauge_transformations_energy_2}\\
\tilde v^{(m)} &= v^{(m)}+\dot L^{(m)}\:,\ \ \ m = 0,\pm 1\label{eq:gauge_transformations_energy_3}\\
\tilde \Pi^{(m)} &= \Pi^{(m)}\:,\ \ \ m = 0,\pm 1,\pm 2\:.\label{eq:gauge_transformations_energy_4}
\end{empheq}

\subsection{Particular gauges}

Since tensor perturbations are gauge invariant, there is no gauge choice for them. On the other hand, since vector modes are very unlikely to be generated in the early universe (see the footnote in Sect.~\ref{sec:canonical_quantization}), we will not consider them any more. We will only consider scalar perturbations and thus drop the $(0)$ superscripts for ease of notation. Although there are many scalar gauges discussed in the literature, we will only introduce two of them, because we will need them in the course of our analysis. The Newtonian gauge is particularly useful for analytical treatment inside the horizon and allows a very simple interpretation in terms of Newtonian physics. The comoving gauge will be used for the study of perturbations outside the horizon in the context of the generation of initial perturbations (see Ch.~\ref{sec:generation_of_initial_perturbations}). 

\subsubsection{Newtonian Gauge}

The Newtonian gauge is defined by the condition
\begin{equation}
	B = H = 0\:,
\end{equation}
and we will rename the non-vanishing perturbations by
\begin{equation}
	\Psi \equiv A\:,\ \ \ \ \ \Phi \equiv -H_{\rm L}\:.
\end{equation}
Using Eqs.~(\ref{eq:gauge_transformation_2}) and (\ref{eq:gauge_transformation_4}) and setting $\tilde B = \tilde H = 0$ we see that the Newtonian gauge is obtained from a general gauge by the transformation
\begin{equation}
	T = \frac{B}{k}+\frac{\dot H}{k^2}\:,\ \ \ \ \ \ \ L = - \frac{H}{k}\:.
\end{equation}
Since the transformation is uniquely determined, the Newtonian gauge is entirely fixed. The field equations (\ref{eq:scalar_field_equations}) become\footnote{Eqs.~(\ref{eq:second_field_equations_Newtonian_gauge}) and (\ref{eq:last_field_equations_Newtonian_gauge}) correspond to the momentum and anisotropic stress equation, respectively. Eq.~(\ref{eq:first_field_equations_Newtonian_gauge}) is the density equation minus $3\mathcal{H}$ times the momentum equation. Eq.~(\ref{eq:third_field_equations_Newtonian_gauge}) is the pressure equation minus $1/3$ times Eq.~(\ref{eq:first_field_equations_Newtonian_gauge}).}
\begin{empheq}[box=\fbox]{align}
	-k^2 \Phi - 3 \mathcal{H}\left(\dot \Phi + \mathcal{H}\Psi \right) &= 4\pi G a^2 \delta \rho \label{eq:first_field_equations_Newtonian_gauge}\\
	\dot \Phi + \mathcal{H}\Psi &= 4\pi G a^2 \frac{v}{k}\left(\bar \rho + \bar p\right)\label{eq:second_field_equations_Newtonian_gauge}\\
	\ddot \Phi + \mathcal{H}\left(\dot \Psi + 2 \dot \Phi \right)+\left(2 \dot{\mathcal{H}}+\mathcal{H}^2\right)\Psi+\frac{k^2}{3}\left(\Phi-\Psi\right) &= 4\pi Ga^2 \delta p\label{eq:third_field_equations_Newtonian_gauge}\\
	k^2\left(\Phi-\Psi\right) &= 8\pi Ga^2 \Pi\label{eq:last_field_equations_Newtonian_gauge}
	\end{empheq}
and the equations of motion (\ref{eq:scalar_equations_of_motion})
\begin{equation}\label{eq:equations_of_motion_Newtonian_gauge}
\boxed{\begin{aligned}
\left(\partial_\tau+3\mathcal{H}\right)\delta \rho + 3 \mathcal{H}\delta p &= -\left(\bar \rho+\bar p\right)\left(kv-3 \dot \Phi\right)\\
\left(\partial_\tau+4\mathcal{H}\right)\left[\frac{v}{k}\left(\bar\rho+\bar p\right)\right] &= \delta p - \frac{2}{3}\Pi+\left(\bar \rho+\bar p\right)\Psi\:.
\end{aligned}}
\end{equation}
These equations are still exact aside from the use of first-order perturbation theory. Since $D_{\rm H} = 1/\mathcal{H}$ is the comoving Hubble length (see Sect.~\ref{sec:horizons}), $\mathcal{H}/k\ll 1$ means that the length scale of a given Fourier mode $k$ is well within the horizon. In this limit Eq.~(\ref{eq:first_field_equations_Newtonian_gauge}) becomes
\begin{equation}\label{eq:generalized_poisson_equation}
	-k^2 \Phi = 4\pi G a^2 \delta \rho
\end{equation}
being identical to the Poisson equation (see Eq.~\ref{eq:hydrodynamical3}) derived by Newtonian fluid dynamics, if $\Phi$ is interpreted as the perturbation of the Newtonian gravitational potential. Thus, $\Phi$ has a simple physical interpretation and, well within the horizon, the Newtonian gauge reduces to Newtonian mechanics. We will call $\Phi$ the ``generalized Newtonian potential''.\footnote{Note that the gauge invariant ``Bardeen potentials'' defined by \cite{bardeen1980} as
\begin{equation}
		\Phi_A \equiv A-\frac{1}{k}\dot B-\frac{1}{k}\mathcal{H}B-\frac{1}{k^2}\left(\ddot H+\mathcal{H}\dot H\right)\:,\ \ \ \ \ \Phi_H \equiv H_{\rm L}+\frac{1}{3}H-\frac{1}{k}\mathcal{H}B-\frac{1}{k^2}\mathcal{H}\dot H\:,
\end{equation}
reduce in the Newtonian gauge to
\begin{equation}
	\Psi = \Phi_A\:, \ \ \ \ \ \ \Phi = -\Phi_H\:.
\end{equation}
This means that the simple physical interpretation of $\Phi$ is automatically conveyed to $-\Phi_H$ in all gauges, since $\Phi_H$ is gauge invariant.} Another advantage of the Newtonian gauge is that the metric tensor $g_{\mu\nu}$ is diagonal, which makes analytic calculation convenient.

Eq.~(\ref{eq:last_field_equations_Newtonian_gauge}) yields a simple algebraic formula connecting $\Psi$ and $\Phi$ by the anisotropic stress $\Pi$. During the matter dominated era in which independent, ideal fluids (DM and baryonic matter) dominate the energy density, we may neglect the anisotropic stress. With this approximation we obtain $\Psi \simeq \Phi$ and there remains only a single parameter in metric perturbation being the generalized Newtonian potential $\Phi$.

\subsubsection{Comoving gauge}

The comoving gauge is defined by
\begin{equation}
	B + v = 0\:,\ \ \ \ \ \ \ H = 0\:.
\end{equation}
Using Eqs.~(\ref{eq:gauge_transformation_2}), (\ref{eq:gauge_transformation_4}) and (\ref{eq:gauge_transformations_energy_3}) and setting $\tilde B+\tilde v = 0$ and $\tilde H=0$ we see that the transformation from a general gauge into comoving gauge is given by
\begin{equation}\label{eq:transformation_to_comoving_gauge}
	T = \frac{1}{k}\left(B+v\right)\:,\ \ \ \ \ \ L = -\frac{H}{k}\:.
\end{equation}
This transformation is uniquely determined, so the comoving gauge is entirely fixed. We rename the remaining metric perturbations as
\begin{equation}\label{eq:xi_and_zeta}
	\xi \equiv A\:,\ \ \ \ \ \ \ \zeta \equiv H_{\rm L}\:.
\end{equation}

There is a relation between $\zeta$ and the 3-curvature perturbation $ {}^{(3)}\delta\mathcal{R}$ on a slice of constant comoving time $\tau$. In a general gauge the 3-scalar curvature perturbation is given by\footnote{This is easily seen by using the expression for the 4-scalar curvature perturbation in the Appendix D of \cite{kodama1984} and setting all terms containing $A$ or a time derivative to zero.}
\begin{equation}
	{}^{(3)}\delta \mathcal{R} = 4 \frac{k^2}{a^2}\left(H_{\rm L}+\frac{1}{3}H\right)\:,
\end{equation}
so it reduces in the comoving gauge to
\begin{equation}\label{eq:curvature_and_zeta}
	 {}^{(3)}\delta\mathcal{R} = 4\frac{k^2}{a^2}\zeta\:.
\end{equation}
As is shown in Section \ref{sec:conservation_of_perturbations}, the 3-scalar curvature perturbation is a useful quantity, since, under certain conditions, it stays constant outside the horizon if multiplied by $a^2$.
 
\section{Evolution of the perturbations}

In this section we try to find solutions to the field equations and equations of motion from the previous section. After the electron-positron annihilation at a temperature of $T \sim 0.5\ {\rm MeV}/k_{\rm B} \sim 10^{10}$ K, the universe contained only four different components: cold DM, baryons, photons, and neutrinos. We refer to these components with the subscripts d, b, $\gamma$, and $\nu$, respectively. While cold DM and neutrinos interacted with all components only by means of gravitation, the baryons and the photons were tightly coupled to each other until the epoch of decoupling.\footnote{So Eq.~(\ref{eq:equation_of_motion1}) is not satisfied separately for photons and baryons during this time.} Again we will only consider scalar perturbations and thus drop the $(0)$ superscripts for ease of notation.

\subsection{Dark matter and baryons}\label{sec:dark_matter_and_baryons}

Since cold DM interacts with the baryons, photons, and neutrinos only by means of gravity, the conservation equation (\ref{eq:energy_momentum_conservation}) is satisfied separately for the DM fluid $[T_{\rm d}]^{\mu\nu}$, i.e., $\nabla_\mu [T_{\rm d}]^{\mu\nu} = 0$. In the Newtonian gauge, with $p_{\rm d}=\Pi_{\rm d} = 0$ for DM, the equations of motion (\ref{eq:equations_of_motion_Newtonian_gauge}) become
\begin{align}
	\big(\partial_\tau + 3 \mathcal{H}\big)\delta \rho_{\rm d} &= \bar \rho_{\rm d}\big(3 \dot \Phi - k v_{\rm d}\big)\:,\label{eq:dm_newtonian1}\\
	\big(\partial_\tau+4 \mathcal{H}\big)\bar \rho_{\rm d} v_{\rm d} &= k \bar \rho_{\rm d} \Psi\:.\label{eq:dm_newtonian2}
\end{align}
Photons, neutrinos, and baryons affect the evolution of the DM only by means of $\Psi$ and $\Phi$. By introducing the DM overdensity
\begin{equation}
	\delta_{\rm d} = \frac{\delta \rho_{\rm d}}{\bar \rho_{\rm d}}
\end{equation}
and by using the equation of motion (\ref{eq:equation_of_motion_total_2}) for DM, Eqs.~(\ref{eq:dm_newtonian1}) and (\ref{eq:dm_newtonian2}) simply become
\begin{align}
	\dot \delta_{\rm d} &= - k v_{\rm d} + 3\dot \Phi \label{eq:dm_newtonian3}\:,\\
	\dot v_{\rm d} &= - \mathcal{H}v_{\rm d} + k\Psi\:.\label{eq:dm_newtonian4}
\end{align}
So by differentiating Eq.~(\ref{eq:dm_newtonian3}) and replacing $\dot v_{\rm d}$ by Eq.~(\ref{eq:dm_newtonian4}) we can eliminate $v_{\rm d}$ and obtain a relation between $\delta_{\rm d}$ and the metric perturbations:
\begin{equation}\label{eq:cdm_grwoth_gr}
	\ddot \delta_{\rm d} + \mathcal{H} \dot \delta_{\rm d} = -k^2\Psi +3 \mathcal{H}\dot \Phi + 3 \ddot \Phi\:.
\end{equation}
In order to solve this equation, we have to know the metric perturbations $\Psi$ and $\Phi$, which depend on the total energy budget of the universe. However, in the matter dominated era we can assume $T^{\mu\nu} \simeq T^{\mu\nu}_{\rm d}$, and so we have $\Psi \simeq \Phi$ due to $\Pi \simeq \Pi_{\rm d} = 0$ (see Eq.~(\ref{eq:last_field_equations_Newtonian_gauge})). Then for modes well inside the horizon we can replace $k^2\Phi$ by the Poisson equation (\ref{eq:generalized_poisson_equation}) yielding
\begin{equation}
	\ddot \delta_{\rm d} + \mathcal{H} \dot \delta_{\rm d} = 4\pi G a^2 \bar \rho_{\rm d} \delta_{\rm d}+ 3 \mathcal{H} \dot \Phi + 3 \ddot \Phi\:.
\end{equation}
If we can further ignore the time derivatives of $\Phi$, we obtain
\begin{equation}\label{eq:growth_delta_d}
\boxed{\ddot \delta_{\rm d} + \mathcal{H} \dot \delta_{\rm d} - 4\pi G a^2 \bar \rho_{\rm d} \delta_{\rm d} = 0\:,}
\end{equation}
which is equivalent to the evolution equation of DM in the Newtonian regime (see Eq.~(\ref{eq:linear_perturbation_equation_simplified})). However, neglecting the time derivatives of $\Phi$ is only acceptable, if the solutions lead to more or less time independent $\Phi$. Using Eqs.~(\ref{eq:dat}) and (\ref{eq:a_tau}), the growing mode during matter domination is given by $\delta_{\rm d} \propto a(\tau) \propto \tau^2$. Thus, according to the Poisson equation (\ref{eq:generalized_poisson_equation}) this leads indeed to a constant $\Phi$ inside the horizon during matter domination and, thus, justifies Eq.~(\ref{eq:growth_delta_d}) \emph{a posteriori}.

While Eq.~(\ref{eq:cdm_grwoth_gr}) was derived for cold DM, it applies for baryons as well after the epoch of decoupling and on scales larger than the Jeans mass $M_{\rm J}$ (see Sect.~\ref{sec:linear_theory}), i.e.,
\begin{equation}\label{eq:baryon_grwoth_gr}
	\ddot \delta_{\rm b} + \mathcal{H} \dot \delta_{\rm b} = -k^2\Psi +3 \mathcal{H}\dot \Phi + 3 \ddot \Phi\:.
\end{equation}
Subtracting Eq.~(\ref{eq:cdm_grwoth_gr}) from Eq.~(\ref{eq:baryon_grwoth_gr}), we obtain
\begin{equation}
		\boxed{\frac{d^2}{d\tau^2} \left(\delta_{\rm d}-\delta_{\rm b}\right) + \mathcal{H} \frac{d}{d\tau}\left(\delta_{\rm d}-\delta_{\rm b}\right) = 0\:.}
\end{equation}
The growing solution for the difference $\delta_{\rm d}-\delta_{\rm b}$ is a constant, so the difference between cold DM and baryon density perturbation does not change with time. Thus, during matter domination it holds
\begin{equation}
	\frac{\delta_{\rm d}-\delta_{\rm b}}{\delta_{\rm d}} = 1-\frac{\delta_{\rm b}}{\delta_{\rm d}}  \propto \tau^{-2}\:. 
\end{equation}
This means that $\delta_{\rm b}$ catches up with $\delta_{\rm d}$ after decoupling and grows at an equal rate.

\begin{figure}[t]\centering
   \includegraphics[scale = 1]{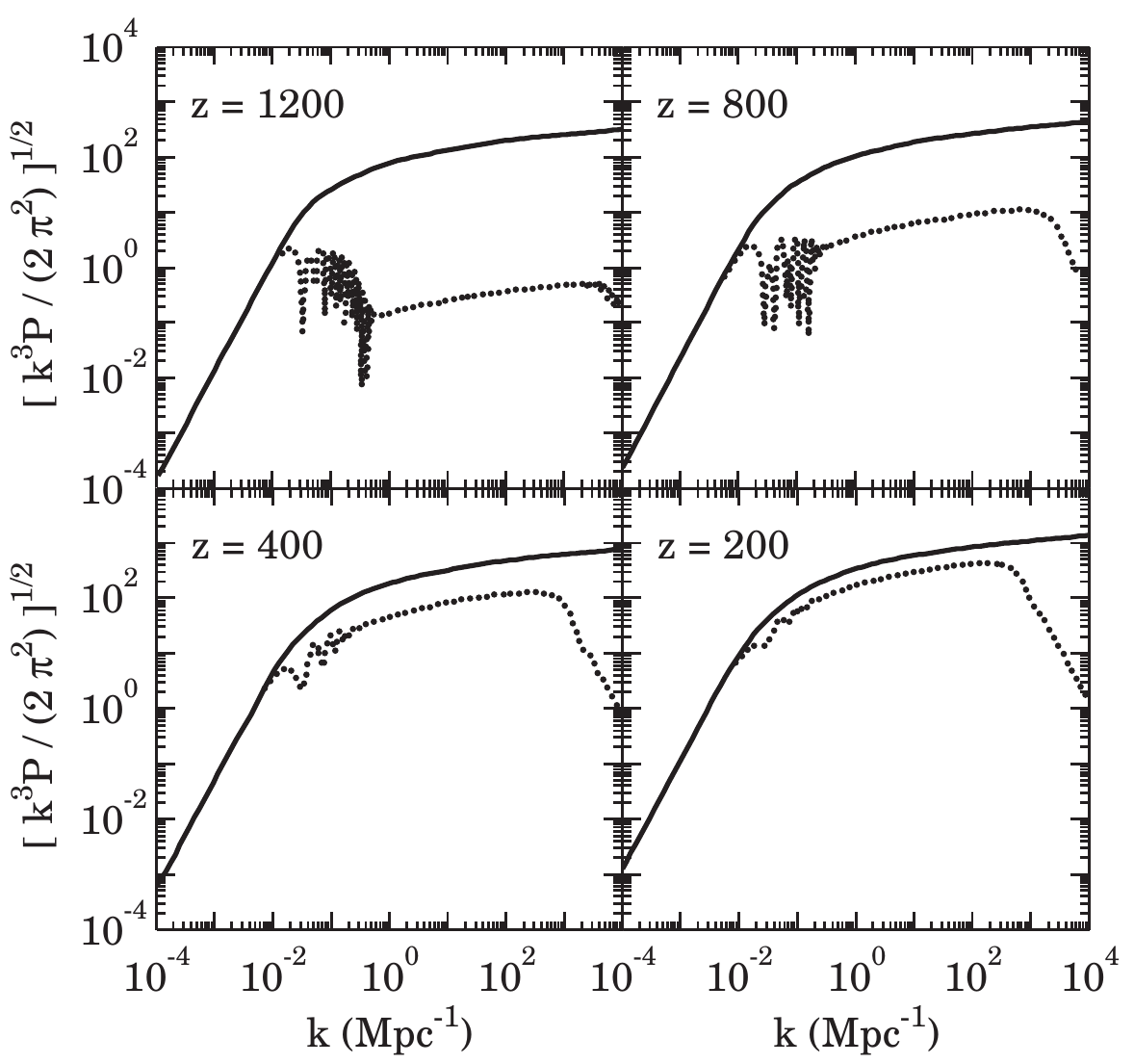}
     \caption{Power spectra (in dimensionless form $k^3 \mathcal{P}$) of the density fluctuations of DM and baryons at different redshifts. The solid curve corresponds to DM perturbations and the dotted curve to baryon perturbations. The horizon scale at $z = 1000$ is about $k \sim 0.01$. Inside the horizon the baryonic acoustic oscillations in the power spectrum of the baryons are clearly visible and it is shown how over the timespan of $z = 1200$ to $z = 200$ the baryon power spectrum slowly catches up with that of DM, as the baryons fall into the gravitational potential wells of previously formed DM structures. \citep[Adapted from][]{naoz2005}}\label{fig:baryon_growth}
\end{figure}

This result can be summarized in simple terms as follows. During matter domination but before recombination ($z_{\rm dec}  \simeq 1089$), the baryons are tightly coupled to the photons and thus are prevented from growing due to the photon pressure, while the DM perturbations grow unimpeded. After recombination, the photons basically propagate freely and the baryons start to feel the gravitational pull of the DM structures that have evolved in the meantime. However, since the photon-to-baryon ratio in the universe is huge, the residual ionization of cosmic gas keeps the temperature of the baryons close to the temperature of the CMB and the radiation drag prevents the baryons to fall into the DM potential wells (see, e.g., Sect.~17.3 of \citealt{peacock1999}, \citealt{naoz2005}). It is only for $z \lesssim 100$ that the baryons entirely decoupled from the photons and catched up with the DM structures.
This is shown in Figure \ref{fig:baryon_growth}, where calculations from \cite{naoz2005} for the power spectra of DM and the baryon density are shown at different epochs using an extension of the CMBfast code (see the following section).
While for scales outside the horizon ($k \lesssim 0.01$ at $z = 1000$) both power spectra are equal, we see inside the horizon the result of the photon-baryon plasma before recombination in form of acoustic oszillations. The figure nicely shows how the baryons slowly catch up by falling into the DM structures over the timespan of $z = 1200$ to $z = 200$.

This scenario is also a strong indication for the existence of some sort of non-baryonic DM, since the DM perturbations help the baryonic perturbations to grow. We saw in Section \ref{eq:lin_growth} that during matter domination the perturbations grow proportionally to the scale factor $a$. In fact, it can be shown \citep{lifshitz1946} that in the general relativistic framework perturbations can maximally grow proportionally to the cosmic time $t$. However, since the baryonic perturbations were prevented from growing at least until the epoch of decoupling, which took place around $z_{\rm dec} \simeq 1089$, and since the perturbations at that time were of the order of $10^{-5}$, they could in a baryon-only scenario (without any DM) maximally grow about a factor of $a_0/a(t_{\rm dec}) = (z_{\rm dec}+1) \sim 1000$ until the present epoch and would be of the order $10^{-2}$ today. Obviously, this is not enough to reach the nonlinear regime and to form the galaxies and clusters we observe today. On the other hand, since the DM perturbations were not coupled to the radiation, they could start to grow much earlier and thus enable us to reconcile the small perturbations at the epoch of decoupling as observed in the CMB with the nonlinear LSS today. 

\subsection{Complete treatment}\label{sec:complete_treatement}

The results in the last section are only approximately valid under certain conditions and during certain cosmological eras. To obtain an accurate model for the transfer function $T(k)$ of DM (see Sect.~\ref{sec:transferfunction}) at the present epoch or of the CMB multipoles $C_l$, it is unavoidable to numerically solve the complete set of equations taking into account all existing components of the universe (cold DM, baryons, photons, and neutrinos). To accurately describe the evolution of photon and neutrino perturbations, one has to apply the general relativistic Boltzmann equation, i.e.,
\begin{equation}\label{eq:boltzmann_equation}
	\left(p^{\mu}\partial_{\mu} - \Gamma^i_{\mu\nu}p^\mu p^\nu \frac{\partial}{\partial p^\mu}\right) f = C[f]\:,
\end{equation}
to the 1-particle distribution functions $f(x,\bv p)$ of photons and neutrinos, where $p^\mu = (p^0,\bv p)$ is the momentum and $C[f]$ the collision term, which accounts for Thomson scattering between electrons and photons and vanishes for neutrinos. The energy-momentum tensor for a particle of mass $m$ is then given by (cf.~Eq.~(\ref{eq:gas}))
\begin{equation}
	T^{\mu\nu}(x) = \int_{P_m(x)} \:\frac{p^\mu p^\nu}{p^0} \:f(x,\bv p)\:\sqrt{-g(x)}\:dp^3\:,
\end{equation}
where ${P_m(x)}$ is the configuration space of particles with mass $m$ at position $x$, i.e., all $p^\mu$ with $g_{\mu\nu}(x)p^\mu p^\nu = -m^2$, and $g$ is the determinant of $g_{\mu\nu}$.
Since Thomson scattering can produce a net polarization of the photons, the complete analysis has to take into account polarization perturbations as well. The derivation of the full set of equations from first principles is given, for instance, in \cite{weinberg2008} in a self-contained way. We refer the interested reader to this comprehensive review for more details.

After having derived the complete set of equations, the remaining challenge is to find a solution, which can only be achieved numerically. But even finding a numerical solution is extremely challenging. Since the 1-particle distribution function $f(x,\bv p)$ is not only a function of spacetime $x$, but also of direction $\bv p$, it is usually expanded into Legendre polynomials $P_l(\mu)$, i.e.,\footnote{Since in both, the collisionless Boltzmann equation and the collision term $C[f]$ for Thomson scattering, the directional dependence enters only by means of the scalar product $\hat{\bv k} \cdot \hat{\bv p} = \mu$,  the 1-particle distribution functions for photons and neutrinos also depend only on $\mu$, if this was initially the case. So we may assume here that the initial momentum dependence is in fact axially symmetric.
Thus the Legendre polynomials $P_l(\mu)$ constitute a complete set for representing $f(\tau,\bv k,\mu)$ for each $\bv k$ and $\tau$. Furthermore, since the field equations (\ref{eq:scalar_field_equations}) and equations of motions (\ref{eq:energy_momentum_conservation}) are linear differential equations depending only on $k = |\bv k|$, two modes $\bv k$ and $\bv k'$ with $|\bv k| = |\bv k'|$ obey the same time evolution up to the amplitude of the initial conditions. Thus, it suffices to solve the field equations and equations of motion for each $k$, i.e., we can write $f(\tau,k,\mu)$.}
\begin{equation}
	f(\tau,k,\mu) = \sum_{l=0}^\infty i^{-l}(2l+1)P_l(\mu)f_l(\tau,k)\:, \ \ \ \ \mu = \hat{\bv k} \cdot \hat{\bv p}\:.
\end{equation}
The corresponding Boltzmann equation is then converted into a ``Boltzmann hierarchy'' using the relation
\begin{equation}
	\left(2l+1\right)\mu P_l(\mu) = \left(l+1\right) P_{l+1}(\mu) + l P_{l-1}(\mu)
\end{equation}
and the orthogonality of the Legendre polynomials. This is a set of coupled differential equations where for each $l$ there is an equation which couples the moments $f_{l+1}(\tau,k)$, $f_{l}(\tau,k)$, and $f_{l-1}(\tau,k)$. To compute the perturbations on scales appropriate for current CMB observations, one needs all Legendre moments $f_{l}$ up to $l \sim 1000$. Thus, one has to solve a system of about 3000 coupled differential equations: 1000 for photon perturbations, 1000 for photon polarization, 1000 for neutrino perturbations (e.g., \citealt{ma1995}, \citealt{seljak1996}). Furthermore, this system of equations has to be solved for every Fourier mode $k$, and since the solutions are rapidly oscillating functions of time, the integration has to proceed in small time steps. This demands a lot of computation time even on present day computers.

Fortunately, \cite{seljak1996} found a way to compute all these Legendre moments without solving this huge system of differential equations. Following their method, one only has to solve this system for $l \lesssim 20$ and all the higher moments are then obtained by means of a line-of-sight integral. This is the heart of the \textbf{CMBfast} code \citep{seljak1996}. The computation with CMBfast was about two orders of magnitude faster than the standard Boltzmann methods, while preserving the same accuracy which was about 1\%-2\% at that time. In the meantime, the code has been continuously extended. The initial code was developed for scalar quantities in flat FLRW background cosmologies, but now it includes open and closed background cosmologies, vector and tensor modes, weak lensing etc.
Today, for parameter sets around the present concordance cosmology, the accuracy of the CMBfast code is about 0.1\% for the CMB multipoles $C_l$ up to $l = 3000$ and even better for the DM transfer function $T(k)$ \citep{seljak2003}. 
Today the CMBfast-package is not suported anymore, but there are other publicly available CMB-computation-packages, such as CAMB \citep{lewis2000} and CMBeasy \citep{doran2005}, which are based on CMBfast.\footnote{For further information about these codes we refer to the corresponding websites \href{http://camb.info/}{http://camb.info/} and \href{http://www.cmbeasy.org}{http://www.cmbeasy.org} respectively.}


\chapter{Generation of primordial perturbations}\label{sec:generation_of_initial_perturbations}

The generation of perturbations in the very early universe is important insofar as it produces the initial conditions for the theories of structure formation as described in the Chapters \ref{sec:structure_formation} and \ref{sec:general_relativistic_treatment}. The basic idea is that during inflation
quantum perturbations are created and stretched to scales outside the horizon, where they are conserved until they reenter the horizon and become observable (see Sect.~\ref{sec:Generation of the primordial perturbations}, where this process is described in simple terms). We will consider only the simplest inflationary scenarios which are driven by a single scalar field $\phi$ (the ``inflaton'') and obey the ``slow roll'' conditions (see Eq.~(\ref{eq:slow_roll_conditions})).

In this chapter, we develop the basic theory behind the generation of perturbations in the very early universe. In Section \ref{sec:quantization_of_perturbations}, we quantize the initial perturbation during inflation by means of the canonical quantization, then in Section \ref{sec:conservation_of_perturbations} we derive the behavior of perturbations outside the horizon, and finally in Section \ref{sec:primordial_power_spectrum} we derive the primordial DM power spectrum for scales inside the horizon.

Since for modes outside the horizon we are in the relativistic regime, this chapter is based on the general relativistic theory of linear perturbations, which was described in the previous chapter. The quantization of the perturbations is performed in Newtonian gauge, the constancy of perturbations outside the horizon is shown for comoving gauge, and finally for the power spectrum within the horizon we transform back to Newtonian gauge. Since we have to deal with perturbed scalar fields $\phi$, we also need certain relations from Appendix \ref{sec:scalar_field_perturbed_universe}. Readers that are not familiar with the theory of scalar fields in the context of general relativity are encouraged to first study the full Appendix \ref{sec:classical_scalar_field}. Also some general knowledge on ``canonical quantization'' of a scalar field in Minkoswski spacetime is required and we refer the reader to textbooks of quantum field theory for an introduction \citep[see, e.g.,][]{mandl1993}. In this section, we will mainly follow the discussion in Seljak (unpublished lecture notes), \cite{liddle2000}, and \cite{mukhanov2005}. 

Like in Chapter \ref{sec:general_relativistic_treatment} the background FLRW universe is assumed to be flat and the time coordinate is conformal time $\tau$ (see Eq.~(\ref{eq:conformal_time})). We consider only scalar perturbations so that the superscript $(0)$ can be omitted in the following.  We choose natural units, i.e., $c = \hbar = 1$.

\section{Quantization of perturbations}\label{sec:quantization_of_perturbations}

To quantize the initial perturbations in the universe, we need basic concepts from quantum field theory. In general, the quantization of a system in curved spacetime is rather complicated and the meaning of a particle or even of the vacuum is subtle (see \cite{birrell1982} for a general discussion). Fortunately, the Lagrangian for the scalar field perturbations reduce to those of a scalar field in Minkowski spacetime (up to an effective time dependent mass). This simplifies our treatment a lot.

\subsection{Classical equation of motion}

The proper quantization of a system has to start from its action $S$. It is dangerous to simply compute the classical equation of motion and try to interpret it in the context of quantum field theory. This could lead to a wrong normalization and thus to an incorrect result as demonstrated by \cite{deruelle1992}. The action for the field $\phi$ in Newtonian gauge is given by Eqs.~(\ref{eq:larangian_gr})-(\ref{eq:action}):
\begin{equation}\label{eq:action2}
	S_\phi = \int_{\mathcal D} \mathcal{L}\: \sqrt{-g}\: dx^4 = \int_{\mathcal D} \left(\frac{1}{16\pi G}\mathcal{R} -\frac{1}{2}\partial_\mu \phi \partial^\mu \phi - V(\phi)\right)\sqrt{-g}\: dx^4\:.
\end{equation}
In order to find the action for the field perturbation $\delta \phi$, the action (\ref{eq:action2}) needs to be expanded to \textit{second} order in perturbations. This is a straightforward, but lengthy calculation and will not be reproduced here. The result is \citep[Sect.~10.3]{mukhanov1992}
\begin{equation}\label{eq:action_second_order}
	S_q = \int_{\mathcal D} \mathcal L_q \:dx^4 = \frac{1}{2} \int_{\mathcal D} \left(-\eta^{\alpha\beta}\partial_\alpha q \:\partial_\beta q + \frac{\ddot z}{z} q^2 + \text{total derivatives}   \right) dx^4\:,
\end{equation}
where we have introduced the notation
\begin{equation}\label{eq:action_v}
	q(x) = \delta \phi + \frac{\dot{\bar \phi}}{\mathcal H}\Phi\:,\quad\quad\quad z(\tau) = \frac{a\dot{\bar \phi}}{\mathcal H}\:.
\end{equation}
The action (\ref{eq:action_second_order}) effectively describes a Klein-Gordon field $v$ with time-dependent mass $m^2(\tau) = -\ddot z/z$ in the Minkowski spacetime $\eta_{\alpha\beta} = {\rm diag}(-1,+1,+1,+1)$ (cf.~Eq.~(\ref{eq:klein_gordon})). This becomes obvious by deriving the equation of motion for $q$. Varying Eq.~(\ref{eq:action_second_order}) with respect to $q$ yields the classical equation of motion by means of the Euler-Lagrange equation (\ref{eq:euler_lagrange})
\begin{equation}\label{eq:equation_of_motion_v}
	-\eta^{\alpha\beta}\partial_\alpha \partial_\beta q - \frac{\ddot z}{z}q = \ddot q - \Delta q - \frac{\ddot z}{z}q = 0\:.
\end{equation}
The total derivatives vanish due to the usual condition $\delta q = 0$ on $\partial \mathcal D$.

Before quantizing the system, we want to find analytic solutions to Eq.~(\ref{eq:equation_of_motion_v}). Such solutions are found by introducing the following approximations. Since $\dot {\bar \phi}/\mathcal H$ is approximately constant during slow roll inflation, we obtain at first order
\begin{equation}
	\frac{\ddot z}{z}q \supeq{(\ref{eq:action_v})}{\simeq} \frac{\ddot a}{a}q = \left( \dot{\mathcal H}+\mathcal H^2 \right) q \simeq 2 \mathcal H^2 q + \mathcal O\left(\mathcal H^2\epsilon q\right)\:.
\end{equation}
In the last step we used the expression (\ref{eq:slow_roll_parameters}) for the slow roll parameter $\epsilon$ in terms of the conformal Hubble parameter, i.e.,
\begin{equation}
\epsilon = 1 - \frac{\dot{\mathcal{H}}}{\mathcal{H}^2}\:,
\end{equation}
and $\epsilon \ll 1$ during slow-roll inflation. If we neglect the term of order $\mathcal O\left(\mathcal H^2\epsilon q\right)$, the equation of motion becomes
\begin{equation}\label{eq_equation_of_motion_v2}
	\ddot q - \left(\Delta + 2\mathcal H^2 \right)q = 0\:.
\end{equation}
This can be further simplified by measuring our time with respect to the end of inflation $\tau_{\rm f}$. That is, we introduce a new time variable $\tilde \tau$ such that
\begin{equation}\label{eq:new_time}
	\tilde \tau (\tau) = \int^{\tau}_{\tau_{\rm f}} d \tau = \int^{a(\tau)}_{a_{\rm f}} \frac{da}{\mathcal H a}  = \frac{a(\tau)}{\mathcal H(\tau)} \int^{a(\tau)}_{a_{\rm f}} \frac{da}{a^2} = - \frac{a(\tau)}{\mathcal H(\tau)} \left( \frac{1}{a(\tau)} - \frac{1}{a_{\rm f}} \right) \simeq -\frac{1}{\mathcal H(\tau)}\:,
\end{equation}
where we have assumed that the Hubble parameter $H = \mathcal H/a$ is approximately constant during inflation and the scale parameter at the end of inflation $a_{\rm f}$ is much larger than at the time $\tau$, i.e., $a(\tau)/a_{\rm f} \ll 1$. This means that $\tau$ and $\tilde \tau$ are identical up to a constant shift and we will use this new time variable until the end of this section. For ease of notation we will also denote it by $\tau$ and we will use the same symbol for $\mathcal{H}(\tau)$ and $a(\tau)$ irrespective of which time coordinate is used. Note that the relation (\ref{eq:new_time}) allows us to express the comoving Hubble length (see Sect.~\ref{sec:horizons}) as $D_{\rm H} = 1/\mathcal H(\tau) = -\tau$. So the condition for a comoving Fourier mode $\bv k$ to cross the horizon is $-k \tau = k/\mathcal H(\tau) = 1$.\footnote{Recall that for a universe that is dominated by an equation of state $w = -1$, the event horizon $D_{\rm e}$ is identical with the Hubble length $D_{\rm H}$. Moreover, here we interpreted the mode that corresponds to a scale $\lambda$ by $k = 1/\lambda$.} The approximate equation of motion (\ref{eq_equation_of_motion_v2}) then becomes in Fourier space
\begin{equation}\label{eq_equation_of_motion_v3}
	\ddot q_{\bv  k} + \left(k^2 - \frac{2}{\tau^2} \right)q_{\bv  k} = 0\:,
\end{equation}
which allows analytical solutions. Since it is a linear differential equation of second order, it has two independent solutions for each $\bv k$ mode, and since with $q_{\bv k}(\tau)$ also $q_{\bv  k}^\ast(\tau)$ is a solution, the solutions
\begin{equation}\label{eq:v_k}
	q_{\bv k}(\tau) = \frac{1}{\sqrt{2k}}\left(1-\frac{i}{\tau k}\right)e^{-ik\tau}\:,
\end{equation}
along with $q_{\bv  k}^\ast(\tau)$ for all $\bv k$ constitute a complete set of independent solutions to Eq.~(\ref{eq_equation_of_motion_v3}).

\subsection{Canonical quantization}\label{sec:canonical_quantization}

As already mentioned, the action (\ref{eq:action_second_order}) describes a Klein-Gordon field $q$ with time-dependent mass in Minkowski spacetime. Since we can quantize this field as in standard quantum field theory (up to the time dependent mass), we will quantize $q$ directly rather than $\delta \phi$.\footnote{\label{foot:deg}This is no loss of generality, since for a single scalar field $\phi$ the system has only one degree of freedom. That is, the metric perturbations are determined as soon as $\delta \phi$ is determined and vice versa. So once we have quantized $q$ all the other perturbations such as $\delta \phi$ and the metric perturbations follow through the constraints which relate them to $q$. As a consequence there are no scalar metric perturbations without a scalar field. The same holds for metric vector perturbations, i.e., there are no metric vector perturbations present without a vector source. Since a scalar field does not exhibit any vector perturbations (see App.~\ref{sec:scalar_field_perturbed_universe}), no vector perturbations are generated during inflation. However, this is not true for tensor perturbations (``gravitational waves''). Tensor perturbations have their own degrees of freedom which might get excited during inflation even without tensor sources. This is why gravitational waves but no vector perturbations can be generated in inflationary scenarios.}

The first step of the canonical quantization is determining the canonical conjugate momentum field to $q$ defined by
\begin{equation}
	\pi(x) \equiv \frac{\partial \mathcal L_q}{\partial \dot q} = \dot q(x)\:,
\end{equation}
and interpret these variable as operators
\begin{equation}
	q(x) \rightarrow \hat q(x)\:,\quad\quad \pi(x) \rightarrow \hat \pi(x) = \dot{\hat q}(x)\:,
\end{equation}
subject to the equal-time commutation relations
\begin{equation}\label{eq:commutation_relation}
	\big[\hat q(\tau,\bv x),\hat \pi(\tau,\bv x')\big] = i \delta(\bv x-\bv x')\:, \quad\quad \big[\hat q(\tau,\bv x),\hat q(\tau,\bv x')\big] = \big[\hat \pi(\tau,\bv x),\hat \pi(\tau,\bv x')\big]= 0\:.
\end{equation}
Since $\phi$ is a real scalar field, the operators $\hat q$ and $\hat \pi$ are hermitian and we can expand them into Fourier integrals in the following way
\begin{equation}
\begin{aligned}
	\hat q(\tau,\bv x) &= \frac{1}{(2\pi)^3} \int \left(q_{\bv k}(\tau)\:\hat a_{\bv k} e^{i \bv k \bv x} + q_{\bv k}^\ast(\tau)\: \hat a_{\bv k}^\dagger\  e^{-i \bv k \bv x} \right)dk^3\:,\\
	\hat \pi(\tau,\bv x) &= \frac{i}{(2\pi)^3} \int \left(\dot q_{\bv k}(\tau)\:\hat a_{\bv k} e^{i \bv k \bv x} + \dot q_{\bv k}^\ast(\tau)\: \hat a_{\bv k}^\dagger\  e^{-i \bv k \bv x} \right)dk^3\:,
	\end{aligned}
\end{equation}
where the functions $q_{\bv k}(\tau)$ are the complete set of solutions to the equation of motion given by Eq.~(\ref{eq:v_k}). Note that these functions are normalized such that it follows with the commutation relations (\ref{eq:commutation_relation}) for the operators $\hat a_{\bv k}$ and $\hat a_{\bv k}^\dagger$
\begin{equation}\label{eq:commutation_relation_ak}
	\big[\hat a_{\bv k},\hat a_{\bv k'}^\dagger \big] = \left(2\pi\right)^3 \delta(\bv k-\bv k')\:, \quad\quad \big[\hat a_{\bv k},\hat a_{\bv k'} \big] = \big[\hat a_{\bv k}^\dagger,\hat a_{\bv k'}^\dagger \big]= 0
\end{equation}
as required by the canonical quantization. Finally, we define the vacuum state $\left|0\right\rangle$ by\footnote{As discussed in Chapter 11 of \cite{mukhanov1992}, the vacuum $\left|0\right\rangle$ changes for different times $\tau$ due to the time dependence of the effective mass $m^2(\tau) = -\ddot z/z$. This is a common feature of the vacuum in curved space (see \citealt{birrell1982}). However, at least for modes well inside the horizon, i.e., $\mathcal H/ k \ll 1$, the short-wavelength part of the initial vacuum spectrum should be independent of the choice of the vacuum.
This condition is satisfied to high precision if inflation lasts long enough. So there is no problem computing the power spectrum of initial perturbations as long as a considered perturbation was initially well inside the horizon.}
\begin{equation}
	\hat a_{\bv k} \big|0\big\rangle = 0
\end{equation}
for all $\bv k$.

\subsection{Expectation values}\label{sec:quantum_power_spectrum}

Now we will investigate the expectation values of the perturbations. The simplest and most natural assumption is that the state of the universe during inflation is the vacuum $|0\rangle$.\footnote{This is not the only possibility as discussed in Section 10.1 of \cite{weinberg2008}. A reason to expect the system to be in the the vacuum state during inflation is that the vacuum is the energetically lowest state and so any other state should decay into the vacuum. However, it is not clear whether this would happen fast enough.} In this state, the expectation value $\mu_q(x)$ of the field $q(x)$ is
\begin{equation}\label{eq:quantum_mean}
	\mu_q(x) = \big\langle 0 \big|\hat q(\tau,\bv x) \big|0\big\rangle = \frac{1}{(2\pi)^3} \int\left( q_{\bv k} (\tau)\big\langle 0 \big|\hat a_{\bv k} \big|0\big\rangle  e^{i\bv k \bv x} + q_{\bv k}^\ast (\tau)\big\langle 0 \big|\hat a_{\bv k}^\dagger \big|0\big\rangle  e^{-i\bv k \bv x}\right)dk^3 = 0
\end{equation}
and the correlation function $\xi_q(\tau,\bv x,\bv x')$ of $q(x)$ is given by (cf.~Eq.~(\ref{correlationfunction}))
\begin{align}
	\xi_q(\tau,\bv x,\bv x') &= \big\langle 0 \big|\hat q(\tau,\bv x)\: \hat q(\tau,\bv x') \big|0\big\rangle\\
	&= \frac{1}{(2\pi)^6} \iint q_{\bv k}(\tau)\: q_{\bv k'}^\ast(\tau) \:\big\langle 0 \big|\hat a_{\bv k}\: \hat a_{\bv k'}^\dagger \big|0\big\rangle \:e^{i(\bv k \bv x - \bv k'\bv x')}\:dk^3\: {dk'}^3\label{eq:power_spectrum_phi}\\
&=	\frac{1}{(2\pi)^3} \int \big | q_{\bv k} (\tau)\big |^2 e^{i\bv k (\bv x-\bv x')} \:dk^3\:,
\end{align}
where we have used the relation
\begin{equation}
	\big\langle 0 \big|\hat a_{\bv k} \:\hat a_{\bv k'}^\dagger \big|0\big\rangle = \big\langle 0 \big|\hat a_{\bv k}^\dagger \:\hat a_{\bv k'} \big|0\big\rangle+\left(2\pi\right)^3\delta(\bv k-\bv k') = \left(2\pi\right)^3\delta(\bv k-\bv k')\:.
\end{equation}
Thus, the power spectrum (see Eq.~(\ref{eq:power_spectrum_definition})) can be directly read off from Eq.~(\ref{eq:power_spectrum_phi}) as
\begin{equation}\label{eq:power_spectrum_v_definition}
	q_{\bv k}(\tau)\: q_{\bv k'}^\ast(\tau) \:\big\langle 0 \big|\hat a_{\bv k}\: \hat a_{\bv k'}^\dagger \big|0\big\rangle = \left(2\pi\right)^3\delta(\bv k-\bv k') \big | q_{\bv k} (\tau)\big |^2 = \left(2\pi\right)^3\delta(\bv k-\bv k') \mathcal P_q(\tau,k)\:,
\end{equation}
where $\mathcal P_q(\tau,k)$ is rotational invariant due to the rotational invariance of $q_{\bv k}$ and different $\bv k$ modes are decoupled due to the canonical commutation relations (\ref{eq:commutation_relation_ak}). The latter property of the power spectrum makes the correlation function to depend only on the difference $\bv x-\bv x'$, and the former property leads to rotational invariance for the correlation function. That is, it holds in fact $\xi_q(\tau,|\bv x-\bv x'|) = \xi_q(\tau,\bv x,\bv x')$ as was assumed in Section \ref{sec:statistics_overdensity_field} as a consequence of the cosmological principle.

Starting with a perturbation $\bv k$ well inside the horizon so that its vacuum $|0\rangle$ is well defined, and waiting until the mode is well outside the horizon, i.e., $k \tau \ll 1$, we have
\begin{equation}
q_{\bv k} (\tau) q_{\bv k}^\ast(\tau) \supeq{(\ref{eq:v_k})}{=} \frac{1}{2k} \left( 1 + \frac{1}{k^2\tau^2}\right) \simeq \frac{1}{2 k^3 \tau^2} \supeq{(\ref{eq:new_time})}{=} \frac{\mathcal{H}^2}{2 k^3} 
\end{equation}
and thus the power spectrum well outside the horizon is
\begin{equation}\label{eq:power_spectrum_inflation}
		\boxed{\mathcal P_q (k) = \frac{\mathcal H^2}{2 k^3}\:.}
\end{equation}

\subsection{Gaussianity}

What is the probability distribution function of the field $q(x)$ in the ground state $\left|0\right\rangle$? To answer this question, we express the operator $\hat q(\tau,\bv x)$ in terms of its Fourier transformed operator $\hat q(\tau,\bv k)$ as
\begin{equation}
	\begin{aligned}
	\hat q(\tau,\bv x) &= \frac{1}{(2\pi)^3} \int \left(q_{\bv k}(\tau)\:\hat a_{\bv k} e^{i \bv k \bv x} + q_{\bv k}^\ast(\tau)\: \hat a_{\bv k}^\dagger\  e^{-i \bv k \bv x} \right)dk^3\\
	&= \frac{1}{(2\pi)^3} \int \left(q_{\bv k}(\tau)\:\hat a_{\bv k} + q_{\bv k}^\ast(\tau)\: \hat a_{-\bv k}^\dagger\ \right)e^{i \bv k \bv x}\:dk^3\\
	&= \frac{1}{(2\pi)^3} \int \hat q(\tau,\bv k)\: e^{i \bv k \bv x} \:dk^3\:,
	\end{aligned}
\end{equation}
where in the second step we performed a change of variables for the second term and used that $q_{-\bv k}^\ast(\tau) = q_{\bv k}^\ast(\tau)$ (see Eq.~(\ref{eq:v_k})). The last term corresponds to the definition of the Fourier transformed operator $\hat q(\tau,\bv k)$ and we can read off
\begin{equation}\label{eq:fourier_transform_v_and_pi}
	\hat q(\tau,\bv k) = q_{\bv k}\hat a_{\bv k} + q_{\bv k}^\ast \hat a_{-\bv k}^\dagger\:, \quad\quad \hat \pi(\tau,\bv k) = \dot{\hat q}(\tau,\bv k) = \dot q_{\bv k}\hat a_{\bv k} + \dot q_{\bv k}^\ast \hat a_{-\bv k}^\dagger\:.
\end{equation}
We can now introduce four hermitian operators
\begin{equation}
\begin{aligned}
	\hat q_{\rm Re}(\tau,\bv k) &= q_{\bv k} \hat a_{\bv k} + q_{\bv k}^\ast \hat a_{-\bv k}^\dagger + q_{\bv k}^\ast \hat a_{\bv k}^\dagger + q_{\bv k}\hat a_{-\bv k}\:, \quad\quad \hat \pi_{\rm Re}(\tau,\bv k) = \dot{\hat q}_{\rm Re}(\tau,\bv k)\:,\\
		\hat q_{\rm Im}(\tau,\bv k) &= q_{\bv k} \hat a_{\bv k} + q_{\bv k}^\ast \hat a_{-\bv k}^\dagger - q_{\bv k}^\ast \hat a_{\bv k}^\dagger - q_{\bv k} \hat a_{-\bv k}\:, \quad\quad \hat \pi_{\rm Im}(\tau,\bv k) = \dot{\hat q}_{\rm Im}(\tau,\bv k)\:,\\
	\end{aligned}
\end{equation}
such that
\begin{equation}
	\hat q(\tau,\bv k) = \hat q_{\rm Re}(\tau,\bv k) + i \:\hat q_{\rm Im}(\tau,\bv k)\:,\quad \quad	\hat \pi(\tau,\bv k) = \hat \pi_{\rm Re}(\tau,\bv k) + i \:\hat \pi_{\rm Im}(\tau,\bv k)\:.
\end{equation}
These are the real and imaginary parts of $\hat q(\tau,\bv k)$ and $\hat \pi(\tau,\bv k)$, respectively. Their equal time commutation relations are as follows:
\begin{equation}
\begin{aligned}
	\big [\hat q_{\rm Re}(\tau,\bv k), \hat \pi_{\rm Re}(\tau,\bv k') \big ] &= \frac{i}{2} (2 \pi)^3 \left[ \delta(\bv k-\bv k')+\delta(\bv k+\bv k') \right]\:,\\
	\big [\hat q_{\rm Im}(\tau,\bv k), \hat \pi_{\rm Im}(\tau,\bv k') \big ] &= \frac{i}{2} (2 \pi)^3 \left[ \delta(\bv k-\bv k')-\delta(\bv k+\bv k') \right]\:,
	\end{aligned}
\end{equation}
while all other pairwise commutators vanish.

The non-vanishing commutation relations between modes with $\bv k$ and $-\bv k$ are due to the reality of the field $q(x)$ leading to $q(\tau,\bv k) = q^\ast(\tau,-\bv k)$. Hence with $q(\tau,\bv k)$ also $q^\ast(\tau,-\bv k)$ is fully determined and so they must exhibit the same commutation relations with respect to $\pi(\tau,\bv k)$ and $\pi^\ast(\tau,-\bv k)$ respectively. However, if we restrict ourselves to the upper half of the Fourier space, i.e., $k_z > 0$, the real as well as the imaginary parts of $q(\tau,\bv k)$ are simultaneously measurable and behave (together with their canonical momenta) like independent harmonic oscillators (up to the factor $1/2$). So analog to the simple harmonic oscillator in quantum mechanics, the probability density functions of the real and imaginary parts of $q(\tau,\bv k)$ are independent Gaussian distributions for every $\bv k$ with the constraint $q(\tau,\bv k) = q^\ast(\tau,-\bv k)$.

\subsection{Transition from quantum perturbations to classical perturbations}

So far, we treated the universe entirely quantum mechanical. But when and how does the transition to the classical universe that we can observe today take place? For modes well outside the horizon, i.e., $k \tau \ll 1$, the Fourier transforms (\ref{eq:fourier_transform_v_and_pi}) become
\begin{equation}
	\hat q (\tau,\bv k) \simeq -\frac{1}{\sqrt{2 k}} \: \frac{i}{\tau k}\left(\hat a _{\bv k}e^{-i k \tau} - \hat a_{-\bv k}^\dagger e^{i k \tau} \right)\:, \quad \quad \hat \pi (\tau, \bv k) \simeq \sqrt{\frac{k}{2}} \: \frac{i}{\tau^2 k^2} \left(\hat a _{\bv k}e^{-i k \tau} - \hat a_{-\bv k}^\dagger e^{i k \tau} \right)\:.
\end{equation}
That is $\hat q(\tau, \bv k)$ and $\hat \pi (\tau,\bv k)$ are proportional to the same operator and thus they started to commute. This means that the system behaves classical outside the horizon.

This simple consideration, however, does not explain how a particular realization --- our universe --- is chosen out of the quantum ensemble during this process (see \citealt{mukhanov2005}, Sect.~8.3.3, for a brief discussion). The way this happens lies at the very base of quantum mechanics and is not yet fully understood in general. Here we will just assume that outside the horizon the field $q$ became classical and also the power spectrum (\ref{eq:power_spectrum_inflation}) can be interpreted as a power spectrum of an ensemble of classical universes.

Thus, we can summarize this section with the statement that during inflation quantum mechanical processes produce tiny fluctuations in the universe, which become classical as they leave the horizon and can be regarded a realization of a Gaussian random field with zero mean and with a translational and rotational invariant correlation function.

\section{Conservation of perturbations outside the horizon}\label{sec:conservation_of_perturbations}

One of the key points in the context of inflation is the behavior of perturbations outside the horizon. We will show that in comoving gauge the 3-curvature perturbation ${}^{(3)}\delta \mathcal R$ times $a^2$ (or equivalently the perturbation $\zeta$, see Eq.~(\ref{eq:curvature_and_zeta})) is conserved outside the horizon if perturbations are adiabatic. Without this feature, it would be basically impossible to make any robust prediction from inflation, since we practically know nothing about the fundamental physics associated with inflation and the transition from inflation to the radiation dominated universe.

Before going into the details of the calculation, we want to point out that outside the horizon, i.e., for $k/\mathcal{H} \ll 1$, it holds
\begin{equation}\label{eq:pi_zero}
0 = \frac{\bar \rho}{3} \left( \frac{k}{\mathcal{H}}\right)^2 (\Phi - \Psi) \supeq{(\ref{eq:last_field_equations_Newtonian_gauge})}{=} \frac{8\pi G a^2 \bar \rho}{3\mathcal{H}^2}\:\Pi^{\rm N} \supeq{(\ref{eq:friedmann_equations_conformal})}{=} \Pi^{\rm N} \supeq{(\ref{eq:gauge_transformations_energy_4})}{=} \Pi^{\rm (any\: gauge)}\:,
\end{equation}
where we denoted the anisotropic stress in the Newtonian gauge by $\Pi^{\rm N}$. From the last step we can conclude that outside the horizon the anisotropic stress is zero for every constituent of the universe in every gauge. That is, outside the horizon we can neglect the dissipative effects of $T_{\mu\nu}$ and we can essentially regard the constituents of the universe as ideal fluids.

\subsection{Adiabaticity}\label{sec:adiabaticity}

After the inflationary slow-roll stage, the inflaton $\phi$ decays by a process that is not very well understood and the unverse becomes radiation dominated. What can we say about the perturbations of the decay products of the inflaton? To answer this question we consider a scale $\bv k$ well outside the horizon and smooth the universe on this scale at every point. Since $\bv k$ is well outside the horizon, each smoothed patch is causally disconnected from any other smothed patch and evolves like a homogeneous and isotropic universe. These ``universes'' have slightly different mean densities and become identical if we synchronize them on slices of constant $\phi$. Since the inflaton is the only field, such a synchronized universe would be absolutely homogeneous. So whatever happens after inflation, it happens everywhere the same, and any decay product of $\phi$ will be homogeneous as well. Then transforming back to the old or any other time variable yields (see Eq.~(\ref{eq:gauge_transformation_scalar}))
\begin{equation}\label{eq:transformation_s}
	\delta s(\tau,\bv k) = -\dot{\bar s}(\tau) T(\tau,\bv k)
\end{equation}
with the same $T(\tau,\bv k)$ for any 4-scalar $s(\tau,\bv k)$. That is, we find for any constituent $I$
\begin{equation}
	 \boxed{\frac{\delta \rho}{\dot{\bar \rho}} = \frac{\delta \rho_I}{\dot{\bar \rho}_I}}
\end{equation}
in any gauge as long as the scale $\bv k$ is outside the horizon. This is the \textbf{generalized adiabatic condition}. Moreover, with Eq.~(\ref{eq:transformation_s}) it follows for each constituent separately (we omit the index $I$ for the ease of notation)
\begin{equation}\label{eq:adiabatic2}
	\frac{\delta \rho}{\dot{\bar \rho}} = \frac{\delta p}{\dot{\bar p}}\:,
\end{equation}
which is important for the following reason. Since it holds for the pressure
\begin{equation}
	p(\tau,\bv k) = p(\bar \rho + \delta \rho,\bar S + \delta S) = \bar p + \delta p
\end{equation}
with $S = \bar S + \delta S$ the entropy of the fluid, we can express $\delta p$ at first order as
\begin{equation}\label{eq:dp_expressed}
	\delta p = c_{\rm s}^2 \delta \rho + \sigma \delta S\:, \quad\quad c_{\rm s}^2 = \left(\frac{\partial \bar p}{\partial\bar\rho}\right)_{\!\!\bar S}\:, \quad \sigma = \left(\frac{\partial \bar p}{\partial \bar S}\right  )_{\!\!\bar \rho}\:,
\end{equation}
where $c_{\rm s}$ is the speed of sound. It follows
\begin{equation}\label{eq:delta_p_over_delta_rho}
	\frac{\delta p}{\delta \rho} \supeq{(\ref{eq:adiabatic2})}{=} \frac{\dot{\bar p}}{\dot{\bar \rho}} = \frac{1}{\dot{\bar \rho}} \left[ \left( \frac{\partial \bar p}{\partial \bar \rho} \right)_{\!\!\bar S} \dot{\bar \rho} + \left( \frac{\partial \bar p}{\partial \bar S} \right)_{\!\!\bar \rho} \dot{\bar S}\right] = \left(\frac{\partial \bar p}{\partial \bar \rho}\right)_{\!\!\bar S} = c_{\rm s}^2\:,
\end{equation}
where we have used that the entropy is conserved, i.e., $\dot{\bar S} = 0$, in a FLRW universe.\footnote{The conservation of entropy in a FLRW universe is easily seen. Let $V$ be a comoving volume, so that its corresponding proper volume is $V_{\rm pr} = a^3 V$. The equation of motion (\ref{eq:equation_of_motion_total}) can be written as $d(\bar \rho a^3)/d(a^3) = -\bar p$. The change of the internal energy $d\bar U$ in this comoving volume during the expansion of the universe is then related to the change of the proper volume $dV_{\rm pr}$ by
\begin{equation}
	d\bar U = d\!\left(\bar\rho a^3 V\right) = -\bar p \: d\!\left(a^3 V\right) = -\bar p \:dV_{\rm pr}\:.
\end{equation}
Thus, during the expansion of the universe, the change of entropy in this comoving volume is
\begin{equation}
	d \bar S = \frac{d\bar U + \bar p\:dV_{\rm pr} - \sum_i \mu_i dN_i}{\bar T} = -\frac{\sum_i \mu_i dN_i}{\bar T}\:,
\end{equation}
where $N_i$ is the number of particles of species $i$ in this volume and $\mu_i$ the corresponding chemical potential. If $\mu_i \simeq 0$, i.e., negligible particle-antiparticle asymmetries, it follows $d \bar S = 0$ within any comoving volume $V$.} Thus, with Eqs.~(\ref{eq:dp_expressed}) and (\ref{eq:delta_p_over_delta_rho}) the entropy perturbations vanish for each constituent separately, i.e.,
\begin{equation}\label{eg:entropy_perturbation}
	 \boxed{\delta S = \frac{1}{\sigma}\left(\delta p - c_{\rm s}^2 \delta \rho\right) = 0\:.}
\end{equation}
This is why these perturbations are called adiabatic. The adiabaticity of the perturbation is important for the conservation of perturbations outside the horizon as is shown in the next section.

\subsection{$\zeta$ outside the horizon}\label{sec:zeta_outside_the_horizon}

In this section, we show that in comoving gauge $\zeta$ is conserved well outside the horizon. As we deal with different gauges, we denote quantities in the Newtonian gauge with a superscript ``N'' and in the comoving gauge with a superscript ``com''. Using Eq.~(\ref{eq:transformation_to_comoving_gauge}) the gauge transformation from Newtonian gauge to comoving gauge is 
\begin{equation}\label{eq:t_L_newtonian_comoving}
	T = \frac{v^{\rm N}}{k}\:, \quad\quad L = 0\:.
\end{equation}
So we can express $\delta \rho^{\rm com}$ in terms of $\Phi$ as
\begin{equation}\label{eq:delta_rho_com}
	\delta \rho^{\rm com} \supeq{(\ref{eq:gauge_transformations_energy_1})}{=}  \delta \rho^{\rm N} - \dot{\bar\rho} \frac{v^{\rm N}}{k} \supeq{(\ref{eq:equation_of_motion_total_2})}{=} \delta \rho^{\rm N} + 3 \mathcal H \left( \bar \rho + \bar p \right) \frac{v^{\rm N}}{k}\supeq{(\ref{eq:second_field_equations_Newtonian_gauge})}{=}\delta \rho^{\rm N} + 3 \mathcal H\frac{\dot \Phi + \mathcal{H}\Psi}{4 \pi G a^2}  \supeq{(\ref{eq:first_field_equations_Newtonian_gauge})}{=} - \frac{k^2 \Phi}{4 \pi G a^2}\:.
\end{equation}
To derive the next relation, we express the momentum equation (\ref{eq:scalar_field_equations}) and the Euler equation (\ref{eq:scalar_equations_of_motion}) in comoving gauge, i.e.,
\begin{equation}\label{eq:zeta_comovin_gauge}
\mathcal{H} \xi = \dot \zeta \:, \qquad \qquad \delta p^{\rm com} = -\left(\bar \rho + \bar p \right)\xi + \frac{2}{3}\Pi^{\rm com}\:.
\end{equation}
With $\Pi^{\rm com} = 0$ outside the horizon (see Eq.~(\ref{eq:pi_zero})), we then obtain the relation
\begin{equation}
	\dot \zeta \supeq{(\ref{eq:zeta_comovin_gauge})}{=} \mathcal H \xi \supeq{(\ref{eq:zeta_comovin_gauge})}{=} - \mathcal H \: \frac{\delta p^{\rm com}}{\bar \rho + \bar p} = - \frac{c_{\rm s}^2 \mathcal H \delta \rho^{\rm com}}{\bar \rho + \bar p} \supeq{(\ref{eq:delta_rho_com})}{=} \frac{1}{4 \pi G a^2}\frac{c_{\rm s}^2 k^2 \mathcal H \Phi}{\bar \rho + \bar p} \supeq{(\ref{eq:friedmann_equations_conformal})}{=} \frac{2}{3} \left(\frac{c_{\rm s}k}{\mathcal H}\right)^2 \frac{\mathcal H \Phi \bar \rho}{\bar \rho + \bar p}\:,
\end{equation}
where in the third step we used that for slow roll inflation with a single scalar field, the perturbations are adiabatic, i.e., $\delta p^{\rm com} = c_{\rm s}^2\delta \rho^{\rm com}$ (see Eq.~(\ref{eq:dp_expressed}) for $\delta S = 0$). Thus, $|\dot \zeta|$ is of order
\begin{equation}\label{eq:zeta_dot}
	 \boxed{\big|\dot \zeta \big|\sim \mathcal O \left(\frac{k^2}{\mathcal H^2}\big|\Phi\big| \right)}
\end{equation}
and can be neglected for $k/\mathcal H \ll 1$. So $\zeta$ is (at first order) constant outside the horizon if the perturbations are adiabatic.\footnote{\cite{weinberg2008} gives in his Section 5.4 a proof of a more general theorem stating that irrespective of the contents of the universe there are always two independent adiabatic physical scalar solutions for which $\zeta$ is time independent outside the horizon. Since for an inflationary scenario with only one inflaton there is only one degree of freedom and since the equations of motion are ordinary second order differential equations, during inflation the two independent solutions must be adiabatic and $\zeta$ must be constant outside the horizon. Moreover, since there are always two independent solutions with these properties, the perturbations remain adiabatic and $\zeta$ remains constant even after the inflaton decays. However, if inflation is governed by more than one inflaton field, this reasoning is no longer valid and there might by entropy perturbations produced during inflation.}

\subsection{$\Phi$ outside the horizon}

There is also a similar theorem for the generalized Newtonian potential $\Phi$ as for $\zeta$. Using the transformation (\ref{eq:gauge_transformation_3}) from Newtonian gauge to comoving gauge, we have
\begin{equation}\label{eq:zeta_phi_relation}
	\zeta \supeq{(\ref{eq:t_L_newtonian_comoving})}{=} -\Phi - \mathcal H \frac{v^{\rm N}}{k}\:.
\end{equation}
Then we can connect the Newtonian perturbations $\Psi$ and $\Phi$ with $\zeta$ by means of
\begin{equation}
	\dot \Phi + \mathcal H \Psi \supeq{(\ref{eq:second_field_equations_Newtonian_gauge})}{=} 4 \pi G a^2 \frac{v^{\rm N}}{k} \left(\bar \rho + \bar p \right) \supeq{(\ref{eq:friedmann_equations_conformal})}{=} \frac{3}{2} \mathcal H^2 \frac{v^{\rm N}}{k} \left(1 + w\right) \supeq{(\ref{eq:zeta_phi_relation})}{=} - \frac{3}{2} \mathcal H \left(1 + w\right)\left(\zeta + \Phi \right)\:.
\end{equation}
Solving this for $\zeta$ yields the following relation for any equation of state $w$:
\begin{equation}\label{eq:phi_zeta_relation_general}
	\zeta = -\Phi - \frac{2}{3}\:\frac{\Psi + \mathcal H^{-1} \dot \Phi}{1+w}\:.
\end{equation}
Using Eqs.~(\ref{eq:last_field_equations_Newtonian_gauge}) and (\ref{eq:pi_zero}), i.e., $\Phi = \Psi$, and allowing only adiabatic perturbations, i.e., $\delta p^{\rm N} = c_{\rm s}^2 \delta \rho^{\rm N}$, we obtain
\begin{equation}
	\ddot \Phi + 3 \mathcal H \dot \Phi + \left(2 \dot{\mathcal H} + \mathcal H^2\right) \Phi   \supeq{(\ref{eq:third_field_equations_Newtonian_gauge})}{=}      4 \pi G a^2 \delta p^{\rm N} = 4 \pi G a^2 c_{\rm s}^2 \delta \rho^{\rm N} \supeq{(\ref{eq:first_field_equations_Newtonian_gauge})}{=} -c_{\rm s}^2 \left[k^2 \Phi+ 3 \mathcal H \left(\dot \Phi + \mathcal H \Phi\right)\right]
\end{equation}
yielding the following differential equation for $\Phi$:
\begin{equation}\label{eq:bardeen_equation}
	 \boxed{\ddot \Phi + 3 \mathcal H \dot \Phi \left(1 + c_{s}^2\right) + \left[2 \dot{\mathcal H} + \mathcal H^2 \left(1 + 3 c_{\rm s}^2\right)\right] \Phi + c_{\rm s}^2k^2 \Phi = 0\:.}
\end{equation}
For a constant equation of state $w(t) \equiv w$, we can find an analytic solution to this equation for scales well outside the horizon. If $w$ is constant, it holds
\begin{equation}\label{eq:cah}
	c_{\rm s}^2 \supeq{(\ref{eq:delta_p_over_delta_rho})}{=} \frac{\partial \bar p}{\partial \bar \rho} \supeq{(\ref{eq:equation_of_state})}{=} w\:, \quad\quad a(\tau) \supeq{(\ref{eq:a_tau})}{\propto} \tau^{2/(1+3w)}\:, \quad\quad \mathcal H(\tau) = \frac{2}{1+3w} \tau^{-1}\:,
\end{equation}
so that Eq.~(\ref{eq:bardeen_equation}) reduces to
\begin{equation}
	\ddot \Phi + \frac{6(1+w)}{1+3w} \: \frac{1}{\tau} \dot \Phi + w k^2 \Phi = 0\:.
\end{equation}
For scales well outside the horizon, i.e., $k \tau \propto k/\mathcal H \ll 1$, we can neglect the last term and the general solution in this limit is then given by
\begin{equation}
	\Phi(\tau) = c_1 + c_2 \tau^{- \nu}, \quad\quad \nu = \frac{6(1+w)}{1+3w} - 1 = \frac{5+3w}{1+3w} > 1\:,
\end{equation}
where $c_1$ and $c_2$ are two constants. Thus, for a constant equation of state, the growing mode of $\Phi(\tau)$ outside the horizon is just a constant, i.e.,
\begin{equation}
\Phi \simeq \rm const\:.
\end{equation}
Thus, in this case the relation (\ref{eq:phi_zeta_relation_general}) between $\zeta$ and $\Phi$ simplifies to the constant expression
\begin{equation}\label{eq:relation_zeta_phi_simple}
	 \boxed{\zeta = - \frac{5+3w}{3+3w}\:\Phi\:.}
\end{equation}

\section{Primordial power spectrum}\label{sec:primordial_power_spectrum}

After the end of inflation, the universe finally turned into the radiation and then matter dominated eras, during which the universe is decelerating. So the modes that left the horizon during inflation will reenter it after a certain time (see Fig.~\ref{fig:inflation}). In the previous section we studied the behavior of the perturbation when they are outside the horizon. In this section we use these results along with the power spectrum (\ref{eq:power_spectrum_inflation}) that was created during inflation to compute the explicit form of the primordial DM power spectrum inside the horizon for modes that enter the horizon during matter domination. The power spectrum for modes that enter during radiation domination can then be obtained by using the transfer function $T(k)$ (see Sect.~\ref{sec:transferfunction}).

\subsection{Scale invariant power spectrum}

Due to the constancy of $\zeta$ outside the horizon, we can evolve the power spectrum (\ref{eq:power_spectrum_inflation}) until the time when a given mode $\bv k$ reenters the horizon. To express the power spectrum in terms of $\zeta$ we need a relation between $\zeta$ and the variable $q$ that was quantized during inflation, which is obtained by transforming the right hand side of Eq.~(\ref{eq:action_v}), i.e.,
\begin{equation}
q = \delta \phi^{\rm N} +  \frac{\dot{\bar \phi}}{\mathcal H}\: \Phi\:,
\end{equation}
from Newtonian gauge to comoving gauge. With equation Eq.~(\ref{eq:t_L_newtonian_comoving}) we have
\begin{equation}\label{eq:delta_phi_com}
\delta \phi^{\rm com} \supeq{(\ref{eq:gauge_transformation_scalar})}{=} \delta \phi^{\rm N} -  \frac{\dot{\bar \phi}}{k} v^{\rm N}\supeq{(\ref{eq:gauge_transformations_energy_3})}{=}  \delta \phi^{\rm N} -  \frac{\dot{\bar \phi}}{k}\: v^{\rm com} \:,
\end{equation}
and so with the density equation (\ref{eq:scalar_field_equations}) in comoving gauge, i.e.,
\begin{equation}\label{eq:k_zeta}
k^2 \zeta + \mathcal{H}kv^{\rm com} = 4 \pi G a^2 \delta \rho^{\rm com} \supeq{(\ref{eq:delta_rho_com})}{=} -k^2 \Phi\:,
\end{equation}
we have 
\begin{equation}
q =  \delta \phi^{\rm N} +  \frac{\dot{\bar \phi}}{\mathcal H}\: \Phi \supeq{(\ref{eq:delta_phi_com})}{=} \delta \phi^{\rm com} + \frac{\dot {\bar \phi}}{k}v^{\rm com}-  \frac{\dot{\bar \phi}}{\mathcal H}\: \Phi \supeq{(\ref{eq:k_zeta})}{=} \delta \phi^{\rm com} - \frac{\dot{\bar \phi}}{\mathcal{H}}\: \zeta \:.
\end{equation}
Using Eq.~(\ref{eq:phi_expressions}) it holds in comoving gauge
\begin{equation}\label{eq:comoving_gauge_phi}
	\dot{\bar \phi}\:\delta \phi^{\rm com} = 0
\end{equation}
and thus $\delta \phi^{\rm com} = 0$, since $\bar \phi$ is generally a nonzero function determined by the background cosmology.\footnote{Only for a constant potential $V(\bar \phi)$ is a constant $\bar \phi$ a solution to the equation of motion (\ref{eq:equation_of_motion_of_unperturbed_phi}). This case is not considered here. During slow roll inflation $\bar \phi$ is only approximately constant and so $\delta \phi^{\rm com}$ must be exactly zero to satisfy the relation (\ref{eq:comoving_gauge_phi}).} So we obtain the relation between $q$ and $\zeta$ as
\begin{equation}
	\zeta= \frac{\mathcal H}{\dot{\bar \phi}} \left(  \delta \phi^{\rm com} - q \right)   = -\frac{\mathcal H}{\dot{\bar \phi}}\:q\:.
\end{equation}

The power spectrum well outside the horizon is then given by constant expression
\begin{align}
	\big \langle \Phi(\bv k) \Phi^\ast(\bv k')  \big \rangle &= C^2 \big \langle \zeta(\bv k) \zeta^\ast(\bv k')  \big \rangle = C^2 \left(\frac{\mathcal H}{\dot{\bar \phi}}\right)^2 \big \langle q(\bv k) q^\ast (\bv k') \big \rangle\\
&= C^2 (2 \pi)^3 \delta(\bv k-\bv k') \left(\frac{\mathcal H}{\dot{\bar \phi}}\right)^2 \left(\frac{\mathcal H^2}{2 k^3 a^2}\right)\:,\label{eq:power_spectrum_relation}
	\end{align}
where in the first step we have used Eq.~(\ref{eq:relation_zeta_phi_simple}) for a constant $w$ and in the last step power spectrum at horizon exit given by the Eqs.~(\ref{eq:power_spectrum_definition}) and (\ref{eq:power_spectrum_inflation}). The constant $C$ is $-2/3$ during radiation domination and $-3/5$ during matter domination. Since $\mathcal H/\dot{\bar \phi}$ and $\mathcal H/a^2$ are roughly constant during slow-roll inflation, we can evaluate the right-hand-side of Eq.~(\ref{eq:power_spectrum_relation}) for each mode at the time $\tau_{\rm out}(\bv k)$ when it leaves the horizon.

The power spectrum of DM is defined by $\langle \delta_{\rm d}(\bv k)\delta_{\rm d}^\ast(\bv k')\rangle$, where $\delta_{\rm d}(\bv k) = \delta \rho_{\rm d}/\bar \rho_{\rm d}$ and $\rho_{\rm d} = \bar \rho_{\rm d}+\delta \rho_{\rm d}$ is the matter density of DM. If the universe is dominated by DM, i.e., $\rho \simeq \rho_{\rm d}$, then we can relate $\delta_{\rm d}$ and $\Phi$ well inside the horizon by means of the Poisson equation (\ref{eq:generalized_poisson_equation})
\begin{equation}\label{eq:generalized_poisson_equation2}
	-k^2 \Phi = 4 \pi G a^2 \delta \rho_{\rm d} = 4 \pi G a^2 \bar \rho_{\rm d} \delta_{\rm d} \supeq{(\ref{eq:friedmann_equations_conformal})}{=} \frac{3}{2}\mathcal H^2 \delta_{\rm d}\:.
\end{equation}
With this relation and with Eq.~(\ref{eq:power_spectrum_relation}) we get for the power spectrum inside the horizon
\begin{equation}\label{eq:power_spectrum_relation2}
	\big \langle \delta_{\rm d}(\bv k)\delta_{\rm d}^\ast(\bv k')  \big \rangle = (2 \pi)^3 \delta(\bv k-\bv k')\left(\frac{2}{5}\right)^2  \left.\left(\frac{\mathcal H^4}{2 \dot{\bar \phi}^2 k^3 a^2}\right)\right|_{\tau_{\rm out}(\bv k)}\left(\frac{k}{\mathcal H}\right)^4\:.
\end{equation}
Note that to properly compute the DM power spectrum inside the horizon, we would have to evolve $\Phi$ from outside into the horizon. But since during matter domination $\Phi$ is also constant inside the horizon (see Sect.~\ref{sec:dark_matter_and_baryons}) we will accept the approximation (\ref{eq:power_spectrum_relation2}). Moreover, for simplicity we will evaluate the term $(k/\mathcal H)^4$ at horizon entry, so that it just becomes unity for all $\bv k$. The DM power spectrum within the horizon then becomes at horizon entry $\tau_{\rm in}(\bv k)$ for each mode
\begin{equation}
	\big \langle \delta_{\rm d}(\bv k)\delta_{\rm d}^\ast(\bv k')  \big \rangle = (2 \pi)^3 \delta(\bv k-\bv k') \mathcal P_{\rm d}(\tau,k)\:, \quad\quad \mathcal P_{\rm d}(\tau_{\rm in}(\bv k),k) \propto \frac{1}{k^3}\:,
\end{equation}
since during slow roll inflation $\mathcal H/\dot{\bar \phi}$ and $\mathcal H/a^2$ are roughly constant and thus are the same for all $\bv k$. It is convenient to evaluate $\mathcal P_{\rm d}(\tau,k)$ at the same time $\tau$ for all modes $\bv k$. To achieve this, we express $P_{\rm d}(\tau,k)$ as
\begin{equation}
	\mathcal P_{\rm d}(\tau,k) = \mathcal P_{\rm d}(\tau_{\rm in}(\bv k),k)\left(\frac{D_+(\tau)}{D_+(\tau_{\rm in}(\bv k))}\right)^2\:,
\end{equation}
where $D_+(\tau)$ is the linear growth function (see Sect.~\ref{sec:linear_theory}) being independent of $\bv k$ at first order. With the Eqs.~(\ref{eq:dat}) and (\ref{eq:a_tau}), we immediately obtain that during matter domination $D_+(\tau) \propto a(\tau) \propto \tau^2$ and that the condition of horizon entry is $k\tau_{\rm in}(\bv k) \propto k/\mathcal H(\tau_{\rm in}(\bv k)) \propto 1$. Thus, it holds
\begin{equation}
	D_+(\tau_{\rm in}(\bv k)) \propto k^{-2}
\end{equation}
and finally the primordial power spectrum inside the horizon at a given time is
\begin{equation}\label{eq:initial_power_spectrum}
\boxed{\mathcal P_{\rm d}(\tau,k) \propto k^{n_{\rm s}} D_+(\tau)\:, \quad\quad n_{\rm s} \simeq 1\:,}
\end{equation}
where $n_{\rm s}$ is the spectral index. Interestingly, this form of the power spectrum was proposed even ten years before inflation was introduced (see Sect.~\ref{sec:Initial_conditions}).

For some applications it is useful to express the power spectrum in dimensionless form. For a random field $\delta$, the \textbf{dimensionless power spectrum} is defined by
\begin{equation}
\Delta_{\delta}^2(\tau,\bv k) \equiv \frac{1}{2 \pi^2} \: k^3\: \mathcal P_{\delta}(\tau,\bv k)\:,
\end{equation}
where $P_{\delta}(\tau,\bv k)$ is its (dimensioned) power spectrum (\ref{eq:power_spectrum_definition}). For DM, i.e., $\delta_{\rm d}$, this becomes
\begin{equation}\label{eq:dimensionless_DM_spectrum}
	\Delta_{\rm d}^2(\tau,k) = \frac{1}{2 \pi^2} \: k^3 \: \mathcal P_{\rm d}(\tau,k) \supeq{(\ref{eq:initial_power_spectrum})}{\propto} k^{n_{\rm s}+3}
\end{equation}
and thus, for $n_{\rm s} = 1$ the dimensionless power spectrum for $\Phi$ is independent of $\bv k$, i.e.,
\begin{equation}\label{eq:delta_phi1}
	\Delta_{\Phi}(k) \supeq{(\ref{eq:generalized_poisson_equation2})}{\propto} \frac{\Delta_{\rm d}(k)}{k^4} \supeq{(\ref{eq:dimensionless_DM_spectrum})}{\propto} k^{n_{\rm s}-1} \propto \rm const\:.
\end{equation}
For this reason, the spectral index $n_{\rm s} = 1$ is called \textbf{scale invariant}. However, this scale invariance applies only approximately and in the next section we will compute the deviation from it.

\subsection{Deviation from scale invariance}

The deviation of the power spectrum from scale invariance is produced by the amount the right hand side of Eq.~(\ref{eq:power_spectrum_relation}) changes during inflation for the different modes. It is obtained by computing
\begin{equation}
	\frac{d \ln \Delta_{\Phi}}{d \ln k} \supeq{(\ref{eq:delta_phi1})}{=} k \frac{d\left[(n_{\rm s}-1) \ln k\right]}{dk} = n_{\rm s}-1\:.
\end{equation}
For our calculation, we again adopt the time coordinate which is defined with respect to the end of inflation (see Eq.~(\ref{eq:new_time})), so that it holds $-k \tau_{\rm out}(\bv k) = k/\mathcal{H}(\tau_{\rm out}(\bv k)) = 1$ at the time of horizon exit for a given mode $\bv k$. Then we can express the derivative after $\ln k$ as
\begin{equation}\label{eq:d_dk}
	\frac{d}{d \ln k} = k \frac{d}{dk} = k \frac{d\tau_{\rm out}}{dk}\:\frac{d \bar \phi}{d\tau_{\rm out}}\:\frac{d}{d \bar \phi} \quad \Rightarrow \quad \left.\frac{d}{d \ln k}\right|_{\tau_{\rm out}(\bv k)} = \frac{\dot{\bar \phi}(\tau_{\rm out}(\bv k))}{k}\:\left.\frac{d}{d\bar \phi}\right|_{\tau_{\rm out}(\bv k)}\:.
\end{equation}
Combining the two Eqs.~(\ref{eq:equations_slow_roll}) yields
\begin{equation}\label{eq:dot_bar_phi_dev}
	\dot{\bar \phi} = - \frac{V'}{V}\:\frac{\mathcal H}{8\pi G}
\end{equation}
and we obtain
\begin{equation}\label{eq:delta_phi}
	\Delta_{\Phi} \propto k^3\big\langle \Phi(\bv k)\Phi^\ast(\bv k') \big\rangle \supeq{(\ref{eq:power_spectrum_relation})}{\propto} \frac{\mathcal H^4}{\dot{\bar \phi}^2a^2}  \supeq{(\ref{eq:dot_bar_phi_dev})}{\propto} \left( \frac{\mathcal{H}}{a}\right)^2  \left(\frac{V}{V'}\right)^2 \supeq{(\ref{eq:equations_slow_roll})}{\propto}  V \left(\frac{V}{V'}\right)^2 \propto \frac{V^3}{V'^2} \:.
\end{equation}
With
\begin{equation}
\frac{\dot {\bar \phi}(\tau_{\rm out}(\bv k))}{k} \supeq{(\ref{eq:dot_bar_phi_dev})}{=} - \frac{1}{8 \pi G} \frac{V'}{V} \frac{\mathcal{H}(\tau_{\rm out}(\bv k))}{k} = - \frac{1}{8 \pi G} \frac{V'}{V}
\end{equation}
and with the Eqs.~(\ref{eq:d_dk}) and (\ref{eq:delta_phi}), the deviation from scale invariance is approximately given by
\begin{equation}
\begin{aligned}
	n_{\rm s} - 1 &= \left.\frac{d \ln \Delta_{\Phi}}{d\ln k}\right|_{\tau_{\rm out}(\bv k)} = - \frac{1}{8\pi G}\:\frac{V'}{V} \left.\frac{d}{d\bar \phi}\left(3 \ln V-2 \ln V'\right)\right|_{\tau_{\rm out}(\bv k)}\\
	&= \left.\left(-\frac{6}{16 \pi G}\left(\frac{V'}{V}\right)^2+\frac{2}{8\pi G}\:\frac{V'}{V''}\right)\right|_{\tau_{\rm out}(\bv k)}\:.
		\end{aligned}
\end{equation}
Thus, with the definition of the slow-roll parameters (\ref{eq:slow_roll_parameters}) we obtain the final result
\begin{equation}\label{eq:n_s_3}
\boxed{n_{\rm s} - 1 = \left.\left(- 6 \epsilon + 2 \eta\right)\right|_{\tau_{\rm out}(\bv k)}\:.}
\end{equation}
If the spectral index $n_{\rm s}$ is unequal 1, the spectrum is called ``tilted'', where $n_{\rm s}>1$ is called ``blue spectrum'' and $n_{\rm s}<1$ ``red spectrum''. The measured value of $n_{\rm s}$ in the universe is about 0.97 (see Tab.~\ref{tab:cosmological_parameters}), so the actual spectrum in the universe is almost scale invariant, but slightly tilted to the red side. This is an excellent confirmation of the phenomenology of slow roll inflation. If $n_{\rm s}$ has an additional (weak) dependence on $\bv k$, $d n_{\rm s}/d \ln k$ is called the ``running'' of the spectral index.

We conclude this chapter with the statement that as a result of the simplest models of inflation the DM perturbation $\delta_{\rm d}$ on large scales (linear regime) but well inside the horizon can be regarded as a realization of a homogeneous and isotropic Gaussian random field with zero mean and an almost scale invariant power spectrum. Additionally the perturbations are adiabatic, i.e., outside the horizon every energy contribution is subjected to the generalized adiabatic condition (\ref{eq:adiabatic2}) and there are no entropy perturbations.


\appendix


\chapter{Classical scalar field theory}\label{sec:classical_scalar_field}

The simplest kind of field in classical (and quantum) field theory is a real scalar field $\phi(x)$. This is also the kind of field that is the dominant energy component in the simplest models of cosmic inflation (see Sect.~\ref{sec:inflation}). The theory of the scalar field is usually formulated by means of the Lagrangian (density) $\mathcal L$ for the field $\phi$. In the first section of this appendix, we briefly introduce the scalar field theory in the context of general relativity and apply it in the second section to cosmology. We choose natural units, i.e., $c = \hbar = 1$.

\section{Scalar field theory in general relativity}

The Lagrangian $\mathcal L_\phi$ for a real scalar field $\phi$ moving in a potential $V(\phi)$ and in the presence of gravity is given by
\begin{equation}\label{eq:larangian_gr}
	\mathcal{L}_\phi = -\frac{1}{2}\nabla_\mu \phi \nabla^\mu \phi - V(\phi) = -\frac{1}{2}\partial_\mu \phi \partial^\mu \phi - V(\phi)\:.
\end{equation}
The second equality follows from the fact that applying the covariant derivative $\nabla_\mu$ to a scalar field reduces to the normal derivative $\partial_\mu$, i.e., $\nabla_\mu \phi = \partial_\mu \phi$. Note that the gravitational interaction enters this formalism merely through the metric $g_{\mu\nu}$ as always in general relativity. For $g_{\mu\nu} \rightarrow \eta_{\mu\nu} = {\rm diag}(-1,1,1,1)$ gravity is ``switched off'' and we are in the regime of special relativity. The Lagrangian for the metric field $g_{\mu\nu}$ is
\begin{equation}
	\mathcal L_{\rm H} = \frac{1}{16\pi G}\mathcal{R}\:,
\end{equation}
where $\mathcal R$ is the curvature for $g_{\mu\nu}$. This is the Lagrangian of the \textbf{Einstein-Hilbert action}. So the total Lagrangian becomes
\begin{equation}
	\mathcal L = \mathcal L_{\rm H} + \mathcal L_\phi
\end{equation}
with the associated action
\begin{equation}\label{eq:action}
		S = \int_{\mathcal D} \mathcal{L}\: \sqrt{-g}\: dx^4\:,
\end{equation}
where $\mathcal D$ is a compact region with smooth boundary $\partial \mathcal D$ and $g = \det(g_{\mu\nu})$.

The variation of the action with respect to the field $\psi$, where $\psi$ stands for either the scalar field $\phi$ or the metric component $g_{\mu\nu}$, is defined by
\begin{equation}
	\delta S = \left. \frac{d}{d \varepsilon} S(\psi_\varepsilon) \right| _{\varepsilon = 0}
\end{equation}
with $\psi_\varepsilon$ being a 1-parameter family of fields satisfying $\psi_{\varepsilon = 0}(x) = \psi(x)$. Note that we can manipulate with $\delta$ in a similar way as with normal derivatives (e.g., chain rule). Now, the Lagrangian $\mathcal L$ is defined such that the condition
\begin{equation}
	\delta S = 0
\end{equation}
for $\delta \psi = 0$ on the boundary $\partial \mathcal D$ leads to the equation of motion for $\psi$. Thus, the variation of the action (\ref{eq:action}) with respect to $\phi$ leads to the equation of motion for $\phi$, and the corresponding variation with respect to $g_{\mu\nu}$ yields the Einstein field equations. In the following two sections we will compute these two variations explicitly.

\subsection{Variation with respect to the scalar field}

Varying the action (\ref{eq:action}) with respect to $\phi$ yields
\begin{equation}\label{eq:euler_lagrange_help}
\begin{aligned}
		\delta S &= \delta \int_{\mathcal D} \mathcal{L}\: \sqrt{-g}\: dx^4 = \int_{\mathcal D} \big(\delta \mathcal{L}\big)\: \sqrt{-g}\: dx^4 = \int_{\mathcal D} \big(\delta \mathcal{L}_\phi\big)\: \sqrt{-g}\: dx^4\\
		&= \int_{\mathcal D}\bigg(\frac{\partial \mathcal L_\phi}{\partial \phi}\:\delta \phi + \frac{\partial \mathcal L_\phi}{\partial (\partial_\mu \phi)}\:\delta(\partial_\mu \phi) \bigg) \sqrt{-g}\: dx^4\:.\\
\end{aligned}
\end{equation}
Then with $\delta (\partial_\mu \phi) = \partial_\mu (\delta \phi) = \nabla_\mu (\delta \phi)$ it holds
\begin{equation}
	\frac{\partial \mathcal L_\phi}{\partial (\partial_\mu \phi)}\:\delta (\partial_\mu \phi) = \nabla_\mu \left(\frac{\partial \mathcal L_\phi}{\partial (\partial_\mu \phi)}\:\delta \phi\right) - \left(\nabla_\mu\frac{\partial \mathcal L_\phi}{\partial (\partial_\mu \phi)}\right)\delta \phi\:,
\end{equation}
and inserting this into Eq.~(\ref{eq:euler_lagrange_help}) we obtain
\begin{equation}
	\delta S = \int_{\mathcal D} \left(\frac{\partial \mathcal L_\phi}{\partial \phi}-\nabla_\mu \frac{\partial \phi}{\partial(\partial_\mu \phi)}\right)   \delta \phi\: \sqrt{-g}\: dx^4 + \int_{\mathcal D}\nabla_\mu \left(\frac{\partial \mathcal L_\phi}{\partial (\partial_\mu \phi)}\:\delta \phi\right)\sqrt{-g}\: dx^4\:.
\end{equation}
By means of Gauss theorem the second integral is equivalent to a surface integral over $\partial \mathcal D$, so it vanishes due to the boundary condition $\delta \phi = 0$ on $\partial \mathcal D$. Since $\delta \phi$ and $\mathcal D$ are arbitrary, the condition $\delta S = 0$ leads to
\begin{equation}\label{eq:euler_lagrange}
	\boxed{\frac{\partial \mathcal L_\phi}{\partial \phi} - \nabla_\mu \frac{\partial \mathcal L_\phi}{\partial(\partial_\mu \phi)} = -\nabla_\mu \nabla^\mu \phi + V' = 0}
\end{equation}
with $V' = \partial V/\partial \phi$. This is the \textbf{Euler-Lagrange equation} being the equation of motion for the scalar field $\phi$.\footnote{Note that in the special relativistic limit, i.e., $g_{\mu\nu} \rightarrow \eta_{\mu\nu}$, the equation of motion reduces for the potential $V(\phi) = m^2\phi^2/2$ to the familiar \textbf{Klein-Gordon equation}
\begin{equation}\label{eq:klein_gordon}
\left(-\eta_{\mu\nu}\partial^\mu \partial^\nu + m^2\right)\phi = \ddot \phi - \bv \nabla^2 \phi + m^2 \phi = 0\:.
\end{equation}
}

\subsection{Variation with respect to the metric}

On the other hand, varying the action (\ref{eq:action}) with respect to $g_{\mu\nu}$ such that $\delta g_{\mu\nu} = 0$ on $\partial \mathcal D$ leads to the Einstein field equations and thus defines the energy-momentum tensor for the field $\phi$ being
\begin{equation}\label{eq:energy_momentum_tensor_definition}
	\delta \int_D \mathcal{L}_\phi \sqrt{-g}\: dx^4 = -\frac{1}{2}\int_{\mathcal D} [T_\phi]_{\mu\nu}\delta g^{\mu\nu} \sqrt{-g}\: dx^4\:.
\end{equation}
To see see this, we need the identity (see, e.g., \citealt{straumann2004}, Sect.~2.3)
\begin{equation}\label{eq:einstein_hilbert_action}
	\delta \int_{\mathcal D} \mathcal R \sqrt{-g}\: dx^4 = \int_{\mathcal D} G_{\mu\nu}\delta g^{\mu\nu} \sqrt{-g}\: dx^4\:,
\end{equation}
where $G_{\mu\nu}$ is the Einstein tensor. Then using the Eqs.~(\ref{eq:einstein_hilbert_action}) and (\ref{eq:energy_momentum_tensor_definition}), the variation of $S$ becomes
\begin{equation}
\begin{aligned}
	\delta S &= \delta \int_{\mathcal D} \mathcal{L}\sqrt{-g}\: dx^4 \\
	&= \delta \int_{\mathcal D} \mathcal L_{\rm H}\sqrt{-g}\: dx^4 + \delta \int_{\mathcal D} \mathcal{L}_{\phi}\sqrt{-g}\: dx^4\\
	&= \int_{\mathcal D} \frac{1}{16\pi G}G_{\mu\nu}\delta g^{\mu\nu}\sqrt{-g}\: dx^4 - \frac{1}{2}\int_{\mathcal D} [T_\phi]_{\mu\nu}g^{\mu\nu}\sqrt{-g}\:dx^4\\
	&= \frac{1}{2}\int_{\mathcal D} \left(\frac{1}{8\pi G}G_{\mu\nu}-[T_\phi]_{\mu\nu}\right) g^{\mu\nu}\sqrt{-g}\: dx^4\:.
	\end{aligned}
\end{equation}
Together with the condition $\delta S = 0$ for arbitrary $\delta g _{\mu\nu}$ and $\mathcal D$ we obtain the field equations
\begin{equation}
	G_{\mu\nu} = 8\pi G \:[T_\phi]_{\mu\nu}\:.
\end{equation}
This justifies the definition of the energy-momentum tensor (\ref{eq:energy_momentum_tensor_definition}).

In order to compute the energy-momentum tensor explicitly, we need the variation of $\sqrt{-g}$. To compute it, we need the auxiliary relation that every real, invertible, differentiable square matrix $M(x)$ satisfies (see, e.g., \citealt{weinberg1972}, Sect.~4.7)\footnote{This is a special formulation of Jacobi's formula
\begin{equation}
\frac{d }{dx} \det(M(x))= {\rm tr}\left( {\rm adj}(M(x)) \frac{dM(x)}{dx}  \right),
\end{equation}
where $M$ is a real, differentiable square matrix and ${\rm adj}(M)$ its adjugate. If $M$ is invertible, it holds ${\rm adj}(M) = \det(M)M^{-1}$, and thus it follows Eq.~(\ref{eq:jacobi}).}
\begin{equation}\label{eq:jacobi}
	{\rm tr}\Big[M^{-1}(x)\frac{\partial}{\partial x} M(x)\Big] = \frac{\partial}{\partial x} \ln \big| \det M(x) \big|\:,
\end{equation}
so that it holds for the metric $\delta g/g = g^{\mu\nu}\delta g_{\mu\nu}$. This yields
\begin{equation}\label{eq:auxiliary_relation}
	\delta (\sqrt{-g}) = -\frac{1}{2}\frac{1}{\sqrt{-g}}\delta g = -\frac{1}{2}\frac{1}{\sqrt{-g}} \:g\: g^{\mu\nu}\delta g_{\mu\nu} = \frac{\sqrt{-g}}{2} \:g^{\mu\nu}\delta g_{\mu\nu} = -\frac{\sqrt{-g}}{2} \:g_{\mu\nu}\delta g^{\mu\nu}\:,
\end{equation}
where in the last step we have used
\begin{equation}
	0 = \delta\left( \delta^\mu_{\phantom \mu \nu}\right) = \delta \left(g^{\mu\alpha}g_{\alpha \nu}\right) = \delta g^{\mu\alpha} g_{\alpha \nu}+ g^{\mu\alpha}\delta g_{\alpha \nu}\:.
\end{equation}
With the relation (\ref{eq:auxiliary_relation}), the variation of the left hand side of Eq.~(\ref{eq:energy_momentum_tensor_definition}) with respect to $g_{\mu\nu}$ becomes
\begin{equation}
\begin{aligned}
	\delta \int_{\mathcal D} \mathcal L_\phi \sqrt{-g}\: dx^4 &= \int_{\mathcal D} \Big[\big(\delta \mathcal L_\phi\big)\sqrt{-g} + \mathcal L_\phi \:\big(\delta \sqrt{-g}\big)\Big]dx^4\\
	&= \int_{\mathcal D}\Big[-\frac{1}{2} \delta g^{\mu\nu} \partial_\mu \phi \partial_\nu \phi \sqrt{-g} - \frac{1}{2}\mathcal L_\phi g_{\mu\nu}\sqrt{-g}\:dg^{\mu\nu}\Big]dx^4\\
	&= \frac{1}{2}\int_{\mathcal D}\Big[-\partial_\mu \phi \partial_\nu \phi-  \mathcal L_\phi g_{\mu\nu}\Big]dg^{\mu\nu}\:\sqrt{-g}\:dx^4\:,
\end{aligned}
\end{equation}
and it follows by comparison with the right hand side of Eq.~(\ref{eq:energy_momentum_tensor_definition}) the explicit form of the energy-momentum tensor:\footnote{In special relativity, the energy-momentum tensor $T_{\mu\nu}$ for a field $\psi$ is usually derived by a symmetry argument (see, e.g., \citealt{mandl1993}, Sect.~2.4). If the Lagrangian of the field $\mathcal L(\psi, \partial_\mu \psi)$ is invariant under spacetime translations, then the quantity 
\begin{equation}\label{eq:noether_theorem}
	 T_{\mu\nu} = \frac{\partial \mathcal L}{\partial (\partial^\mu \psi)}\: \partial_\nu \psi + \mathcal L\: \eta_{\mu\nu}
\end{equation}
is conserved as a consequence of Noether's theorem and thus is interpreted as energy-momentum tensor. This criterion of translational invariance is satisfied by the Lagrangian (\ref{eq:larangian_gr}) of the scalar field $\phi$ and leads with the formula (\ref{eq:noether_theorem}) after generalizing to curved spacetime, i.e., $\eta_{\mu\nu} \rightarrow g_{\mu\nu}$, to the same energy-momentum tensor as in Eq.~(\ref{eq:energy_momentum_tensor_field}):
\begin{equation}
	 T_{\mu\nu} = \frac{\partial \mathcal L_\phi}{\partial (\partial^\mu \phi)} \partial_\nu \phi + \mathcal L_\phi g_{\mu\nu} = g_{\mu\gamma} \partial^\gamma \phi \partial_\nu \phi + \mathcal L_\phi g_{\mu \nu} = \partial_\mu \phi \partial_\nu \phi + \mathcal L_\phi g_{\mu \nu}\:.
\end{equation}
This confirms our general relativistic approach by means of an action variation.
}
\begin{equation}\label{eq:energy_momentum_tensor_field}
	\boxed{[T_\phi]_{\mu\nu} = \partial_\mu \phi \partial_\nu \phi+  \mathcal L_\phi g_{\mu\nu}\:.}
\end{equation}
If $\phi^\mu \phi_\mu < 0$, the energy-momentum tensor takes the form of an ideal fluid
\begin{equation}\label{eq:energy_momentum_tensor_fluid_field}
	[T_\phi]_{\mu\nu} = \left(\rho_\phi + p_{\phi}\right)[u_\phi]_\mu [u_\phi]_\nu + p_\phi g_{\mu\nu}
\end{equation}
with the effective energy density $\rho_\phi$, effective pressure $\rho_\phi$, and 4-velocity $[u_\phi]^\mu$ being
\begin{equation}\label{eq:field_fluid_expressions}
	\boxed{\rho_\phi = -\frac{1}{2}\partial^\mu \phi \partial_\mu \phi + V(\phi)\:,\quad\quad p_\phi = -\frac{1}{2}\partial^\mu \phi \partial_\mu \phi - V(\phi)\:,\quad\quad [u_\phi]^\mu =  \frac{\partial^\mu \phi}{\sqrt{-\partial^\mu \phi \partial_\mu \phi}}\:.}
\end{equation}

\section{Scalar field in cosmology}\label{sec:scalar_field_in_cosmology}

In this section, we apply the results of the previous section to cosmology. First, we consider a homogeneous and isotropic FLRW universe and then a linearly perturbed universe as discussed in Chapter \ref{sec:general_relativistic_treatment}.

\subsection{FLRW universe}

In an unperturbed FLRW universe using comoving coordinates, the scalar field can only depend on time, i.e., $\phi(t,\bv x) \equiv \phi(t)$, due to the homogeneity of the universe. Moreover, since in a FLRW universe the energy-momentum tensor always takes the form of an ideal fluid, our energy-momentum tensor (\ref{eq:energy_momentum_tensor_fluid_field}) has already the right form and we just have to evaluate the expressions (\ref{eq:field_fluid_expressions}) using the Robertson-Walker metric (\ref{eq:robertson_walker_metric3}) and $\partial_i \phi = 0$:
\begin{equation}\label{eq:fluid_for_scalar_field}
	\boxed{\rho_\phi = \frac{1}{2}\dot \phi^2 + V(\phi)\:,\quad\quad p_\phi = \frac{1}{2}\dot \phi^2 - V(\phi)\:, \quad\quad [u_\phi]^\mu = [u_\phi]_\mu = (1,0,0,0)\:.}
\end{equation}
So a scalar field has the time dependent equation of state
\begin{equation}
	w_\phi(t) = \frac{\frac{1}{2}\dot \phi^2 - V(\phi)}{\frac{1}{2}\dot \phi^2 + V(\phi)}
\end{equation}
with the bounds $-1 \leq w_\phi(t) \leq 1$. If the kinetic energy of the field is small compared to the potential $V$, i.e., $\dot \phi^2 \ll V$, it follows $w_\phi \simeq -1$. Thus, such a fluid can mimic a cosmological constant $\Lambda$ and leads, if it is the dominant energy component of the universe, to an exponential expansion of the universe (see Sect.~\ref{sec:equation_of_state}). Both kinds of accelerations, inflation as well as the recent acceleration by dark energy, could in principle be caused by a scalar field.

If the scalar field $\phi$ does not interact with any other energy component in the universe, its equation of motion is either given by the Euler-Lagrange equation (\ref{eq:euler_lagrange}) or by the energy-momentum conservation $\nabla_\mu [T_\phi]^{\mu\nu} = 0$. Both approaches yield the same result. Since we have already computed $\nabla_\mu [T_\phi]^{\mu\nu} = 0$ (see Eq.~(\ref{eq:equation_of_motion1})), we can just insert the expressions (\ref{eq:fluid_for_scalar_field}) into Eq.~(\ref{eq:equation_of_motion1}) yielding
\begin{equation}\label{eq:equation_of_motion_of_unperturbed_phi}
	\boxed{\ddot \phi + 3 H \dot \phi + V' = 0\:.}
\end{equation}
This is the equation of motion for a scalar field in a FLRW universe.

\subsection{Perturbed universe}\label{sec:scalar_field_perturbed_universe}

In the following, we extend our discussion by considering a perturbed universe like in Chapter \ref{sec:general_relativistic_treatment}. We adopt the corresponding notation and use conformal time $\tau$ (see Eq.~(\ref{eq:conformal_time})) instead of cosmic time $t$. The perturbed scalar field is
\begin{equation}\label{eq:phi_perturbed}
	\phi = \bar \phi + \delta \phi\:,
\end{equation}
where $\bar \phi(\tau)$ denotes the field of the background FLRW universe and $\delta \phi(x)$ is a small perturbation to be treated at first order. We want to express the perturbations in the energy-momentum tensor (\ref{eq:energy_momentum_tensor}) in terms of $\delta \phi$. Since the energy-momentum tensor (\ref{eq:energy_momentum_tensor_fluid_field}) has the form of an ideal fluid, we immediately obtain $\Pi_\phi = 0$. The other perturbations in Eq.~(\ref{eq:energy_momentum_tensor}) are computed by inserting Eq.~(\ref{eq:phi_perturbed}) into the expressions (\ref{eq:field_fluid_expressions}) and evaluating them at first order. With the inverse (\ref{eq:g_inverse}) of the perturbed metric $g_{\mu\nu}$ and with $\partial_\mu \phi = (\dot{\bar\phi}+\delta \dot \phi,\partial_i \delta \phi)$ we have
\begin{equation}
\partial^\mu \phi \partial_\mu \phi = g^{\mu\nu}\partial_\mu \phi \partial_\nu \phi = - \frac{1}{a^2} \left(1-2A \right)\left(\dot{\bar \phi} + \delta \dot \phi\right)^2 = \frac{1}{a^2}\left(- \dot{\bar \phi}^2 - 2\dot{\bar \phi} \:\delta \dot \phi + 2A \dot{\bar \phi}^2\right)\:.
\end{equation}
Thus, we obtain for the first two expressions in Eq.~(\ref{eq:field_fluid_expressions})
\begin{equation}\label{eq_rho_and_p_of_phi}
\begin{aligned}
	\rho_\phi &= -\frac{1}{2}\partial^\mu \phi \partial_\mu \phi + V(\phi) = \frac{1}{2a^2}\dot{\bar \phi}^2 + V(\bar \phi) + \frac{1}{a^2}\left(\dot{\bar \phi}\delta \dot \phi -A \dot{\bar \phi}^2\right) + V'(\bar \phi)\delta \phi\:,\\
	p_\phi &= -\frac{1}{2}\partial^\mu \phi \partial_\mu \phi - V(\phi) = \frac{1}{2a^2}\dot{\bar \phi}^2 - V(\bar \phi) + \frac{1}{a^2}\left(\dot{\bar \phi}\delta \dot \phi -A \dot{\bar \phi}^2\right) - V'(\bar \phi)\delta \phi\:,
	\end{aligned}
\end{equation}
where we used the first order expansion $V(\bar \phi+\delta \phi) = V(\bar \phi) + V'(\bar \phi)\delta \phi$. Furthermore, with
\begin{equation}
	[u_\phi]_\mu [u_\phi]_\nu = \frac{\partial_\mu \phi \partial_\nu \phi}{\sqrt{-\partial^\mu \phi \partial_\mu \phi}} = \frac{\partial_\mu \phi \partial_\nu \phi}{\rho_\phi + p_\phi}
\end{equation}
it follows
\begin{equation}
	[T_\phi]_{0i} = \left(\rho_\phi + p_\phi \right)[u_\phi]_0 [u_\phi]_i + g_{0i} p_\phi = \dot \phi \:\partial_i\delta \phi  + a^2 B_i \bar p_{\phi} = \dot{\bar \phi} \:\partial_i\delta \phi  + a^2 B_i \bar p_{\phi}\:.
\end{equation}
A comparison of this expression to the $0i$-component of Eq.~(\ref{eq:energy_momentum_tensor}) yields
\begin{equation}\label{eq:v_of_phi}
	\left(\bar \rho_\phi + \bar p_\phi\right)\left(B_i+[v_\phi]_i\right) = -\frac{1}{a^2} \dot{\bar \phi} \:\partial_i \delta \phi\:.
\end{equation}
We can decompose the energy-momentum perturbations $\delta [T_\phi]_{\mu\nu}$ into scalar, vector, and tensor helicity states (see Sect.~\ref{sec:SVT}). Since $\phi$ is a scalar field, $\delta \phi$ is a 3-scalar under spatial rotations due to the rotational invariance of the background FLRW universe (see the discussion in Sect.~\ref{sec:energy_momentum_tensor}). Therefore, the scalar field perturbation consists only of a scalar helicity state, i.e., $\delta \phi \equiv \delta \phi^{(0)}$, with the consequence that a scalar field does not produce vector and tensor modes (see also the discussion in the Footnote \ref{foot:deg} of Ch.~\ref{sec:generation_of_initial_perturbations}). With the Eqs.~(\ref{eq_rho_and_p_of_phi}) and (\ref{eq:v_of_phi}) we obtain after subtracting the corresponding background quantities $\bar \rho_{\phi}$ and $\bar p_{\phi}$ the following explicit expressions for the scalar modes in Fourier space
\begin{empheq}[box=\fbox]{align}
\delta \rho_\phi^{(0)} &= \frac{1}{a^2}\left(\dot{\bar \phi}\:\delta \dot \phi^{(0)} -A^{(0)} \dot{\bar \phi}^2\right) + V'(\bar \phi)\delta \phi^{(0)} \\
	\delta p_\phi^{(0)} &= \frac{1}{a^2}\left(\dot{\bar \phi}\:\delta \dot \phi^{(0)} -A^{(0)} \dot{\bar \phi}^2\right) - V'(\bar \phi)\delta \phi^{(0)} \\
	\left(\bar \rho_\phi + \bar p_\phi\right)\left(B^{(0)}+v_\phi^{(0)}\right) &= -\frac{k}{a^2} \dot{\bar \phi}\:\delta \phi^{(0)} \label{eq:phi_expressions}\\
	\Pi_\phi^{(0)} &= 0\:.
\end{empheq}
Inserting these expressions into the equations of motion (\ref{eq:scalar_equations_of_motion}) yields
\begin{equation}\label{eq:equation_of_motion_perturbed_scalar_field}
\boxed{\text{continuity:}\quad \delta \ddot \phi + 2 \mathcal{H} \delta \dot \phi + \left(k^2 + a V'\right) \delta \phi = \left(\dot A^{(0)}-3 \dot H_{\rm L}^{(0)}+k B^{(0)}\right)\dot{\bar\phi} - 2 a^2 V' A^{(0)}\:,}
\end{equation}
whereas the Euler equation just reduces to Eq.~(\ref{eq:equation_of_motion_of_unperturbed_phi}) for the unperturbed field $\bar \phi$.


\setlength{\bibsep}{.045\baselineskip}

\cleardoublepage
\phantomsection
\addcontentsline{toc}{chapter}{Bibliography}
\bibliographystyle{apj3}
\bibliography{apj-jour,bibliography}

\end{document}